\documentclass{ucdthesis}

\usepackage{exscale}
\usepackage{graphicx}
\usepackage{latexsym}
\usepackage{amsmath}
\usepackage{amsfonts}
\usepackage{mathtools}
\usepackage{listings}
\usepackage{textcomp}
\usepackage[usenames]{color}
\usepackage{pdfpages}

\makeatletter
\renewcommand{\section}{\@startsection%
{section}%
{1}%
{0mm}%
{-\baselineskip}%
{0.5\baselineskip}%
{\normalfont\large\bfseries}%
} \makeatother
\renewcommand{\theequation}{\arabic{chapter}.\arabic{equation}}
\renewcommand{\baselinestretch}{1}
\makeatletter
\renewcommand{\subsection}{\@startsection%
{subsection}%
{2}%
{0mm}%
{-\baselineskip}%
{0.5\baselineskip}%
{\normalfont\normalsize\bfseries}}%

\makeatletter

\newcommand{\Rmnum}[1]{\expandafter\@slowromancap\romannumeral #1@}
\makeatother

\newtheorem{remark}{Remark}
\newtheorem{proposition}{Proposition}[chapter]
\newtheorem{theorem}{Theorem}[chapter]
\newtheorem{lemma}{Lemma}[chapter]
\newtheorem{definition}{Definition}
\newtheorem{corollary}{Corollary}[chapter]
\newenvironment{proof}{{\bf Proof:}}{\,\hfill$\square$\vskip.5cm}

\newcommand{\astr}{^{\phantom *}}

\usepackage{color}

\def\*{{\phantom *}}

\usepackage[pdftitle={Infinite Cycles in Boson Lattice Models},pdfpagemode=UseOutlines,colorlinks=true,linkcolor=black,citecolor=black,pdfauthor={G. Boland},bookmarks=true,pdftex=true,linktocpage]{hyperref}

\hypersetup{
   bookmarksnumbered,
   pdfstartview={FitH},
   urlcolor={black},
   pdfpagemode={UseOutlines}
}
\makeatletter
\newcommand\org@hypertarget{}
\let\org@hypertarget\hypertarget
\renewcommand\hypertarget[2]{%
\Hy@raisedlink{\org@hypertarget{#1}{}}#2%
} \makeatother

\newcommand{\sLambda}{{\hbox {\tiny{$\Lambda$}}}}
\newcommand{\sV}{{\hbox {\tiny{$V$}}}}
\newcommand{\sbeta}{{\hbox {\tiny{$\beta$}}}}

\newcommand{\trace}{\text{\rm{trace}}\,}
\newcommand{\sym}[1]{\sigma_{+}^{(#1)}\,}
\newcommand{\Prob}{\mathbb{P}}
\newcommand{\Ex}{\mathbb{E}}

\newcommand{\Htilde}{\widetilde{H}}
\newcommand{\ii}{\mathbf{i}}
\newcommand{\jj}{\mathbf{j}}
\newcommand{\kk}{\mathbf{k}}
\newcommand{\boldl}{\mathbf{l}}
\newcommand{\e}{\mathrm{e}}

\newcommand{\CC}{\mathbb{C}}
\newcommand{\RR}{\mathbb{R}}

\newcommand{\KK}{\mathbb{K}}
\newcommand{\NN}{\mathbb{N}}
\newcommand{\II}{\mathbb{I}}
\newcommand{\ee}{\mathbf{e}}
\newcommand{\g}{\mathbf{g}}
\newcommand{\cc}{\mathbf{c}}

\newcommand{\bigbar}{\bigg|}
\newcommand{\la}{\langle}
\newcommand{\ra}{\rangle}
\newcommand{\lla}{\left\langle\!\left\langle}
\newcommand{\rra}{\right\rangle\!\right\rangle}

\newcommand{\HH}{\mathcal{H}_\sLambda}

\newcommand{\HHcansym}[1]{{\mathcal{H}_{\sLambda,+}^{(#1)}}}
\newcommand{\HHcan}[1]{{\mathcal{H}_{\sLambda}^{(#1)}}}
\newcommand{\FFH}{\mathcal{F}(\HH)}
\newcommand{\FFHsym}{\mathcal{F}_{+}(\HH)}

\newcommand{\cHEx}{\Ex^n_\sLambda}   
\newcommand{\gHEx}{\Ex^\mu_\sLambda}   
\newcommand{\cHProb}{\Prob^n_\sLambda}   
\newcommand{\gHProb}{\Prob^\mu_{\phantom{l}\!\!\sLambda}}   

\newcommand{\cHHam}{H_\sLambda^{n\phantom{j}}}		
\newcommand{\gHHam}{H_\sLambda}			
\newcommand{\cHPart}{Z_\sLambda(n)}	
\newcommand{\gHPart}{\Xi_\sLambda(\mu)}           
\newcommand{\gHC}{c_\sLambda^\mu}		
\newcommand{\cHC}{c_\sLambda^n}		

\newcommand{\lthermlim}{ \lim_{\Lambda}}
\newcommand{\clthermlim}{ \lim_{\stackrel{n,|\Lambda| \to \infty}{n/|\Lambda|=\rho}}}

\newcommand{\glHC}{c_l^\mu}		
\newcommand{\glPart}{\Xi_l(\mu)}           
\newcommand{\glPartF}{\Xi^0_l(\mu)}           
\newcommand{\dthermlim}{ \lim_{l \to \infty}}
\newcommand{\Hlone}{\mathcal{H}_l}
\newcommand{\Hlcan}[1]{{\mathcal{H}_l^{(#1)}}}

\newcommand{\FFlsym}{\mathcal{F}_{+}(\Hlone)}

\newcommand{\mucrit}{\mu_\infty}

\newcommand{\Hone}{\mathcal{H}_\sV}

\newcommand{\FF}{\mathcal{F}(\Hone)}
\newcommand{\FFsym}{\mathcal{F}_{+}(\Hone)}

\newcommand{\Hcansym}[1]{{\mathcal{H}_{\sV,+}^{(#1)}}}
\newcommand{\Hcan}[1]{{\mathcal{H}_{\sV\phantom{l}}^{(#1)}}\!}
\newcommand{\Happ}[1]{H_{#1,\sV}^{\textrm{\tiny{APP}}}}
\newcommand{\Hv}[1]{{H_{#1,\sV}}}
\newcommand{\qFFsym}{\mathcal{H}_{q, \sV}}
\newcommand{\mcC}{\widetilde{c}_\sV^{\,n}}

\newcommand{\gHam}{H_\sV}      
\newcommand{\cPart}{Z_\sV(n)}           
\newcommand{\gPart}{\Xi_\sV^\mu}           
\newcommand{\gC}{c_\sV^\mu}			
\newcommand{\cC}{c_\sV^n}		

\newcommand{\thermlim}{ \lim_{V \to \infty}}
\newcommand{\cthermlim}{ \lim_{\stackrel{n,\sV \to \infty}{n/\sV=\rho}}}

\newcommand{\Hhccan}[1]{{\mathcal{H}_{#1,\sV}^{\mathsf{hc}}}}
\newcommand{\Hhccansym}[1]{{\mathcal{H}_{#1,\sV,+}^{{\mathsf{hc}}}}}
\newcommand{\Hhccyclespace}{{\mathcal{H}_{q,n,\sV}^{{\mathsf{hc}}}}}

\newcommand{\hcHam}{H^{\mathsf{hc}}_{n,\sV}}      

\newcommand{\Phc}[1]{\mathcal{P}_{#1}^{\mathsf{hc}}}
\newcommand{\ccPart}{Z_{V-q}\left(\tfrac{V-q}{V},\, n\!-\!q \right)}

\newcommand{\mgC}{\widetilde{c}_\sV^{\,\mu}}

\newcommand{\rholong}{{\varrho^\mu_{{\rm long}}}}
\newcommand{\rhoshort}{{\varrho^\mu_{{\rm short}}}}
\newcommand{\rhocond}{{\varrho^\mu_{{\rm cond}}}}
\newcommand{\rhocondfree}{{\varrho_{{\rm cond}}}}
\newcommand{\rhodens}{{\rho(\mu)}}
\newcommand{\rhocrit}{{\rho_{{\rm cr}}}}

\newcommand{\rhoblong}{{\varrho^\rho_{{\rm long}}}}         
\newcommand{\rhobshort}{{\varrho^\rho_{{\rm short}}}}
\newcommand{\rhobcond}{{\varrho^\rho_{{\rm cond}}}}
\newcommand{\rhobcrit}{\rho_{\rm cr}}             

\begin{document}

\title{Infinite Cycles in Boson Lattice Models}
\author{Gerard Gordon Boland}
\newcommand{\supervisor}{Professor Joseph V.\ Pul\'e}
\newcommand{\institution}{University College Dublin}
\newcommand{\department}{School of Mathematical Sciences}
\newcommand{\college}{University College Dublin}

\newcommand{\draftimages}{false}

\begin{spacing}{1.5}

\setlength{\parskip}{5mm plus3mm minus2mm}
\setlength{\parindent}{0in}

\setlength{\headheight}{14.5pt}

\frontmatter
\thispagestyle{empty}

\begin{doublespacing}
\begin{center}

\begin{scshape}
\begin{LARGE}
\makeatletter
\MakeLowercase{\@title}
\makeatother
\end{LARGE}
\end{scshape}

\vspace{2cm}

\begin{large}
\begin{scshape}
a thesis submitted to\\
the national university of ireland\\
for the degree of\vspace{0.175cm}\\
doctor of philosophy
\end{scshape}
\end{large}

\vspace{0.5cm}

by

\makeatletter
\textbf{\@author}
\makeatother

\vspace{4cm}

Based on research carried out in the\vspace{0.175cm}\\
{\department},\\
College of Engineering, Mathematical and Physical Sciences, \\
{\institution}

\vspace{0.5cm}

under the direction of \textbf{\supervisor}\\
Head of School: Dr. M\'icheal \'O Searc\'oid

\vspace{1.5cm}

\textit{\institution}\hfill\textit{\todaymonth}

\end{center}
\end{doublespacing}

\pagebreak
\thispagestyle{empty}
\addcontentsline{toc}{chapter}{Abstract} 

\begin{center}
	{\large\textbf{Abstract}}
\end{center}
We study the relationship between long cycles and Bose-Einstein condensation (BEC)
in the case of several models. A convenient expression for the density of particles on cycles
of length $q$ is obtained, in terms of $q$ unsymmetrised particles coupled
with a boson field. 

Using this formulation we reproduce known results on the Ideal Bose Gas, Mean-Field 
and Perturbed Mean-Field Models, where the condensate 
density exactly equals the long cycle density.

Then we consider the Infinite-Range-Hopping Bose-Hubbard Model:
\[
	H^{\rm BH}_V
=	\frac{1}{2V}\!\sum_{x,y = 1}^V(a^\ast_x-a^\ast_y)(a\astr_x-a\astr_y)
	+\lambda\sum_{x=1}^V n_x(n_x-1)
\]
in two cases, first for $\lambda = +\infty$, otherwise known as the hard-core boson model;
and secondly for $\lambda$ finite, representing a finite on-site repulsion interaction.

For the hard-core case, we find we may disregard the hopping contribution of the 
$q$ unsymmetrised particles, allowing us to calculate an exact expression for the density of
particles on long cycles. It is shown that only the cycle of length one contributes to the cycle density.
We conclude that while the existence of a non-zero long cycle density coincides with the occurrence of
Bose-Einstein condensation, the respective densities are not necessarily equal.

For the case of a finite on-site repulsion, we obtain an expression for the cycle density 
involving the partition function for a Bose-Hubbard Hamiltonian with a single-site 
correction again by neglecting the $q$ unsymmetrised hop. Inspired by the Approximating 
Hamiltonian method we conjecture a simplified 
expression for the short cycle density as a ratio of single-site partition functions. 
In the absence of condensation we prove that this simplification is exact and use it to 
show that in this case the long-cycle density vanishes. In the presence of condensation 
we can justify this simplification when a gauge-symmetry breaking term is introduced 
in the Hamiltonian. Assuming our conjecture is correct, we compare numerically the 
long-cycle density with the condensate and again find that though they coexist, in general 
they are not equal.

\pagebreak
\thispagestyle{empty}
\pdfbookmark[0]{Acknowledgements}{acknowledgements} 

\begin{center}
	{\large\textbf{Acknowledgements}}
\end{center}

I am extremely indebted to Professor Joseph V. Pul\'e for his guidance, encouragement
and patience. He never failed to be a source of inspiration, happy to explain everything
from the deepest methodologies to the slightest of mathematical details, and when leaving
me with the impression he was several steps ahead, always waited patiently for me to catch up.
My introduction to Quantum Statistical Mechanics has been thorough and exciting due to
his tireless efforts, for which I am eternally grateful.

I thank the external examiner Professor Valentin Zagrebnov for a careful and thorough
reading of the thesis and several helpful comments and suggestions for its improvement.

I also wish to thank Professor Teunis C. Dorlas for his many inputs in my education,
always eager to suggest ideas, offer help and propose a new topic for discussion.

To all the friends I have made in the process of studying for this Ph.D., 
especially those of us who all began together; Alex Byrne, Ian Harris, Anna Heffernan, Liquin Tao, Barry Wardell,
and in particular Thomas Jaeck, Ciara Morgan and Anne Ghesquiere, I want to say
it is a pleasure to know you all. 
And thanks to those who have known me even longer, especially Cormac, Eleanor, John, Odhran, 
Terry and Harriet. I am lucky to have such good friends.

Also I wish to thank the administrative staff of UCD, who helped me navigate through the mire
of paperwork with ease, and went out of their way to ensure everything ran smoothly.

I acknowledge the support of the Irish Research Council for Science, Engineering and Technology.

Finally I must thank my parents, my sisters Leonie and Maureen, my uncle Michael and my niece Kate, whom I always know 
I can depend on when I need some encouragement, distraction, comfort and humility. 

\begin{flushright}\textit{Go raibh m\'ile maith agaibh go leor.}\end{flushright}

\pagebreak
\renewcommand{\contentsname}{Table of Contents}
\pdfbookmark[0]{Table of Contents}{contents} 
\tableofcontents

\mainmatter


\chapter{Introduction}									\label{chapter1}
\markboth{Introduction}{Introduction}

\hrule
\textbf{Summary}\\
\textit{To begin it is useful to look back through the literature to trace the progression of
discoveries made by many great minds on the phenomenon of Bose-Einstein condensation in
quantum statistical mechanics. We discuss the order parameters that have been postulated to indicate
the presence of this condensate, and give details on all known about their relationships to date.
Finally we briefly detail the layout of this thesis.}
\vspace{15pt}
\hrule
\vspace{35pt}


\section{Bose-Einstein Condensation}
In 1925, the phenomenon of Bose-Einstein condensation was predicted by 
Einstein following correspondence with S.N. Bose, who had found a novel derivation of
photon statistics and the Planck distribution \cite{einstein1}. Einstein extended these ideas to an ideal
monatomic gas \cite{einstein2} and observed that below some critical temperature, a number
of molecules which grows with the total density, transition to an ``unmoving state,'' and
this condensed substance coexists with the monatomic gas. These papers appeared about a
year before the development of Quantum Mechanics
and was the first time the wave nature of matter was explored, with particles obeying
the same statistics as photons.

Since then we now know that all particles in nature can be classified into 
two classes according to their intrinsic angular momentum or spin.
Those with half-integer spin are called fermions, those with integral spin are bosons.
Fermions are restricted by the Pauli-Exclusion principle, which prevent multiple
fermions from existing in the same quantum state, however bosons have no such restriction.
The aforementioned ``unmoving state'' is now known to consist of bosons at
(or near) the bosonic ground state, with particles in this state denoted as the 
(generalised) Bose-Einstein condensate (BEC).

Nevertheless at the time, Einstein's claim of a phase transition in the ideal bose gas
was dismissed, largely due to the criticism in 1927 of G. Uhlenbeck \cite{uhlenbeck1} 
(a graduate student of P. Ehrenfest) who
argued that a gas of finitely many particles can exhibit no discontinuity, which
was technically correct at the time, as it was yet to be shown that singularities
in thermodynamic functions are a consequence of the thermodynamic limit
(i.e. second-order phase transitions). This opinion prevailed until 1937 when H.A. Kramers
pointed out the importance of the limit, causing Uhlenbeck \cite{uhlenbeck2} to withdraw his objections
and assert that Einstein's formula for the particle number-density is correct in the thermodynamic
limit.

\begin{figure}[hbt]
\begin{center}
\includegraphics[width=13cm]{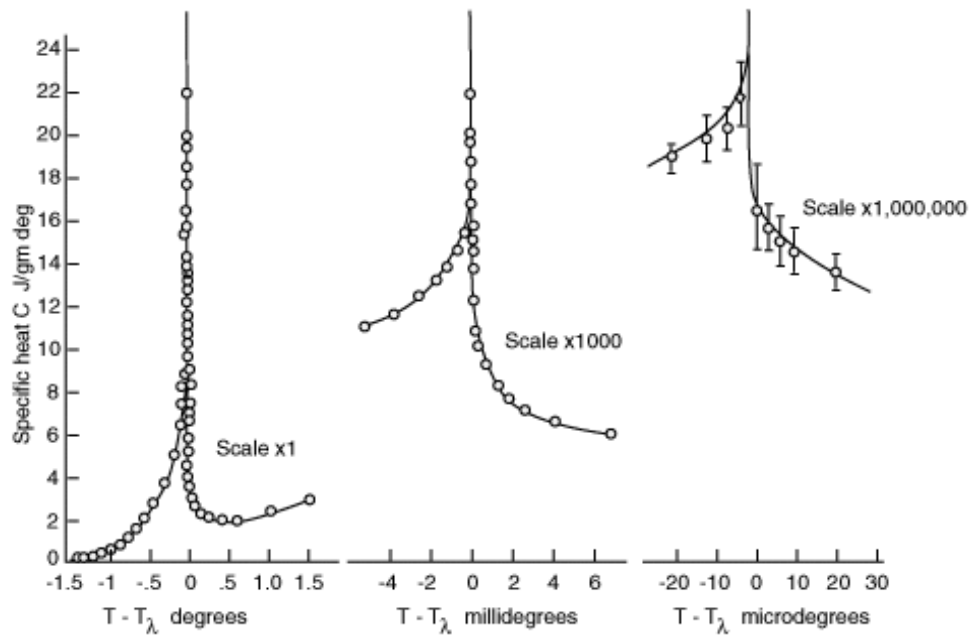}
\end{center}
\caption[The $\lambda$-transition in $^4$He. Image from Buckingham and Fairbank \cite{buckfair}]{\it The 
$\lambda$-transition in $^4$He.}
\label{fighe4}
\end{figure}

Meanwhile although the experimentalist H. Kamerlingh Onnes \cite{onnes} had succeeded in liquifying 
Helium ($^4$He) in 1908, it was not until 1928 that his student W.H. Keesom with M. Wolfke \cite{keesomwolfke} 
discovered  that at temperatures below 2.18K, a second phase of liquid helium appears, and christened it Helium II. 
Four years later Keesom and K. Clusius \cite{keesomclusius} measured the famous peak in the specific heat of helium,
whose shape inspired the name $\lambda$-transition for the behaviour at the critical temperature 
$T_\lambda = 2.18K$ (see Figure \ref{fighe4}). Further research \cite{kapitza,allenmisener} in 1938 into Helium II found that 
a portion of the liquid flows without any apparent viscosity, a phenomenon now called superfluidity.
In 1937 F. London \cite{london1} connected this curious phase transition in $^4$He (a bosonic gas) with 
the disputed ideas of Einstein on the ideal gas. He noted that Einstein's formula for the transition temperature
was a good estimate for $T_\lambda$ and consequently introduced the concept of macroscopic occupation
of the ground state and related it to the coherence properties of the condensate. That
the superfluidity of $^4$He is due to the presence of Bose-Einstein condensation is 
now a fact which is largely accepted today. 

Cowley and Woods \cite{cowleywoods} in 1968 were the first to experimentally detect condensate at $T=1.1K$. 
Later work \cite{alekzag_etal,dokukoz_etal} repeated and improved these experiments and found the critical 
temperature to be $2.24\pm0.04 K$, closely matching the corresponding value for the lambda-point $T_\lambda$,
lending weight to a correspondence between condensation and superfluidity.
A more thorough historical account of these discoveries can be found in \cite{griffin}. 
While BEC has been proved experimentally, nevertheless it remains a valuable concept to the quantum 
statistical mechanics community as it is the only known phase-transition taking place in the
absence of interactions, its origin being purely quantum mechanical.\phantom{\cite{anderson_etal}}

A general theory for condensation of interacting bosonic gasses remain to this day a
notoriously difficult problem. The theory of the ideal gas has led to the formulation that 
``Bose-Einstein condensate'' labels those particles which occupy lowest kinetic energy state, i.e.
\[
	\varrho^\rho_0 = \thermlim \frac{\la N_0 \ra}{V}.
\]
A rigourous proof that $\varrho^\rho_0 > 0$
for systems with realistic inter-particle interactions is an ongoing pursuit.
In 1947 N.N. Bogoliubov computed the excitation spectrum for the weakly interacting Bose gas
and pointed out the similarities with the spectrum of Helium II. This implied  
that the inclusion of interactions in the model had little effect upon the condensate,
but instead had a pronounced impact upon the wave function.

Until recently in fact, BEC had not been proved
for many-body Hamiltonians with genuine interactions except for one special case:
hard core bosons on a lattice at half-filling, i.e., number of particles is exactly half the number of sites,
in 1978 \cite{DysonLiebSimon78, KennedyLiebSriram88}. Five years later in 1983,
E. Buffet, Ph. de Smedt and J.V. Pul\'e \cite{BPdSP83} showed the existence of BEC for a class
of homogeneous bosonic systems with interactions 
close to the van~der~Walls limit with a gap in the one-particle excitation spectrum.
Independently in 2002, E. Lieb and R. Seiringer \cite{liebseiringer1}
considered a model of an interacting system of a trapped (and thus inhomogeneous) gas
in the Gross-Pitaevski limit. J. Lauwers, A. Verbeure and V.A. Zagrebnov 
\cite{LVZ} in 2003, proved the occurrence of zero-mode Bose-Einstein 
condensation for a class of continuous homogeneous systems of boson particles with 
super-stable interactions and a one-particle energy gap.

Because of the difficulty of treating the boson gas interacting through a pair potential,
various approximations have been proposed.
One approach is to isolate the portion of the interaction which depends only upon the occupation numbers,
separating the interaction it into a \textit{diagonal} part (with respect to the occupation number
basis) and an \textit{off-diagonal} part (see \cite{leeyang57}). K. Huang, C.N. Yang and
J.M. Luttinger in 1957 \cite{HYL57} argued that some of the diagonal terms
can enhance the amount of condensate, while Bogoliubov and his school claimed that
it contains terms which deplete it. C.N Yang and C.P. Yang \cite{yangyang69} considered
a special case in $d=1$ of a repulsive delta potential and prove that it exhibits no
phase transition (see also \cite{DLP89-yang, Dorlas93}). However considering the diagonal portion 
of this interaction alone, the Hamiltonian reduces to that of the Huang-Yang-Luttinger model
which has a guaranteed phase transition at $d=1$. This points to the idea that the off-diagonal 
terms must suppress this condensate, which indicates a more complex interplay of the
diagonal and off-diagonal portions of the interaction than previously thought.
Select models considering only the diagonal terms include the Mean-Field model \cite{Dav72, BLS},
the Huang-Yang-Luttinger model \cite{vdBLP88, bergdorlaslewispule_hyl},
the Perturbed Mean-Field model \cite{vdBDLP, dlp91}, and the
Full-Diagonal model \cite{fulldiagmodel_dorlaslewispule}, all of which can possess a critical phase.

Another model which has been considered includes non-diagonal BCS-type interaction terms
in addition to the diagonal terms \cite{PZ93, PZ07}.

Analysis of further models was made possible by two Ans\"atze by Bogoliubov, 
the first to truncate the full Hamiltonian of the interacting bosons to produce the 
Weakly Imperfect Bose Gas (WIBG)\cite{BruZagrebnov01}, the second to perform apt substitutions 
of some operators by $c$-numbers, known as the Bogoliubov approximation. Condensation is shown
to appear in the WIBG (with some peculiarities). Investigations into the experimentally
observed \textit{Mott insulator-superfluid} (or condensate) phase transition \cite{freemon} 
resulted in the analysis of the so-called Bose-Hubbard model, proposed by
Fisher et. al. \cite{Fisher}, and successfully analysed using the Bogoliubov approximation in \cite{BruDorlas,DPZ}
to exhibit BEC.

\section{The Feynman-Kac Integral and Cycle Lengths}
An alternative parameterisation to predict the occurrence of the condensate 
was postulated by R.P. Feynman in 1953 \cite{Fey_1953_and_stat._mech.}. He
was discouraged by the fact that energies of a complex system are ``hopelessly'' difficult
to calculate, so he tried combine his path integral formulation with Bose statistics to see if a 
transition could be found by other means. 

The idea behind the Feynman path integral can be traced back to a paper by Dirac \cite{dirac}
in 1933 on an alternative formulation of quantum mechanics. Initially the
quantum theories concurrently developed by Heisenberg and Schr\"odinger and others in the 1920
were built as a close analogy to classical mechanics where they considered the Hamiltonian $H$,
an energy function of coordinates and momenta. Dirac found that the Lagrangian $L$, a function
of the coordinates $x(t)$ and velocities $\dot{x}(t)$, to be more intuitive than the 
Hamiltonian approach. His reasoning is as follows, in classical dynamics the action 
is given by
\begin{equation}							\label{theaction}
	S[x(\,\cdot\,)] = \int L(\dot{x}(t), x(t)) dt
\end{equation}
where the integral is taken between initial and final times $0$ and $t$. The 
action must depend on the path $x(\,\cdot\,)$ taken by the particles, and hence is a 
functional of these paths. The \textit{principle of least action} states that for small
variations of these paths, the action $S$ is minimised. He hoped that these classical
paths could somehow be useful to describe the evolution of a quantum particle.

In his article, Dirac considers the time propagation operator $\e^{-i t H /\hbar}$.
For two points $x$ and $y$ in a region $\Lambda$, quantum mechanics 
tells us that we can write the probability $P_{t}(x,y)$ that a particle with initial position $x$
propagates to position $y$ after time $t$, as $P_t(x,y) = |K_{t}(x,y)|^2$, the square of the amplitudes
\[
	K(x,y; t) = \la y | \e^{-i t H /\hbar} | x \ra.
\]
This amplitude
is the sum of the amplitudes of all possible paths which start at $x$ and end at $y$ after time $t$.
We will denote such a path by $\omega^t_{x,y}: [0,t] \to \Lambda$ and denote its amplitude 
by $K[\omega^t_{x,y}]$. Let $\Omega^t_{x,y}$ be the set of all these paths. Then
\[
	K(x,y; t) = \sum_{\omega \in \Omega^t_{x,y}} K[\omega^t_{x,y}].
\]
We would like to see how each path $\omega^t_{x,y}$ contributes to the total amplitude. Dirac
suggests that the amplitude $K[\omega^t_{x,y}]$ ``corresponds to'' the quantity
\[
	\exp\left\{ \frac{i}{\hbar} S[\omega^t_{x,y}(t)] \right\}
\]
and later writes that it is the ``classical analogue'' of the propagator. 

By 1948 Feynman \cite{feynman48} had developed this suggestion and showed it was more than an analogy:
it was an equality. He succeeded in formulating his ``space-time approach,'' deriving a third formulation 
of Quantum Mechanics using this expression for the propagator in terms of the action (a formulation later
shown to be equivalent to the Heisenberg and Schr\"odinger pictures). And while Dirac 
considered only the classical path, Feynman showed that all paths contribute: in a sense, the quantum 
particle takes all paths, and the amplitudes for each path add according to the usual quantum mechanical rule 
for combining amplitudes. What was lacking was a rigourous definition of summing over these paths.

Surprisingly a well-defined calculus of such paths was already well known to mathematicians 
by the 1920s, due to N. Wiener and his study of stochastic processes and Brownian motion
(extending the earlier work of Einstein and M. Smoluchowski). He introduced the concept of integration in 
functional spaces that now bears his name: the Wiener measure and Wiener integral.
In 1949, Kac \cite{kac} considered Feynman's path integral and realised that if one replaces the time 
parameter in the Dirac-Feynman expression for the action, equation (\ref{theaction}), with a purely 
imaginary time, the Feynman path integral becomes a Wiener integral which (with some work) can be written 
in terms of the Hamiltonian $H = H_0 + V$:
\[
	K(x,y; t \to -i \hbar\beta) = \la y | \e^{-\beta H} | x \ra
=	\int W^\beta_{xy}(d\omega) \exp\left\{ - \int_0^\beta V(\omega(s)) ds \right\}
\]
which is now called the Feynman-Kac path-integral.

Returning to Feynman's proposition of an order parameter for BEC, he started out by analysing the 
partition function for a system of $n$ particles interacting via a pair potential of
the form $V(x,y) = V(|x-y|)$. Using the Feynman-Kac formula, one can write
the $\e^{-\beta H}$,
\begin{multline}									\label{feynmankac}
	K(x_1, \dots, x_n; y_1, \dots , y_n)
=	\int W^\beta_{x_1, y_1}(d\omega_1) \cdots \int W^\beta_{x_n, y_n}(d\omega_n)
\\
	\times \exp\left\{  -  \sum_{j < k} \int_0^\beta V(\omega_j(s) - \omega_k(s)) ds \right\}
\end{multline}
where the Wiener measure $W^\beta_{x,y}(\omega)$ describes a Brownian path
that starts at $x$ and ends at $y$ with ``time'' $\beta$, i.e. $\omega(0)=x$ and $\omega(\beta)=y$. 

The Feynman path integral has become a very powerful tool in Statistical Mechanics. For a 
system of bosons with Hamiltonian $H$, the canonical partition function can be 
written as
\begin{equation}									\label{intor_partfn}
	\cPart 
=	\trace\left[  \sigma_+ \e^{-\beta H}\right]
=	\frac{1}{n!} \sum_{\pi \in S_n} \int dx_1\dots dx_n K(x_1, \dots, x_n; x_{\pi(1)}, \dots , x_{\pi(n)})
\end{equation}
where the sum over permutations appears because we take the trace over the symmetric states
(as indicated by $\sigma_+$), and we integrate over all particle positions.
(See Appendix A of \cite{ueltschi_fyncyclesbosegas} for a clear derivation of this expression).

\begin{figure}[hbt]
\begin{center}
\includegraphics[width=13cm]{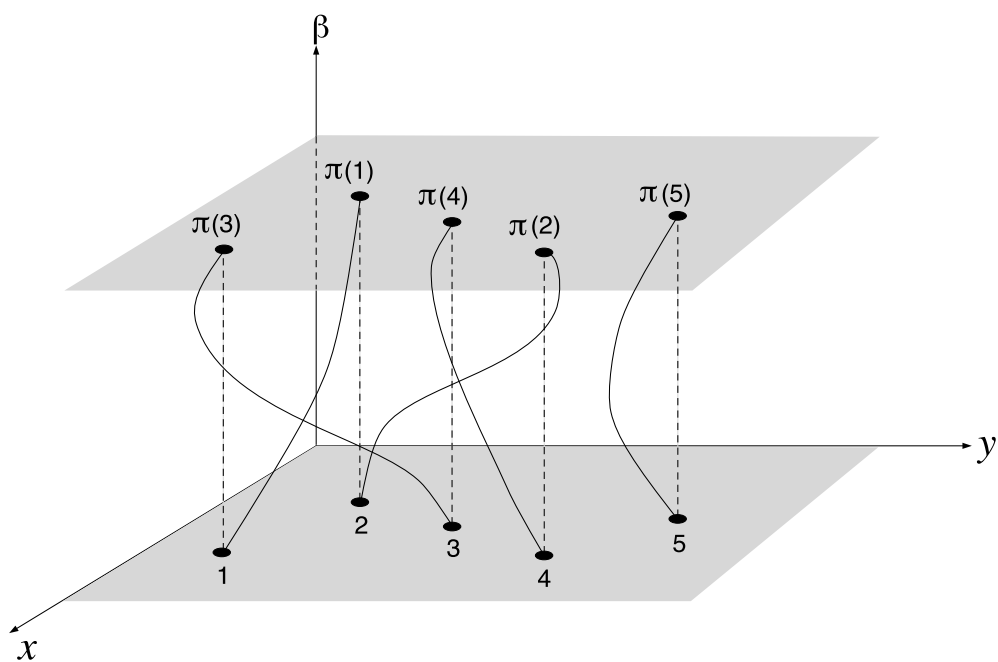}
\end{center}
\caption[Interpretation of Feynman-Kac integral. Image copyright: D. Ueltschi]{\it An interpretation
of the Feynman-Kac representation of the partition function for a gas of bosons. The horizontal
plane represents the $d$ spatial dimensions, and the vertical axis is the imaginary time direction.
This picture considers five particles on two cycles, one of length 1 and the other of length 4
(Picture from \cite{ueltschi_fyncyclesbosegas}).}
\label{figfeykac}
\end{figure}

In Statistical Mechanics, $\beta$ is the inverse temperature. However if one were consider
it as a time parameter for the Feynman-Kac formula, then one 
returns to the intuitive picture of moving particles following paths. Then equation (\ref{feynmankac}) 
can be thought of as $n$ particles with initial positions $x_i$ and 
final positions $y_i$ at ``time'' $\beta$.

Now if one considers the integrand of the partition function expression (\ref{intor_partfn}), 
one sees that the initial and final positions for the particles are fixed, but a particle
with initial position $x_i$ will have final position $x_{\pi(i)}$, thus its path
will depend upon the permutation $\pi$.

Figure \ref{figfeykac} provides an example, where five particles propagate in
the imaginary time direction on paths with endpoints specified by the permutation
\[
	\pi = \left(\begin{array}{ccccc}1&2&3&4&5\\2&4&1&3&5\end{array}\right)
	 = (1 2 4 3 )(5).
\]
The particle labelled five is on a closed Brownian bridge, starting and ending at the point $x_5$. 
We shall say this is a cycle of length one, whereby it took one timestep $\beta$ for the 
particle to return to its original position.

However the particle labelled one however lands on a different site, $x_2$, after time $\beta$. 
Should it continue its journey, it will land at $x_4$ at time $2\beta$, $x_3$ at time $3\beta$
and back to its starting position $x_1$ after four timesteps. We shall call this a cycle
of length four.

Feynman considered the probability that a given particle
belongs to a cycle of length, say $q$, and questioned if there was a strictly positive probability that,
after taking the thermodynamic limit, a cycle of infinite length exist? He postulated that
the occurrence of infinite (long) cycles is indicative of the presence of BEC.

\begin{figure}[hbt]
\begin{center}
\includegraphics[width=13cm]{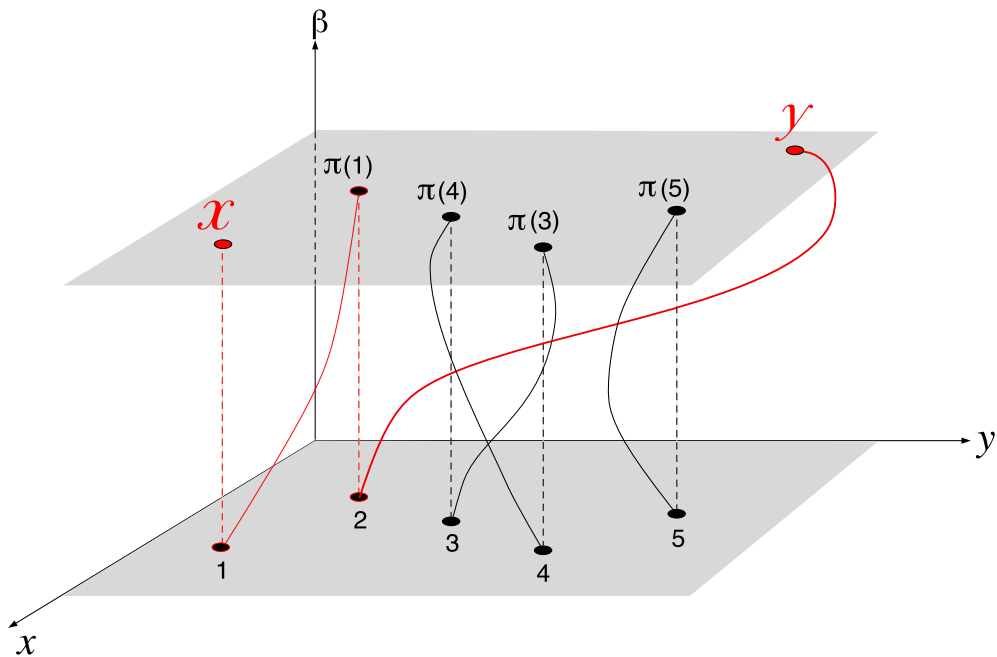}
\end{center}
\caption[Interpretation of Off-Diagonal Long-Range Order in terms of Feynman-Kac integrals]{Interpretation 
of Off-Diagonal Long-Range Order in terms of Feynman-Kac integrals}
\label{figfeykacoldro}
\end{figure}

Three years after Feynman's conjecture, O. Penrose and L. Onsager \cite{Pen.Onsa.} introduced a 
different order parameter for BEC,  related to the concept of ``off-diagonal long-range order'' (OLDRO) \cite{Yang}
which investigates the correlation between two particles at positions $x$ and $y$, i.e.
\[
	D_\rho(x,y) = \cthermlim \la a^\ast_x a\astr_y \ra.
\]
We say the system displays OLDRO when $D_\rho(x,y) > 0$ when $|x-y| \to \infty$.
This correlation can be expressed in the Feynman-Kac representation, which involves
an open cycle starting at $x$ and finishing at $y$ which may wind many times around the 
imaginary time direction (see Figure \ref{figfeykacoldro}):
\begin{multline}
	D_\rho(x,y) = \cthermlim \frac{1}{\cPart V} \frac{1}{(n-1)!} \int dx_2\dots dx_n \sum_{\pi \in S_n}
	\int dW^\beta_{x_1, \hat{x}_{\pi(1)}}(\omega_1) \cdots \int dW^\beta_{x_n, \hat{x}_{\pi(n)}}(\omega_n)
\\
	\times \exp\left\{  -  \sum_{j < k} \int_0^\beta V(\omega_j(s) - \omega_k(s)) ds \right\}
\end{multline}
where we have set $x_1 = x, \hat{x}_1 = y$ and $\hat{x}_j = x_j$ for $j=2,\dots,n$.
As $|x-y| \to \infty$ there corresponds a notion of infinite
winding which is reminiscent of Feynman's approach. Penrose and Onsager 
observed that there should be BEC when the fraction of the total particle number belonging to infinite 
cycles is strictly positive. 

It was generally agreed at the time that this parameter is the correct one for BEC, however
infinite cycles were still considered noteworthy, with for instance a possible role in superfluidity \cite{pollock},
and showing analogous phase transitions to various physical processes such as percolation,
gelation and polymerisation, see for example \cite{Chandler, Schakel, Sear-Cuesta, deGennes}.

However it was not until 1993 that A. S\"ut\H{o} \cite{Suto1,Suto2} gave
a precise mathematical and quantitative formulation relating the occurrence
of BEC (and hence OLDRO) with the existence of infinite cycles. 
He proved in the cases of the free bose gas and 
the mean field gas that in fact infinite cycles do occur below the transition temperature for BEC.
In 2001, D. Ueltschi \cite{ueltschi_cycles_offdiag} considered a simple lattice model in which the 
integral over Brownian paths is replaced by an effective weight. He proved that infinite cycles do not 
occur for high temperatures. Later in 2003, T.C. Dorlas, Ph. Martin and J.V. Pul\'e \cite{DMP} considered 
a model of a Bose gas with some genuine short range interactions between modes
using the theory of Large Deviations developed by Varadhan \cite{varadhan}, and proved that the 
average fraction of infinite cycles is equal to the density of condensate. 

However more recent investigations cast doubt upon the suitability of infinite cycles as an
order parameter for BEC. Ueltschi \cite{ueltschi_cycles_offdiag} argues that both order 
parameters are identical in the case of a weakly interacting
Bose gas, but in the presence of a strong inter-particle interaction this correlation breaks
down. The example offered is a crystalline system. The off-diagonal correlation
function decays exponentially as the inter-particle distance increases so there in no OLDRO. 
However Brownian motion occurs as a consequence of the interchange of particles via
quantum tunnelling. Due to the strong interaction, these Brownian paths need to avoid each other,
which for dimensions greater than two and for low enough temperatures
can result in a macroscopic number of particles on random walks of infinite length.

As a result there exists uncertainty as to exactly what connection there is between
Bose-Einstein condensation and infinite cycles, which motivates the subject of this
thesis. Here we compare the long cycle density with that of the condensate for two models
of interacting bosons, both of which involve an on-site repulsive interaction.

We find a weaker result than \cite{DMP}, that the existence of long cycles coincides with
the presence of BEC, but their respective densities are not necessarily equal.
In fact, the long cycle density can be both greater than or less than the 
condensate density. It appears that the single-site repulsion term can either
increase or decrease the quantity of long cycles, depending upon the 
thermodynamic parameters of the system. Nevertheless this analysis does
lend weight to Feynman's conjecture in the absence of strong interactions.

\section*{Outline of Thesis}
The object of this thesis is to study the relation between the occurrence of BEC and 
the existence of infinite cycles for some Boson lattice models, in particular 
the infinite-range-hopping Bose-Hubbbard model with both hard cores and without.
These models correspond to bosonic systems that describe particles hopping on a lattice 
and interacting among themselves with the constraint 
that their wavefunctions are invariant under permutations of the particles. The partition function
can induce a probability measure upon the set of permutations, with which one can calculate
expectations of cycle lengths. 

In Chapter 2 we apply the techniques for cycle statistics described in \cite{DMP}, following
\cite{Martin}, to establish the general framework for evaluating the density of particles on cycles of
particular lengths, and show that this formulation uniquely partitions the particles into
their cycle structures.

In Chapter 3 we formulate a relation between occupation numbers and the 
cycle structure of models which satisfy particular conditions, which implies
the condensate density equals the infinite cycle density. We then check the conditions
in the case of three simple models, the Ideal Bose Gas, the Mean-Field Bose Gas
and the Perturbed Mean-Field Bose gas, to reproduce the results of S\"ut\H{o} \cite{Suto1,Suto2}
and Dorlas, Martin and Pul\'e \cite{DMP}.

In Chapter 4 we consider the Infinite-Range-Hopping Bose-Hubbard Model
with hard-cores, i.e. infinite on-site repulsion to prevent multiple particles occupying 
individual sites. We calculate the density of particles on cycles on infinite length
and compare the result with the condensate density.

In Chapter 5 we consider the Infinite-Range-Hopping Bose-Hubbard Model
(without hard-cores) where now there is a finite on-site repulsion to discourage multiple particles
per site. Again we consider infinite cycle density
and compare the result with the condensate density.

\chapter{Cycle Statistics}								\label{chapter2}

\hrule
\textbf{Summary}\\
\textit{In this chapter we will introduce a general framework for cycle statistics
by expanding the bosonic partition function into a sum over permutations of an unsymmetrised
trace term and inducing a probability measure upon these
permutations. By decomposing these permutations into their cycle structures, we
derive a convenient expression for the density of particles on cycles of
particular lengths, and show that this formulation uniquely partitions the particles depending upon
the cycle lengths they live on. In taking the thermodynamic limit we define the 
long cycles to be those which are of infinite length, and short cycles as those which
are finite in length. We find that we can uniquely decompose the total density of
particles into those on long cycles and those on short cycles, i.e.
$\rho = \varrho_{\rm short} + \varrho_{\rm long}$.
Finally we also obtain an expression for the expectation of
certain operators in terms of cycle statistics, which will be useful for analysing
occupation number densities and off-diagonal long-range order in later chapters.
}
\vspace{15pt}
\hrule

\section{Cycle Representation of the Partition Function}

In his paper investigating the low-temperature properties of liquid He$^4$, R. Feynman \cite{Fey_1953_and_stat._mech.}
re-expressed the partition function of a boson gas in terms of the cycle structures
of permutations induced by Bose-Statistics. He postulated that at low temperatures
(or high densities), bosons belong to cycles of different lengths, and below some
critical temperature cycles of infinite length appear. Feynman associated this critical behaviour
with the phase transition of He$^4$ known as Bose-Einstein condensation.

In this chapter we will establish a general framework for cycle statistics as described in
\cite{DMP}, following \cite{Martin}. This framework will be applied throughout this script
so it is valuable to lay down the notational foundation for this thesis before applying
the techniques to various models.

To begin, we shall consider a system of $n$ identical bosons enclosed in a region $\Lambda$
with volume $|\Lambda|$, in thermal equilibrium at temperature $T$ (i.e. in the canonical ensemble). 
Denote the inverse temperature $\beta \vcentcolon= k_B T$, where $k_B$ is Boltzmann's constant. For this 
ensemble we take the thermodynamic limit at a fixed temperature $\beta$ by letting the quantities $n$ 
and $|\Lambda|$ tend to infinity in such a way that $n/|\Lambda| = \rho$ for a fixed $\rho$. 
For the entirety of this text we shall fix $\beta$, and drop it from our notation where possible.

Denote the single particle Hilbert space by $\HH$. Let $I$ be the identity operator on $\HH$. 
For an operator $A$ on $\HH$
denote $A^{(n)}$ as the operator on an \textit{unsymmetrised} (i.e. no Bose-statistics) $n$-particle Hilbert 
space $\HHcan{n} \vcentcolon= \underbrace{\HH \otimes \HH \otimes \cdots \otimes \HH}_{\text{$n$ times}}$ by
\begin{equation*}
	A^{(n)} = A \otimes I \otimes \dots \otimes I + I \otimes A \otimes \dots \otimes I
	+ \dots + I \otimes \dots  \otimes I \otimes A.
\end{equation*}
When considering bosons, we need to restrict the system to the \textit{symmetric} subspace of $\HHcan{n}$.
Let $S_n$ be the set of all permutations of $\{1,\dots,n\}$.
The symmetrisation projection $\sym{n}$ on $\HHcan{n}$ is defined by 
\begin{equation*}
	\sym{n} \vcentcolon= \frac{1}{n!} \sum_{\pi \in S_n} U_\pi
\end{equation*}
where $U_\pi : \HHcan{n} \mapsto \HHcan{n}$ is a unitary representation of the permutation group 
$S_n$ on $\HHcan{n}$ defined by
\begin{equation*}
	U_\pi(\phi_1 \otimes \phi_2 \otimes \cdots \otimes \phi_n) 
	= \phi_{\pi(1)} \otimes \phi_{\pi(2)} \otimes \cdots
			 \otimes \phi_{\pi(n)}
\end{equation*}
with $\phi_j \in \HH$, $j=1, \dots, n$ and any $\pi \in S_n$. Thus we denote the 
symmetric or bosonic $n$-particle Hilbert space by $\HHcansym{n} \vcentcolon= \sym{n} \HHcan{n}$. 
We assume that there is
a symmetric Hamiltonian $\cHHam$ on $\HHcan{n}$ whose restriction to $\HHcansym{n}$ is the
Boson Hamiltonian for our system.
The canonical partition function for the $n$-particle Hamiltonian $\cHHam$ can be written as
\begin{equation*}
	\cHPart 
=	\trace_{\HHcansym{n}} \left[ \e^{-\beta \cHHam} \right] 
	= \trace_{\HHcan{n}} \left[ \sym{n} \e^{-\beta \cHHam} \right]
=	\frac{1}{n!} \sum_{\pi \in S_n} \trace_{\HHcan{n}} \left[ U_\pi \e^{-\beta \cHHam} \right].
\end{equation*}
This leads us naturally to define a probability measure on $S_n$ by
\begin{equation}									\label{cProb_defn} 
	\cHProb(\pi) = \frac{1}{\cHPart} \frac{1}{n!} \trace_{\HHcan{n}} \left[ U_\pi \e^{-\beta \cHHam} \right]
\end{equation}
provided that the following holds for all $\pi \in S_n$:
\begin{equation}									\label{Cycle_Cond1} 
	\trace_{\HHcan{n}} \left[ U_\pi \e^{-\beta \cHHam} \right] \ge 0.
\end{equation}
Note that in general this condition is not obviously satisfied. However if $\e^{-\beta \cHHam}$
has a Feynman-Kac representation, then (\ref{Cycle_Cond1}) must be true.

In the grand-canonical ensemble, the number of particles is not fixed, so we must be prepared to
deal with an arbitrary number of particles. To do so we construct the Fock space $\FFH$ which consists of
the tensor product of each canonical $n$-particle Hilbert space $\HHcan{n}$ for every $n \ge 0$. 
Then the thermodynamic limit simply consists of letting the region $\Lambda$ tend to $\RR^d$. So
\[
	\FFH = \bigoplus_{n \ge 0} \HHcan{n}
\]
where $\HHcan{0} \vcentcolon= \CC$ and $\HHcan{1} \vcentcolon= \HH$. 
Define the second quantization of the single-particle operator $A$, denoted by $d\Gamma(A)$, as the operator 
which acts upon $\FFH$ by
\[
	d\Gamma(A) = \overline{\bigoplus_{n \ge 0} A^{(n)}}
\]
where $A^{(0)} = 0$ and $A^{(1)} = A$. The number of particles is represented by the operator
$N_\sLambda \vcentcolon= d\Gamma(I)$.

Again for the case of bosons, we need to restrict the system to the symmetric subspace of $\FFH$,
denoted $\FFHsym$, constructed by applying the symmetrisation operator to each $\HHcan{n}$ in 
the tensor product:
\[
	\FFHsym
\vcentcolon=	\bigoplus_{n \ge 0} \HHcansym{n}
=	\bigoplus_{n \ge 0} \sym{n} \HHcan{n} 
= 	\bigoplus_{n \ge 0} \frac{1}{n!} \sum_{\pi \in S_n} U_\pi \HHcan{n}.
\]
The grand-canonical partition function for a system with Hamiltonian $\gHHam$ defined
on the symmetric Fock space $\FFHsym$ and with chemical potential $\mu$ can be written as 
\[
	\gHPart 
= 	\trace_{\FFHsym} \left[ \e^{-\beta (\gHHam - \mu N_\sLambda) }\right]
= 	\sum_{n=0}^\infty \frac{1}{n!} \sum_{\pi \in S_n} \trace_{\HHcan{n}} 
			\left[ U_\pi \e^{-\beta (\gHHam - \mu n)} \right].
\]
We define the corresponding probability measure on $\bigcup_{n=0}^\infty S_n$ by
\begin{equation*}
	\gHProb(\pi) = \frac{1}{\gHPart} \sum_{n=0}^\infty \frac{1}{n!} \trace_{\HHcan{n}} 
			\left[ U_\pi \e^{-\beta (\gHHam - \mu n)} \right] \mathcal{I}_{S_n}(\pi)
\end{equation*}
(true if (\ref{Cycle_Cond1}) is satisfied) where $\mathcal{I}$ is the indicator function. 
We take $S_0 = \{ 1 \}$.

Each permutation $\pi \in S_n$ can be decomposed into a number of cyclic permutations of lengths 
$q_1, q_2, \dots, q_r$ with $r<n$ and $q_1 + q_2 + \dots + q_r = n$. We consider the set 
$\Omega = \bigcup_{r \in \mathbb{N}} \Omega_r$ of unordered $r$-tuples of natural numbers 
$\mathbf{q} = [q_1, q_2, \dots ,q_r] \in \Omega_r$ for $r=0,1,2,\dots$, and let 
$|\mathbf{q}| = q_1 + q_2 + \dots + q_r$ for $\mathbf{q} \in \Omega$. Then a decomposition of 
$\pi \in S_n$ into cycles is labelled by $\mathbf{q} \in \Omega$ with $|\mathbf{q}| = n$. We recall the
following facts on the permutation group.
\begin{itemize}
\item The decomposition into cycles leads to a partition of $S_n$ into equivalence classes of permutations 
$C_{\mathbf{q}}$, $|\mathbf{q}| = n$.
\item Two permutation $\pi'$ and $\pi''$ belong to the same class if and only if they are conjugate in 
$S_n$, i.e. if there exists a $\pi \in S_n$ such that
\begin{equation} 								\label{conjclass} 
	\pi'' = \pi^{-1} \pi' \pi.
\end{equation}
\item The number of permutations belonging to the class $C_{\mathbf{q}}$ is 
\begin{equation} 								\label{conjclasscount} 
	\frac{n!}{n_{\mathbf{q}}! (q_1 q_2\dots q_r)}
\end{equation}
with $n_\mathbf{q}! = n_1! n_2! \dots n_j! \dots$ and $n_j$ is the number of cycles of length $j$ 
in $\mathbf{q}$.
\end{itemize}

We observe for a symmetric Hamiltonian (i.e. $[\gHHam,U_\pi] = 0, \forall\, \pi \in S_n$), for 
$\pi', \pi'' \in C_{\mathbf{q}}$, and using (\ref{conjclass}) one has that
\begin{equation}
  \begin{split}								\label{invar_under_cycles} 
	\trace_{\HHcan{n}} \left[ U_{\pi''} \e^{-\beta \cHHam} \right] 
		&= \trace_{\HHcan{n}} \left[ U_\pi^{-1} U_{\pi'} U_\pi \e^{-\beta \cHHam} \right] \\
		&= \trace_{\HHcan{n}} \left[ U_\pi^{-1} U_{\pi'} \e^{-\beta \cHHam} U_\pi \right] \\
		&= \trace_{\HHcan{n}} \left[ U_{\pi'} \e^{-\beta \cHHam} \right].
   \end{split}
\end{equation}
Hence in the case of a symmetric Hamiltonian we may express the symmetrisation
operator $\sym{n}$ as a sum (with some scaling) over all unitary representations of
permutations with unique cycle structures of $S_n$
i.e.
\begin{align*}
	\trace_{\HHcansym{n}} \left[ \e^{-\beta \cHHam} \right]
&=	\sum_{C_\mathbf{q}} \sum_{\pi \in C_\mathbf{q}} \trace_{\HHcan{n}} \left[ U_{\pi} \e^{-\beta \cHHam} \right]
\\
&=	\sum_{C_\mathbf{q}} \frac{n!}{n_{\mathbf{q}}! (q_1 q_2\dots q_r)} 
	\trace_{\HHcan{n}} \left[ U_{\pi} \e^{-\beta \cHHam} \right].
\end{align*}
This motivates a more careful analysis of the class of cycle structures,
specifically the expectation of cycles of particular lengths in $C_\mathbf{q}$.

\section{Cycle Densities}

For $q \in \mathbb{N}$, define the random variable $N_q(\pi)$ to be the number of cycles of length $q$ in $\pi$.
\[
	\cHProb[ N_q(\pi)\!=\!r ] \,=\!\sum_{N_q(\pi) = r} \cHProb(\pi).
\]
Then we can define the expectation of cycles of length $q$ in the canonical ensemble by
\begin{equation*}
	\cHEx(N_q)
=	\sum_{r=1}^n r\cHProb(N_q\!=\!r)
=	\sum_{r=1}^n \sum_{\pi \in S_n} r \cHProb[N_q(\pi)\!=\!r]
\end{equation*}
and hence the average density of particles on $q$-cycles for a system of $n$ bosons is
\begin{equation}								\label{canCycleDens}
	\cHC(q) = \frac{q \; \cHEx(N_q) }{|\Lambda|}.
\end{equation}
We note here that if $n/|\Lambda|=\rho$ then the sum of the densities of particles 
on all cycle lengths is exactly the density of the system, i.e.
\begin{equation}									\label{canDensity}
	\sum_{q=1}^n \cHC(q)=\rho.
\end{equation}
In the grand-canonical ensemble, we similarly obtain that the expectation of cycles of length $q$
is
\begin{equation*}
	\gHEx(N_q) = \sum_{r=1}^\infty r\gHProb(N_q\!=\!r),
\end{equation*}
and so the average density of particles on $q$-cycles is
\begin{equation}									\label{gcanCycleDens}
	\gHC(q) = \frac{q \; \gHEx(N_q) }{|\Lambda|}.
\end{equation}

Define the grand-canonical expectation of the operator $d\Gamma(A)$ as
\begin{equation}										\label{gcanExpect}
	\la d\Gamma(A) \ra_{\gHHam}
=	\frac{1}{\gHPart} \trace_{\FFHsym} \big[ d\Gamma(A) \e^{-\beta(\gHHam - \mu N_\sLambda) } \big]
=	\frac{1}{\gHPart} \sum_{n=0}^\infty \e^{\beta \mu n} \cHPart \la A^{(n)} \ra_{\cHHam}.
\end{equation}
where
\begin{equation}										\label{canExpect}
	\la A^{(n)} \ra_{\cHHam} = \frac{1}{\cHPart} \trace_{\HHcansym{n}} \left[ A^{(n)} \e^{-\beta \cHHam}\right].
\end{equation}
The density of particles for finite volume is the expected number of particles per unit volume
\[
	\rho_\sLambda(\mu) = \frac{\la N_\sLambda \ra_{\gHHam}}{|\Lambda|}.
\]
Then the density of the system is the sum of all the cycle densities:
\begin{equation}										\label{gcanDensity}
	\sum_{q=1}^\infty \gHC(q)=\rho_\sLambda(\mu). 
\end{equation}

Hence with this formulation, equations (\ref{canDensity}) and (\ref{gcanDensity}) imply that
we can uniquely partition the ensemble of the particles into their respective cycle structures.

However we would like to observe the behaviour of these cycles in the thermodynamic limit.
To do so we shall decompose the total system density into the density of particles belonging 
to cycles of finite length ($\varrho_{{\rm short}}$) and to infinitely long cycles ($\varrho_{{\rm long}}$) 
in the thermodynamic limit. We are thus led to the following definition:

\begin{definition}									\label{rholongshort}
The expected density of particles on cycles of \textbf{infinite} length is given
\begin{itemize}
\item in the canonical ensemble by
\begin{equation*}
	\rhoblong = \lim_{Q \to \infty} \;\clthermlim \; \sum_{q=Q+1}^\infty \cHC(q),
\end{equation*}

\item in the grand-canonical ensemble by
\begin{equation*}
	\rholong = \lim_{Q \to \infty} \;\lim_{|\Lambda| \to \infty} \; \sum_{q=Q+1}^\infty \gHC(q).
\end{equation*}
\end{itemize}
\end{definition}

For notational convenience, we shall henceforth define $\lthermlim$ as the limit as 
$\Lambda \uparrow \RR^d$, where $d$ is the dimension of the system. The limited
density in the grand-canonical ensemble is defined by $\rhodens \vcentcolon= \lthermlim \rho_\sLambda(\mu)$.

Note that if we define the corresponding densities of short cycles as
\[
	\rhobshort = \lim_{Q \to \infty} \;\clthermlim \; \sum_{q=1}^Q \cHC(q),
\qquad\qquad
	\rhoshort = \lim_{Q \to \infty} \;\lthermlim \; \sum_{q=1}^Q \gHC(q),
\]
then using (\ref{canDensity}) and (\ref{gcanDensity}) one easily sees that
\[
	\rho = \rhobshort + \rhoblong 
\qquad\quad\text{and}\qquad\quad
	\rhodens = \rhoshort + \rholong.
\]
Hence in the limit, all particles can be classified as those belonging to cycles of
finite length and those belonging to cycles of infinite length.

\section{Expression for the \texorpdfstring{$q$}{q}-cycle Density}
Here we shall obtain a convenient expression for the density of particles on cycles of specific length
in the following theorem. The idea is that since the trace of the partition function only
depends upon the cycle structure of the symmetrisation operator, 
we can decompose this operator into its cycle distribution, isolate one $q$ cycle
and then recombine the remaining cycles into a symmetrisation operator over $n-q$
particles leaving us with a trace over $q$ distinguishable particles coupled with a boson 
field (with some normalisation).
 
Before stating the theorem we first shall fix some notation.
To indicate that a permutation $\pi$ contains at least one cycle of length $q$, we shall say $q \in \pi$.

Let $U_q: \HHcan{q} \to \HHcan{q}$ be the unitary representation of a $q$-cycle on 
$\HHcan{q}$ defined by
\begin{equation*}
	U_q(\phi_1 \otimes \phi_2 \otimes \phi_3 \otimes  \dots \otimes \phi_q) 
		= \phi_2 \otimes \phi_3 \otimes \dots \otimes \phi_q \otimes \phi_1.
\end{equation*}
Denote the $n$-space identity operator by $I_n = \underbrace{I \otimes I \otimes \cdots 
\otimes I}_{\text{$n$ times}}$ and the corresponding Fock space identity operator by $\II$.
When there is no ambiguity we shall simply write $U_q$ for $U_q \otimes 
\mathbb{I}: \FFsym \to \FFsym$. Note that $[U_q, \sigma^n_{+}] = 0$. 

Also we shall define the operator $M^\mu_{q,\sLambda}$ on $\HHcan{q}$ as the partial trace of 
$\e^{-\beta(\gHHam-\mu N_\sLambda)}$ over $\FFHsym$, i.e.
\begin{equation}									\label{partial-trace}
	M^\mu_{q,\sLambda} = \trace_{\FFHsym} \left[ \e^{-\beta(\gHHam-\mu N_\sLambda)} \right].
\end{equation}

\par

\begin{theorem}								\label{thm1-cycledens}
	The expectation of the density of particles on cycles of length $q$ can be expressed\vspace{-4mm}
	\begin{enumerate}
		\item in the canonical case as
	\begin{equation}							\label{can-cycle-dens}
		\cHC(q)
		=  \frac{1}{\cHPart |\Lambda|} \;
			\trace_{\HHcan{q} \otimes \HHcansym{n-q}} \left[ ( U_q \otimes I_{n-q} )
			\e^{-\beta \cHHam} \right],
	\end{equation}
	\item in the grandcanonical ensemble as
	\begin{equation}							\label{grandcan-cycle-dens}
	   \begin{split}
		\gHC(q) &= \frac{1}{\gHPart |\Lambda|} \; \trace_{ \HHcan{q} \otimes \FFHsym }
				\left[ ( U_q \otimes \II ) \e^{-\beta(\gHHam -\mu N_\sLambda)} \right]   \\
			&= \frac{1}{\gHPart |\Lambda|} \; \trace_{\HHcan{q} }
				\Big[   U_q \, M^\mu_{q,\sLambda}  \Big].
	   \end{split}
	\end{equation}
	\end{enumerate}
\end{theorem}

\begin{proof}
Define $C_n(q_1,n_1; \, q_2,n_2; \dots; q_r,n_r)$, $q_i$ distinct, $q_1 n_1 + q_2 n_2 +\dots + q_r n_r =n$ as 
the class of permutations with the following cycle structure: $n_1$ cycles of length $q_1$, $n_2$ cycles of length 
$q_2$, etc. By (\ref{conjclasscount}) the number of elements in each class is
\begin{equation*}
	\#C_n(q_1,n_1; \, q_2,n_2; \dots; q_r,n_r) = \frac{n!}{q_1^{n_1} \dots q_r^{n_r} \, n_1! n_2! \dots n_r!}.
\end{equation*}

From (\ref{cProb_defn})
\begin{equation*}
	\cHProb(N_q\!=\!r) 
= 	\frac{1}{\cHPart} \frac{1}{n!}
	\sum_{\substack{{q_1,\dots q_k \ge 0; \,q_i \ne q} \choose {\sum q_i n_i = n-rq}}}
	\sum_{\pi \in C_n(q,r; \, q_1,n_1; \dots; q_k,n_k)} \!\!\!\!
	\trace_{\HHcan{n}} \left[ U_\pi \e^{-\beta \cHHam} \right].
\end{equation*}
Using (\ref{invar_under_cycles}) we see that
\begin{multline*}
\qquad
	\cHProb(N_q\!=\!r)
= 	\frac{1}{\cHPart} \frac{1}{n!}
	\sum_{\substack{{q_1,\dots q_k \ge 0;\, q_i \ne q} \choose {\sum q_i n_i = n-rq}}}
	\#C_n(q,r; \, q_1,n_1; \dots; q_k,n_k) 
\\
	\times \trace_{\HHcan{n}} \left[ U_{\tilde{\pi}} \e^{-\beta \cHHam} \right]
\qquad
\end{multline*}
where $\tilde{\pi}$ is any fixed element in $C_n(q,r; \, q_1,n_1; \dots; q_k,n_k)$. Therefore
\begin{equation*} \begin{split}
	\cHProb(N_q\!=\!r)
&=  \frac{1}{\cHPart} \frac{1}{n!}\sum_{\substack{{q_1,\dots q_k \ge 0; \,q_i \ne q} 
	\choose {\sum q_i n_i = n-rq}}}  
	\tfrac{\#C_n(q,r; \, q_1,n_1; \dots; q_k,n_k)}{\#C_{n-q}(q,r-1; \, q_1,n_1; \dots; q_k,n_k)} 
\\
&	\hspace{40pt} \times \#C_{n-q}(q,r-1; \, q_1,n_1; \dots; q_k,n_k) \;
	\trace_{\HHcan{n}} \left[ U_{\tilde{\pi}} \e^{-\beta \cHHam} \right] 
\\[0.4cm]
&= \frac{1}{\cHPart} \frac{1}{n!}\sum_{\substack{{q_1,\dots q_k \ge 0; \,q_i \ne q}
	\choose {\sum q_i n_i = n-rq}}} \frac{n!}{rq(n-q)!}
\\
&	\hspace{40pt} \times \#C_{n-q}(q,r-1; \, q_1,n_1; \dots; q_k,n_k) \;
	\trace_{\HHcan{n}} \left[ U_{\tilde{\pi}} \e^{-\beta \cHHam} \right].
\end{split}\end{equation*}
We choose one $q$-cycle from $\tilde{\pi}$, since $\tilde{\pi} \in C_n(q,r; \, q_1,n_1; \dots; q_k,n_k)$. 
Thus write $\tilde{\pi} = q \circ \pi'$, where $\pi' \in  C_{n-q}(q,r-1; \, q_1,n_1; \dots; q_k,n_k)$ and 
$q \in \tilde{\pi}$. The corresponding representations can be written as 
$U_{\tilde{\pi}} = U_q \otimes U_{\pi'}$, so summing over $\pi'$ 
\begin{multline*}
\qquad
	\cHProb(N_q\!=\!r) 
=	\frac{1}{\cHPart} \frac{1}{rq(n-q)!} 
	\sum_{\substack{{q_1,\dots q_k \ge 0; \,q_i \ne q} \choose {\sum q_i n_i = n-rq}}}
\\
	\times\hspace{-6pt}\sum_{\pi' \in C_{n-q}(q,r-1; \, q_1,n_1; \dots; q_k,n_k)} \hspace{-16pt}
	\trace_{\HHcan{n}} \left[ (U_q \otimes U_{\pi'}) \e^{-\beta \cHHam} \right] 
\qquad
\end{multline*}
and by using (\ref{invar_under_cycles}) again we have
\begin{multline*}
\qquad
	\cHProb(N_q\!=\!r)
=	\frac{1}{\cHPart} \frac{1}{rq(n-q)!} 
	\sum_{\substack{{q_1,\dots q_k \ge 0; \,q_i \ne q} \choose {\sum q_i n_i = n-rq}}}
\\
	\times\hspace{-6pt}\sum_{\pi' \in C_{n-q}(q,r-1; \, q_1,n_1; \dots; q_k,n_k)} \hspace{-16pt}
	\trace_{\HHcan{n}} \left[ (U_q \otimes U_{\pi'}) \e^{-\beta \cHHam} \right].
\qquad
\end{multline*}
Then the canonical expectation of the number of $q$-cycles is found to be
\begin{align*}
	\cHEx(N_q)
&=	\sum_{r=1}^{\infty} r\cHProb(N_q\!=\!r)
\\
&=	\frac{1}{\cHPart} \frac{1}{q(n-q)!} \sum_{r=1}^{\infty} 
	\sum_{\substack{{q_1,\dots q_k \ge 0; \,q_i \ne q} \choose {\sum q_i n_i = n-rq}}}
\\*
&	\hspace{3.5cm} \times\hspace{-6pt}\sum_{\pi' \in C_{n-q}(q,r-1; \, q_1,n_1; \dots; q_k,n_k)}\!\!\!
	\trace_{\HHcan{n}} \left[ (U_q \otimes U_{\pi'}) \e^{-\beta \cHHam} \right]
\\[0.3cm]
&=	\frac{1}{\cHPart} \frac{1}{q(n-q)!} \sum_{\pi \in S_{n-q}}
	\trace_{\HHcan{n}} \left[ (U_q \otimes U_{\pi}) \e^{-\beta \cHHam} \right] 
\\[0.3cm]
&=	\frac{1}{\cHPart} \frac{1}{q} 
	\trace_{\HHcan{q} \otimes \HHcansym{n-q}} 
	\left[ (U_q \otimes I_{n-q}) \e^{-\beta \cHHam} \right]
\end{align*}
and the corresponding cycle density expression follows from (\ref{canCycleDens}).
Going to the grand-canonical ensemble, using (\ref{gcanCycleDens}) we have:
\begin{align*}
	\gHC(q)
&=	\frac{q\,\gHEx(N_q)}{|\Lambda|}
=	\frac{q}{\gHPart |\Lambda|} \sum_{n=q}^\infty \e^{\beta \mu n} \cHPart \cHEx(N_q)
\\[0.2cm]
&=	\frac{1}{\gHPart |\Lambda| } \sum_{n=q}^\infty 
	\trace_{ \HHcan{q} \otimes \HHcansym{n-q}} 
	\left[ (U_q \otimes I_{n-q}) \e^{-\beta(\gHHam - \mu N_\sLambda)} \right]
\\[0.2cm]
&=	\frac{1}{ \gHPart |\Lambda| } \trace_{ \HHcan{q} \otimes \FFHsym }
	\left[ (U_q \otimes \II) \e^{-\beta(\gHHam - \mu N_\sLambda)} \right]
\end{align*}
as required.
\end{proof}

\section{\texorpdfstring{Expectation of Operators on $q$-cycles}{Expectation of Operators on q-cycles}}
											\label{02cycle-expect}

For an arbitrary one particle operator $A$, define the following cycle-dependent quantities:
\begin{itemize}
\item In the canonical ensemble,
\begin{equation}									\label{cycle-expt-can}
	\cHC(q,A)
= 	\frac{1}{\cHPart |\Lambda|}\trace_{\HHcan{q} \otimes \HHcansym{n-q}} \bigg[
	\underbrace{ \left( A \otimes I \otimes \dots \otimes I \right) }_{\text{$q$-terms}} 
	(U_q \otimes I_{n-q}) \e^{-\beta \cHHam} \bigg].
\end{equation}
It is easily seen that $\cHC(q,I) = \cHC(q)$. Also notice that $\cHC(q,A)$ is linear in $A$, since the trace is linear.
\item In the grand-canonical ensemble
\begin{equation}									\label{cycle-expt-gcan}
	\gHC(q,A)
= 	\frac{1}{\gHPart |\Lambda|}\trace_{\HHcan{q}} \bigg[
	\underbrace{ \left( A \otimes I \otimes \dots \otimes I \right) }_{\text{$q$-terms}}
	U_q M^\mu_{q,\sLambda} \bigg].
\end{equation}
which again is linear in $A$, and $\gHC(q,I) = \gHC(q)$.
\end{itemize}

Then the following theorem shows that we can express the expectation of those multi-particle operators
of the form
\begin{equation*}
	A^{(n)} = A \otimes I \otimes \dots \otimes I + I \otimes A \otimes \dots \otimes I 
	+  \dots +  I \otimes \dots \otimes I \otimes A
\end{equation*}
on $\HHcan{n}$ in the canonical ensemble, or 
\begin{equation*}
	d\Gamma(A) = \overline{\bigoplus_{n \ge 0} A^{(n)}}
\end{equation*}
(where $A^{(0)} \vcentcolon= 0$ and $A^{(1)} \vcentcolon= A$) on $\FFHsym$ in the grand-canonical ensemble, 
in terms of cycle statistics. Specifically we reduce the expectation of a multi-particle operator 
to an expression where the single-particle operator is applied to one of the $q$ distinguishable
particles, all coupled with the boson field (with some scaling).

\begin{theorem}									\label{expectation-cycled}
	In the canonical ensemble
	\begin{equation*}
		\frac{\la A^{(n)} \ra}{|\Lambda|} = \sum_{q=1}^n \cHC(q,A)
	\end{equation*}
	and similarily in the grand-canonical ensemble
	\begin{equation*}
		\frac{\la d\Gamma(A)\ra}{|\Lambda|} = \sum_{q=1}^\infty \gHC(q,A).
	\end{equation*}
\end{theorem}

\begin{proof}
Consider the canonical case
\begin{align*}
	\langle A^{(n)} \rangle 
&= 	\frac{1}{\cHPart} \trace_{\HHcansym{n}} \left[ A^{(n)} \e^{-\beta \cHHam} \right] 
\\
&= 	\frac{1}{\cHPart} \frac{1}{n!} \sum_{\pi \in S_n} \trace_{\HHcan{n}} \left[ U_\pi A^{(n)} \e^{-\beta \cHHam} \right] 
\\
&= 	\frac{1}{\cHPart} \frac{1}{n!} \sum_{\pi \in S_n} \trace_{\HHcan{n}} \left[ A^{(n)} U_\pi \e^{-\beta \cHHam} \right] 
\end{align*}
using the cyclicity of the trace and the fact that $[U_\pi,\cHHam]=0$. Note that we can simplify this 
expression by the following method:
{\allowdisplaybreaks
\begin{align*}
& 	\trace_{\HHcan{n}} \left[ A^{(n)} U_\pi \e^{-\beta \cHHam} \right] 
\\
&= 	\trace_{\HHcan{n}} \Bigg[ \sum_{i=1}^n 
	(I \otimes \cdots \otimes \!\!\!\underbrace{A}_{i^\text{th}\text{-position}}\!\!\! \otimes \cdots \otimes I) 
	U_\pi \e^{-\beta \cHHam} \Bigg]
\\
&= 	\trace_{\HHcan{n}} \left[ \sum_{i=1}^n U_{(1\,i)} (A \otimes I \otimes \cdots \otimes I)
	U_{(1\,i)} U_\pi \e^{-\beta \cHHam} \right]
\intertext{where $U_{(1\,i)}$ represents the two-cycle $(1 \, i)$, so again using cyclicity of the trace}
		&= 	\trace_{\HHcan{n}} \left[ \sum_{i=1}^n (A \otimes I \otimes \cdots \otimes I)
				U_{(1\,i)} U_\pi \e^{-\beta \cHHam} U_{(1\,i)} \right] \\
		&= 	\trace_{\HHcan{n}} \left[ \sum_{i=1}^n (A \otimes I \otimes \cdots \otimes I)
				U_{(1\,i)} U_\pi U_{(1\,i)} \e^{-\beta \cHHam} \right] \\
		&= 	n \; \trace_{\HHcan{n}} \left[ (A \otimes I \otimes \cdots \otimes I) U_{\pi'}  \e^{-\beta \cHHam} \right]
\end{align*}}
by (\ref{invar_under_cycles}). Thus
\begin{equation}										\label{sum_perms}
	\langle A^{(n)} \rangle 
=	\frac{1}{\cHPart} \frac{1}{(n-1)!} \sum_{\pi \in S_n}
	\trace_{\HHcan{n}} \Big[ (A \otimes I \otimes \cdots \otimes I) U_\pi \e^{-\beta \cHHam} \Big] .
\end{equation}

Define $S_r(\{ a_1, \dots a_r \})$ as the set of all permutations of $\{ a_1, \dots a_r \}$.
For some $q$ choose $i_2, i_3, \dots i_q$ to be distinct elements of the set $\{2,3, \dots n\}$ and define
\begin{equation*}
	S_n^q(i_2, i_3, \dots i_q) = \Big\{ (1,i_2, i_3, \dots i_q) \circ \tau :
						\tau \in S_{n-q}(\{2,3, \dots q\} \setminus \{i_2, i_3, \dots i_q\} )\Big\}.
\end{equation*}
Using this construction, we see that the sets $S_n^q(i_2, i_3, \dots i_q)$ are disjoint for each $q$, since for each 
case $q$ is the length of the cycle containing 1, the element never touched by $S_{n-q}$. Sets are also distinct
by choice of $(i_2, i_3, \dots i_q)$, evident when comparing two different choices of indices $(i_2, i_3, \dots i_q)$ 
and $(i'_2, i'_3, \dots i'_q)$ the element 1 is in a different cycle. 

Considering the cardinality of these sets, for fixed $(i_2, i_3, \dots i_q)$, 
$\#S_{n}^q(i_2, i_3, \dots i_q) = \#S_{n-q} = (n-q)!$ by construction. We pick the $q-1$ indices $(i_2, i_3, \dots i_q)$ 
from $n-1$. Thus summing over all possibilities
\begin{multline*}
\qquad
	\# \bigcup_{q=1}^n \bigcup_{(i_2, i_3, \dots i_q)} S_{n}^q(i_2, i_3, \dots i_q) 
= 	\sum_{q=1}^n \frac{(n-1)!}{(n-q)!} \# S_{n}^q
\\
=	\sum_{q=1}^n \frac{(n-1)!}{(n-q)!} (n-q)!
=	\sum_{q=1}^n (n-1)! 
= 	n!
\qquad
\end{multline*}
where $S_n^q = S_n^q(2,3, \dots ,n)$. Due to the facts that $S_n^q(i_2, i_3, \dots i_q)$ are distinct with respect to 
both the value of $q$ and the choices of $(i_2, i_3, \dots i_q)$, and summing all possible choices gives $n!$, we 
may conclude the identity
\begin{equation*}
	S_n = \bigcup_{q=1}^n \bigcup_{i_2, \dots i_q} S_n^q(i_2, \dots , i_q).
\end{equation*}
For $\pi \in S_n^q(i_2, i_3, \dots i_q)$, we may find $\sigma$ a permutation of $(2,3,\dots ,n)$ such that
\begin{equation*}
	\sigma \pi \sigma^{-1} = \tilde{\pi} \in S_n^q
\end{equation*}
and therefore
\begin{align*}
	\trace_{\HHcan{n}} &\Big[ (A \otimes I \otimes \cdots \otimes I) U_{\pi}  \e^{-\beta \cHHam} \Big] 
\\
&=	\trace_{\HHcan{n}} \Big[ (A \otimes I \otimes \cdots \otimes I) 
	U_\sigma^{-1} U_{\tilde{\pi}} U_\sigma  \e^{-\beta \cHHam} \Big]
\\
&= 	\trace_{\HHcan{n}} \Big[ U_\sigma (A \otimes I \otimes \cdots \otimes I) U_\sigma^{-1}
	U_{\tilde{\pi}}  \e^{-\beta \cHHam} \Big]
\\
&=	\trace_{\HHcan{n}} \Big[ (A \otimes I \otimes \cdots \otimes I) 
	U_{\tilde{\pi}}  \e^{-\beta \cHHam}  \Big] 
\end{align*}
since $U_\sigma (A \otimes I \otimes \cdots \otimes I) U_\sigma^{-1} = A \otimes I \otimes \cdots \otimes I$ and 
$[\cHHam, U_\sigma] = 0$.

Thus instead of summing over all elements in $S_n$ in equation ($\ref{sum_perms}$), we may sum
over all $S_n^q$ from $q=1, \dots ,n$ as shown below
{\allowdisplaybreaks
\begin{align*}
	\la A^{(n)} \ra
&=	\frac{1}{\cHPart} \frac{1}{(n-1)!} \sum_{\pi \in S_n}   
	\trace_{\HHcan{n}} \Big[ (A \otimes I \otimes \cdots \otimes I) U_\pi \e^{-\beta \cHHam} \Big]
\\[0.2cm]
&=	\frac{1}{\cHPart} \frac{1}{(n-1)!} \sum_{q=1}^n \sum_{(i_2, i_3, \dots i_q)}  
	\sum_{\pi \in S_n^q(i_2, i_3, \dots i_q)}
\\*
	&\hspace{130pt}\trace_{\HHcan{n}} \Big[ (A \otimes I \otimes \cdots \otimes I) 
	U_{\pi}  \e^{-\beta \cHHam} \Big] 
\\[0.2cm]
&= 	\frac{1}{\cHPart} \frac{1}{(n-1)!} \sum_{q=1}^n \frac{(n-1)!}{(n-q)!}  
	\sum_{\tilde{\pi} \in S_n^q} 
	\trace_{\HHcan{n}} \Big[ (A \otimes I \otimes \cdots \otimes I) 
	U_{\tilde{\pi}}  \e^{-\beta \cHHam} \Big]
\\[0.2cm]
&=	\frac{1}{\cHPart} \sum_{q=1}^n \frac{1}{(n-q)!}  \sum_{\pi' \in S_{n-q}} 
	\trace_{\HHcan{n}} \Big[ (A \otimes I \otimes \cdots \otimes I) 
	U_q \circ U_{\pi'}  \e^{-\beta \cHHam} \Big]
\\[0.2cm]
&=	\frac{1}{\cHPart} \sum_{q=1}^n \trace_{\HHcan{q} \otimes \HHcansym{n-q}} 
	\Big[ (A \otimes I \otimes \cdots \otimes I) U_q  \e^{-\beta \cHHam} \Big]
\end{align*}}
as desired. The grand-canonical case follows with ease.
\end{proof}

In the following chapters we consider Feynman's conjecture for a selection of models
and investigate the relationship, if any, between a non-zero infinite cycle density 
$\varrho_{\rm long}$ and Bose-Einstein condensation.

\chapter{Cycle Statistics of Simpler Models}								\label{chapter3}

\hrule
\textbf{Summary}\\
\textit{In this chapter a relation between the ground state
occupation density and cycle statistics is made. This relation rests upon the validity of three conditions,
all of which are verified for three different models: the ideal Bose gas, the Mean-Field model and the Perturbed 
Mean-Field model. From this theory we reproduce the well known fact that the density of long cycles 
for these three models is equal to the density of the condensate. However we note that the core condition 
does not hold for all models.}
\vspace{15pt}
\hrule

\section{Occupation Numbers}									\label{SectionOCCMO}

The appearance of (generalised) Bose-Einstein condensation depends on the low-energy
behaviour of the density of states. We intend to investigate the connection between the cycle 
statistics and the occurrence of condensation by considering the expansion of the expectation
of a low-energy density operator in terms of the formulation described in 
Section \ref{02cycle-expect}, and comparing it with the long-cycle density $\varrho_{\text{long}}$.

Consider a system of bosons enclosed a region $\Lambda \subset \RR^d$.
Let 
\[
	h_\sLambda = -\Delta
\]
act upon a dense domain in the Hilbert space $\mathcal{H}_\sLambda \vcentcolon= L^2(\Lambda)$ with 
suitable boundary conditions on $\partial \Lambda$. Suppose that $h_\sLambda$ has 
eigenvalues $E_k^\sLambda$ with corresponding eigenvectors $\phi_k^\sLambda$,
indexed by $k \ge 1$ such that $0 \le E_1^\sLambda \le E_2^\sLambda \le \dots$.

Define a symmetric Hamiltonian $H_\sLambda$ on the Fock space $\mathcal{F}_{+}(\HH)$ by
\[
	H_\sLambda = d\Gamma(h_\sLambda) + U
\]
where $U$ denotes the inter-particle interaction.
For any state $\phi \in \HH$ define the orthogonal 
projection in $\HH$ onto $\phi$ by $P_{\phi} \vcentcolon= | \phi \ra\la \phi |$. The 
corresponding second quantization 
is the operator $N_\sLambda(\phi) \vcentcolon= d\Gamma(P_\phi)$ which 
counts the number of particles in the state $\phi$. Let
$P_\varepsilon \vcentcolon= \sum_{k: E^\sLambda_k < \varepsilon} P_{\phi_k}$ which is a projection
onto the subspace of $\HH$ with single particle kinetic energy not exceeding $\varepsilon$. Then the operator
\[	
	N_\sLambda(\varepsilon) = d\Gamma(P_\varepsilon) 
=	\sum_{k : E_k^\sLambda < \varepsilon} N_\sLambda(\phi^\sLambda_k)
\]
counts the number of particles with kinetic energy less than $\varepsilon$.

In terms of cycle statistics established in Chapter \ref{chapter2}, the average expectation of this operator can
be expressed by
\[
	\frac{ \la N_\sLambda(\epsilon) \ra_{H_\sLambda} }{ |\Lambda|}
= 	\sum_{q=0}^\infty \gHC(q, P_\varepsilon)
=	\frac{1}{ |\Lambda | \gHPart} \sum_{q=0}^\infty \trace_{\HHcan{q}} 
	\bigg[\underbrace{ \left( P_\varepsilon \otimes I \otimes
	\dots \otimes I \right) }_{q\text{-terms}} U_q M^\mu_{q, \sLambda}\bigg]
\]
where $M^\mu_{q, \sLambda}$ is a partial trace of $\e^{-\beta ( \gHHam - \mu N_\sLambda)}$
over $\FFHsym$, see (\ref{partial-trace}).  We shall consider
this cycle expectation expression to investigate whether cycles of infinite length appear
and if they coincide with a macroscopic occupation of particles near the ground state.

Using this formulation, under certain conditions, we can prove three Theorems
relating the density of the Bose-Einstein condensate with the long-cycle density,
which can be applied to the simpler models discussed in the next section.
These Theorems depend upon the three conditions, which we shall number \Rmnum{1}--\Rmnum{3}.
Later we shall prove the validity of these conditions for a selection of models.

For the next three Theorems let us first assume that the following condition holds:
\newtagform{brackets2}{[\;}{.\;]}
\usetagform{brackets2}
\begin{equation}				\tag{\Rmnum{1}}	\label{positivity2}
	\gHC(q,A) \ge 0 \text{\; for all \;} A \ge 0.
\end{equation}
\usetagform{default}%
The first theorem proves that the occupation density for any single particle state is bounded above by
$\rholong$.

\begin{theorem}								\label{longcycleslowerbound}
Suppose that Condition \ref{positivity2} is satisfied and if
\begin{equation}									\label{condition1}
	\lthermlim \gHC(q, P_{\phi_\sLambda}) = 0
\end{equation}
holds for all $q \ge 1$ and any $\phi_\sLambda \in \HH$ with $||\phi_\sLambda||=1$. 
Then for any single-particle state $\phi_\sLambda \in \HH$ we have the following upper bound:
\begin{equation}									\label{statebound}
		\lthermlim \frac{\la N_\sLambda(\phi_\sLambda) \ra}{ |\Lambda|} 
	\le	\rholong.
\end{equation}
\end{theorem}
\begin{proof}
\begin{align*}
	\lthermlim \sum_{q=1}^\infty \gHC(q,I-P_{\phi_\sLambda})
&\ge 	\lim_{Q \to \infty} \lthermlim \sum_{q=1}^{Q}\gHC(q,I-P_{\phi_\sLambda}) 
\\
&=	\lim_{Q \to \infty} \lthermlim \sum_{q=1}^{Q} \gHC(q) 
	- \lim_{Q \to \infty} \lthermlim\sum_{q=1}^{Q} \gHC(q,P_{\phi_\sLambda}) 
\\
&=	\lim_{Q \to \infty} \lthermlim \sum_{q=1}^{Q} \gHC(q)
\end{align*}
using Condition \ref{positivity2} and (\ref{condition1}). Then since 
$\displaystyle\lthermlim \sum_{q=1}^\infty \gHC(q) = \rho(\mu)$ and
$\gHC(q,A)$ is linear in $A$, we may write
\[
	\lthermlim \frac{\la N_\sLambda(\phi_\sLambda) \ra}{ |\Lambda|} 
=	\lthermlim \sum_{q=1}^\infty \gHC(q,P_{\phi_\sLambda})
\le	\lim_{Q \to \infty} \lthermlim \sum_{q=Q+1}^\infty \gHC(q) 
= \rholong
\]
as desired.
\end{proof}

We note that condition (\ref{condition1}) holds if we assume that
$\e^{-\beta H}$ has a Feynman-Kac representation. In this case we can show that
\begin{equation}									\label{condition1general}
	\gHC(q,P_{\phi_\sLambda})
\;\le\; 	\frac{e^{-\beta q(E_1^\sLambda-\mu)}}{|\,\Lambda|}
\;\le\; 	\frac{e^{\beta q\mu}}{|\,\Lambda|}
\end{equation}
where $E_1^\sLambda$ is the lowest eigenvalue of the single-particle free Hamiltonian
$h_\sLambda$, which immediately implies condition (\ref{condition1}).

Justification of (\ref{condition1general}) is as follows: let $H^n_\sLambda$ be the restriction
of $H_\sLambda$ to $\HHcan{n}$. The kernel of $\exp(-\beta \cHHam)$ on
$\HHcan{q} \otimes \HHcansym{n-q}$ is positive and bounded above by the kernel of
$\exp(-\beta H_\sLambda^q) \otimes \exp(-\beta H_\sLambda^{n-q})$. Thus
the kernel of $M^\mu_{q,\sLambda}$ is positive and bounded
above by $\gHPart \e^{\beta \mu q}\ \times$ the kernel of
$\exp(-\beta h_\sLambda^{(q)})$. Therefore
\[
	\gHC(q,P_{\phi_\sLambda})
\le	 \frac{1}{|\Lambda|} \trace_{\HHcan{q}} \left[
	(P_{|\phi_\sLambda|}\otimes I
	\otimes \ldots \otimes I) U_q \e^{-\beta( h_\sLambda^{(q)} - \mu q)} \right].
\]
Then considering the trace and expanding in terms of the eigenbasis of $h_\sLambda$:
\begin{align*}
&\hspace{-0.7cm}
	\trace_{\HHcan{q}} \left[ (P_{|\phi_\sLambda|}\otimes I \otimes \ldots \otimes I)
	U_q \e^{-\beta h_\sLambda^{(q)}} \right]
\\
&=	\sum_{k_1,k_2,\ldots,k_q\ge 1} \e^{-\beta(E_{k_1}^\sLambda+ E_{k_2}^\sLambda +\ldots+ E_{k_q}^\sLambda)}
\\
&\hspace{1.5cm}
	\times \big\la \phi_{k_1}^\sLambda \otimes \phi_{k_2}^\sLambda \otimes \ldots \otimes \phi_{k_q}^\sLambda
	\big| (P_{|\phi_\sLambda|}\otimes I \otimes \ldots \otimes I)U_q \big|
	\phi_{k_1}^\sLambda  \otimes \phi_{k_2}^\sLambda \otimes \ldots \otimes \phi_{k_q}^\sLambda \big\ra
\\[0.2cm]
&=
	\sum_{k_1,k_2,\ldots,k_q\geq 1} \e^{-\beta(E_{k_1}^\sLambda+ E_{k_2}^\sLambda +\ldots+ E_{k_q}^\sLambda)}
	\\
&\hspace{1.5cm}
	\times \big\la \phi_{k_1}^\sLambda \otimes \phi_{k_2}^\sLambda \otimes \ldots \otimes \phi_{k_q}^\sLambda
	\big| (P_{|\phi_\sLambda|}\otimes I \otimes \ldots \otimes I) \big|
	\phi_{k_2}^\sLambda \otimes \phi_{k_3}^\sLambda \otimes \ldots \otimes \phi_{k_1}^\sLambda \big\ra
\\[0.2cm]
&=	\sum_{k_1,k_2,\ldots,k_q\geq 1} \e^{-\beta(E_{k_1}^\sLambda+ E_{k_2}^\sLambda +\ldots+ E_{k_q}^\sLambda)}
	\big\la \phi_{k_1}^\sLambda \big| |\phi_\sLambda| \big\ra
	\big\la |\phi_\sLambda| \big| \phi_{k_2}^\sLambda \big\ra 
	\big\la \phi_{k_2}^\sLambda \big| \phi_{k_3}^\sLambda \big\ra 
	\big\la \phi_{k_3}^\sLambda \big| \phi_{k_4}^\sLambda \big\ra
	\ldots \big\la \phi_{k_q}^\sLambda \big| \phi_{k_1}^\sLambda \big\ra
\\[0.2cm]
&=	\sum_{k\geq 1} \e^{-\beta qE_k^\sLambda}
	\big\la \phi_k^\sLambda \big| |\phi_\sLambda| \big\ra
	\big\la |\phi_\sLambda| \big| \phi_k^\sLambda \big\ra
\\
&\le	\e^{-\beta qE_{1\,\sLambda}}
\end{align*}
and hence (\ref{condition1general}) follows.

In the next theorem we show that the generalised condensate in the lower single particle kinetic energy states,
defined by
\[
	\rhocond \vcentcolon
=	\lim_{\varepsilon \to 0} \lthermlim \frac{\la N_\sLambda(\varepsilon) \ra}{ |\Lambda|}
\]
is bounded above by $\rholong$.

\begin{theorem}								\label{longcyclesotherlowerbound}
If Condition \ref{positivity2} is satisfied and 
\usetagform{brackets2}
\begin{equation}								\tag{\Rmnum{2}}	\label{condition2}
	\lim_{\varepsilon \to 0} \lthermlim \gHC(q, P_\varepsilon) = 0,
\end{equation}
\usetagform{default}%
then  for any single-particle state $\phi_\sLambda \in \HH$ we have the following upper bound:
\[
	\rhocond \le 	\rholong \; .
\]
\end{theorem}

\begin{proof}
Similarly to the previous proof we have that
\begin{align*}
	\lthermlim \sum_{q=1}^\infty \gHC(q,I-P_\varepsilon)
&\ge 	\lim_{Q \to \infty} \lthermlim \sum_{q=1}^{Q}\gHC(q,I-P_\varepsilon) 
\\
&=	\lim_{Q \to \infty} \lthermlim \sum_{q=1}^{Q} \gHC(q) 
	- \lim_{Q \to \infty} \lthermlim\sum_{q=1}^{Q} \gHC(q,P_\varepsilon) 
\\
&=	\lim_{Q \to \infty} \lthermlim \sum_{q=1}^{Q} \gHC(q)
\end{align*}
using Conditions \ref{positivity2} and \ref{condition2} respectively, with which one can derive
the following:
\[
	\lim_{\varepsilon \to 0} \lthermlim \frac{\la N_\sLambda(\varepsilon) \ra }{ |\Lambda|}
=	\lthermlim \sum_{q=1}^\infty \gHC(q,P_\varepsilon)
\le	\lim_{Q \to \infty} \lthermlim \sum_{q=Q+1}^\infty \gHC(q)
=	\rholong.\vspace{-20pt}
\]
\end{proof}


Finally we show that $\rholong$ is also an lower bound for the generalised condensate.

\begin{theorem}								\label{longcyclesupperbound}
Suppose that for each $\varepsilon > 0$, 
\usetagform{brackets2}
\begin{equation}								\tag{\Rmnum{3}}	\label{condition3}
	\lim_{Q\rightarrow \infty} \displaystyle\lthermlim 
	\displaystyle\sum_{q=Q+1}^\infty \gHC(q,I - P_\varepsilon) = 0,
\end{equation}
\usetagform{default}%
then 
\begin{equation*}
 	\rhocond \ge \rholong \; .
\end{equation*}
\end{theorem}

\begin{proof}
Let $Q_\varepsilon = I - P_\varepsilon$ and recall that $\la N_\sLambda(\varepsilon) \ra \vcentcolon= 
\la d\Gamma(P_\varepsilon)\ra$.
Then
\begin{align*}
	\lthermlim \sum_{q=1}^\infty \gHC(q,Q_\varepsilon)
= & 	\lim_{Q \to \infty} \lthermlim \sum_{q=1}^{Q} \gHC(q,Q_\varepsilon)
		+ \lim_{Q \to \infty} \lthermlim \sum_{q=Q+1}^{\infty} \gHC(q,Q_\varepsilon) \\
= & 	\lim_{Q \to \infty} \lthermlim \sum_{q=1}^{Q} \gHC(q,Q_\varepsilon) \\	
		\le &  \lim_{Q \to \infty} \lthermlim \sum_{q=1}^{Q} \gHC(q)
\end{align*}
by Conditions \ref{positivity2} and \ref{condition3}. Hence
\begin{equation*}
	\lim_{\varepsilon \to 0} \lthermlim \frac{\la N_\sLambda(\varepsilon) \ra}{ |\Lambda|} 
= 	\lim_{\varepsilon \to 0} \lthermlim \sum_{q=1}^\infty \gHC(q,P_\varepsilon) 
\ge 	\lim_{Q \to \infty} \lthermlim \sum_{q=Q+1}^\infty \gHC(q)
=	\rholong.
\end{equation*}
\end{proof}

Consequently those models which satisfy Conditions \Rmnum{1}--\Rmnum{3},
i.e. the foundations of Theorems \ref{longcyclesotherlowerbound} and 
\ref{longcyclesupperbound}, are endowed with the property that
\[
	\rholong = \rhocond.
\]
We shall prove these conditions for three models, the Ideal Bose Gas, the
Mean-Field model and the Perturbed Mean-Field model. In the following section
we give a brief introduction to these three models, and prove the conditions
in Section \ref{proveconds}.

\section{The Models}
The regions for these models, $\Lambda_l$, are a sequence of $d$-dimensional cubic boxes
centred at the origin with sides of length $l=1,2,\dots$, i.e. $\Lambda_l \vcentcolon= [ -\tfrac{l}{2} ,\tfrac{l}{2} ]^d$,
and thus with volume $V_l \vcentcolon= |\Lambda_l| = L^d$. We shall relabel all quantities with 
$l$ instead of $\Lambda$.
Thus the single-particle Hamiltonian $h_l = -\Delta$ acts upon 
$\mathcal{H}_l \vcentcolon= L^2(\Lambda_l)$ with Dirichlet boundary conditions
on the boundary, and we label its eigenvalues by $0 \le E^l_1 \le E^l_2 \le \dots$
with corresponding eigenvectors $\phi^l_k$.

The ideal Bose gas Hamiltonian is the 
second quantization of $h_l$ on the bosonic Fock space $\mathcal{F}_{+}(\mathcal{H}_l)$:
\begin{equation}									\label{02freeham}
	H^0_l = d\Gamma(h_l).
\end{equation}
The thermodynamic limit is formed by letting $l \to \infty$, i.e. $V_l \to \infty$.
The phenomenon of Bose-Einstein condensation occurs when eigenstates of the 
Hamiltonian are macroscopically occupied in the thermodynamic limit. To examine this behaviour, 
we define a distribution function $\nu_l$ which counts the number
of eigenvalues per unit volume by
\[
	\nu_l(\lambda) = \frac{1}{V_l} \# \{k : E^l_k \le \lambda \}
\]
for any $\lambda \in \RR_+$. 
One can prove (see \cite{vdB-L-P} and \cite{vdBLP88} for more details) that this sequence of measures 
converges weakly in the thermodynamic limit to a measure $\nu$, known as the integrated 
density of states, which in this case evaluates to
\begin{equation}								\label{densityofstates}
	\nu(\lambda) = C_d \lambda^{d/2}
\end{equation}
where $C_d = (\pi^{d/2} 2^d \Gamma(\tfrac{d}{2}))^{-1}$.

In terms of the single-particle eigenstates $\phi_k$ (dropping the $l$ for clarity), define the usual
creation and annihilation operators $a^\ast(\phi_k)$ and $a(\phi_k)$ which satisfy the canonical
commutation relations $[a(\phi_j) , a^\ast(\phi_k)]=\delta_{j,k}$. The operator $N_l(\phi_k) 
\vcentcolon= a^\ast(\phi_k)a(\phi_k)$ then counts the 
number of particles in $\mathcal{F}_{+}(\mathcal{H}_l)$ with state $\phi_k$. Using this formalism
we may rewrite the grand-canonical Hamiltonian of the ideal Bose gas, equation (\ref{02freeham}), as
\[
	H^0_l(\mu) \vcentcolon= H^0_l - \mu N_l
	= \sum_{k\ge1} (E^l_k - \mu) \,N_l(\phi_k)
\]
where $\mu$ is the chemical potential of the ensemble and $N_l \vcentcolon= 
\sum_{k\ge1}N_l(\phi_k)$ counts the total number of particles. 

For the ideal Bose gas in a finite volume with chemical potential $\mu<0$, at thermal equilibrium 
with inverse temperature $\beta$, we define the grand-canonical partition function by
\begin{equation}									\label{freepartition}
	\glPartF = \trace_{\mathcal{F}_{+}(\mathcal{H}_l)} \left[ \e^{- \beta H^0_l(\mu)}\right].
\end{equation}
This allows us to define the finite-volume pressure and density of the ensemble as 
\[
	p^0_{l}(\mu) = \frac{1}{\beta V_l} \ln \glPartF,
\qquad\qquad
	\rho^0_l(\mu) = \frac{d}{d\mu} p^0_{l}(\mu).
\]
We denote the corresponding limiting quantities by $p^0(\mu) \vcentcolon= \dthermlim p^0_{l}(\mu)$ and 
$\rho^0(\mu) \vcentcolon= \dthermlim \rho^0_{l}(\mu)$ respectively.

The expectation $\la N_l(\phi_k)\ra_{H^0_l(\mu)}$ returns the average number 
of particles occupying a specific quantum state $\phi_k$ 
when the system is in thermal equilibrium. It may be evaluated to yield
\[
	\la N_l(\phi_k)\ra_{H^0_l(\mu)} = \frac{1}{\e^{\beta (E^l_k - \mu)} - 1},
\]
and hence we can re-express the total average density of the system in finite volume as
\begin{equation}									\label{freedensity}
	\rho^0_l(\mu) 
=	\frac{1}{ V_l} \sum_{k \ge 1} \la N_l(\phi_k)\ra_{H^0_l(\mu)}
= 	\int_{\RR_+} \frac{1}{\e^{\beta (\lambda - \mu)} - 1} d\nu_l(\lambda).
\end{equation}
Similarly one can derive the following expression for the pressure
\begin{equation}									\label{freepressure}
	p^0_l(\mu) 
= 	\frac{1}{\beta} \int_{\RR_+} \ln \left[ 1- \e^{-\beta (\lambda - \mu)} \right]^{-1} d\nu_l(\lambda).
\end{equation}
It is useful to define the partial pressure $\pi : (-\infty, 0) \to \RR_+$ by the following
\begin{equation}									\label{partialpressure}
	\pi(s) 
= 	\frac{1}{\beta} \ln \left[ 1 - \e^{\beta s }\right]^{-1}
\end{equation}
which allows us to write
\begin{equation*}	
	p^0_l(\mu) 
= 	\int_{\RR_+} \pi(\mu - \lambda ) d\nu_l(\lambda).
\end{equation*}

To observe the well-known phase transition is this model, we proceed as follows:
fix the mean-density of the ensemble as $\bar{\rho}$. We then require that the chemical potential $\mu$ satisfies
\[
	\rho_l^0(\mu_l) = \bar{\rho}
\]
for all $l$. Set $\mucrit \vcentcolon= \dthermlim \mu_l$, 
and denote the \textit{critical density} by
\[
	\rhocrit = \lim_{\mu \to 0} \rho(\mu)
\]
and hence 
\begin{equation}									\label{freecriticaldensity}
	\rhocrit = \int_{\RR_{+}} (\e^{\beta \lambda}-1)^{-1} d\nu(\lambda).
\end{equation}
Note that the limiting thermodynamic quantities like $\rho(\mu)$ only exist for negative values 
of $\mu$. But for more general models, $\mu$ can be
positive or negative depending on the interaction and the boundary conditions of the system.

If $\rhocrit$ is infinite, then we find that $\mucrit < 0$ always satisfies
the equation 
\begin{equation}									\label{freedensities}
	\bar{\rho} = \rho^0(\mucrit)
\end{equation}
which implies a lack of a phase transition (this occurs for $d=1,2$, see \cite{hohenberg}). If
$\rhocrit$ is finite (here $d\ge3$), then equation (\ref{freedensities}) is again solved by a unique 
non-positive $\mucrit$ when $\bar{\rho} < \rhocrit$. However if $\bar{\rho} \ge \rhocrit$ 
then (\ref{freedensities}) has no solution and $\mucrit = 0$. It can be shown that the corresponding 
grand-canonical pressure $p^0(\mu)$ is constant when $\bar{\rho} \ge \rhocrit$, indicating a phase-transition at 
$\rhocrit$ to which we associate with the appearance of \textit{generalised} Bose-Einstein condensate.
To locate this condensate we focus on the eigenstates with energies in the neighbourhood of the ground-state
before taking the thermodynamic limit. For any $\varepsilon > 0$ let
\begin{equation*}
	N_l(\varepsilon) = \sum_{E^l_k < \varepsilon}  N_l(\phi_k)
\end{equation*}
which counts the number of states with energies less than $\varepsilon$. Then one may calculate
the average number of particles with arbitrarily low energies to find $\rhocondfree$, the density of the condensate:
{\allowdisplaybreaks
\begin{align*}
	\rhocondfree
&\vcentcolon=	\lim_{\varepsilon \to 0} \dthermlim \frac{ \la N_l(\varepsilon)\ra_{H^0_l(\mu)} }{ V_l}
\quad=\quad
	\lim_{\varepsilon \to 0} \dthermlim \int_{[0,\varepsilon)} 
	\frac{1}{\e^{\beta(\lambda-\mu_l)} - 1} d\nu_l(\lambda)
\\
&=	\bar{\rho} - \lim_{\varepsilon \to 0} \dthermlim \int_{[\varepsilon,\infty)} 
	\frac{1}{\e^{\beta(\lambda-\mu_l)} - 1} d\nu_l(\lambda)
\quad=\quad
	\bar{\rho} - \int_0^\infty \frac{1}{\e^{\beta(\lambda-\mucrit)} - 1} d\nu(\lambda)
\end{align*}}
\[
\Rightarrow \quad \rhocondfree
=	\begin{cases}
		0 						& \bar{\rho} < \rhocrit, \\
		\bar{\rho} - \rhocrit 	\qquad	& \bar{\rho} \ge \rhocrit.
	\end{cases}
\]
(Here there is no $\mu$ dependence as we fixed the mean density of the system $\bar{\rho}$. 
In the other models we consider this will not be the case so we shall denote the condensate density 
by $\rhocond$.) It is fascinating that a collection of non-interacting bosons
can yet display a phase transition, purely as a consequence of their quantum mechanical properties.
As a result, this model has been the object of study
for many years, and has been approached in various ways, for
examples see \cite{KacUhlenbeckZiff, ZP, vdB-L, vdB-L-L, vdBLP88.1}.

Note also that here we did not consider the nature of the condensate, but one in fact can show 
that the ground state is macroscopically occupied
\[
	\dthermlim \frac{\la N_l(\phi_1) \ra}{V_l} = \bar{\rho} - \rhocrit.
\]
However it has been shown by van\,den\,Berg, Lewis, Pul\'e \cite{vdB-L-P}
(see also \cite{vdB, vdB-L, vdB-L-L}) that the ground state is not necessarily the only 
state to be macroscopically occupied, but instead observed there are three possible ways the condensate
may form, depending on the shape of the region $\Lambda_l$. Instead of taking $\Lambda_l$
to be a $d$-dimensional cube, they considered rectangular parallelepipeds whose edges go to infinity
at different rates (known as Casimir boxes \cite{casimir}). A consequence of this anisotropy
is that the standard ground-state BEC is converted into \textit{generalised} BEC of three
types. These condensates may be classified as follows:
type I has a finite number of eigenstates macroscopically occupied, type II an infinite number of states occupied
and most surprisingly a type III, which has no macroscopically occupied eigenstates
before the thermodynamic limit, but where a macroscopic number accumulate in the ground state in the limit.

This thesis will only deal with the generalised condensate, but an investigation of the relation
between cycles and these three condensate types is currently being performed by M. Beau \cite{Beau}.

With respect to cycle statistics, S\"ut\H{o} showed in his two papers on the topic \cite{Suto1,Suto2}
that $\rhocond$ exactly equals the long cycle density $\rholong$ for the ideal Bose Gas.

\subsubsection{The Mean-Field Model}
In the mean-field model of a system of interacting bosons (also known as the Imperfect Bose Gas),
an external energy proportional to the term $N_l^2/ V_l$
is added to the Hamiltonian of the ideal boson gas:
\begin{equation*}
	H^{\textsf{mf}}_l
=	H_l^0 + \frac{a}{2 V_l} N_l^2 
\end{equation*}
where $a>0$ controls the strength of the interaction \cite{HYL57}.

This model has been studied extensively, see for instance \cite{Dav72, vdBDLP, PZ04, BLS},
where it is shown that Bose-Einstein condensation persists in the presence
of this external interaction. Condensation only occurs if it occurs in the case $a=0$, 
i.e. the ideal case, with similar dependence on the dimensionality. Note that the
presence of the mean-field interaction stabilises the system so that the thermodynamic
quantities are defined for any $\mu$.
Here the condensate density is
\[
	\rhocond = 
	\begin{cases}
		0 & 								\mu \le a \rhocrit, \vspace{5pt}\\
		\displaystyle{\frac{\mu}{a}} - \rhocrit 	\qquad & 	\mu > a \rhocrit.
	\end{cases}
\]
where $\rhocrit$ is the critical density of the ideal Bose gas. 

However in contast to the ideal Bose gas, the occurrence of the condensate 
is due to an effective attraction between bosons in the zero-mode \cite{BruZagrebnov}.
Moreover at fixed mean-density it is independent of the strength of 
the repulsive interaction and equal to the total amount of condensate in the Ideal 
Boson gas at the same mean-density. 

S\"ut\H{o} again proved the equivalence of long cycles
and the condensate for this model.

\subsubsection{The Perturbed Mean-Field Model}
This model is an extension of the mean-field model with the addition
of some interactions between modes. Its Hamiltonian is
\begin{equation}									\label{PMFham}
	H^{\mathsf{pmf}}_l
=	H_l^0 + \frac{1}{2 V_l} \left[ a N_l^2 
	+ \sum_{k, k' \ge 1} v(k,k') N_k N_{k'} \right]
\end{equation}
where $N_k \vcentcolon= N_l(\phi_k)$ is the occupation number operator of the mode $k$, 
$a>0$ and we write $v(k,k')$ to denote $v(E_k,E_{k'})$, where $v$ has a continuous and positive definite kernel 
with suitable decay properties.

Originally considered in 1990 by van den Berg, Dorlas, Lewis and Pul\'e \cite{vdBDLP}
(c.f also \cite{DLP92}), they prove the existence of the thermodynamic pressure
and find a variational formula for this pressure using Large Deviations
(we shall lay out the rough procedure of this paper as the method will be required later).
Later in 2005, Dorlas, Martin and Pul\'e \cite{DMP}
consider the cycle statistics of this model and prove that the density of long cycles equals 
the condensate density.

To apply Large Deviation Theory, one must first phrase the question in terms of probabilities.
The Hamiltonian (\ref{PMFham}) is diagonal with respect to the occupation number operators $N_k$, so it is
possible to consider the occupation numbers as random variables rather than operators.
Define our probability space $\Omega$ as the countable set of terminating sequences of 
non-negative integers, so that any $\omega \in \Omega$ is of the form $\{\omega_1, \omega_2, \dots |
\ \omega_j \in \NN \ \forall j\ \text{with} \sum \omega_j < \infty \}$. Then the basic random variables
are the occupation numbers $\{N_1, N_2, \dots \}$, which are maps $N_j : \Omega \to \NN$
defined by $N_j(\omega) = \omega_j$ for each configuration $\omega \in \Omega$. The total 
number operator $N(\omega)$ for a configuration $\omega$ is defined by
\[
	N(\omega) = \sum_{ j \ge 1} N_j(\omega)
\]
which is necessarily finite by construction. With respect to this formulation, the Ideal Bose gas Hamiltonian 
in the region $\Lambda_l$ with eigenvalues $0 \le E^l_1 \le E^l_2 \le \dots$ 
(henceforth neglecting the $l$) may be expressed for a configuration $\omega$ as
\[
	H^0_l(\omega) = \sum_{j \ge 1} E_j N_j(\omega)
\]
and thus the Perturbed Mean-Field Hamiltonian is
\begin{equation*}
	H^{\mathsf{pmf}}_l(\omega)
=	H_l^0(\omega) + \frac{1}{2 V_l} \left[ a N(\omega)^2 
	+ \sum_{k, k' \ge 1} v(k,k') N_k(\omega) N_{k'}(\omega) \right].
\end{equation*}
%

In this representation the corresponding finite-volume grand-canonical partition function and pressure 
for a chemical potential $\mu \in \RR$ are (respectively)
\[
	\glPart = \sum_{\omega \in \Omega} \e^{-\beta ( H^{\mathsf{pmf}}_l(\omega) - \mu N(\omega))},
\qquad\qquad
	p_l(\mu) = \frac{1}{\beta V_l} \ln \glPart.
\]
Again we set the limited pressure $p(\mu) = \dthermlim p_l(\mu)$.
Recall for $\mu < 0$ that equations (\ref{freepartition}) and (\ref{freepressure}) define the ideal-gas partition 
function $\glPartF$ and pressure $p^0_l(\mu)$.

Using Large Deviation theory, a variational expression for $p(\mu)$ can
be found with respect to the pressure of the ideal Bose gas. We begin by placing a probability measure
upon $\Omega$ with respect to $H^0_l$:
\[
	\Prob^\mu_l(\omega) 
=	\frac{1}{\glPartF} \e^{-\beta ( H^0_l(\omega) - \mu N(\omega))}.
\]
Introduce the occupation measure $L_l$ by defining for each Borel subset A of $[0,\infty)$
and $\omega \in \Omega$
\[
	L_l[\omega; A] = \frac{1}{ V_l} \sum_{j\ge1} N_j(\omega) \delta_{E_j}(A),
\]
where $\delta_{x}(\,\cdot\,)$ is the Dirac measure concentrated at $x$. Then for every $\omega$ in
$\Omega$, $L_l$ is a bounded positive measure. It is convenient to define the 
following notation for any $m \in \mathcal{M}^b_+(\RR_+)$, the positive bounded measures on $[0,\infty)$:
\[
	|| m || = \int_{\RR_+} m(d\lambda),
\qquad\qquad
	\lla m , Vm \rra = \iint_{\RR_+^2} v(\lambda,\lambda') m(d\lambda) m(d\lambda'),
\]
and put
\[
	G^\mu[m] = \mu ||m|| - \frac{a}{2} || m ||^2  - \frac{1}{2} \lla m , Vm \rra.
\]
In terms of the probability measure $\Prob^\mu_l$ we can express the 
pressure for the interacting model as a ideal gas pressure plus a correction term
which incorporates the inter-particle interactions (given by $G^\mu$):
\begin{equation}									\label{LDPpressure}
	p_l(\mu)
=	p^0_l(\alpha) + \frac{1}{\beta V_l} \ln \sum_{\omega \in \Omega}
	\e^{\beta  V_l G^{\mu - \alpha}[L_l[\omega,\,\cdot\,]]}
	\Prob^\alpha_l(\omega)
\end{equation}
for $\alpha<0$ (since the ideal-gas pressure is only defined for strictly-negative chemical
potentials, we use a trick from \cite{vdBLP88} whereby negative $\alpha$ is used for this purpose, 
for which we compensate by introducing a $\mu - \alpha$ term in the interaction term $G^{\mu-\alpha}$).

Next we rewrite (\ref{LDPpressure}) as an integral over $\mathcal{E} \vcentcolon= \mathcal{M}^b_+(\RR_+)$,
which we equip with the narrow topology to guarantee the continuity of the mapping
$m \to G^\mu[m]$ (see Lemma 4.1 of \cite{vdBDLP}). $G^\mu$ also has an upper bound
since $v$ is positive definite, $\lla m,vm\rra$ is non-negative and $a>0$.
Let $\KK_l$ be the probability measure induced on $\mathcal{E}$ by $L_l$:
\[
	\KK_l^\alpha = \Prob^\alpha_l \circ L_l^{-1},
\]
with which (\ref{LDPpressure}) can be re-expressed as
\begin{equation}									\label{LDPpressure1}
	p_l(\mu)
=	p^0_l(\alpha) + \frac{1}{\beta V_l} \ln \int_{\mathcal{E}}
	\e^{\beta  V_l G^{\mu - \alpha}[m]}
	\KK^\alpha_l[dm].
\end{equation}
The form of the above integral suggests the use of Laplace's method to evaluate it.
In fact, we shall use a variant called Varadhan's theorem \cite{varadhan}, which depends upon
the measures $\KK_l$ satisfying certain criteria so that 
for large $V_l$, they exhibit the following behaviour
\begin{equation*}
	\KK_l[dm] \approx \exp\big( - V_l I[m]\big)dm
\end{equation*}
where $I[m]$ is called the rate function. The criteria are stated as follows:
we say that the sequence of measures $\KK_l$
satisfy the Large Deviation Principle \cite{ellis85, varadhan} with constants $ V_l$ (which diverge to $+\infty$)
and rate function $I : \mathcal{E} \to [0,\infty]$ if the following hold:
\setlength{\leftmargini}{50pt}\vspace{-20pt}
\begin{enumerate}
\setlength{\itemsep}{1pt}
\setlength{\parskip}{0pt}
\setlength{\parsep}{0pt}
\renewcommand{\labelenumi}{(LD\arabic{enumi})}
\item $I$ is lower semi-continuous,
\item for each $b<\infty$, the set $\{ m : I[m] \le b \}$ is compact,
\item For each closed set $C$
\[
	\limsup_{l \to \infty} \frac{1}{ V_l} \ln \KK_l[C] \le -\inf\{ I[m] : m\in C \},
\]
\item For each open set $O$
\[
	\liminf_{l \to \infty} \frac{1}{ V_l} \ln \KK_l[O] \ge -\inf\{ I[m] : m\in O \}.
\]
\end{enumerate}

Now we may state Varadhan's Theorem:

\textbf{Theorem: Varadhan \cite{varadhan}}\\
\textit{
Let $\KK_l$ be a sequence of probability measures on the 
Borel subsets of $\mathcal{E}$ satisfying the Large Deviation Principle with constants $ V_l$ and 
rate-function $I$. Then, for any continuous function $G: E \to \RR$ which is bounded above, 
we have 
\[
	\dthermlim \frac{1}{ V_l} \ln \int_\mathcal{E} \e^{ V_l G[m]}
	\KK_l[dm] = \sup_{ m \in \mathcal{E}} \left\{ G[m] - I[m] \right\}.
\]
}

It is shown in a paper by R.S. Ellis, J. Gough and J.V. Pul\'{e}\cite{goughpuleellis_lpdrandomweights}
that there is a large class of measures with random weights which satisfy the large deviation principle
and they give a general procedure to obtain the corresponding rate function in an explicit form. It is easy to check that 
$\KK^\alpha_l$ is of this class, so by taking the thermodynamic limit, Varadhan's Theorem implies that 
(\ref{LDPpressure1}) reduces to
\begin{equation}									\label{LDPpressures}
	p(\mu) 
= 	p^0(\alpha) + \sup_{ m \in \mathcal{E}} \left\{ G^{\alpha-\mu}[m] - I^\alpha[m] \right\}.
\end{equation}
Then to find an expression for the rate function $I^\alpha[m]$, 
Ellis, Gough and Pul\'{e} describe the following procedure:
first single out the part of the measure $m$ which is singular with respect
to the integrated density of states $d\nu$, i.e. let $m = m_s + m_a$ be the Lebesgue 
decomposition of $m$ with respect to $d\nu$, where $m_s$ is the singular part and $m_a$ 
the absolutely continuous part.
Let $\rho$ be the density of $m_a$, i.e. $m_a(d\lambda) = \rho(\lambda) d\nu(\lambda)$.
Define $U : \mathcal{E} \to [0,\infty]$ by
\[
	U^\alpha[m] = - \int_{[0,\infty)} (\alpha - \lambda) m_s(d\lambda).
\]
Let $\pi^\ast : \RR \to (-\infty, \infty]$ be the Legendre-Fenchel transform of $\pi$, that is
\[
	\pi^\ast(t) = \sup_{s<0}\left\{ ts - \pi(s) \right\}
\]
where $\pi$ is defined in equation (\ref{partialpressure}).

For $t \in \RR$ and $r < 0$, denote
\[
	J(t,r) = \pi^\ast(t) - rt + \pi(r).
\]
Then Theorem 3 of \cite{goughpuleellis_lpdrandomweights} gives an explicit expression for the 
rate function of the form 
\begin{align*}
	I^\alpha[m] &= U^\alpha[m_s] + \int_{[0,\infty)} J(\rho(\lambda), \alpha-\lambda) m_a(d\lambda)
\\
&=	 - \int_{[0,\infty)} (\alpha - \lambda) m_s(d\lambda) 
	+ \int_{[0,\infty)} \pi^\ast(\rho(\lambda)) d\nu(\lambda) + p^0(\alpha)
\end{align*}
and hence from (\ref{LDPpressures}) get an expression for the limiting pressure in terms of
a variational problem
\[
	p(\mu) = -\inf_{m\in \mathcal{E}} \left\{ ( E_1 - \mu) ||m|| + \frac{a}{2} || m ||^2  
	+ \frac{1}{2} \lla m , Vm \rra + \int_{[0,\infty)} \pi^\ast(\rho(\lambda)) d\nu(\lambda)  \right\}.
\]
The existence of a minimizer is proved by Dorlas, Lewis and Pul\'e \cite{vdBDLP}, but an analytical
expression remains to be found. However they succeeded in showing that if $v(\cdot, \cdot)$
is continuously differentiable and $\partial v/\partial \lambda$ is bounded, then the minimising measure 
has an atom at the origin, i.e. $m_s$ is concentrated at $\lambda=0$.
With this Dorlas, Martin and Pul\'e \cite{DMP} prove that the density of long cycles equals 
the condensate density. We shall reproduce this result in the next section by a different method.

%
\section{Checking the Three Conditions for Simple Models}				\label{proveconds}

The technique detailed in Chapter \ref{chapter2} depends upon finding the probabilities of particles on
cycles of particular lengths. The validity of these probabilities rests upon (\ref{Cycle_Cond1}), 
ensuring for the model chosen that the Hamiltonian $H_l$ satisfies
\[
	\trace_{\Hlcan{n}} \left[ U_\pi \e^{-\beta H_l} \right] \ge 0
\]
for any $n\in\NN$, and for all $\pi \in S_n$. 

We shall begin by checking this for the Perturbed Mean-Field model.
As will often be the case, in considering the Perturbed Mean-Field model, the result will also
stand for the Mean-Field by taking $v=0$, and the Ideal Gas with $a=v=0$.

To prove (\ref{Cycle_Cond1}) for the Perturbed Mean-Field model, we first shall denote the basis 
of unsymmetrised Hilbert space $\Hlcan{n}$ by $| \kk \ra
\equiv | k_1, \dots, k_n \ra$, for all $\kk \in \NN^n$. Then it is easy to see that
$N | \kk \ra = n | \kk \ra$ and $N_j | \kk \ra = n_j(\kk) | \kk\ra$, where 
$n_j(\kk) \in \{0,1,\dots,n\}$ such that $\sum_j n_j(\kk) = n$.
We expand the (unsymmetrised) trace in terms of this basis as follows:
{\allowdisplaybreaks
\begin{align*}
&	\trace_{\Hlcan{n}} \left[ U_\pi \e^{-\beta H^{\mathsf{pmf}}_l} \right]
=	\sum_{\kk} \la k_1, k_2, \dots, k_n | \e^{-\beta H^{\mathsf{pmf}}_l } U_\pi
	|  k_1, k_2, \dots, k_n \ra
\\[0.2cm]
&=	\sum_{\kk} \la k_1, k_2, \dots, k_n | \e^{-\beta H^{\mathsf{pmf}}_l }
	|  k_{\pi(1)}, k_{\pi(2)}, \dots, k_{\pi(n)} \ra
\\[0.2cm]
&=	\sum_{\kk} \big\la k_1, k_2, \dots, k_n \big| \exp\left\{-\beta \left[ \sum_{j=1}^n E_{k_j} +
	\frac{a}{2 V_l} N_l^2 + \frac{1}{2 V_l} \sum_{j,j'=1}^n v(j, j') N_j N_{j'} \right] \right\}
\\*
&	\hspace{2cm} \big|  k_{\pi(1)}, k_{\pi(2)}, \dots, k_{\pi(n)} \big\ra
\\[0.3cm]
&=	\sum_{\kk} \exp\left\{-\beta \left[ \sum_{j=1}^n E_{k_j} +
	\frac{a}{2 V_l} n^2 + \frac{1}{2 V_l} \sum_{j,j'=1}^n v(j, j') 
	n_j(\kk) n_{j'}(\kk) \right] \right\}
	\prod_{j=1}^n \delta_{k_j, k_{\pi(j)}}
\end{align*}}
which is a sum of exponentials of real numbers, the resulting sum of which must be non-negative.

With this foundation in place, Theorems \ref{longcycleslowerbound} and \ref{longcyclesupperbound}
state that if the symmetric grand-canonical Hamiltonian $H_l$ satisfies the following conditions:
\begin{enumerate}
\renewcommand{\labelenumi}{\Roman{enumi}.}
\item $\glHC(q,A) \ge 0$\; for all $A \ge 0$ and $q \in \NN$,
\item for any $q \in \NN$, \;$\displaystyle\lim_{\varepsilon \to 0} \displaystyle\dthermlim \glHC(q, P_\varepsilon) = 0$,
\item for some $\varepsilon > 0$,\; $\displaystyle\lim_{Q\to \infty} \displaystyle\dthermlim 
	\displaystyle\sum_{q=Q+1}^\infty \glHC(q,I - P_\varepsilon) = 0$,
\end{enumerate}
then the model is endowed with the property that the density
of cycles of infinite length equals that of the Bose-Einstein condensate, i.e. 
\[
	\rholong = \rhocond \;. 
\]

We shall consider the following three models introduced above with this 
framework: the Ideal Bose Gas, the Mean-Field Model and the Perturbed Mean-Field Model, 
to reproduce the results of S\H{u}t\"o \cite{Suto1,Suto2} and Dorlas, Martin and Pul\'e \cite{DMP}, i.e.
that the condensate density equals the long cycle density. Again it suffices to assert Conditions 
\Rmnum{1}--\Rmnum{3} for the Perturbed Mean-Field Model, since setting $v=0$ 
returns the Mean-Field Model, which also implies the result for the Ideal Bose Gas.

However before considering these three conditions, we shall first state and prove a useful lemma.

\begin{lemma}										\label{cycexpectlemma}
For any non-negative single-particle operator $A$ we have the following identity:
\[
	\glHC(q, A) = \frac{1}{ V_l \glPart} \sum_{k \ge 1} \la \phi_k | A \phi_k \ra
	\mathop{\trace_{\FFlsym}}_{N_k \ge q} \left[ 
	\e^{-\beta(H^{\mathsf{pmf}}_l -\mu N_l)}\right].
\]
\end{lemma}

\begin{proof}
%
Before we begin we need to establish some notation.
Denote an ordered set of $q$ (not necessarily distinct) positive numbers by $\kk \vcentcolon=
( k_1, k_2, \dots, k_q ) \in \NN^q$ and let
\[
	| \kk \ra = \phi_{k_1} \otimes \phi_{k_2} \otimes \cdots \otimes \phi_{k_q}.
\]
Then $\{ | \kk \ra | \kk \in \NN^q \}$ is an orthonormal basis of $\Hlcan{q}$, and
a summation over $\kk$ will sum over all possible $q$-tuples.

Let $\{ \psi_n \}_{n=1}^\infty$ be an orthonormal basis for $\FFlsym$. Then a basis for
$\Hlcan{q} \otimes \FFlsym$ may be formed by taking the tensor product of the bases of 
$\Hlcan{q}$ and $\FFlsym$, thus the set $\{ \ |\kk \ra \otimes \psi_n \vcentcolon=
| \kk, \psi_n \ra \, |\, \kk \in \NN^q; n=1,2,\dots \}$
is an orthonormal basis of $\Hlcan{q} \otimes \FFlsym$.

Since the expression for the grand-canonical cycle density, equation (\ref{grandcan-cycle-dens}),
involves a Hilbert space of the form $\Hlcan{q} \otimes \FFlsym$, it is
convenient to establish a notation for operators acting on $\Hlcan{q}$ and $\FFlsym$ individually.

Define $N^q_k$ on $\Hlcan{q}$ as the operator which counts the number of $q$
unsymmetrised particles in the state $\phi_k$. This may be expressed as 
\begin{equation*}
	N^q_i | \kk \ra
=	N^q_i | ( k_1, k_2, \dots ,k_q ) \ra 
= 	\sum_{j=1}^q \delta_{i, k_j} |  ( k_1, k_2, \dots ,k_q ) \ra
= 	| \kk \ra \sum_{j=1}^q \delta_{i, k_j} 
= 	| \kk \ra n_i(\kk).
\end{equation*}
where $n_i(\kk) \in \NN$. Note that $\sum_{i\ge1} n_i(\kk) = q$.
The $q$-unsymmetrised free-particle Hamiltonian is $h_l^{(q)}$ which has eigenvectors
$|\kk\ra$ for all $\kk \in \NN^q$, with corresponding eigenvalues
\[
	\sum_{i=1}^q E_{k_i}.
\]
Therefore the operator corresponding to $N_k$ applied to $\Hlcan{q} \otimes \FFlsym$ is
$N^q_k + N_k$, and the total number operator is $q + N$.

For the remainder of this proof we discard the $l$ subscripts from the Hamiltonian $H$,
the volume $V$ and the number operator $N$ for clarity.
{\allowdisplaybreaks
By definition%
\begin{align*}
&	\glHC(q,A)
= 	\frac{1}{ V \glPart} \trace_{\Hlcan{q} \otimes \FFlsym}
	\left[ (A \otimes I \otimes \dots \otimes I) (U_q \otimes \II )
	\e^{-\beta(H^{\mathsf{pmf}} - \mu N)} \right]
\intertext{which when expanding the trace in terms of the basis of $\Hlcan{q} \otimes \FFlsym$ becomes}
&= 	\frac{1}{ V \glPart} \sum_{\kk} \sum_{n=1}^\infty 
	\big\la \kk, \psi_n \big|
	(A \otimes I \otimes \dots \otimes I) (U_q \otimes \II)
\\*
&	\hspace{2cm}  \times \exp\bigg\{-\beta \bigg[ d\Gamma(h) + 
	\frac{a}{2V} N^2 + \frac{1}{2 V} 
	\sum_{i,j = 1}^\infty v( k_i, k_j ) N_i N_j
	- \mu \, N \bigg] \bigg\} \big| \kk, \psi_n \big\ra.
\intertext{Now split the operators $N_j$ and $H^0$ into their $\Hlcan{q} \otimes \FFlsym$ representations,
$N^q_j + N_j$ and $h^{(q)} + d\Gamma(h)$ respectively, to obtain}
&=	\frac{1}{ V \glPart} \sum_{\kk} \sum_{n=1}^\infty 
	\big\la \kk, \psi_n
	\big|  (A \otimes I \otimes \dots \otimes I) (U_q \otimes \II)
	\exp\bigg\{-\beta \bigg[h^{(q)} + d\Gamma(h) + \frac{a}{2 V}(q+N)^2 
\\*
&	\hspace{2cm} + \frac{1}{2 V} \sum_{i,j=1}^\infty v( i, j ) 
	(N^q_i + N_i)( N^q_j + N_j )  - \mu (q+N)
	\bigg] \bigg\}  \big| \kk, \psi_n \big\ra.
\intertext{All the operators in the exponential commute, and as $N^q_k |\kk\ra = n_k(\kk) |\kk\ra$, this is equal to}
&=	\frac{1}{ V \glPart} \sum_{\kk} \sum_{n=1}^\infty 
	\big\la \kk, \psi_n
	\big|  (A \otimes I \otimes \dots \otimes I) (U_q \otimes \II)
	\exp\bigg\{-\beta \bigg[h^{(q)} + d\Gamma(h) + \frac{a}{2 V}(q+N)^2 
\\*
&	\hspace{2cm} + \frac{1}{2 V} \sum_{i,j=1}^\infty v( i, j ) 
	(n_i(\kk) + N_i)( n_j(\kk) + N_j )  - \mu (q+N)
	\bigg] \bigg\}  \big| \kk, \psi_n \big\ra.
\end{align*}
Since
\[
	\la \kk |  (A \otimes I \otimes \dots \otimes I) U_q | \kk \ra = 0
\]
unless $k_1 = k_2 = \ldots = k_q \vcentcolon= k$, we can write the same expression over one $k$
and replace the $n_i(\kk)$ with $q\delta_{i \cdot}$ as follows:
\begin{align}
	\glHC(q,A)
&=	\frac{1}{ V \glPart} \sum_{k \ge1}  
	\la k | Ak \ra \sum_{n=1}^\infty \big\la \psi_n \big|
	\exp\bigg\{-\beta \bigg[h^{(q)} + d\Gamma(h) + \frac{a}{2 V}(q+N)^2 	\notag
\\*
&	\hspace{2cm} + \frac{1}{2 V} \sum_{i,j=1}^\infty v( i, j ) 
	(q\delta_{ki} + N_i)( q\delta_{kj} + N_j )  - \mu (q+N)
	\bigg] \bigg\}  \big| \psi_n \big\ra 						\label{trivialeh}
\\
&=	\frac{1}{ V \glPart} \sum_{k \ge1}  
	\la k | Ak \ra \mathop{\trace_{\FFlsym}}_{N_k \ge q} \Bigg[ 
	\exp\bigg\{ -\beta \bigg[ 
	\sum_{j\ge1} N_j (E_j - \mu) 							\notag
\\*
&\hspace{6.5cm}
	+ \frac{a}{2 V} N^2
	+ \frac{1}{ 2V}\sum_{i, j =1}^\infty v(i,j) N_i N_j
	\bigg] \bigg\} \Bigg]								\notag
\\[0.3cm]
&=	\frac{1}{ V \glPart} \sum_{k \ge1}  
	\la k | Ak \ra \mathop{\trace_{\FFlsym}}_{N_k \ge q} \left[ 
	\e^{-\beta(H^{\mathsf{pmf}} -\mu N)}\right].				\notag
\end{align}}
where we mean that the trace is taken over the subspace of Fock space with at least $q$ particles
in the state $\phi_k$. 
\end{proof}

\begin{corollary}
Let $\phi_k$ be an eigenstate of the ideal Bose gas Hamiltonian, and $P_{\phi_k}$ the orthogonal
projection upon this state, then
\[
	\glHC(q, P_{\phi_k}) = \frac{1}{ V_l \glPart}
	\mathop{\trace_{\FFlsym}}_{N_k \ge q} \left[ 
	\e^{-\beta(H^{\mathsf{pmf}}_l -\mu N_l)}\right].
\]
\end{corollary}

Now we may proceed to consider the three conditions proposed in Section \ref{SectionOCCMO}.

\subsection{Condition \Rmnum{1}}
Now we shall check the non-negativity of $\glHC(q,A)$ for any single-particle operator $A \ge0$
for the case of the three models described above: the Ideal Bose Gas, the Mean Field model 
and the Perturbed Mean-Field model. 
By Lemma \ref{cycexpectlemma} we have that
\[
	\glHC(q, A) = \frac{1}{ V_l \glPart} \sum_{k \ge 1} \la \phi_k | A \phi_k \ra
	\mathop{\trace_{\FFlsym}}_{N_k \ge q} \left[ 
	\e^{-\beta(H^{\mathsf{pmf}}_l -\mu N_l)}\right].
\]
Since $A \ge 0$, then $\la \phi_k | A\phi_k \ra \ge 0$ for all $k \ge 1$. The trace term 
in the summation is positive as it involves the exponential of positive operators.
Hence $\glHC(q,A) \ge 0$ for any $q$ and \mbox{Condition \Rmnum{1}} is satisfied.

\subsection{Condition \Rmnum{2}}

Here we wish to prove for the Perturbed Mean-Field model that
\[
	\lim_{\varepsilon \to 0} \dthermlim \glHC(q, P_\varepsilon) = 0
\]
where $P_\varepsilon \vcentcolon= \sum_{k: E_k < \varepsilon} P_{\phi_k}$.

By Lemma \ref{cycexpectlemma}, for a state $\phi \in \HH$ we immediately obtain
\[
	\glHC(q, P_{\phi}) 
= 	\frac{1}{ V_l \glPart} \sum_{k \ge 1} | \la \phi | \phi_k \ra |^2
	\mathop{\trace_{\FFlsym}}_{N_k \ge q} \left[ 
	\e^{-\beta(H^{\mathsf{pmf}}_l -\mu N_l)}\right].
\]
The trace with the restriction $N_k \ge q$ is bounded above by the same trace
but with $N_k \ge 0$, which actually is the usual partition function for the Perturbed
Mean-Field model. Hence
\begin{equation*}
	\glHC(q, P_{\phi}) 
\le	\frac{1}{ V_l \glPart} \sum_{k \ge 1} | \la \phi | \phi_k \ra |^2
	\mathop{\trace_{\FFlsym}}_{N_k \ge 0} \left[ 
	\e^{-\beta(H^{\mathsf{pmf}}_l -\mu N_l)}\right]
\le	\frac{1}{ V_l } \sum_{k \ge 1} | \la \phi | \phi_k \ra |^2 = \frac{1}{ V_l }
\end{equation*}
which tends to zero as $l \to \infty$. Then using this
\begin{multline*}
	\lim_{\varepsilon \to 0} \dthermlim \glHC(q, P_\varepsilon)
=	\lim_{\varepsilon \to 0} \dthermlim 
	\sum_{k: E_k < \varepsilon} \glHC(q, P_{\phi_k})
\le	\lim_{\varepsilon \to 0} \dthermlim 
	\frac{1}{ V_l} \sum_{k: E_k < \varepsilon} 1
\\*
\le	\lim_{\varepsilon \to 0} \dthermlim 
	\int_{[0,\varepsilon)} d\nu_l(\lambda)
\le	\lim_{\varepsilon \to 0} 
	\int_{[0,\varepsilon)} d\nu(\lambda)
\\*
=	\lim_{\varepsilon \to 0} 
	C_d \int_{[0,\varepsilon)} \lambda^{d/2} d\lambda
=	\frac{2 C_d}{d+2}\, \lim_{\varepsilon \to 0} \varepsilon^{d/2+1}
=	0.
\end{multline*}
for any $d \in \NN$, recalling (\ref{densityofstates}), the expression for the limiting integrated density of states.

\subsection{Condition \Rmnum{3}}
Here we shall prove that
for some $\varepsilon > 0$, with $P_\varepsilon \vcentcolon= \sum_{k:E_k < \varepsilon} P_{\phi_k}$ that \vspace{-0.2cm}
\[ 
	\lim_{Q\to \infty} \dthermlim \sum_{q=Q+1}^\infty \glHC(q,I - P_\varepsilon) = 0.
\]
It is convenient to note that we may interchange the summations using Fubini's Theorem, i.e.
\begin{equation}										\label{noteworthy}
	\sum_{q=Q+1}^\infty \glHC(q,I - P_\varepsilon)
=	\sum_{k:E_k \ge \varepsilon} \sum_{q=Q+1}^\infty \glHC(q, P_{\phi_k})
\end{equation}
since the total sum is bounded above by the total density $\rho(\mu)$.

The approach for the Perturbed Mean-Field case will involve Large Deviation theory.
Certainly while this calculation will imply that Condition \Rmnum{3} holds for the case of the Mean-Field model too, there
is a short and intuitive alternative proof for the Mean-Field case alone which we shall give. 
Also to address possible concerns with guaranteeing the negativity of the chemical potential for the Ideal
Bose Gas, we will consider each model individually.

\subsubsection{Ideal Bose Gas case}
Due to the absence of interactions between particles, it is trivial to deduce from equation (\ref{trivialeh})
the following expression for the cycle expectation of an operator $A \ge 0$:
\[
	\glHC(q, A) = \frac{1}{ V_l } \sum_{k\ge1} \la \phi_k | A \phi_k \ra \e^{-\beta q (E_k - \mu)}
\]
for $\mu < 0$, and hence 
\[
	\glHC(q, P_{\phi_k}) = \frac{1}{ V_l } \e^{-\beta q (E_k - \mu)}.
\]
Then taking (\ref{noteworthy}), one immediately obtains
\[
	\sum_{q=Q+1}^\infty \sum_{E_k \ge \varepsilon} \glHC(q, P_{\phi_k})
=	\frac{1}{ V_l } \sum_{k:E_k \ge \varepsilon} \sum_{q=Q+1}^\infty \e^{-\beta q (E_k - \mu)}
\le	\frac{1}{ V_l } \sum_{k:E_k \ge \varepsilon} \frac{\e^{-\beta (Q+1) E_k}}{\e^{\beta E_k} - 1}.
\]
For large enough $l$ there exists an $\varepsilon > 0$ 
satisfying $E_k > \varepsilon$ such that
\[
	\frac{1}{ V_l } \sum_{k:E_k \ge \varepsilon} \frac{\e^{-\beta (Q+1) E_k}}{\e^{\beta E_k} - 1}
\le	\frac{\e^{-\beta (Q+1) \varepsilon}}{ V_l } \sum_{k:E_k \ge \varepsilon} \frac{1}{\e^{\beta E_k} - 1}
=	\e^{-\beta Q \varepsilon} \int_{[\varepsilon, \infty)} \frac{1}{{\e^{\beta \lambda} - 1}} \nu_l(d\lambda).
\]
In taking the thermodynamic limit, one thus obtains the upper bound
\[
	\dthermlim \sum_{q=Q+1}^\infty \sum_{E_k \ge \varepsilon} \glHC(q, P_{\phi_k})
\le	\e^{-\beta Q \varepsilon}
	\int_{[\varepsilon, \infty)} \frac{1}{{\e^{\beta \lambda} - 1}} \nu(d\lambda)
\le	\e^{-\beta Q \varepsilon} \rhocrit
\]
which tends to zero in the limit $Q \to \infty$.

\subsubsection{Mean-Field case}
One can easily deduce an equivalent form of Lemma (\ref{cycexpectlemma}):
\begin{equation*}
\glHC(q,P_{\phi_k})
= 	\frac{1}{ V_l\glPart} \mathop{\trace_{\FFlsym}}_{N_k \ge q} \big[
	 \e^{-\beta(H^{\textsf{mf}} - \mu N)}
	\big].
\end{equation*}
Expand the Hamiltonian and incorporate the $N_k \ge q$ restriction to obtain
\begin{align*}
\glHC(q,P_{\phi_k})
&=	\frac{1}{ V_l\glPart} \trace_{\FFlsym} \bigg[
	 \exp\bigg\{\!\!-\!\beta\bigg[ \sum_{j\ge1} (E_j + \delta_{jk} q) N_j + \frac{a}{2 V_l} (N+q)^2 - 
	(N+q) \mu \bigg]\bigg\} 
	\bigg]	
\\[0.2cm]
&=	\frac{\e^{-\beta q(E_k - E_1)}}{ V_l\glPart} \trace_{\FFlsym} \bigg[
	 \exp\bigg\{\!\!-\!\beta\bigg[ \sum_{j\ge1} (E_j + \delta_{1j} q) N_j 
\\*&\hspace{7.5cm}
	+ \frac{a}{2 V_l} (N+q)^2 - (N+q) \mu \bigg]\bigg\} 
	\bigg]	
\\[0.2cm]
&=	\frac{\e^{-\beta q(E_k - E_1)}}{ V_l\glPart} \mathop{\trace_{\FFlsym}}_{N_1 \ge q} \big[
	 \e^{-\beta(H^{\textsf{mf}} - \mu N)}
	\big]	
\\[0.2cm]
&\le	\frac{\e^{-\beta q(E_k - E_1)}}{ V_l\glPart} \mathop{\trace_{\FFlsym}}_{N_1 \ge 0} \big[
	 \e^{-\beta(H^{\textsf{mf}} - \mu N)}
	\big]	
=	\frac{\e^{-\beta q(E_k - E_1)}}{ V_l}.
\end{align*}
Then taking (\ref{noteworthy}), for large enough $l$ there exists an $\varepsilon > 0$ such that 
$E_1 < \varepsilon/2$ and
\[
	\sum_{k : E_k \ge \varepsilon } \sum_{q=Q+1}^\infty \glHC(q,P_{\phi_k})
\le	\sum_{k : E_k \ge \varepsilon } \frac{\e^{\beta Q \varepsilon/2}}{ V_l} 
	\sum_{q=Q+1}^\infty \e^{-\beta q E_k}
=	\frac{ \e^{\beta Q \varepsilon/2} }{V_l} \sum_{k : E_k \ge \varepsilon } 
	\frac{\e^{-\beta Q E_k}}{\e^{\beta E_k} - 1}
\]
which in taking the thermodynamic limit, is bounded above by
\[
	\dthermlim \frac{ \e^{\beta Q \varepsilon/2} }{V_l} \sum_{k : E_k \ge \varepsilon } 
	\frac{\e^{-\beta Q E_k}}{\e^{\beta E_k} - 1}
\le	\e^{-\beta Q \varepsilon /2 } \rhocrit
\]
which goes to zero in the limit $Q \to \infty$.

\subsubsection{Perturbed Mean-Field case}

Recall Lemma \ref{cycexpectlemma} gave us, 
\[
	\glHC(q,P_{\phi_k}) 
=	\frac{1}{ V_l \glPart} \mathop{\trace_{\FFlsym}}_{N_k \ge q} \left[ 
	\e^{-\beta(H^{\mathsf{pmf}}_l -\mu N_l)}\right].
\]
For all $n \in \NN$, define the projection $P(N_k\!=\!n)$ which projects onto the 
subspace of $\FFlsym$ with exactly $n$ particles in the $k^{th}$ mode. With this
we can write
\[
	\glHC(q,P_{\phi_k}) 
=	\frac{1}{ V_l \glPart} \sum_{n=q}^\infty \trace_{\FFlsym} \left[ P(N_k = n)
	\e^{-\beta(H^{\mathsf{pmf}}_l -\mu N_l)}\right].
\]
Then performing the $q$ summation of (\ref{noteworthy}) first we have
{\allowdisplaybreaks
\begin{align*}
	\sum_{q=Q+1}^\infty \glHC(q, P_{\phi_k})
&=	\frac{1}{ V_l \glPart} \sum_{q=Q+1}^\infty \sum_{n=q}^\infty \trace_{\FFlsym} \left[ P(N_k\!=\!n)
	\e^{-\beta(H^{\mathsf{pmf}}_l -\mu N_l)}\right]
\intertext{and by rearranging the sums, get}
&=	\frac{1}{ V_l \glPart} \sum_{n=Q+1}^\infty \sum_{q=Q+1}^n \trace_{\FFlsym} \left[ P(N_k\!=\!n)
	\e^{-\beta(H^{\mathsf{pmf}}_l -\mu N_l)}\right]
\\[0.1cm]
&=	\frac{1}{ V_l \glPart} \sum_{n=Q+1}^\infty (n-Q) \:\trace_{\FFlsym} \left[ P(N_k\!=\!n)
	\e^{-\beta(H^{\mathsf{pmf}}_l -\mu N_l)}\right]
\\[0.1cm]
&=	\frac{1}{ V_l \glPart} \sum_{n=0}^\infty (n-Q)  \Theta(n-Q-1) \:\trace_{\FFlsym} \left[ P(N_k\!=\!n)
	\e^{-\beta(H^{\mathsf{pmf}}_l -\mu N_l)}\right] 
\\[0.1cm]
&=	\frac{1}{ V_l } \sum_{n=0}^\infty \big\la (n-Q) \Theta(n-Q-1) P(N_k\!=\!n) \big\ra_{H^{\mathsf{pmf}}_l}
\\[0.1cm]
&=	\frac{1}{ V_l } \big\la (N_k-Q) \Theta(N_k-Q-1) \big\ra_{H^{\mathsf{pmf}}_l}
\end{align*}
where $\Theta$ is the Heaviside step function. With this (\ref{noteworthy}) becomes
\begin{align*}
	\sum_{q=Q+1}^\infty \glHC(q, I - P_\varepsilon )
&=	\frac{1}{ V_l } \sum_{k : E_k \ge \varepsilon} \big\la (N_k-Q) \Theta(N_k-Q-1) \big\ra_{H^{\mathsf{pmf}}_l}
\\[0.1cm]
&=	\frac{1}{ V_l } \sum_{k\ge1} \Theta(E_k - \varepsilon) \big\la (N_k-Q) \Theta(N_k-Q-1) \big\ra_{H^{\mathsf{pmf}}_l}
\\[0.1cm]
&=	\frac{1}{ V_l }  \left\la\sum_{k\ge1} (N_k-Q) \Theta(N_k-Q-1) \Theta(E_k - \varepsilon) \right\ra_{H^{\mathsf{pmf}}_l}.
\end{align*}}

This inspires us to 
define the following modified ``free'' partition function 
\begin{multline*}
	\Xi^0_l(\mu, \tau, Q,\varepsilon) 
=\\
 	\trace_{\FFlsym} \left[ \exp\left\{ -\beta \left( H^0_l - \mu N_l 
	- \tau \left[ \sum_{k \ge 1} (N_k - Q) \Theta(N_k - Q-1) \Theta(E_k - \varepsilon) \right]
	\right) \right\}\right]
\end{multline*}
and denote its corresponding pressure by $p^0_l(\mu, \tau, Q, \varepsilon)$.
Similarly, define a modified Perturbed Mean-Field partition function 
\begin{multline*}
	\Xi_l(\mu, \tau, Q, \varepsilon) 
=\\
 	\trace_{\FFlsym} \left[ \exp\left\{ -\beta \left( H^\mathsf{pmf}_l - \mu N_l 
	- \tau \left[ \sum_{k \ge 1} (N_k - Q) \Theta(N_k - Q-1) \Theta(E_k - \varepsilon) \right]
	\right) \right\}\right]
\end{multline*}
with pressure $p_l(\mu, \tau, Q, \varepsilon)$. These pressures must satisfy the following relation:
\[
	\frac{\partial}{\partial \tau} \bigbar_{\tau=0} p_l(\mu, \tau, Q, \varepsilon)
=	\frac{1}{ V_l} \left\la \sum_{k \ge 1} (N_k - Q) \Theta(N_k - Q-1) \Theta(E_k - \varepsilon) 
	\right\ra_{H^{\mathsf{pmf}}_l}
\!\!\!\!\vcentcolon=\,\,
	\sum_{q=Q+1}^\infty \glHC(q, I - P_\varepsilon).
\]
We shall apply the technique of Large Deviation Theory to write $p_l(\mu,\tau, Q, \varepsilon)$ in terms of 
the ``free'' pressure $p^0_l(\mu, \tau, Q, \varepsilon)$
and find its asymptotic behaviour, which will allow us to derive an expression for 
the derivative of $p_l(\mu, \tau, Q, \varepsilon)$ wrt $\tau$.

The ``free'' partition function can be written as
\[
	\Xi^0_l(\mu, \tau, Q, \varepsilon)
=	\prod_{k : E_k < \varepsilon} \frac{1}{1-\e^{-\beta (E_k - \mu)}}
	\prod_{k: E_k \ge \varepsilon} \frac{1 - \e^{-\beta( E_k - \mu - \tau)}
	+ (\e^{\beta \tau} - 1)\e^{-\beta(Q+1)(E_k - \mu)}}
	{(1-\e^{-\beta (E_k - \mu)})(1 - \e^{-\beta( E_k - \mu - \tau )})}.
\]
%
We can rewrite the limited ``free'' pressure as
\[
	p^0_l(\mu\equiv\alpha, \tau, Q, \varepsilon) 
= 	\int_{[0,\varepsilon)} \pi
	(\alpha - \lambda)d\nu_l(\lambda)
	+ \int_{[\varepsilon, \infty)} \pi_{\tau, Q } (\alpha - \lambda)d\nu_l(\lambda)
\]
with $\tau \ge 0$, where
\begin{equation}							\label{LDPpi2}
	\pi_{ \tau, Q} ( s )
=			\beta^{-1} \ln \left[ \displaystyle\frac{1 - \e^{\beta(s+\tau)}
		+ (\e^{\beta \tau} - 1)\e^{\beta s(Q+1)} }
		{(1-\e^{\beta s})(1 - \e^{\beta( s + \tau )})} \right] 
\end{equation}
for $s < \min\{ 0 ,-\tau \}$.
Note that for $\tau = 0$, $\pi_{\tau, Q} ( s )$ reduces to the 
ideal gas partial pressure $\pi(s)$, c.f. (\ref{partialpressure}).

Define the following probability measure on $\Omega$:
\[
	\Prob^{\mu, \tau, Q, \varepsilon}_l(\omega) 
=	\frac{1}{\Xi^0_l(\mu, \tau, Q,\varepsilon)} 
	\e^{-\beta \left( H^0_l(\omega) - \mu N(\omega) + 
	\tau \left[ \sum_{k \ge 1} (N_k(\omega) - Q) \Theta(N_k(\omega) - Q-1) 
	\Theta(E_k - \varepsilon)  \right] \right)}.
\]

We shall consider the cases $E_k < \varepsilon$ and 
$E_k \ge \varepsilon$ individually. To do so, define a pair of 
occupation measures as follows:
\[
	L^{(1)}_{l,\varepsilon}[\omega; A] 
= 	\frac{1}{ V_l}\sum_{j: E_j < \varepsilon } N_j(\omega) \delta_{E_j}(A)
\qquad\qquad
	L^{(2)}_{l,\varepsilon}[\omega; A] 
= 	\frac{1}{ V_l} \sum_{j: E_j \ge \varepsilon } N_j(\omega) \delta_{E_j}(A)
\]
so that both $L$ measures are positive and bounded. Now define
$\mathcal{E}_1 \vcentcolon= \mathcal{M}^b_{+}([0,\varepsilon))$ and 
$\mathcal{E}_2 \vcentcolon= \mathcal{M}^b_{+}([\varepsilon, \infty))$ as two spaces of positive
bounded measures on $[0,\varepsilon)$ and $[\varepsilon, \infty)$ respectively. We equip
both these spaces with the narrow topology. Then set $\mathcal{E} \vcentcolon= \mathcal{E}_1
\oplus \mathcal{E}_2$. 

Let $\KK^{\alpha, (i)}$ (for $i=1,2$) be the probability measure induced on $\mathcal{E}_i$ by $L$:
\[
	\KK_{\tau,l}^{\alpha, (1)} 
=	\Prob_l^{\alpha, \tau, Q, \varepsilon} \circ (L_{l,\varepsilon}^{(1)})^{-1}
\qquad\qquad
	\KK_{\tau,l}^{\alpha, (2)} 
=	\Prob_l^{\alpha, \tau, Q, \varepsilon} \circ (L_{l,\varepsilon}^{(2)})^{-1}
\]
where for brevity we will not write all the arguments, and set \vspace{-0.2cm}
\[
	\KK_{\tau, l}^{\alpha} = \KK_{\tau,l}^{\alpha, (1)} \oplus \KK_{\tau,l}^{\alpha, (2)}\vspace{-0.2cm}.
\]
Then define the Legendre transform\vspace{-0.2cm}
\[
	\pi^\ast(t) = \sup_{s\le \min\{0, -\tau \}}\big\{ts - \pi(s)\big\}
= 	ts^\ast(t) - \pi(s^\ast(t)) \vspace{-0.2cm}
\]
where $s^\ast(t) < 0$ is the maximiser satisfying\vspace{-0.2cm}
\[
	t = \pi'(s^\ast(t)).\vspace{-0.2cm}
\]
Similarly set\vspace{-0.2cm}
\[
	\pi^\ast_{\tau,Q}(t) 
= 	\sup_{s\le \min\{0, -\tau \}}\big\{ts(t) - \pi_{\tau,Q}(s(t))\big\}
=	ts^\ast_{\tau, Q}(t) - \pi_{\tau,Q}(s^\ast_{\tau, Q}(t)) 
\]
where $s^\ast_{\tau,Q}(t) < 0$ satisfies
\begin{equation}										\label{Legendident}
	t = \pi_{\tau,Q}'(s^\ast_{\tau,Q}(t)).
\end{equation}
As above for $i=1,2$ and every $m^{(i)} \in \mathcal{E}_i$ let $m^{(i)} = m^{(i)}_s + m^{(i)}_a$ 
be the Lebesgue decomposition of $m^{(i)}$ with respect to $d\nu$, where $m^{(i)}_s$ is the singular part 
and $m^{(i)}_a$ the absolutely continuous part.
Define $U_i : \mathcal{E}_i \to [0,\infty]$ by
\[
	U^\alpha_1[m] = - \int_{[0,\varepsilon)} (\alpha - \lambda) m^{(1)}_s(d\lambda),
\qquad\qquad
	U^\alpha_2[m] = - \int_{[\varepsilon,\infty)} (\alpha - \lambda) m^{(2)}_s(d\lambda).
\]
As above, fix\vspace{-0.2cm}
\begin{equation}											\label{LDP-J}
	J_1(t,r) = \pi^\ast(t) - rt + \pi(r),
\qquad\qquad
	J_{2,\tau}(t,r) = \pi^\ast_{\tau,Q}(t) - rt + \pi_{\tau,Q}(r).
\end{equation}
Then by Theorem 2 of \cite{goughpuleellis_lpdrandomweights}, the sequence of measures 
$\KK^{\alpha}_{\tau, l}$ satisfies the Large Deviation
Principle with constants $ V_l$ and  rate function $I$ of the form
\[
	I^{\alpha}_{\tau}[m] = I^\alpha_1[m] + I^\alpha_{2,\tau}[m]
\]
where\vspace{-0.4cm}
\begin{align}
	I^\alpha_1[m]									\notag
&=	U^\alpha_1[m] + \int_{[0,\varepsilon)} J_1(\rho(\lambda), \alpha - \lambda ) d\nu(\lambda),
\\
	I^\alpha_{2,\tau}[m]								\label{LDP-I2}
&=	U^\alpha_2[m] + \int_{[\varepsilon,\infty)}J_{2,\tau}(\rho(\lambda), \alpha - \lambda ) d\nu(\lambda).
\end{align}
Then we may use Varadhan's Theorem to show that the pressure of the modified Perturbed Mean-Field model 
(using the $\alpha <0$ trick) in the thermodynamic limit can be expressed as
\begin{align*}
	p(\mu, \tau, Q, \varepsilon)
&=	p^0(\alpha, \tau, Q, \varepsilon) + \sup_{m \in \mathcal{E}} \left\{ G^{\mu-\alpha}[m] - I^{\alpha}_{\tau}[m] \right\}
\\[0.2cm]
&=	p^0(\alpha, \tau, Q, \varepsilon) + \mathop{\sup_{m_i \in \mathcal{E}_i}}_{m=m_1+m_2} 
	\left\{ G^{\mu-\alpha}[m_1] - I^\alpha_1[m_1] 
	+  G^{\mu-\alpha}[m_2] - I^\alpha_{2,\tau}[m_2] \right\}
\\
&=	p^0(\alpha, \tau, Q,\varepsilon) + G^{\mu-\alpha}[m_\tau^\ast] 
	- I^\alpha_1[m_\tau^\ast] - I^\alpha_{2,\tau}[m_\tau^\ast]
\end{align*}
for some $m_\tau^\ast \in \mathcal{E}$. The Euler-Lagrange equation for the variational problem implies 
\begin{multline}										\label{LDP-EulerLag}
	\frac{\partial}{\partial \tau} p(\mu, \tau, Q, \varepsilon)
=	\frac{\partial}{\partial \tau} p^0(\alpha, \tau, Q, \varepsilon)
+
\\[0.1cm]
	\lla \frac{\delta}{\delta \tau} G[m_\tau^\ast] - \frac{\delta}{\delta \tau} I_1^{\alpha}[m_\tau^\ast] 
	- \frac{\delta}{\delta \tau} I_{2,\tau}^{\alpha}[m_\tau^\ast], \frac{dm_\tau^\ast}{d\tau}  \rra 
	- \frac{\partial}{\partial\tau} I_{2,\tau}^{\alpha}[m_\tau^\ast]
\end{multline}
of which the inner product must be zero because $\frac{\delta}{\delta \tau} \left( G[m_\tau^\ast] 
- I_1^{\alpha}[m_\tau^\ast] - I_2^{\alpha}[m_\tau^\ast]\right) = 0$. Then taking the remaining term,
and using (\ref{LDP-J}) and (\ref{LDP-I2}),
\begin{align*}
	\frac{\partial}{\partial\tau} I_{2,\tau}^\alpha[m^\ast] 
&=	\int_{[\varepsilon, \infty)} \frac{\partial}{\partial\tau} J_{2,\tau}(\rho(\lambda), \alpha - \lambda) d\nu(\lambda)
\\[0.2cm]
&=	\int_{[\varepsilon, \infty)} \frac{\partial}{\partial\tau} \pi_{\tau, Q}^\ast(\rho(\lambda))d\nu(\lambda) +
	\int_{[\varepsilon, \infty)} \frac{\partial}{\partial\tau} \pi_{\tau, Q}(\alpha - \lambda )d\nu(\lambda)
\\[0.2cm]
&=	\int_{[\varepsilon, \infty)} \frac{\partial}{\partial\tau} \pi_{\tau, Q}^\ast(\rho(\lambda))d\nu(\lambda) +
	\frac{\partial}{\partial\tau} p_l^0(\alpha,Q,\tau,\varepsilon)
\end{align*}
since $\pi$ has no $\tau$ dependence. Inserting this into equation (\ref{LDP-EulerLag}) 
causes the free pressures to cancel, leaving one with 
\[
	\frac{\partial}{\partial \tau} p(\mu, \tau, Q, \varepsilon)
=	\int_{[\varepsilon, \infty)} \frac{\partial}{\partial\tau} \pi_{\tau, Q}^\ast(\rho(\lambda))d\nu(\lambda).
\]
But note that 
\[
	\frac{\partial}{\partial \tau} \pi_{\tau, Q}^\ast(t) 
=	\left( t - \pi_{\tau, Q}' (s^\ast_{\tau,Q}(t)) \right) \frac{\partial s^\ast}{\partial \tau} 
	- \frac{\partial }{\partial \tau} \pi_{\tau,Q}(s^\ast_{\tau,Q}(t))
=	- \frac{\partial }{\partial \tau} \pi_{\tau,Q}(s^\ast_{\tau,Q}(t)) 
\]
since $t=\pi_{\tau,Q}'(s^\ast_{\tau,Q}(t))$ from (\ref{Legendident}), which returns the following expression for 
(\ref{LDP-EulerLag}),
the derivative of the interacting pressure with respect to $\tau$:
\begin{equation}								\label{secondlast}
	\frac{\partial}{\partial \tau} p(\mu,Q,\tau,\varepsilon)
=	- \int_{[\varepsilon, \infty)} \frac{\partial}{\partial\tau} 
	\pi_{\tau, Q}\big(s^\ast_{\tau,Q}(\rho(\lambda))\big) d\nu(\lambda).
\end{equation}
where $s^\ast_{\tau, Q}$ satisfies $t = \pi'_{\tau, Q}(s^\ast_{\tau,Q}(t))$.

From (\ref{LDPpi2}), the definition of $\pi_{\tau, Q}$, one can can immediately calculate that
\begin{equation}								\label{almostlast}
	\frac{\partial }{\partial \tau} \bigbar_{\tau=0} \pi_{\tau, Q}(s^\ast)
=	\frac{\e^{\beta Q s^\ast}}{\e^{-\beta s^\ast}-1}.
\end{equation}
Note that when $\tau=0$, $\pi_{\tau=0, Q}$ reduces to the ideal gas partial 
pressure (\ref{partialpressure}) for which we have the explicit expression:
\[
	\pi_{\tau=0, Q}(s) 
=	\pi(s)
=	\frac{1}{\beta} \ln\big[ 1-\e^{\beta s} \big]^{-1}
\]
which is independent of $Q$ (so we drop its index). Differentiating this with respect to $s$ implies that
\begin{equation*}
	t
=	\frac{1}{\e^{-\beta s^\ast_0(t)}-1}
\quad\Rightarrow\quad
	s^\ast_0(t) = \frac{1}{\beta} \ln \left[ \frac{t}{1+t} \right]
\end{equation*}
and thus the integrand of the non-zero term of (\ref{almostlast}) in terms of $\lambda$ is
\[
	\frac{\e^{\beta Q s^\ast_0(\rho(\lambda))}}{\e^{-\beta s^\ast_0(\rho(\lambda))}-1} 
=	\left(\frac{\rho(\lambda)}{\rho(\lambda)+1}\right)^Q \rho(\lambda).
\]
Combining this with (\ref{secondlast}),
\[
	\dthermlim \sum_{q=Q}^\infty \glHC(q, I - P_\varepsilon)
=	\frac{\partial}{\partial \tau} \bigbar_{\tau=0} p(\mu,Q,\tau,\varepsilon)
=	\int_{[\varepsilon, \infty)} \left(\frac{\rho(\lambda)}{\rho(\lambda)+1}\right)^Q \rho(\lambda) d\nu(\lambda)
\]
which goes to zero at $Q \to \infty$ by Lebesgue's Dominated Convergence Theorem.
\hfill$\square$

\vspace{0.75cm}
Finally we note that when $\e^{-\beta H}$ has a Feynman Kac representation, Condition \ref{condition3} follows
from (\ref{statebound}) since by (\ref{condition1general})
\[
	\lim_{\varepsilon \to 0} \lthermlim \gHC(q, P_\varepsilon)
\;\le\;	\lim_{\varepsilon \to 0} \lthermlim  \e^{\beta \mu q} \int_{[0,\varepsilon)} d\nu_\sLambda(\lambda)
\;\le\;	C_d\e^{\beta \mu q}  \lim_{\varepsilon \to 0} \int_{[0,\varepsilon)} \lambda^{d/2} d\lambda = 0
\]
using the expression for the integrated density of states $(\ref{densityofstates})$.

However for more complex models not all these conditions are not necessarily true. In fact we know
that the vital underlying assumption of the non-negativity of
the cycle expectation for a positive operator (Condition \ref{positivity2}) is not satisfied in general.
In the following two chapters, we shall consider the Infinite-Range-Hopping Bose-Hubbard model in two forms,
first with a hard-core repulsion explicitly forbidding more than a single particle per site and
second with a finite on-site repulsion to discourage multiple particles per site. 
In Appendix \ref{appendixE}, by direct computation we find that the positivity
condition does not hold for the hard-core model, and therefore an alternative approach is required.

\chapter{I.R.H Bose-Hubbard Model with Hard-Cores} 								\label{chapter4}

\hrule
\textbf{Summary}\\
\textit{In this chapter we study the relation between long cycles and Bose-Condensation in the 
Infinite-Range-Hopping Bose-Hubbard Model with a hard core interaction \cite{BolandPule}. Unfortunately the technique 
derived in Chapter \ref{chapter3} cannot be applied to this model (as Condition \Rmnum{1}
fails to hold, see Appendix \ref{appendixE}).
Nonetheless we succeed in calculating the density of particles on long cycles in the 
canonical thermodynamic limit by the following argument. 
We express the density of particles on cycles of length $q$ for $n$ particles 
in terms of $q$ distinguishable particles coupled to a field of $n-q$ bosons.
We can prove that in the thermodynamic limit
we can neglect the hopping of the $q$ particles so that bosons have to avoid
each other and the fixed positions of the
distinguishable particles. This is equivalent to a reduction of the
lattice by $q$ sites. Moreover the $q$ particles are on a cycle of length $q$.
It then is shown that in fact only the single-cycle density contributes
which means that in the thermodynamic limit
the sum of the long cycle densities is the particle density less the one-cycle contribution,
which can be calculated. We find that the existence
of a non-zero long cycle density coincides with the occurrence of Bose-Einstein
condensation but this density is not equal to that of the Bose condensate.}
\vspace{15pt}
\hrule


\section{The Model and Previously Derived Results}

As shown in Chapter \ref{chapter3}, the density of long cycles is equal to the density of the Bose-Einstein
condensate in the cases of the Bose gas, the mean-field model and the perturbed
mean-filed model. We wish to check if this is the case for the selected model: the 
Infinite-Range-Hopping Bose-Hubbard Model with hard-cores, also simply referred to as the
Hard-Core Boson model.

The Bose-Hubbard Hamiltonian is given by
\begin{equation}												\label{B-H}
    H^{\mathrm{BH}}=J\! \mathop{\sum_{x,y\in\Lambda_\sV}}_{|x-y|=1}(a^\ast_x-a^\ast_y)(a\astr_x-a\astr_y)
    +\lambda\sum_{x\in \Lambda_{\sV}} n_x(n_x-1)
\end{equation}
where $\Lambda_\sV$ is a lattice of $V$ sites, $a^\ast_x$ and $a\astr_x$ are the Bose
creation and annihilation operators satisfying the usual commutation relations
$[a\astr_x,a^\ast_y]=\delta_{x,y}$ and $n_x=a^\ast_x a\astr_x$.
The first term with $J>0$ is the kinetic energy operator and the
second term with $\lambda>0$ describes a repulsive interaction, as it discourages the presence of more
than one particle at each site. This model was originally introduced by
Fisher \textit{et al.} \cite{Fisher}.

The infinite-range-hopping version of this model has Hamiltonian
\begin{equation}												\label{I-R}
    H^{\mathrm{IR}}=\frac{1}{2V}\!\!\!\sum_{x,y\in\Lambda_\sV}(a^\ast_x-a^\ast_y)(a\astr_x-a\astr_y)
    +\lambda\sum_{x\in \Lambda_\sV} n_x(n_x-1).
\end{equation}
This is in fact a mean-field version of (\ref{B-H}) but in terms of the kinetic energy rather
than the interaction. In particular as with all mean-field models, the
lattice structure is irrelevant (due to the infinite-range hopping) and there is no dependence
on dimensionality, so we can take $\Lambda_\sV=\{1,2,3, \ldots, V\}$.
The non-zero temperature properties of this model have
been studied by Bru and Dorlas \cite{BruDorlas} and by Adams and Dorlas \cite{AdamsDorlas}. 
Dorlas, Pastur and Zagrebnov \cite{DPZ} considered the model in the presence of 
an additional random potential.
However this chapter will consider a special case of (\ref{I-R}), introduced by T\'oth \cite{Toth} 
where $\lambda=+\infty$, that is with complete single-site exclusion (hard-core). 
The properties of this hard-core model in the canonical ensemble were first obtained by T\'oth 
using probabilistic methods. Later
Penrose \cite{Penrose} and Kirson \cite{Kirson} obtained equivalent results.
In the grand-canonical
ensemble the model is equivalent to the strong-coupling BCS model
(see for example Angelescu \cite{Angelescu}). 

We recall the thermodynamic properties of the
model in the canonical ensemble as given by Penrose (see Appendix \ref{appendixA} for a 
summary).
\par
\begin{figure}[hbt]
\begin{center}
\includegraphics[width=10cm]{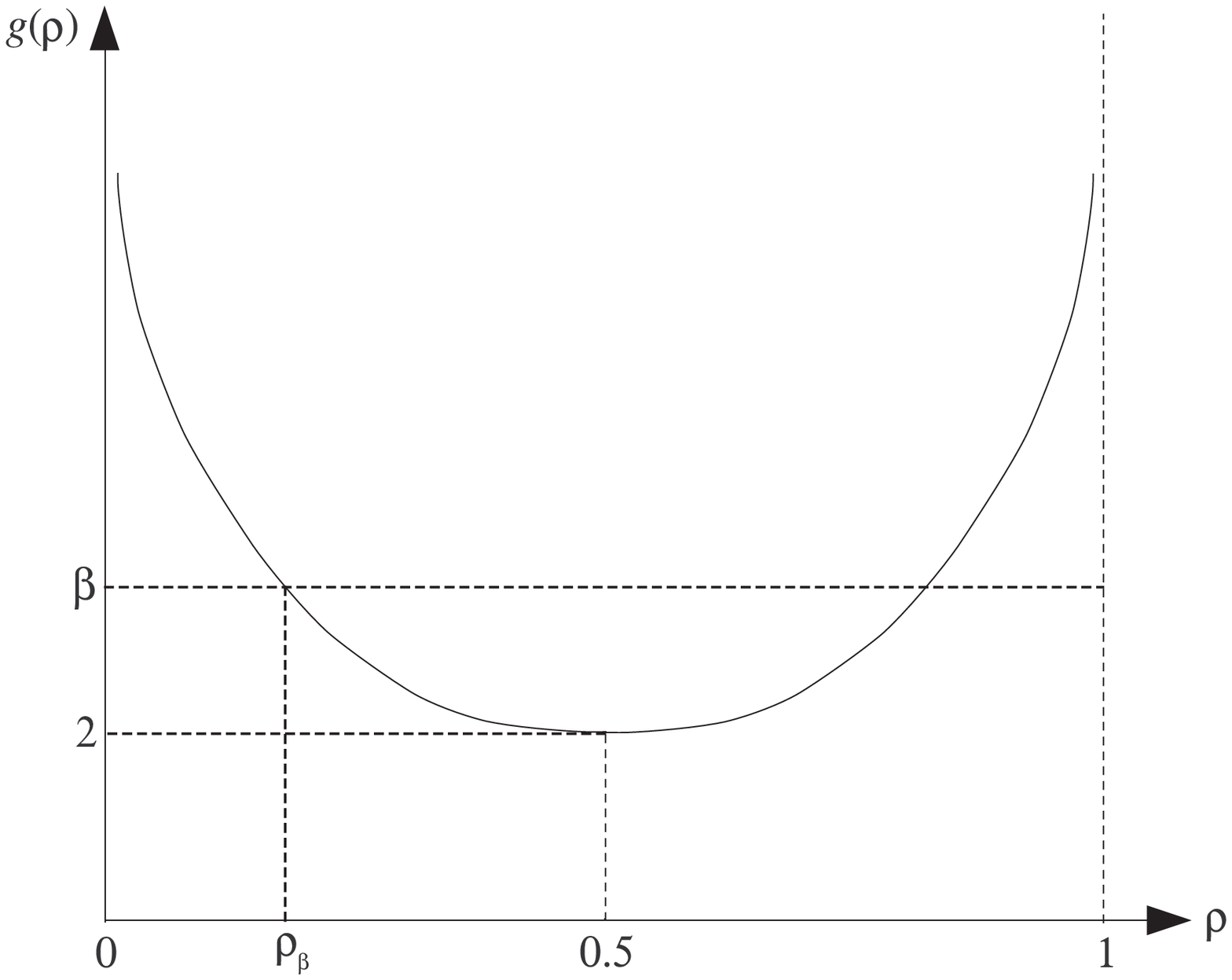}
\end{center}
\caption[Graph of $\rhobcrit$ in presence of hard-cores]{\it Definition of $\rhobcrit$}
\label{fig1}
\end{figure}

For $\rho\in(0, 1)$, let
\begin{equation}\label{gbeta}
    g(\rho)=
    \begin{cases}
    {\displaystyle  \frac{1}{1-2\rho}\ln\left( \frac{1-\rho}{\rho}\right )} \vspace{0.2cm}
    & \mathrm{if}\ \  \rho\neq 1/2,\\
    2 & \mathrm{if}\ \ \rho=1/2.
    \end{cases}
\end{equation}
For each $\beta\geq 2$ the equation $\beta=g(\rho)$ has a unique solution in $(0,1/2]$ denoted by
$\rho_{\rm cr}^\beta$ (see Figure \ref{fig1}). We define $\rho_{\rm cr}^\beta\vcentcolon=1/2$ for $\beta<2$.
The inverse temperature $\beta$ will be fixed for the remainder of this chapter, and so in maintaining 
the notational custom of this document, we shall only write the $\beta$ subscripts when necessary.


\begin{theorem} $\mathrm{(Penrose\ [10],\ Theorem\ 1)}$
The free energy per site at inverse temperature $\beta$ as a function of the particle density
$\rho\in[0, 1]$, $f(\rho)$, is given by
\begin{equation*}
    f(\rho)=
    \begin{cases}
    {\displaystyle \rho+\frac{1}{\beta}\left(\rho\ln\rho+(1-\rho)\ln(1-\rho)\right )}
    & \mathrm{if}\ \  \rho\in[0,\rhobcrit]\cup[1-\rhobcrit,1],\\
    {\displaystyle \rho^2 +\rhobcrit(1-\rhobcrit)+\frac{1}{\beta}
    \left(\rhobcrit\ln\rhobcrit+(1-\rhobcrit)\ln(1-\rhobcrit)\right )}
    & \mathrm{if}\ \ \rho\in[\rhobcrit,1-\rhobcrit].
    \end{cases}
\end{equation*}
\end{theorem}

The density of particles in the ground state in the thermodynamic limit is given by
\begin{equation*}
	\rhobcond=\cthermlim\frac{1}{V^2}\hskip -0.2cm\sum_{x,y\in\Lambda_\sV}\la a^\ast_x a\astr_y\ra
\end{equation*}
where $\la\,\cdot\,\ra$ denotes the canonical expectation for $n$ particles, c.f. (\ref{canExpect}).
Penrose showed that for certain values of $\rho$ and $\beta$, Bose-Einstein condensation occurs,
that is $\rhobcond>0$. The Bose-condensate density is given in the following theorem.
\begin{theorem} $\mathrm{(Penrose\ [10],\ Theorem\ 2)}$\\
The Bose-condensate density, $\rhobcond$ at inverse temperature $\beta$ as a function
of the particle density
$\rho\in[0, 1]$, is given by
\begin{equation}									\label{rho-c}
	\rhobcond=
	\begin{cases}
		0
		& \mathrm{if}\ \  \rho\in[0,\rhobcrit]\cup[1-\rhobcrit,1],\\
		{\displaystyle (\rho-\rhobcrit)(1-\rho-\rhobcrit)}
		& \mathrm{if}\ \ \rho\in[\rhobcrit,1-\rhobcrit].
    	\end{cases}
\end{equation}
\end{theorem}
We note that both $f(\rho)-\rho$ and the condensate density $\rhobcond$ are
symmetric about $\rho=1/2$.
This can easily seen by interchanging particles and holes. The boson states being symmetric can be labelled
unambiguously by the sites they occupy but equivalently they can be labelled by the sites they do not occupy (holes).

It is also worth noting that this model is equivalent to an $XY$ model on a 
complete graph. The mean field theory of the XY model was calculated by 
Matsubara and Matsuda \cite{MatsuMatsu1956}
who found an formula for the critical temperature equivalent to $\rhobcrit$, 
and Zitsel \cite{zitsel} gave a free-energy formula equivalent to 
$f(\rho)$, and a density-temperature graph essentially the same as Figure \ref{fig1}.

We wish to study the cycle statistics of the Infinite-Range-Hopping
Bose-Hubbard model with hard cores and compare $\rhobcond$ with the density of 
long cycles.


\subsection*{Some Notation}\label{sub b}
Before proceeding we need to define the $n$-particle Hamiltonian more carefully. 
The single particle Hamiltonian has eigenvalues 0 and 1. Denote the basis of the single particle Hilbert 
space of $V$ sites, $\Hone\vcentcolon=\CC^V$, by $\{\ee_i\}_{i=1}^V$ 
where $\ee_i=(0,\ldots, 0,1,0,\ldots, 0)$ with 1 in the $i$-th place. 
A Hamiltonian for this model is the following:
\[
	h_\sV | \phi \ra = | \phi \ra - \la \mathbf{g} | \phi \ra| \mathbf{g} \ra
\]
or simply expressed as 
\[
	h_\sV = I - P_\sV
\]
where $P_\sV = \la \mathbf{g}_\sV | \,\cdot\, \ra  \mathbf{g}_\sV $ is the projection onto the ground state
$\mathbf{g}_\sV$ defined by the eigenvector
\[
	\mathbf{g}_\sV = \frac{1}{\sqrt{V}} (1,1,\dots,1) = \frac{1}{\sqrt{V}} \sum_{i=1}^V \mathbf{e}_i
\]
with $V-1$ other possible (excited) states $\mathbf{c}_i$ such that 
$\mathbf{c}_i \perp \mathbf{g} \Leftrightarrow \la \mathbf{c}_i | \mathbf{g} \ra = 0$.
Then $P_\sV$ is given by
\begin{equation*}
	P_\sV  \mathbf{e}_i  = \frac{1}{V} \sum_{j=1}^V \mathbf{e}_j.
\end{equation*}
Thus $h_\sV$ is the orthogonal projection onto the subspace orthogonal to $\mathbf{g}_\sV$.
With this notation we can define the non-interacting $n$-particle Hamiltonian $h^{(n)}_{\sV\phantom{l}}$ acting
on the unsymmetrised Hilbert space $\Hcan{n}$ as:
\begin{align*}
	h^{(n)}_{\sV\phantom{l}}
	&= I^{(n)} - P_\sV^{(n)}
\\
 	&= n - P_\sV \otimes I \otimes \dots \otimes I - I \otimes P_\sV \otimes I \otimes \dots \otimes I
                	- \dots - I \otimes I \otimes \dots \otimes P_\sV.
\end{align*}
For bosons we have to consider the symmetric subspace of $\Hcan{n}$, which we have denoted $\Hcansym{n}$.
When $h_\sV^{(n)}$ is restricted to $\Hcansym{n}$, we obtain the usual free Hamiltonian kinetic energy:
\begin{equation*}
	\frac{1}{2V}\!\!\!\sum_{x,y\in\Lambda_\sV}(a^\ast_x-a^\ast_y)(a\astr_x-a\astr_y).
\end{equation*}

We introduce the hard-core interaction by constraining the Hamiltonian to a subspace of $\Hcan{n}$
where states with more than one particle per site are discarded.
This is done by applying the following hard core projection $\Phc{n}$ to $\Hcan{n}$:
\begin{equation*}
	\Phc{n} ( \mathbf{e}_{i_1} \otimes \mathbf{e}_{i_2} \otimes \dots \otimes \mathbf{e}_{i_n} ) =
	\begin{cases}
 		0   & \text{if } \, \mathbf{e}_{i_k} = \mathbf{e}_{i_{k'}} \; \text{for some }\, k \ne k', \\
		\mathbf{e}_{i_1} \otimes \mathbf{e}_{i_2} \otimes \dots \otimes \mathbf{e}_{i_n}   & \text{otherwise}.
	\end{cases}
\end{equation*}
We shall call $\Hhccan{n} \vcentcolon= \Phc{n} \Hcan{n}$  the unsymmetrised hard-core $n$-particle space
and \linebreak $\Hhccansym{n} \vcentcolon= \Phc{n} \Hcansym{n} $ the symmetrised hard-core $n$-particle space.
Note that as $[U_\pi,\Phc{n} ]=0$ for all $\pi\in S_n$, $\Phc{n}$ commutes with the symmetrisation
and so $\Hhccansym{n} = \sym{n} \Hhccan{n}$.

The hard-core $n$-particle Hamiltonian is then the following operator
\begin{equation}\label{H-hc}
	\hcHam\vcentcolon=\Phc{n} h^{(n)}_{\sV\phantom{l}} \Phc{n}
\end{equation}
acting on the hard-core $n$-particle space $\Hhccan{n}$. Therefore the Hamiltonian for the
infinite-range Bose-Hubbard model with hard-core is (\ref{H-hc}) acting on the \textit{symmetric} hard-core
$n$-particle space $\Hhccansym{n}$.

\section{Applying Cycle Statistics to this Model}

We wish to apply the framework established in Chapter \ref{chapter2} to this model. 
The canonical partition function for this model is
\[
	\cPart
=	\trace_{\Hhccansym{n}} \left[ \e^{-\beta \hcHam } \right]
=	\frac{1}{n!} \sum_{\pi \in S_n} \trace_{\Hhccan{n}} \left[ U_\pi \e^{-\beta \hcHam } \right].
\]
Recall that the validity of this framework rests on (\ref{Cycle_Cond1}), i.e. the condition that
each term in the above summation is non-negative:
\[
	\trace_{\Hhccan{n}} \left[ U_\pi \e^{-\beta \hcHam }\right] \ge 0
\]
for all $\pi \in S_n$.
But from the random walk formulation (see for example \cite{Toth}),
one can see that the kernel of $\e^{-\beta \hcHam}$
is positive and therefore the left-hand side is always positive.

Following the procedure of the proof of Theorem \ref{thm1-cycledens} with the following replacements:
\begin{align*}
	\cHHam &\rightarrow \hcHam 	\hspace{3cm}&
	\HHcan{n} &\rightarrow \Phc{n} \Hcan{n} \vcentcolon= \Hhccan{n}
\\
	\cHPart &\rightarrow \cPart &
	\HHcan{q} \otimes \HHcansym{n-q} &\rightarrow \Phc{n} ( \Hcan{q} \otimes \Hcansym{n-q} ) 
\end{align*}
one may derive the the following more expression for the $q$-cycle density:
\[
	\cC(q) = \frac{1}{\cPart V} \, \trace_{ \Hhccyclespace  }
				\left[ U_q \e^{-\beta \hcHam } \right]
\]
where $\Hhccyclespace \vcentcolon= \Phc{n} ( \Hcan{q} \otimes \Hcansym{n-q} )$.

Then as before we have that the density of particles on cycles of 
\textit{infinite} length is defined by
\begin{equation*}
	\rhoblong = \lim_{Q \to \infty} \;\cthermlim \; \sum_{q=Q+1}^\infty \cC(q)
\end{equation*}
(c.f. Definition \ref{rholongshort}). We wish to compare this quantity
with the condensate density $\rhobcond$.

By defining the density of short cycles as
\[
	\rhobshort = \rho - \rhoblong = \lim_{Q \to \infty} \;\cthermlim \; \sum_{q=1}^Q \cC(q),
\]
then observe that we can interchange the $q$ summation and 
the thermodynamic limit to obtain the expression
\begin{equation*}
	\rhobshort = \lim_{Q \to \infty} \; \sum_{q=1}^Q \cthermlim \;\cC(q),
\end{equation*}
which will make it much easier to calculate later.

%
\section{Neglecting the Hop of the \texorpdfstring{$q$}{q}-unsymmetrised Particles}\label{sub c}

By using cycle statistics we can split our symmetric hard-core Hilbert space $\Hhccansym{n}$
into a tensor product of two spaces, an unsymmetrised $q$-particle space $\Hcan{q}$ and a
symmetric $n-q$ particle space $\Hcansym{n-q}$, with the hard-core projection applied.
However in doing so we explicitly isolate a finite quantity of distinguishable particles.
In the thermodynamic limit, the total number of particles goes to infinity, so one queries
whether the hopping contributions of the $q$ distinguishable particles are significant in the limit.

To isolate the hopping contribution of these $q$ unsymmetrised particles, we need to 
establish some notation to clearly segment the action of an operator onto the two Hilbert spaces
$\Hcan{q}$ and $\Hcansym{n-q}$. Write
\begin{equation*}
	A^{(q)}\vcentcolon= A^{(q)} \otimes I^{(n-q)} \ \ \ \ \mathrm{and}
	\ \ \ \ A^{(n-q)} \vcentcolon= I^{(q)} \otimes A^{(n-q)}
\end{equation*}
for any operator $A$ on $\Hone$.
With this form we can express the Hamiltonian (\ref{H-hc}) on $\Hhccyclespace$ by the following:
\begin{equation*}
	\hcHam = \Phc{n} \left( n - P_\sV^{(q)} - P_\sV^{(n-q)} \right) \Phc{n}.
\end{equation*}
Then the operator $\widetilde{P}^{(q)}_\sV = \Phc{n} P^{(q)}_\sV \Phc{n}$ represents 
the hopping contribution of the $q$ distinguishable particles to the Hamiltonian (\ref{H-hc}).

Now let us define the following modified Hamiltonian
\begin{equation}\label{redHam}
	\widetilde{H}^{\mathsf{hc}}_{q,n,\sV}
	= \Phc{n} \left(n - P_\sV^{(n-q)} \right) \Phc{n}
\end{equation}
so that
\begin{equation*}
	\hcHam
	= \widetilde{H}^{\mathsf{hc}}_{q,n,\sV} - \widetilde{P}^{(q)}_\sV.
\end{equation*}
This Hamiltonian considers $n$ particles on a lattice of $V$ sites, with $q$ particles
in fixed locations, and the remaining $n-q$ particles hopping from site to site
with the hard-core constraint (a particle must avoid landing on any other particle). One 
may observe that this ensemble is similar to a system of $n-q$ hopping particles on
$V-q$ sites with hard-cores, a fact that will be made rigourous shortly.

Now we wish to estimate the effect of neglecting the
action of the $\widetilde{P}^{(q)}_\sV$ term in the unsymmetrised space.
Define the modified cycle density by
\begin{equation*}
	\mcC(q) 
=  	\frac{1}{\cPart V} \trace_{\Hhccyclespace}
	\left[ U_q \mathrm{e}^{-\beta \widetilde{H}^{\mathsf{hc}}_{q,n,\sV} } \right].
\end{equation*}
We want to show that the $q$-unsymmetrised particles' hopping contribution may be 
neglected in the limit. This is expressed in the following theorem
\begin{theorem}										\label{c_theorem}
\[
	\cthermlim \cC(q) = \cthermlim \mcC(q).
\]
\end{theorem}
Proof of this result requires two lemmas. In Lemma \ref{c} we find an upper bound for the difference of the
two cycle densities for fixed $q$, which involves a ratio of partition functions, one of $n$ particles on $V$ sites,
the other of $n-q$ particles on $V-q$ sites with a scaled hopping parameter. Secondly in Lemma \ref{Z} we
determine the limiting behaviour of this ratio.

A little more notation before proceeding, define a scaled partition function by
\begin{equation*}
	Z_\sV(\lambda,n) 
= 	\trace_{\Hhccansym{n}} \left[ \e^{-\beta H^{\mathsf{hc}}_{\lambda, n,\sV} } \right]
\end{equation*}
where the Hamiltonian introduces a weighting parameter $\lambda$ to the hopping action:
\begin{equation}										\label{h-lambda}
	H^{\mathsf{hc}}_{\lambda, n,\sV} = \Phc{n} \left( n - \lambda P_\sV^{(n)} \right) \Phc{n} .
\end{equation}

The first step is determining the following estimate.
\begin{lemma}\label{c}
\begin{equation*}
	\left| \cC(q) -\mcC(q) \right|
\leq	\frac{(1-\e^{-\beta q})}{V}
	\frac{Z_{V-q}(\frac{V-q}{V},n-q)}{\cPart}.
\end{equation*}
\end{lemma}

To prove this Lemma we intend to obtain an upper bound for
\begin{equation*}
	\left |\trace_{\Hhccyclespace}
	\left[ U_q \e^{-\beta H^{\mathsf{hc}}_{n,\sV} } \right]
-	\trace_{\Hhccyclespace}
	\left[ U_q \e^{-\beta \widetilde{H}^{\mathsf{hc}}_{q,n,\sV} } \right]\right |.
\end{equation*}
However before proceeding with a proof we first shall introduce some notation and make some remarks.

\subsection*{Some Preliminaries}
Let $\Lambda_{\sV +}^{(n-q)}$ be the family of sets of $n-q$ distinct points
of $\Lambda_\sV$. For $\mathbf{k}=\{k_1, k_2, \dots , k_{n-q}\}\in \Lambda_{\sV +}^{(n-q)}$
let
\begin{equation*}
	|\mathbf{k}\ra \vcentcolon= \sigma_{+}^{n-q}(\mathbf{e}_{k_1} \otimes \mathbf{e}_{k_2}
	\otimes \dots \otimes \mathbf{e}_{k_{n-q}}).
\end{equation*}
Then $\{|\mathbf{k}\ra\,|\,\mathbf{k}\in \Lambda_{\sV +}^{(n-q)}\}$
is an orthonormal basis for
$\Hhccansym{n-q} \vcentcolon=\mathcal{P}^{\mathrm{hc}}_{n-q} \mathcal{H}_{\sV +}^{(n-q)}$.
\par
Similarly let $\Lambda_{\sV}^{(q)}$ be the set of
ordered $q$-tuples of distinct indices of $\Lambda_\sV$ and
for $\mathbf{i}=(i_1, i_2, \dots , i_q)\in \Lambda_{\sV}^{(q)}$
let
\begin{equation*}
	|\mathbf{i}\ra \vcentcolon= \mathbf{e}_{i_1} \otimes \mathbf{e}_{i_2}
	\otimes \dots \otimes \mathbf{e}_{i_q}.
\end{equation*}
Then $\{|\mathbf{i}\ra\,|\,\mathbf{i}\in \Lambda_{\sV}^{(q)}\}$ is an orthonormal basis for
$\Hhccan{q}\vcentcolon=\mathcal{P}^{\mathrm{hc}}_q \mathcal{H}_{\sV}^{(q)}$.
\par
If $\mathbf{k}\in \Lambda_{\sV +}^{(n-q)}$ and
$\mathbf{i}\in \Lambda_{\sV}^{(q)}$
we shall write $\mathbf{k}\sim\mathbf{i}$
if $\{k_1, k_2, \dots , k_{n-q}\}\cap \{i_1, i_2, \dots , i_q\}=\emptyset$
and we shall use the notation
\begin{equation*}
	|\mathbf{i};\mathbf{k}\ra\vcentcolon=|\mathbf{i}\ra\otimes|\mathbf{k}\ra.
\end{equation*}
Then a basis for $\Hhccyclespace$ may
be formed by taking the tensor product of the bases
of $\Hhccansym{n-q}$ and $\Hhccan{q}$ where we disallow particles from appearing in both
spaces simultaneously.
Thus the set
$\{|\mathbf{i};\mathbf{k}\ra\,|\,\mathbf{k}\in \Lambda_{\sV +}^{(n-q)},\
\mathbf{i}\in \Lambda_{\sV}^{(q)},\ \mathbf{k}\sim\mathbf{i}\}$
is an orthonormal basis for $\Hhccyclespace$.
\par
We shall need also the following facts. For simplicity we shall write $\Htilde$ and
$\widetilde{P}$ for $\widetilde{H}^{\mathsf{hc}}_{q,n,\sV}$ (c.f. equation (\ref{redHam})\!)
and $\widetilde{P}^{(q)}_\sV$ respectively.
\par
Let $\mathcal{P}_\ii^{(n-q)}$ be the projection
of $\mathcal{H}_{(n-q),\sV, +}^{\mathrm{hc}}$ onto a space
with none of the $n-q$ particles at the points $i_1, i_2, \dots , i_q$
(so there are $V-q$ available sites for $n-q$ particles) and not more than
one particle at any site. Then we state the following useful remarks:

\begin{remark}									\label{rem a}
For $\ii \sim \kk$, if $s>0$, then
\begin{equation}									\label{Htilde}
	\e^{-\beta \Htilde s} | \ii; \kk \ra = |\ii; \e^{-\beta H^\ii s} \kk \ra \e^{-\beta q s}
\end{equation}
where $H^\ii = \mathcal{P}_\ii^{(n-q)} ((n-q) - P_\sV^{(n-q)} ) \mathcal{P}_\ii^{(n-q)}$.
\end{remark}
\begin{proof}
This can be seen as follows:
For $\ii \sim \kk$,
\begin{align*}
    \widetilde{H} |\ii; \kk \ra
&=  \Phc{n} ( n - P_\sV^{(n-q)} ) \Phc{n} |\ii; \kk \ra
\\
&=  q\Phc{n} |\ii; \kk \ra + \Phc{n} | \ii; ( (n-q)- P_\sV^{(n-q)} ) \kk \ra
\\
&=  q |\ii; \kk \ra +  |\ii; \mathcal{P}_\ii^{(n-q)} ( (n-q) - P_\sV^{(n-q)} ) \kk \ra
\\
&=  q |\ii; \kk\ra +  |\ii; H^\ii\kk \ra.
\end{align*}
\end{proof}
\begin{remark}									\label{rem b}
For $\ii \sim \kk$,
\begin{equation*}
	H^\ii|\kk \ra
=	(n-q)|\kk \ra-\frac{1}{V}
	\sum_{l=1}^{n-q}\sum_{j\notin\, \ii\,\cup\,\kk\setminus\{k_l\}}
	|(k_1,k_2, \ldots,\widehat{k}_l,j,\ldots,k_{n-q}) \ra
\end{equation*}
where the hat symbol implies the term is removed from the sequence.
\end{remark}
\begin{proof}
This statement follows from the definition of $H^\ii$ and the properties of the ket. Note that
the $l$ summation selects each of the $n-q$ particles in turn, and the $j$ summation is over
all the possible landing-sites for each particle (i.e. that do not violate the hard-core condition).
\end{proof}

Note that from (\ref{h-lambda}), for $\kk\in \mathcal{H}^{(n-q)}_{\sV-q,+}$ we have
\begin{equation*}
	H^{\mathsf{hc}}_{\lambda, n-q,\sV-q} | \kk \ra 
=	(n-q)|\kk \ra-\frac{\lambda}{V-q}
    	\sum_{l=1}^{n-q}
	\sum^{V-q}_{\stackrel{j=1}{j\notin\, \kk\setminus\{k_l\}}}
	|(k_1,k_2, \ldots,\widehat{k}_l,j,\ldots,k_{n-q}) \ra.
\end{equation*}
Thus $H^\ii$ is unitarily equivalent to $H^{\mathsf{hc}}_{(\sV-q)/\sV,\, n-q,\sV-q}$ and
\begin{equation*}
	\trace_{\Hhccyclespace}  \left[
	\mathcal{P}_\ii^{(n-q)} \e^{-\beta \widetilde{H}^\ii}  \mathcal{P}_\ii^{(n-q)}
	\right]
    	=\ccPart.
\end{equation*}
\begin{remark}									\label{rem c}
For $s, \alpha \in \RR$
\[
        \left( \mathcal{P}_\ii^{(n-q)} \e^{-s \widetilde{H}^\ii}  \mathcal{P}_\ii^{(n-q)} \right)^\alpha
        = \mathcal{P}_\ii^{(n-q)} \e^{-s \alpha \widetilde{H}^\ii}  \mathcal{P}_\ii^{(n-q)}.
\]
\end{remark}
\begin{proof}
Let $A =  \mathcal{P}_\ii \e^{-s \widetilde{\mathcal{H}}^\ii}  \mathcal{P}_\ii$, 
$B = \e^{-s \widetilde{\mathcal{H}}^\ii}$. We shall show that for any 
$f$ continuous, $f(A) =  \mathcal{P}_\ii f(B) \mathcal{P}_\ii$. 
These are self-adjoint operators, so using the Spectral
theorem, there exists unique spectral measures $\mu_\phi^A( d\lambda )$ and
$\mu_{ \mathcal{P}_\ii \phi}^B (d\lambda)$ respectively so that
\[
	\la \phi | A^\alpha \phi \ra = \la \mathcal{P}_\ii \phi| B^\alpha  \mathcal{P}_\ii \phi \ra \Rightarrow
	\mu_\phi^A( d\lambda ) = \mu_{ \mathcal{P}_\ii \phi}^B (d\lambda).
\]
Therefore we may write
\[
	\la \phi | f(A) \phi \ra = \int f(\lambda) \mu_\phi^A (d\lambda) 
	= \int f(\lambda) \mu_\phi^{\mathcal{P}_\ii B \mathcal{P}_\ii} (d\lambda)
	= \int f(\lambda) \mu_{\mathcal{P}_\ii \phi}^B (d\lambda)
	= \la \mathcal{P}_\ii \phi | f(B) \mathcal{P}_\ii \phi \ra.
\]
By setting $f(x) = x^\alpha$ we retrieve our result.
\end{proof}

\subsection*{Proof of Lemma \ref{c}}
To proceed we expand
\[
    	\trace_{\Hhccyclespace} \left[U_q \e^{-\beta H^{\mathsf{hc}}_{n,\sV} } \right]
= 	\trace_{\Hhccyclespace} \left[U_q \e^{-\beta (\Htilde - \widetilde{P}) } \right]
\]
in a Dyson series (see \cite{ReedSimon1, ReedSimon2}) in powers of
$\widetilde{P}$. If $m\geq 1$, the $m^\text{th}$ term is
\begin{align*}
	X_m \vcentcolon= \beta^m \int_0^1\hskip -0.3cm ds_1 \int_0^{s_1} \hskip -0.4cmds_2 \dots
& 	\int_0^{s_{m-1}}\hskip -0.8cm ds_m  \;\;
	\trace_{\mathcal{H}^{\mathsf{hc}}_{q,n,\sV}}
	\Bigg[ \e^{-\beta \Htilde (1-s_1)} \widetilde{P}
	\e^{-\beta \Htilde (s_1-s_2)} \widetilde{P}
	\cdots
\\
&	\hskip 6cm\cdots \widetilde{P}
	\e^{-\beta \Htilde (s_{m-1} -s_m)} \widetilde{P}
	\e^{-\beta \Htilde s_m}  U_q \Bigg].
\end{align*}
Recall that $\widetilde{P}\vcentcolon= \Phc{n} P_\sV^{(q)} \Phc{n}$ where
\begin{equation*}
	P_\sV^{(q)} =
	P_\sV \otimes I \otimes \cdots \otimes I + I \otimes P_\sV \otimes I \otimes \cdots \otimes I + \cdots +
	I \otimes \cdots \otimes I \otimes P_\sV
\end{equation*}
has $q$ terms, so in the above trace we have $m$ instances of this form.
Let $P^{(q)}_r = I \otimes \cdots \otimes \underbrace{P_\sV}_{\text{$r^{th}$ place}} \otimes \dots \otimes I$,
and let $\widetilde{P}_r = \Phc{n} P^{(q)}_r \Phc{n}$.

Then we can write
\begin{equation} 	           						\label{dyson_term}   
    	X_m
=	\beta^m \sum_{r_1 = 1}^q \sum_{r_2 = 1}^q \cdots \sum_{r_m = 1}^q X_m(r_1, r_2, \dots, r_q )
\end{equation}
where
\begin{multline*}
	X_m(r_1, r_2, \dots, r_q )
 =	\int_0^1 \hskip -0.3cm ds_1 \int_0^{s_1}\hskip -0.4cm ds_2 \dots
	\int_0^{s_{m-1}}\hskip -0.8cm ds_m
\\
	\times \trace_{\mathcal{H}^{\mathsf{hc}}_{q,n,\sV}}
	\Bigg[ \e^{-\beta \Htilde (1-s_1)} \widetilde{P}_{r_1}
	\e^{-\beta \Htilde (s_1-s_2)} \widetilde{P}_{r_2}
	\cdots \widetilde{P}_{r_{m-1}}
	\e^{-\beta \Htilde (s_{m-1} -s_m)} \widetilde{P}_{r_m}
	\e^{-\beta \Htilde s_m}  U_q \Bigg].
\end{multline*}
In terms of the basis of $\mathcal{H}^{\mathsf{hc}}_{q,n,\sV}$ we may write the expression for 
$X_m(r_1, r_2, \dots, r_q )$ as
\begin{multline} 	           						\label{trace_term}   
	\hspace{1cm} 
	X_m(r_1, r_2, \dots, r_q )
=	\int_0^1\hskip -0.3cm ds_1 \int_0^{s_1} \hskip -0.4cm ds_2
	\dots  \int_0^{s_{m-1}}\hskip -0.8cm ds_m \;\;
	\sum_{\kk^0, \dots\ ,\kk^m} \;\; \sum_{\ii^0 \sim \kk^0} \cdots
	\sum_{\ii^m \sim \kk^m} 
\\
	\la \ii^0; \kk^0 | \e^{-\beta \Htilde (1-s_1)} \widetilde{P}_{r_1}
	| \ii^1; \kk^1 \ra
	\la \ii^1; \kk^1 | \e^{-\beta \Htilde (s_1-s_2)} \widetilde{P}_{r_2}
	| \ii^2; \kk^2 \ra \cdots
\\
	\cdots
	\la \ii^{m-1}; \kk^{m-1} |\e^{-\beta \Htilde (s_{m-1} -s_m)} \tilde{P}_{r_m}
	| \ii^m; \kk^m \ra
	\la \ii^m; \kk^m |\e^{-\beta \Htilde s_m}  U_q  | \ii^0; \kk^0 \ra
\end{multline}
where it is understood that the $\ii$ summations are over $\Lambda_{\sV}^{(q)}$
and the $\kk$ summations are over $\Lambda_{\sV +}^{(n-q)}$.

Note that for $\ii \sim \kk$
\begin{equation}\label{P-r}
	\widetilde{P}_r | \ii ; \kk \ra 
=	\frac{1}{V}\hskip -0.8cm \sum_{\stackrel{l=1...V}
	{l \notin \kk;\ l \neq i_1 \dots \widehat{i}_r \dots i_q}}\hskip -0.8cm
	|(i_1,\cdots , \widehat{i}_r ,l , \cdots , i_q ); \kk \ra
\end{equation}
where again the hat symbol implies that the term is removed from the sequence.

Consider one of the inner products in the expression (\ref{trace_term}) for $X_m(r_1, r_2, \dots, r_q )$, using
(\ref{Htilde}) and (\ref{P-r}) above. For
$\ii \sim \kk$ and $\jj \sim \kk'$:
\begin{align*}
	\la \ii ; \kk\, |\, \e^{-\beta s \widetilde{H}} \widetilde{P}_r | \jj ; \kk' \ra
&= 	\frac{\e^{-\beta q s}}{V}\hskip -0.7cm  \sum_{\stackrel{l=1...V}{l \notin \kk';\, l \neq j_1, \dots
	\widehat{j_r}, \dots j_q}} \hskip -0.7cm\la \ii\,|\,( j_1,
	\cdots ,
	\widehat{j_r}, l, \cdots ,j_q) \ra
	\la \kk\, |\, \e^{-\beta s H^\ii} | \kk' \ra
\\[0.3cm]
&= 	\frac{\e^{-\beta q s}}{V}
	\hskip -0.7cm\sum_{\stackrel{l=1...V}{l \notin \kk';\, l \neq j_1, \dots ,
	\widehat{j_r} , \dots ,  j_q}}\hskip -0.7cm \delta_{i_1 j_1}
	\dots \widehat{\delta_{i_r j_r}} \, \delta_{i_r l} \,
	\dots \delta_{i_q j_q}
	\la \kk | \e^{-\beta s H^\ii} | \kk' \ra.
\end{align*}
In summing over $l$ we replace $l$ by $i_r$ and the result is non-zero
only if $i_r \notin \kk'$ and $i_r \neq j_1, \dots , \widehat{j_r},  \dots , j_q$.
However this last condition is not necessary because if $i_r=j_s\ (s\neq r)$
then $j_s\neq i_s$ and we get zero. Also if for some $s\neq r$, $i_s \in \kk'$ then once again
$j_s\neq i_s$. We can therefore replace the condition $i_r \notin \kk'$ by $\ii \sim \kk'$.
Using $\mathcal{I}$ for the indicator function, we have
\begin{align*}
	\la \ii ; \kk\, |\, \e^{-\beta s \widetilde{H}} \widetilde{P}_r | \jj ; \kk' \ra
&= \frac{\e^{-\beta q s}}{V}  \delta_{i_1 j_1}
    	\dots \widehat{\delta_{i_r j_r}} \, \dots \delta_{i_q j_q}
	\la \kk | \e^{-\beta s H^\ii} | \kk' \ra
	\mathcal{I}_{(\ii \sim \kk')}
\\[0.2cm]
&= 	\frac{\e^{-\beta q s}}{V}  \delta_{i_1 j_1} \dots
	\widehat{\delta_{i_r j_r}} \, \dots \delta_{i_q j_q}
	\la \kk | \mathcal{P}_\ii^{(n-q)} \e^{-\beta s H^\ii}
	\mathcal{P}_\ii^{(n-q)} | \kk' \ra.
\end{align*}
Now if we sum over $\jj \sim \kk'$, with $\ii \sim \kk$ and for a fixed $r$:
\begin{align*}
	\sum_{ \jj \sim \kk'} \la \ii ; \kk | \e^{-\beta \widetilde{H} s}
	\widetilde{P}_r | \jj ; \kk' \ra \la \jj ; \kk' |
&= 	\frac{\e^{-\beta s q}}{V}  \la \kk | \mathcal{P}_\ii^{(n-q)}
	\e^{-\beta H^\ii s} \mathcal{P}_\ii^{(n-q)} | \kk' \ra \\
&	\quad \times \hspace{-0.4cm}
	\sum_{\stackrel{j_r = 1 \dots V}{j_r \notin \kk' \cup \ii\setminus \{ i_r \}}}
	\la (i_1, \dots ,i_{r-1}, j_r, i_{r+1}, \dots,i_q); \kk' |  .
\end{align*}
It is convenient to define the operation $[r,x](\ii)$ which inserts the value of $x$
in the $r^\text{th}$ position of $\ii$ instead of $i_r$.
So for example taking the ordered triplet $\ii = (5,4,1)$,
then $[2,8](\ii) = (5,8,1)$. For brevity we shall denote the composition of these
operators as $[r_k, x_k; \, \dots \, ; r_2, x_2 ; r_1, x_1] \vcentcolon= [r_k, x_k]  \circ \cdots \circ [r_2, x_2] \circ [r_1, x_1]$.

At a very low level, this operation represents the individual hopping of the particles.
To prevent violation of the hard-core condition, one needs to take care when
constructing the set of all possible landing sites.

In this notation the final term in the above expression may be rewritten as $\la [r, j_r](\ii) ; \kk' | $.

Performing two summations for fixed $r_1$ and $r_2$ we get:
{\allowdisplaybreaks
\begin{align*}
	\sum_{ \ii^1 \sim \kk^1} \sum_{ \ii^2 \sim \kk^2} &
	\la \ii^0 ; \kk^0 |
	\e^{-\beta s\widetilde{H} } \widetilde{P}_{r_1} | \ii^1 ; \kk^1 \ra
	\la \ii^1 ; \kk^1 | \e^{-\beta t\widetilde{H} } \widetilde{P}_{r_2}
	| \ii^2 ; \kk^2 \ra
	\la \ii^2 ; \kk^2 |  
\\
&   =  \frac{\e^{-\beta q (s+t)}}{V^2}
	\sum_{i^1_{r_1} \notin \kk^1 \cup \ii^0 \setminus \{ i^0_{r_1} \}} \hspace{0.4cm}
	\sum_{i^2_{r_2} \notin \kk^2 \cup [r_1,i^1_{r_1}](\ii^0) \setminus \{ i^0_{r_2} \}}
	\la \kk^0 | \mathcal{P}_{\ii^0}\
	\e^{-\beta sH^{\ii^0} }\ \mathcal{P}_{\ii^0} | \kk^1 \ra
\\*
&  \qquad \times\la \kk^1 | \mathcal{P}_{[r_1,i^1_{r_1}](\ii^0)}\
\e^{-\beta t H^{[r_1,i^1_{r_1}](\ii^0)} }
	\  \mathcal{P}_{[r_1,i^1_{r_1}](\ii^0)} | \kk^2 \ra
	\la [r_1, i^1_{r_1}; r_2, i^2_{r_2}](\ii^0); \kk^2 |
\\[0.4cm]
&=	\frac{\e^{-\beta q (s+t)}}{V^2}
	\sum_{i^1_{r_1} \notin  \ii^0 \setminus \{ i^0_{r_1} \}} \hspace{0.4cm}
	\sum_{i^2_{r_2} \notin \kk^2 \cup [r_1,i^1_{r_1}](\ii^0) \setminus \{ i^0_{r_2} \}}
	\la \kk^0 |\ \mathcal{P}_{\ii^0}\ \e^{-\beta s H^{\ii^0} }\ \mathcal{P}_{\ii^0}\
	| \kk^1 \ra
\\*
&	\qquad \times \la \kk^1 |\ \mathcal{P}_{[r_1,i^1_{r_1}](\ii^0)}\ \e^{-\beta t H^{[r_1,i^1_{r_1}](\ii^0)}}
	\ \mathcal{P}_{[r_1,i^1_{r_1}](\ii^0)}\ | \kk^2 \ra
	\la [r_2, i^2_{r_2} ; r_1, i^1_{r_1}] (\ii^0); \kk^2|
\end{align*}}
due to the fact that $\mathcal{P}_{[r_1, i^1_{r_1}](\ii^0)}|\kk^1\ra=0$ if $i^1_{r_1}\in \kk^1$.
We may apply this to all inner product terms of (\ref{trace_term}) except the final one.
Note we sum over the $V$ sites of the lattice, with certain points excluded in each case.

For the final inner product of (\ref{trace_term}) we obtain:
{\allowdisplaybreaks
\begin{align*}
&\la	[r_m, i^m_{r_m}; \,  \dots \, ; r_2, i^2_{r_2} ; r_1, i^1_{r_1}](\ii^0) ; \kk^m |
	\e^{-\beta s_m \widetilde{H}  } U_q | \ii^0 ; \kk^0 \ra
\\[0.2cm]
&=	\e^{-\beta q s_m} \la [r_m, i^m_{r_m}; \, \dots \, ; r_2, i^2_{r_2} ; r_1, i^1_{r_1}](\ii^0); \kk^m |
	\e^{-\beta  s_m H^{[r_m, i^m_{r_m}; \, \dots \, ; r_2, i^2_{r_2} ; r_1, i^1_{r_1}](\ii^0)} }
	U_q | \ii^0 ; \kk^0 \ra
\\[0.2cm]
&=	\e^{-\beta q s_m} \la \kk^m |
	\e^{-\beta  s_m H^{[r_m, i^m_{r_m}; \, \dots \, ; r_2, i^2_{r_2} ; r_1, i^1_{r_1}](\ii^0)} }
	| \kk^0 \ra \;
	\la [r_m, i^m_{r_m}; \, \dots \, ; r_2, i^2_{r_2} ; r_1, i^1_{r_1}](\ii^0) | U_q \ii^0 \ra
\\[0.2cm]
&=	\e^{-\beta q s_m} \la \kk^m |
	\mathcal{P}_{[r_m, i^m_{r_m}; \, \dots \, ; r_2, i^2_{r_2} ; r_1, i^1_{r_1}](\ii^0)}
	\e^{-\beta s_m H^{[r_m, i^m_{r_m}; \, \dots \, ; r_2, i^2_{r_2} ; r_1, i^1_{r_1}](\ii^0)}  }
	\mathcal{P}_{[r_m, i^m_{r_m}; \, \dots \, ; r_2, i^2_{r_2} ; r_1, i^1_{r_1}](\ii^0) }
	| \kk^0 \ra
\\*
& \hspace{4cm} 
	\times  \la [r_m, i^m_{r_m};  \, \dots \, ; r_2, i^2_{r_2} ; r_1, i^1_{r_1}](\ii^0) | U_q \ii^0 \ra .
\end{align*}}
Applying this to the whole tracial expression of (\ref{trace_term}) we obtain
{\allowdisplaybreaks
\begin{align*}
   	X_m&(r_1, r_2, \dots, r_q )
\\*[0.2cm]
=& \;  \frac{\e^{-\beta q}}{V^m} \sum_{\kk^0 \dots \kk^m} \sum_{\ii^0}
	\sum_{i^1_{r_1} \notin \ii^0 \setminus \{i^0_{r_1}\}} \hspace{0.3cm}
	\sum_{i^2_{r_2} \notin [r_1,i^1_{r_1}](\ii^0) \setminus \{i^1_{r_2}\}}
	\cdots \sum_{i^m_{r_m} \notin [r_{m-1}, i^{m-1}_{r_{m-1}}; \, \dots \, ; r_2, i^2_{r_2} ; r_1, i^1_{r_1}](\ii^0)
	\setminus \{i^{m-1}_{r_m}\}}
\\*
&	\la \kk^0 | \mathcal{P}_{\ii^0} \e^{-\beta(1-s_1) \Htilde^{\ii^0} } \mathcal{P}_{\ii^0}
	| \kk^1 \ra
\\*
&	\la \kk^1 | \mathcal{P}_{[r_1,i^1_{r_1}](\ii^0)}
	\e^{-\beta (s_1-s_2)\Htilde^{[r_1,i^1_{r_1}](\ii^0)}}
	\mathcal{P}_{[r_1,i^1_{r_1}](\ii^0)} | \kk^2 \ra
\\*
&	\la \kk^2 | \mathcal{P}_{[r_2, i^2_{r_2}; r_1, i^1_{r_1}](\ii^0)} \e^{-\beta (s_2-s_3)
	\Htilde^{[r_2, i^2_{r_2} ; r_1, i^1_{r_1}](\ii^0)} }
	\mathcal{P}_{[r_2, i^2_{r_2} ; r_1, i^1_{r_1}](\ii^0)} | \kk^3 \ra
\\*
&	\cdots
\\*
&	\la \kk^m | \mathcal{P}_{[r_m, i^m_{r_m}; \, \dots \, ; r_2, i^2_{r_2} ; r_1, i^1_{r_1}](\ii^0)}
	\e^{-\beta s_m \Htilde^{ [r_m, i^m_{r_m}; \, \dots \, ; r_2, i^2_{r_2} ; r_1, i^1_{r_1}](\ii^0)} }
	\mathcal{P}_{[r_m, i^m_{r_m}; \, \dots \, ; r_2, i^2_{r_2} ; r_1, i^1_{r_1}](\ii^0)}
	| \kk^0 \ra \\
&   \la [r_m, i^m_{r_m}; \, \dots \, ; r_2, i^2_{r_2} ; r_1, i^1_{r_1}](\ii^0) | U_q \ii^0 \ra
\\[0.5cm]
=& 	\; \frac{ \e^{-\beta q} }{V^m} \sum_{\ii^0}
	\sum_{i^1_{r_1} \notin \ii^0 \setminus \{i^0_{r_1}\}} \hspace{0.3cm}
	\sum_{i^2_{r_2} \notin [r_1,i^1_{r_1}](\ii^0) \setminus \{i^1_{r_2}\}}
	\cdots \sum_{i^m_{r_m} \notin [r_{m-1}, i^{m-1}_{r_{m-1}}; \, \dots \, ; r_2, i^2_{r_2} ; r_1, i^1_{r_1}](\ii^0)
	\setminus \{i^{m-1}_{r_m}\}}
\\*
&	\la [r_m, i^m_{r_m}; \, \dots \, ; r_2, i^2_{r_2} ; r_1, i^1_{r_1}](\ii^0) | U_q \ii^0 \ra
\\*
&	\trace_{\mathcal{H}_{(n-q),\sV, +}^{\mathrm{hc}}} \Bigg[ \mathcal{P}_{\ii^0} \e^{-\beta (1-s_1)\Htilde^{\ii^0} }
	\mathcal{P}_{\ii^0} \mathcal{P}_{i^1_{r_1}(\ii^0)}
	\e^{-\beta (s_1-s_2)\Htilde^{i^1_{r_1}(\ii^0)} }
	\mathcal{P}_{i^1_{r_1}(\ii^0)} \cdots
\\*
&	\qquad \cdots \mathcal{P}_{[r_m, i^m_{r_m}; \, \dots \, ; r_2, i^2_{r_2} ; r_1, i^1_{r_1}](\ii^0)}
	\e^{-\beta s_m\Htilde^{[r_m, i^m_{r_m}; \, \dots \, ; r_2, i^2_{r_2} ; r_1, i^1_{r_1}](\ii^0)} }
	\mathcal{P}_{[r_m, i^m_{r_m}; \, \dots \, ; r_2, i^2_{r_2} ; r_1, i^1_{r_1}](\ii^0)}  \Bigg].
\end{align*}}
From the H\"older inequality (see Manjegani \cite{Manjegani}), for finite dimensional non-negative matrices
$A_1, A_2, \dots , A_{m+1}$ we have the inequality
\[
    \left| \trace \big( A_1 A_2 \dots A_{m+1} \big) \right|
        \le \trace \big| A_1 A_2 \dots A_{m+1} \big|
        \le \prod_{k=1}^{m+1} \big( \trace A_k^{p_k} \big)^{\tfrac{1}{p_k}}
\]
where $\sum_{k=1}^{m+1} \tfrac{1}{p_k} = 1$, $p_i > 0$.

Set $p_1 = \frac{1}{1-s_1},\ p_2 = \frac{1}{s_1 - s_2},\ \dots ,\ p_m = \frac{1}{s_{m-1}-s_m},\ p_{m+1} = \frac{1}{s_m}$.
Taking the modulus of the above trace
\begin{align*}
	\Bigg| &\trace_{\mathcal{H}_{(n-q),\sV, +}^{\mathrm{hc}}}  \Bigg[ \mathcal{P}_{\ii^0}
	\e^{-\beta \Htilde^{\ii^0} (1-s_1)} \mathcal{P}_{\ii^0}
	\mathcal{P}_{[r_1, i^1_{r_1}](\ii^0)} \e^{-\beta \Htilde^{[r_1, i^1_{r_1}](\ii^0)} (s_1-s_2)}
	\mathcal{P}_{[r_1, i^1_{r_1}](\ii^0)} \; \cdots 
\\
&	\qquad \qquad \qquad \cdots \mathcal{P}_{[r_m, i^m_{r_m}; \, \dots \, ; r_2, i^2_{r_2} ; r_1, i^1_{r_1}](\ii^0)}
	\e^{-\beta \Htilde^{[r_m, i^m_{r_m}; \, \dots \, ; r_2, i^2_{r_2} ; r_1, i^1_{r_1}](\ii^0)} (s_m)}
	\mathcal{P}_{[r_m, i^m_{r_m}; \, \dots \, ; r_2, i^2_{r_2} ; r_1, i^1_{r_1}](\ii^0)}
	\Bigg]    \Bigg|
\\[0.4cm]
\le&\quad
	\trace_{\mathcal{H}_{(n-q),\sV, +}^{\mathrm{hc}}} \bigg[ \mathcal{P}_{\ii^0}
	\e^{-\beta \Htilde^{\ii^0}} \mathcal{P}_{\ii^0} \bigg]^{1-s_1}
\\
&\quad 
	\trace_{\mathcal{H}_{(n-q),\sV, +}^{\mathrm{hc}}} \bigg[ \mathcal{P}_{[r_1, i^1_{r_1}](\ii^0)}
	\e^{-\beta \Htilde^{[r_1, i^1_{r_1}](\ii^0)}} \mathcal{P}_{[r_1, i^1_{r_1}](\ii^0)} \bigg]^{s_1-s_2}
\\
&\quad 
	\cdots
\\
&\quad 
	\trace_{\mathcal{H}_{(n-q),\sV, +}^{\mathrm{hc}}}
	\bigg[ \mathcal{P}_{[r_m, i^m_{r_m}; \, \dots \, ; r_2, i^2_{r_2} ; r_1, i^1_{r_1}](\ii^0)}
	\e^{-\beta \Htilde^{[r_m, i^m_{r_m}; \, \dots \, ; r_2, i^2_{r_2} ; r_1, i^1_{r_1}](\ii^0)}}
	\mathcal{P}_{ [r_m, i^m_{r_m}; \, \dots \, ; r_2, i^2_{r_2} ; r_1, i^1_{r_1}](\ii^0)} \bigg]^{s_m} \!\!\!.
\end{align*}
But since the trace is independent of the $V-q$ sites 
$\{\ii^0, [r_1,i^1_{r_1}](\ii^0),\, [r_2, i^2_{r_2} ; r_1,i^1_{r_1}](\ii^0),\,\dots$ 
$\dots, \, [r_m, i^m_{r_m}; \, \dots\ \, ; r_2, i^2_{r_2} ; r_1, i^1_{r_1}](\ii^0)\}$,
therefore using Remark \ref{rem c}, the product of all the trace terms above is equal to
\[
	\trace_{\mathcal{H}_{(n-q),\sV, +}^{\mathrm{hc}}} \left[ \mathcal{P}_{\boldl}
		\e^{-\beta \Htilde^{\boldl}} \mathcal{P}_{\boldl} \right]
\]
with $\boldl = \{V-q+1, V-q+2, \dots ,V\}$ and from Remark \ref{rem b},
\[
	\trace_{\mathcal{H}_{(n-q),\sV, +}^{\mathrm{hc}}} \left[ \mathcal{P}_{\boldl}
		\e^{-\beta \Htilde^{\boldl}} \mathcal{P}_{\boldl} \right]
	=\ccPart.
\]
The integrand is now independent of $s_1, \dots, s_m$, so we may integrate to obtain a factor of
$1/m!$. What remain is to consider the unpleasant looking sum
\begin{multline}  							\label{sum_of_cycled_inner_products} 
	\sum_{\ii^0} \; \sum_{i^1_{r_1} \notin \ii^0 \setminus \{i^0_{r_1}\}} \hspace{0.2cm}
	\sum_{i^2_{r_2} \notin [r_1, i^1_{r_1}](\ii^0) \setminus \{i^1_{r_2}\}}
	\cdots
\\
	\cdots \sum_{i^m_{r_m} \notin [r_{m-1}, i^{m-1}_{r_{m-1}}; \, \dots \, ; r_2, i^2_{r_2} ; r_1, i^1_{r_1}](\ii^0)
	\setminus \{i^{m-1}_{r_m}\}}  \hspace{-0.5cm}
	\la [r_m, i^m_{r_m}; \, \dots \, ; r_2, i^2_{r_2} ; r_1, i^1_{r_1}](\ii^0) | U_q \ii^0 \ra.
\end{multline}
To analyse this, first fix the numbers $r_1, r_2, \dots , r_m$.
If $\{r_1, r_2, \dots , r_m\} \ne \{1,2,\dots, q\} $, then the ket $| [r_m, i^m_{r_m}; \, \dots $ 
$\dots; r_2, i^2_{r_2} ; r_1, i^1_{r_1}](\ii^0) \ra$
is of the form
\[
	| j_1, j_2, \dots ,j_{n_1},i^0_{n_1+1}, \dots ,i^0_{n_2}, j_{n_2+1},
	\dots, j_{n_3}, i^0_{n_3+1}, \dots  ,
	i^0_{n_4}, j_{n_4+1}, \dots \dots\ra
\]
where $\{n_1, n_2, \dots \}$ is a non-empty ordered set of distinct integers between 0 and $q$. This state is
clearly orthogonal to $U_q \ii^0$ for any $q$. Note that this situation does not arise if $q=1$.
Note also that  this is always the case if $ m < q$.
\par
We may bound the remaining sum corresponding to terms for which
$\{r_1, r_2, \dots , r_m\} = \{1,2,\dots, q\}$ by
\[
	\le \sum_{\ii^0} \;\;
	\underbrace{ \sum_{i^1_{r_1}=1}^V \;\;\;
	\sum_{i^2_{r_2}=1}^V \;\;\;
	\cdots  \;\;\;
	\sum_{i^m_{r_m} = 1}^V }_{\stackrel{\text{where $[r_m, i^m_{r_m}; \, \dots \,
		; r_2, i^2_{r_2} ; r_1, i^1_{r_1}]$}}{\text{\tiny{has distinct indices}}}} \;\;
	\la [r_m, i^m_{r_m}; \, \dots \, ; r_2, i^2_{r_2} ; r_1, i^1_{r_1}](\ii^0) | U_q \ii^0 \ra .
\]
Observe that in this case $| [r_m, i^m_{r_m}; \, \dots \, ; r_2, i^2_{r_2} ; r_1, i^1_{r_1}](\ii^0) \ra$ is independent of
$\ii^0$ so we may take it to be
\[
	|[r_m, i^m_{r_m}; \, \dots \, ; r_2, i^2_{r_2} ; r_1, i^1_{r_1}](\mathbf{s}^0) \ra
\]
where $\mathbf{s}^0 = (1,2,3,\dots,q)$. Then we can interchange the $\ii^0$ summation with the others, and
for each choice of $i^1_{r_1}, i^2_{r_2}, \dots , i^m_{r_m}$ there exists only one possible $\ii^0 \in \Lambda_\sV^{(q)}$
such that
\[
	\la [r_m, i^m_{r_m}; \, \dots \, ; r_2, i^2_{r_2} ; r_1, i^1_{r_1}](\mathbf{s}^0) | U_q \ii^0 \ra \ne 0.
\]
So we may conclude that
\begin{multline*}
	\sum_{\ii^0} \; \sum_{i^1_{r_1} \notin \ii^0 \setminus \{i^0_{r_1}\}} \hspace{0.2cm}
	\sum_{i^2_{r_2} \notin [r_1, i^1_{r_1}](\ii^0) \setminus \{i^1_{r_2}\}}
	\cdots
\\
	\cdots \sum_{i^m_{r_m} \notin [r_{m-1}, i^{m-1}_{r_{m-1}}; \, \dots \, ; r_2, i^2_{r_2} ; r_1, i^1_{r_1}](\ii^0)
	\setminus \{i^{m-1}_{r_m}\}}  \hspace{-0.5cm}
	\la [r_m, i^m_{r_m}; \, \dots \, ; r_2, i^2_{r_2} ; r_1, i^1_{r_1}](\ii^0) | U_q \ii^0 \ra \le V^m.
\end{multline*}

\;

Applying this, we see that the modulus of the integrated $m^\text{th}$ term of the
Dyson series may bounded above by
\begin{align*}
    |X_m(r_1, &r_2, \dots , r_m)| 
\\
&\le	\frac{\e^{-\beta q}}{V^m m!}
    	\ccPart \sum_{\ii^0} \;
	\sum_{i^1_{r_1} \notin \ii^0 \setminus \{i^0_{r_1}\}} \hspace{0.2cm}
	\sum_{i^2_{r_2} \notin [r_1, i^1_{r_1}](\ii^0) \setminus \{i^1_{r_2}\}}
	\cdots
\\
&	\qquad \cdots \sum_{i^m_{r_m} \notin [r_{m-1}, i^{m-1}_{r_{m-1}}; \, \dots
		 \, ; r_2, i^2_{r_2} ; r_1, i^1_{r_1}](\ii^0)
	\setminus \{i^{m-1}_{r_m}\}}  \hspace{-0.5cm}
	\la [r_m, i^m_{r_m}; \, \dots \, ; r_2, i^2_{r_2} ; r_1, i^1_{r_1}](\ii^0) | U_q \ii^0 \ra.
\\
&\le	\frac{\e^{-\beta q}}{m!}\ccPart.
\end{align*}
Hence the modulus of (\ref{dyson_term}), the $m^\text{th}$ term of the Dyson series may be bounded above by
\[
	| X_m |
\le	\beta^m \sum_{r_1=1}^q \cdots \sum_{r_q=1}^q |X_m(r_1, r_2, \dots , r_m)|
\le	\e^{-\beta q} \ccPart \frac{q^m \beta^m }{m!} .
\]
Noting that the zeroth term of the Dyson series is
\[
	X_0 = \trace_{\mathcal{H}^{\mathsf{hc}}_{q,n,V}} \left[U_q \e^{-\beta \Htilde } \right],
\]
we may re-sum the series to obtain
\[
	\Bigg| \trace_{\mathcal{H}^{\mathsf{hc}}_{q,n,V}}
	\left[U_q \e^{-\beta H^{\mathsf{hc}}_{n,V} } \right]
-	\trace_{\mathcal{H}^{\mathsf{hc}}_{q,n,V}}
	\left[U_q \e^{-\beta \Htilde } \right]  \Bigg|
\le	\e^{-\beta q}\ccPart\ \sum_{m=1}^\infty \frac{q^m \beta^m}{m!}.
\]
Thus
\begin{align}
	\left|\cC(q)-\widetilde{c}\,_\sV^n(q)\right |
&=	\frac{1}{V} \left| \frac{
	\trace_{\mathcal{H}^{\mathsf{hc}}_{q,n,V}}
	\left[U_q \e^{-\beta H^{\mathsf{hc}}_{n,V} } \right]
-	\trace_{\mathcal{H}^{\mathsf{hc}}_{q,n,V}}
	\left[U_q \e^{-\beta \Htilde } \right]}{\cPart} \right|			\notag
\\
&\le	\frac{\e^{-\beta q}}{V} \,
	\frac{ \ccPart}{\cPart} \sum_{m=1}^\infty \frac{q^m \beta^m}{m!}	\notag
\\
&=	\frac{\e^{-\beta q}}{V} (\e^{\beta q} - 1)
	\frac{\ccPart}{\cPart}.								\label{difference1}
\end{align}
Obviously we want this to tend to zero in the limit, but this is far from clear. The 
next section will consider the behaviour of the troublesome ratio of partition functions
in the thermodynamic limit.

\section{The Variational Argument}\label{sub Z}
\textit{For this section we shall reinsert the $\beta$ dependence in all quantities where required.}

In his paper on the Hard-Core boson model \cite{Penrose}, Penrose gave an explicit expression 
for $Z_\sV(n,\beta)$:
\[
	Z_\sV(n,\beta)=\hskip -0.3cm\sum_{r=0}^{\min(n, V-n)} z(r,n,V,\beta),
\]
where
\[
	z(r,n,V,\beta)
=	\left( \frac{V-2r+1}{V-r+1} \right) \binom{V}{r}
	\exp\left\{ -\frac{\beta}{V} \left[ Vr - r^2 + r + n^2 - n \right] \right\}.
\]
He also proved that if $h_\sV:[0,\min(\rho,1-\rho)]\to \RR$ converges uniformly in $[0,\min(\rho,1-\rho)]$
as $V\to\infty$ to a continuous function $h:[0,\min(\rho,1-\rho)]\to \RR$, then
\begin{equation}\label{LD}
    \cthermlim \frac{1}{Z_\sV(n,\beta)} \sum_{r=0}^{\min(n,V-n)} h_\sV(\tfrac{r}{V})\
    z(r,n,V,\beta) =
    \begin{cases}
    h(\rho),
    & \mathrm{if}\ \  \rho\in[0,\rhobcrit^\beta],\\
    h(\rhobcrit),
    & \mathrm{if}\ \ \rho\in[\rhobcrit^\beta,1-\rhobcrit^\beta],\\
    h(1-\rho),
    & \mathrm{if}\ \  \rho\in[1-\rhobcrit^\beta,1].
    \end{cases}
\end{equation}

We shall use this expansion and his Large-Deviation-like result to consider the limiting 
behaviour of the ratio of partition functions in line (\ref{difference1}), to obtain the following:

\begin{lemma}										\label{Z}
\begin{equation*}
   \cthermlim\frac{Z_{V-q}\left(\tfrac{V-q}{V},\, n-q, \beta \right)}{Z_\sV(n,\beta)}=
    \begin{cases}
    \rho^q \, \e^{\beta q}
    & \mathrm{if}\ \  \rho\in[0,\rhobcrit^\sbeta]\cup[1-\rhobcrit^\sbeta,1],\\
    {\displaystyle (\rhobcrit^\sbeta)^q \, \e^{\beta q(1+\rho-\rhobcrit^\sbeta)}}
    & \mathrm{if}\ \ \rho\in[\rhobcrit^\sbeta,1-\rhobcrit^\sbeta].
    \end{cases}
\end{equation*}
\end{lemma}

\begin{proof}
Recall that we have
\begin{equation}								\label{Z1}
	Z_{V-q}(n-q,\beta)
	= \trace_{\mathcal{H}^{\mathrm{hc}}_{n-q,\sV-q,+}} \left[\e^{-\beta H_{n-q,\sV-q}^{\mathrm{hc}}}\right]
	= \e^{-\beta (n-q)}\ \trace_{\mathcal{H}^{\mathrm{hc}}_{n-q,\sV-q,+}}
	\left[ \e^{\beta \mathcal{P}^{\mathrm{hc}}_{n-q}
		P^{n-q}_{\sV-q}\mathcal{P}^{\mathrm{hc}}_{n-q}} \right]
\end{equation}
while
\begin{align}									\label{Z2}
	Z_{V-q}\left(\tfrac{V-q}{V},\, n-q, \beta \right)
	&= \trace_{\mathcal{H}^{\mathrm{hc}}_{n-q,\sV-q,+}}
  		\left[\e^{-\beta H_{(\sV-q)/V,n-q,\sV-q}^{\mathrm{hc}}}\right]\nonumber
\\
  	&= \e^{-\beta (n-q)}\ \trace_{\mathcal{H}^{\mathrm{hc}}_{n-q,\sV-q,+}}
  		\left[ \e^{\beta(\frac{\sV-q}{\sV}) \mathcal{P}^{\mathrm{hc}}_{n-q} P^{n-q}_{\sV-q}
  		\mathcal{P}^{\mathrm{hc}}_{n-q}} \right].
\end{align}
Comparison of (\ref{Z1}) and (\ref{Z2}) yields an equivalence between the partition functions
\[
	Z_{V-q}\left(\tfrac{V-q}{V},\, n-q, \beta \right)
=	\e^{-\beta \frac{q}{V}(n-q)}\ Z_{V-q}\left(n-q,\beta(\tfrac{V-q}{V})\right)
\]
and thus we have to analyse the following ratio:
\[
	\frac{ \e^{-\beta \frac{q}{V}(n-q)} Z_{V-q}\left(n-q,\beta(\tfrac{V-q}{V})\right)}{Z_\sV(n,\beta)}.
\]
We wish to express this in the form of the lefthand side of
(\ref{LD}). Using Penrose's explicit form for the partition function, we have the following:
\begin{align*}
	Z_{V-q}\left(n-q,\beta(\tfrac{V-q}{V})\right)
= 	\sum_{r=0}^{\min(n-q, V-n)} 
& 	\left( \frac{V-q-2r+1}{V-q-r+1} \right) \binom{V-q}{r} 
\\
&	\times \exp\left\{ -\frac{\beta}{V} \left[ r(V-q) - r^2 + r + (n-q)^2 - (n-q) \right] \right\}
\end{align*}
For the case $\rho > \tfrac{1}{2}$, for large $V$, $n-q> V-n$ we must sum from
zero to $V-n$ and a straightforward calculation then gives
\begin{equation*}
	\e^{-\beta \tfrac{q}{V}(n-q)} Z_{V-q}\left(n-q,\beta(\tfrac{V-q}{V})\right)
=	\sum_{r=0}^{V-n} h_\sV(\tfrac{r}{V})\ z(r,n,V,\beta)
\end{equation*}
where
\begin{multline*}
	h_\sV(x) =  \left( \frac{1-2x-(q-1)/V}{1-2x+1/V} \right)
	\left( \frac{1-x+1/V}{1-x-(q-1)/V} \right)
\\
	\times \prod_{s=0}^{q-1}\left(\frac{1-x-s/V}{1-s/V}\right)
	\exp\left\{ \beta q \left[ x + \rho - 1/V \right] \right\}.
\end{multline*}
Therefore
\begin{equation*}
	h(x)=\lim_{V \to\infty} h_\sV(x)=(1-x)^q \ \e^{q\beta (x+\rho)}.
\end{equation*}
It is clear that the convergence is uniform since $h_\sV(x)$ is a product of terms each of which
converges uniformly on $[0,1-\rho]$ for $\rho>\tfrac{1}{2}$. Thus
\begin{equation*}
	\cthermlim \frac{ \e^{-\beta \frac{q}{V}(n-q)} Z_{V-q}\left(n-q,\beta(\tfrac{V-q}{V})\right) }{Z_\sV(n,\beta)}
=
\begin{cases}
   (1-\rhobcrit^\sbeta)^q\ \e^{q\beta (\rhobcrit^\sbeta+\rho)}
    & \mathrm{if}\ \ \rho\in(1/2,1-\rhobcrit^\sbeta], \vspace{0.1cm}\\
    \rho^q \e^{\beta q}
    & \mathrm{if}\ \  \rho\in(1-\rhobcrit^\sbeta,1] .
    \end{cases}
\end{equation*}
Note that using the relation
\begin{equation*}
	\beta = \frac{1}{1-2\rhobcrit^\sbeta}\ln\left(\frac{1-\rhobcrit^\sbeta}{\rhobcrit^\sbeta}\right )
\end{equation*}
we get
\begin{equation*}
    (1-\rhobcrit^\sbeta)^q\ \e^{q\beta (\rhobcrit^\sbeta+\rho)}
    =(\rhobcrit^\sbeta)^q\ \e^{q\beta (1+\rho-\rhobcrit^\sbeta)}
\end{equation*}
and therefore we have proved Lemma \ref{Z} for $\rho>\tfrac{1}{2}$.

For the case $\rho \le \tfrac{1}{2}$ we have that $n-q < V-n$, the sum
for $\e^{-\beta \frac{q}{V}(n-q)} Z_{V-q}\left(n-q,\beta(\tfrac{V-q}{V})\right)$ is up to $n-q$,
and therefore we need to shift the index by $q$ to get it into the required form.
After shifting we get
\begin{align*}
	\e^{-\beta \frac{q}{V}(n-q)} Z_{V-q}\left(n-q,\beta(\tfrac{V-q}{V})\right)
&=	\sum_{r=q}^{n} z(r,n,V,\beta) \left( \frac{V+q-2r+1}{V-2r+1} \right)\\
&	\hspace{0cm} \times \frac{ r(r-1)(r-2) \cdots (r-q+1)}{V(V-1)(V-2)\cdots (V-q+1)}
	\exp\left\{ \frac{\beta q}{V} \left[ V + n - r \right] \right\}.
\end{align*}
Note that summand is zero if we put $r=0, \dots, q-1$. Thus we may sum
from zero to $n$ to get as before
\begin{equation*}
	\e^{-\beta \frac{q}{V}(n-q)} Z_{V-q}\left(n-q,\beta(\tfrac{V-q}{V})\right)
=	\sum_{r=0}^{n} h_\sV(\tfrac{r}{V})\ z(r,n,V,\beta)
\end{equation*}
where this time
\begin{equation} 									\label{h-rho<=1/2}
	h_\sV(x) 
=	\left( \frac{1-2x +(q+1)/V}{1-2x+1/V} \right)
	\prod_{s=0}^{q-1}\left( \frac{x -s/V}{1-s/V} \right)
	\exp\left\{ \beta q \left[ 1 + \rho - x  \right] \right\}
\end{equation}
so that
\begin{equation*}
	h(x)=\cthermlim h_\sV(x) = x^q \exp\{\beta q(1 + \rho - x) \}.
\end{equation*}
Convergence is again uniform on $[0,\rho]$ for $\rho<\tfrac{1}{2}$
and therefore
\begin{equation*}
	\cthermlim \frac{ \e^{-\beta \frac{q}{V}(n-q)} Z_{V-q}\left(n-q,\beta(\tfrac{V-q}{V})\right) }{ Z_\sV(n,\beta) }
=
\begin{cases}
       \rho^q \e^{\beta q}
    & \mathrm{if}\ \  \rho\in[0,\rhobcrit^\sbeta),  \vspace{0.1cm}\\
    (\rhobcrit^\sbeta)^q\ \e^{q\beta (1+\rho-\rhobcrit^\sbeta)}
    & \mathrm{if}\ \ \rho\in[\rhobcrit^\sbeta,1/2),
    \end{cases}
\end{equation*}
proving Lemma \ref{Z} for $\rho < \tfrac{1}{2}$. The case $\rho = \tfrac{1}{2}$
is more delicate because the first term in (\ref{h-rho<=1/2}) does not converge uniformly.
We can write (taking $V=2n$)
\begin{equation*}
    h_{2n}(r/2n)=\widetilde{h}_{2n}(r/2n)+\frac{q}{2(n-r)+1}\widetilde{h}_{2n}(r/2n)
\end{equation*}
where
\begin{equation*}
    \widetilde{h}_{2n}(x) =
     \prod_{s=0}^{q-1}\left( \frac{x -s/{2n}}{1-s/{2n}} \right)
                        \exp\left\{ \beta q \left[ 3/2 - x  \right] \right\}.
\end{equation*}
Clearly $\widetilde{h}_{2n}(x)$ converges uniformly on $[0,1/2]$ and therefore
\begin{equation*}
    \lim_{n\to\infty}\frac{1}{Z_{2n}(n,\beta)}\sum_{r=0}^{n} \widetilde{h}_{2n}(\tfrac{r}{2n})\
    z(r,n,2n,\beta)= (\rhobcrit^\sbeta)^q \, \mathrm{e}^{\beta q(3/2-\rhobcrit^\sbeta)}.
\end{equation*}
We thus have to show that
\begin{equation*}
    \lim_{n\to\infty}\frac{1}{Z_{2n}(n,\beta)}
    \sum_{r=0}^{n} \frac{\widetilde{h}_{2n}(\tfrac{r}{2n})}{2(n-r)+1}\
    z(r,n,2n,\beta)=0.
\end{equation*}
Since $\widetilde{h}_{2n}(x)$ is bounded, by $C$ say,
\begin{equation*}
    \lim_{n\to\infty}\frac{1}{Z_{2n}(n,\beta)}
    \sum_{r<n-n^{1/4}} \frac{\widetilde{h}_{2n}(\tfrac{r}{2n})}{2(n-r)+1}\
    z(r,n,2n,\beta)\leq \lim_{n\to\infty}\frac{C}{2n^{1/4}}=0.
\end{equation*}
On the other hand one can prove that for $n-2n^{1/2}\leq r\leq n-n^{1/2}$ and $r'\geq n-n^{1/4}$
\begin{equation*}
    \ln z(r,n,2n,\beta)-\ln z(r',n,2n,\beta)>\frac{1}{8}\ln n
\end{equation*}
for $n$ large, so that $z(r',n,2n,\beta)/z(r,n,2n,\beta)<1$.
Therefore
\begin{eqnarray*}
  \lim_{n\to\infty}\frac{1}{Z_{2n}(n,\beta)}
    &&\hskip -1cm\sum_{r\geq n-n^{1/4}} \frac{\widetilde{h}_{2n}(\tfrac{r}{2n})}{2(n-r)+1}\
    z(r,n,2n,\beta)
    \leq
    \lim_{n\to\infty}
    C\frac{\displaystyle{\sum_{r\geq n-n^{1/4}}\hskip -0.2 cm z(r,n,2n,\beta)}}
    {\displaystyle{\sum_{n-2n^{1/2}\leq r\leq n-n^{1/2}}\hskip -0.8cm z(r,n,2n,\beta)}} \\
   &\leq &  \lim_{n\to\infty}\frac{C}{n^{1/4}}
    \frac{\displaystyle{\max_{r\geq n-n^{1/4}}\hskip -0.1 cm z(r,n,2n,\beta)}}
    {\displaystyle{\min_{n-2n^{1/2}\leq r\leq n-n^{1/2}}\hskip -0.8cm z(r,n,2n,\beta)}}
\leq\lim_{n\to\infty}\frac{C}{n^{1/4}}=0.
\end{eqnarray*}
\end{proof}

\section{The Single-Cycle Contribution}\label{sub c-q}
So far we have shown that in the thermodynamic limit
we can neglect the hopping of the $q$ particles so that bosons have to avoid
each other and the fixed positions of the
distinguishable particles. This is equivalent to a reduction of the
lattice by $q$ sites. Moreover the $q$ particles are on a cycle of length $q$.

Now let us consider these $q$ particles in terms of Brownian motion. These 
$q$ particles are distinguishable and are all members of the same $q$-cycle.
For $q > 1$, this means for example, that the position of the second particle 
at the beginning of its path is same as the position of the first particle at the end of its path. 
But since they do not hop this is impossible by the hard core condition.
Therefore among the short cycles only the cycle of unit length can contribute. 

Since the sum of all the cycle densities gives the particle density, this means
that in the thermodynamic limit the sum of the long cycle densities is the particle 
density less the one-cycle contribution, which we can calculate.

These ideas are mode more concrete in the following lemma.
\begin{lemma}										\label{c-q}
$\mcC(q)=0$ if $q>1$ and
\begin{equation*}
   \cthermlim\mcC(1)=
    \begin{cases}
    \rho
    & \mathrm{if}\ \  \rho\in[0,\rhobcrit]\cup[1-\rhobcrit,1],\\
    {\displaystyle \rhobcrit \, \e^{\beta(\rho-\rhobcrit)}}
    & \mathrm{if}\ \ \rho\in[\rhobcrit,1-\rhobcrit].
    \end{cases}
\end{equation*}
\end{lemma}

\begin{proof}
Recall that
\begin{align*}
	\mcC(q)
&=	\frac{1}{\cPart V} \trace_{\Hhccyclespace}
	\left[ U_q \e^{-\beta \widetilde{H}^{\mathsf{hc}}_{q,n,\sV} } \right].
\end{align*}
Considering the trace over $\Hhccyclespace$, expanding it in
terms of its basis $\{|\ii; \kk\ra\}$ and using Remark \ref{rem a} above,
where $\ii \sim \kk$
\begin{align*}
	\trace_{\Hhccyclespace}
	\left[ U_q \e^{-\beta \widetilde{H}^{\mathsf{hc}}_{q,n,\sV} } \right]
&=	\sum_{\kk} \sum_{\ii \sim \kk}
	\la \ii; \kk | U_q \e^{-\beta \Phc{n} (n - P_\sV^{(n-q)} ) \Phc{n}}
	| \ii; \kk \ra\\
&=	\e^{-\beta q} \sum_{\kk} \sum_{\ii \sim \kk} \la U_q \ii; \kk | \e^{-\beta H^{\ii}}
	| \ii; \kk \ra
=	\e^{-\beta q} \sum_{\kk} \sum_{\ii \sim \kk} \la U_q \ii | \ii \ra
	\la \kk | \e^{-\beta H^{\ii}} | \kk \ra.
\end{align*}
For $q > 1$, an element of the basis of the unsymmetrised $q$-space $\Hcan{q}$ may
be written as an ordered $q$-tuple $\ii = (i_1, i_2, \dots , i_q)$ where the $i_l$'s are
all distinct.
Then we may write
\begin{align*}
	\la U_q \ii | \ii \ra
&=	\la U_q (\mathbf{e}_{i_1} \otimes \mathbf{e}_{i_2} \otimes \dots \otimes \mathbf{e}_{i_q} ) \, |\,
	\mathbf{e}_{i_1} \otimes \mathbf{e}_{i_2} \otimes \dots \otimes \mathbf{e}_{i_q} \ra
\\
&=	\la \mathbf{e}_{i_2} \otimes \mathbf{e}_{i_3} \otimes \dots \otimes \mathbf{e}_{i_q} \otimes \mathbf{e}_{i_1}\, |\,
	\mathbf{e}_{i_1} \otimes \mathbf{e}_{i_2} \otimes \dots \otimes \mathbf{e}_{i_q}\ra=0.
\end{align*}
Hence $\mcC(q)$ is non-zero only if $q=1$.

For the second statement, note that we may re-express $\mcC(1)$ as follows:
{\allowdisplaybreaks
\begin{align*}
	\mcC(1)
&=	\frac{1}{\cPart V} \trace_{\Phc{n}(\Hcan{1} \otimes \Hcansym{n-1})}
	\left[ \e^{-\beta \widetilde{H}^{\mathsf{hc}}_{1,n,\sV} } \right]
\\
&=	\frac{\e^{-\beta}}{\cPart V} \sum_{i=1}^V \sum_{\kk\, /\hskip-0.17cm \ni\, i}
	\la \kk | \e^{-\beta H^i } | \kk \ra
\\
&=	\frac{\e^{-\beta}}{\cPart V} \sum_{i=1}^V \trace_{\Hhccansym{n-1}}
	\left[ \mathcal{P}_i \e^{-\beta H^i} \mathcal{P}_i \right]
\\
&=	\e^{-\beta} \, \frac{ Z_{V-1}(\tfrac{V-1}{V}, n-1 )}{ \cPart }
\end{align*}}
and the result follows from Lemma \ref{Z}.
\end{proof}

\section{Conclusion}
\begin{figure}[hbt]
\begin{center}
\includegraphics[width=12cm]{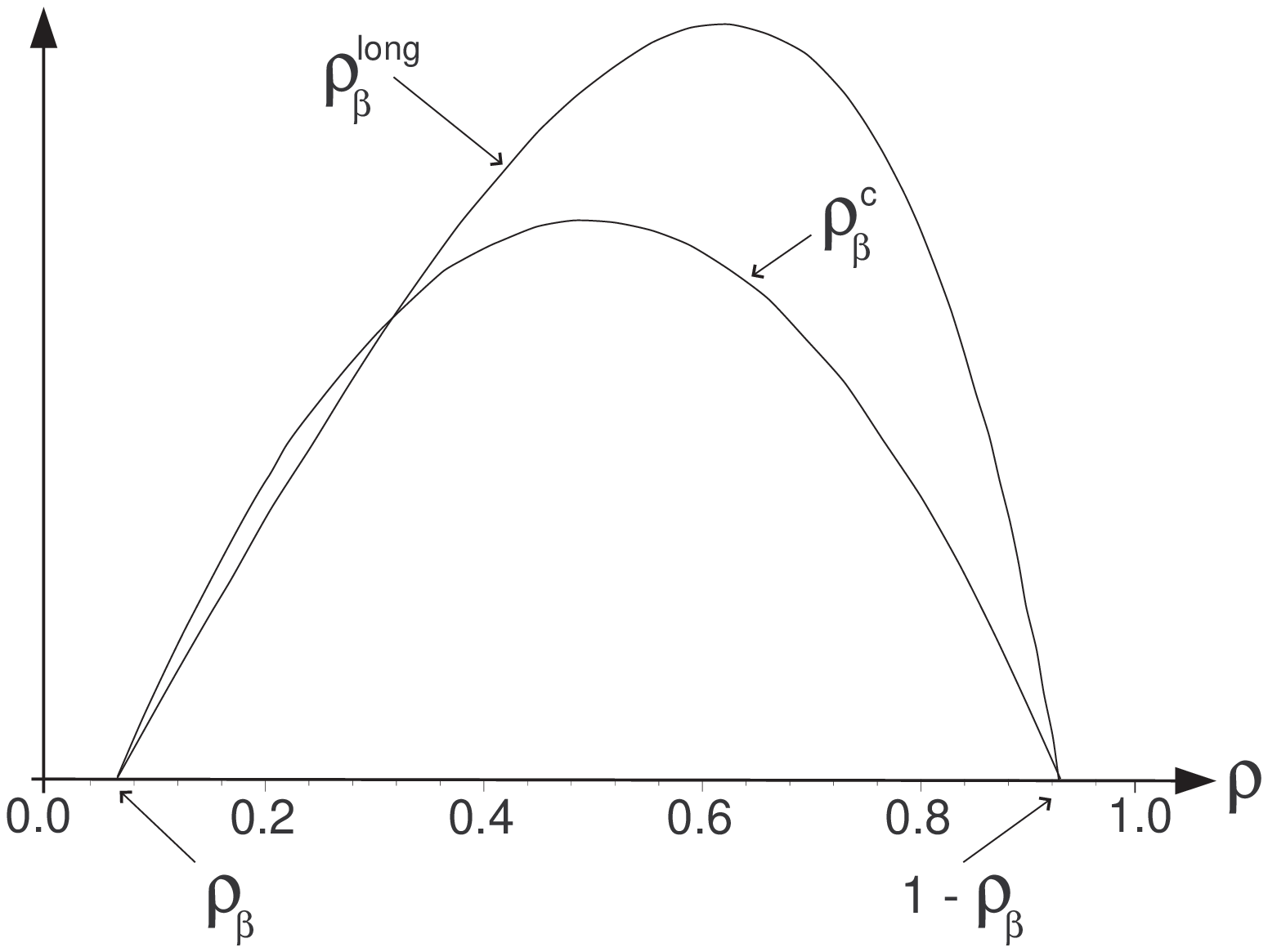}
\caption[Plot of $\rhobcond$ and $\rhoblong$ for $\beta>2$]{\it $\rhobcond$ and $\rhoblong$ for $\beta>2$}
\end{center}
\label{fig2}
\end{figure}

From Lemmas \ref{c} and \ref{Z} we found that
\[
    \cthermlim \cC(q)=\cthermlim\mcC(q).
\]
which implies that in we may neglect the hopping of the $q$ distinguishable particles in the 
thermodynamic limit. Therefore the density of short cycles for the Infinite-Range-Hopping Bose-Hubbard Model
may be expressed in terms of the modified cycle density, where those $q$ particles are fixed:
\[
	\rhobshort = \lim_{Q\to\infty} \sum_{q=Q+1}^\infty \cthermlim \mcC(q)
\]
Since by Lemma \ref{c-q} only cycles of unit length contribute to the summed cycle density, it follows that
\begin{equation*}
   \rhobshort=\cthermlim \cC(1)=
    \begin{cases}
    \rho
    & \mathrm{if}\ \  \rho\in[0,\rhobcrit]\cup[1-\rhobcrit,1],\\
    {\displaystyle \rhobcrit \e^{\beta(\rho-\rhobcrit)}}
    & \mathrm{if}\ \ \rho\in[\rhobcrit,1-\rhobcrit]
    \end{cases}
\end{equation*}
and therefore allows us to state the main result of this chapter:

\begin{theorem}\label{Th-rho-long}
The expected density of particles on cycles of infinite length,
$\rhoblong$, at inverse temperature $\beta$ as a function of the particle density
$\rho\in[0, 1]$, is given by
\begin{equation*}
   \rhoblong=
    \begin{cases}
    0
    & \mathrm{if}\ \  \rho\in[0,\rhobcrit]\cup[1-\rhobcrit,1],\\
    {\displaystyle \rho-\rhobcrit \e^{\beta(\rho-\rhobcrit)}}
    & \mathrm{if}\ \ \rho\in[\rhobcrit, 1-\rhobcrit].
    \end{cases}
\end{equation*}
\end{theorem}
In comparing this result with the expression for the density of the condensate (\ref{rho-c}),
we may form the following conclusions (see Figure \ref{fig2}):
\begin{itemize}
    \setlength{\parskip}{0pt}
    \setlength{\parsep}{0pt}
    \item $\rhoblong=0$ if and only if $\rhobcond=0$.
    \item $\rhoblong$ is not symmetric with respect to $\rho=1/2$. As mentioned above the symmetry of the model about
    $\rho=1/2$ is due to the particle-hole symmetry. But the equivalent labelling of states by sets of occupied or unoccupied
    sites (particles and holes) cannot be used for distinguishable particles. As we derived in Chapter \ref{chapter2}, the
    $q$-cycle occupation density $\cC(q)$ involves $q$ distinguishable particles and $n-q$ bosons and
    therefore the particle-hole symmetry is broken.
    \item When $\rhobcond>0$, $\rhoblong$ starts below $\rhobcond$ since its
    slope at $\rhobcrit$ is equal to $1-2\rhobcrit$ while $\rhobcond$ has slope $1-\beta\rhobcrit$
    and $\beta>2$. Conversely, $\rhoblong$ finishes above
    $\rhobcond$ since its slope at $1-\rhobcrit$ is less than that of $\rhobcond$.
    \end{itemize}

Therefore in contrast to the Ideal Boson Gas, and the Mean-Field and Perturbed Mean-Field Models,
as can be seen in the above diagram, the condensate density is not always equal to the long 
cycle density for the case of the Infinite-Range-Hopping Bose-Hubbard Model with Hard-Cores. 
However it remains true that the absence of condensation implies only finitely long cycles and visa-versa.

In the following chapter we will consider again the Infinite-Range-Hopping Bose-Hubbard Model
but this time without hard-cores to see if a similar result holds. However before doing so
it is of interest to briefly consider the notion of Off-Diagonal Long-Range Order and its
connection to Bose-Einstein condensation using cycle statistics.

\section{\texorpdfstring{Off-Diagonal Long-Range Order}{OLDRO}}
The one-body reduced density matrix for $x,x'\in \Lambda_\sV $ may be defined as
\begin{equation*}
	D_{n,\sV}(x,x')\vcentcolon=\la a^\ast_x a\astr_{x'} \ra_{\hcHam}=\frac{1}{\cPart}
	\trace_{\Hhccansym{n}} \left[ K^{(n)}_{x,\,x'} \e^{-\beta \hcHam} \right].
\end{equation*}
where for $\phi\in \mathcal{H}_\sV$, $K_{x,\,x'}\phi=\la \mathbf{e}_{x'}|\phi \ra \mathbf{e}_x$.

Penrose showed that for $x\neq x'$,
\begin{equation*}
	\cthermlim D_{n,\sV}(x,x')=\rhobcond ,
\end{equation*}
that is, whenever Bose-Einstein condensation occurs, there is
\textit{Off-diagonal long-range order} as defined by Yang \cite{Yang}.
It has been argued and proved in some cases (see for example \cite{ueltschi_fyncyclesbosegas} and 
\cite{DMP}) that in the expansion of $D_{n,\sV}(x,x')$ in terms of 
permutation cycles, only infinite cycles contribute to long-range order. Here we are able to show this 
explicitly.

Using Theorem \ref{expectation-cycled} from Chapter \ref{chapter2}, we have the following
\begin{equation*}
	D_{n,\sV}(x,x') = \sum_{q=1}^n \cC(q;K_{x,\,x'})
\end{equation*}
where
\begin{equation}								\label{hc_cycle_expectation}
	\cC(q;K_{x,\,x'})
	= \frac{1}{\cPart V} \, \trace_{\Hhccyclespace}
	\left[ (K_{x,\,x'}\otimes I\otimes I\otimes \ldots\otimes I)
 	U_q \e^{-\beta \hcHam } \right].
\end{equation}
Note that this is equivalent to the expansion of $\sigma_\rho(x)$ in equations (2.14) and (2.16) from
\cite{ueltschi_fyncyclesbosegas}.

Applying the same argument as in Sections \ref{sub c} and \ref{sub Z}, one may show that
\begin{equation}								\label{proj_in_qspace_irrel}
	\cthermlim \cC(q;K_{x,\,x'})=\cthermlim \mcC(q;K_{x,\,x'})
\end{equation}
where we take
\begin{equation*}
	\mcC(q;K_{x,\,x'}) = \frac{1}{\cPart V} \, \trace_{ \Hhccyclespace  }
	\left[ (K_{x,\,x'}\otimes I\otimes I\otimes \ldots\otimes I)
 	U_q \e^{-\beta \widetilde{H}^{\mathsf{hc}}_{q,n,\sV} } \right].
\end{equation*}
The only difference is that instead of equation (\ref{sum_of_cycled_inner_products}), we obtain
\begin{multline}								\label{summs}
	\qquad \sum_{\ii^0} \sum_{i^1_{r_1} \notin \ii^0 \setminus \{i^0_{r_1}\}} \hspace{0.1cm}
	\sum_{i^2_{r_2} \notin [r_1, i^1_{r_1}](\ii^0) \setminus \{i^1_{r_2}\}} \!\!
	\cdots
	 \!\! \sum_{i^m_{r_m} \notin [r_{m-1}, i^{m-1}_{r_{m-1}}; \, \dots \, ; r_1, i^1_{r_1}](\ii^0)
	\setminus \{i^{m-1}_{r_m}\}}
\\
	\la [r_m, i^m_{r_m}; \, \dots \, ; r_2, i^2_{r_2} ; r_1, i^1_{r_1}](\ii^0) |
		(K_{x,x'} \otimes I \otimes \cdots \otimes I)U_q \ii^0 \ra \qquad
\end{multline}
whose treatment is similar but slightly more complicated, as detailed below.

Let $q > 1$ and consider the case $\{r_1, r_2, \dots , r_m\} \neq \{1,2,\dots, q\} $.
When $1 \notin \{r_1, r_2, \dots , r_m\}$ we obtain inner products of the form:
\[
	\la i^0_1 | K_{x,x'} i^0_2 \ra \la j_2, j_3, \dots , j_q | i^0_3, i^0_4, \dots , i^0_q, i^0_1 \ra
\]
where $j_k \ne i^0_1$ for all $k$ by the hard-core condition, implying the second term is zero
as $j_q \ne i^0_1$. On the other hand, when $1 \in \{r_1, r_2, \dots , r_m\}$, then there exists at least one
$l \notin \{r_1, r_2, \dots , r_m\}$, yielding an inner product of the form
\[
	\la j_1 | K_{x,x'} i^0_2 \ra \la j_2, \dots , j_{l-1}, i_l, j_{l+1}, \dots, j_q | i^0_3, i^0_4, \dots , i^0_q, i^0_1 \ra
\]
which also results in the second term being zero as $\la i_l | i_{l+1} \ra = 0$. Note that the above cases do
not occur for $q=1$.

For the case $\{r_1, r_2, \dots , r_m\} = \{1,\dots, q\} $, as before, the remaining sum may be bounded by a similar
expression whose summations have slightly relaxed restrictions. Also the left hand side of the inner product is
independent of $\ii^0$, so again denoting $\mathbf{s}^0 = (1,2,3,\dots,q)$, we have
\begin{align*}
(\ref{summs})&
	\le
	\underbrace{ \sum_{i^1_{r_1}=1}^V \;\;
	\sum_{i^2_{r_2}=1}^V \;\;
	\cdots  \;\;
	\sum_{i^m_{r_m} = 1}^V }_{\stackrel{\text{where $[r_m, i^m_{r_m}; \, \dots
		\, ; r_2, i^2_{r_2} ; r_1, i^1_{r_1}](\mathbf{s}^0)$}}{\text{\tiny{has distinct indices}}}} \!\!
	\sum_{\ii^0} \;
	\la[r_m, i^m_{r_m}; \, \dots \, ; r_2, i^2_{r_2} ; r_1, i^1_{r_1}]
		(\mathbf{s}^0) | (K_{x,x'} i^0_2), i^0_3, \dots , i^0_q, i^0_1 \ra
\intertext{and as there is only one possible value for each $i^0_1, i^0_3, i^0_4, \dots , i^0_q$
giving a non-zero summand, we can bound above by}
&\le 	\sum_{i^1_{r_1}=1}^V \;
	\sum_{i^2_{r_2}=1}^V \;
	\cdots  \;
	\sum_{i^m_{r_m} = 1}^V
	\sum_{i^0_2=1}^\sV \la i^k_{r_k} | K_{x,x'} i^0_2 \ra \,
=	\, V^{m-1} \sum_{i^k_{r_k} = 1}^\sV \sum_{i^0_2=1}^\sV \la i^k_{r_k} | K_{x,x'} i^0_2 \ra \,
=	\,V^{m-1}
\end{align*}
where $k \in [1,m]$ is the smallest number such that $r_k = 1$, and for any $x, x' \in \Lambda_\sV$. Thus
the entire sum (\ref{summs}) is bounded above by $V^{m-1}$. Therefore one can conclude the argument
of Subsection \ref{sub Z}, proving (\ref{proj_in_qspace_irrel}).

Moreover, following the reasoning in Subsection \ref{sub c-q},
we can then check that for $q \ge 1$ and $x\neq x'$,
$\mcC(q;K_{x,\,x'})=0$, since for $q=1$, $\la \mathbf{e}_i | K_{x,x'} \mathbf{e}_i \ra = 0$, and for $q > 1$
\[
	\la (K_{x,x'} \otimes I \otimes \dots \otimes I) U_q \ii | \ii \ra
=	\la (K_{x,x'} \mathbf{e}_{i_2}) \otimes \mathbf{e}_{i_3} \otimes \dots
	\otimes \mathbf{e}_{i_q} \otimes \mathbf{e}_{i_1}\, |\,
	\mathbf{e}_{i_1} \otimes \mathbf{e}_{i_2} \otimes \dots \otimes \mathbf{e}_{i_q}\ra=0
\]
as the $i_l$'s are all distinct. So we have that
\begin{equation*}
	\cthermlim \cC(q;K_{x,\,x'})=0
\end{equation*}
and that
\begin{equation*}
	\lim_{Q\to \infty} \cthermlim \frac{1}{V} \sum_{q=Q+1}^\infty \cC(q;K_{x,\,x'})=
	\cthermlim D_{n,\sV}(x,x')=\rhobcond
\end{equation*}
thus proving the long-range order coincides with the condensate density.

\phantom{\cite{BratelliRobinson}}

\chapter[I.R.H. Bose-Hubbard Model]{The Infinite-Range-Hopping Bose-Hubbard Boson Model}
												\label{chapter5}
\hrule
\textbf{Summary}\\
\textit{In this chapter we study the relation between long cycles and Bose-Condensation in 
the Infinite-Range-Hopping Bose-Hubbard Model (without hard-cores) \cite{Boland1} and
calculate the density of particles on long cycles in the thermodynamic limit.
The argument is as follows, first we shall obtain an expression for the density of particles
on cycles of length $q$ in the grand-canonical ensemble, isolating $q$ unsymmetrised 
particles. Secondly we prove that in the thermodynamic limit we may neglect the hopping 
of these $q$ particles in the cycle density expression. Third we attempt to justify the application 
of the Approximating Hamiltonian Method upon the cycle density, proving results in the absence
of condensation, and with a gauge-breaking term, in its presence also. 
Finally we apply simple numerical methods to the resulting cycle density expression to 
indicate our conclusion.
We find, as in the hard-core case, that the absence of long cycles implies the absence of condensation and visa versa.
In the presence of condensation, we argue that while the occurrence of Bose-Einstein condensation coincides with
the existence of long cycles, their corresponding densities are not necessarily equal.}
\vspace{15pt}
\hrule

\section{The Model and Previously Derived Results}
The Infinite-Range-Hopping Bose-Hubbard model (without hard-cores) is given by the Hamiltonian 
\begin{equation}							\label{I-R2}
	H_\sV^{\rm IR}
=	\frac{1}{2V}\!\sum_{x,y = 1}^V(a^\ast_x-a^\ast_y)(a\astr_x-a\astr_y)
	+\lambda\sum_{x=1}^V n_x(n_x-1)
\end{equation}
where the magnitude of repulsion is controlled by the parameter $\lambda > 0$.
The zero-temperature properties of this model have been analysed 
for instance in \cite{Fisher, freemon, sheesh, elsemon},
mostly by applying perturbation theory to the ground state wave function
or via computational methods.

To observe the low temperature properties of this model, Bru and Dorlas \cite{BruDorlas} applied the 
``Approximating Hamiltonian'' technique developed by Bogoliubov \cite{Bog4, Bog5} (see also 
\cite{Bog1, Bog2, Bog3} and \cite{Bog-Sr1, Bog-Sr2}) to this model.
In this method one performs the following substitution for the Laplacian term of the Hamiltonian:
\[
	\frac{1}{V}\!\sum_{x,y=1}^V a^\ast_x a\astr_y \rightarrow 
	\sum_{x=1}^V ( \bar{r} a\astr_x + r a^\ast_x ) - V|r|^2
\]
(some $r \in \CC$) to obtain the approximating Hamiltonian:
\begin{equation}							\label{Happrox}
	H^{\textrm{APP}}_\sV(r)
=	\sum_{x=1}^V n_x - 
	\sum_{x=1}^V ( \bar{r} a\astr_x + r a^\ast_x ) + V |r|^2
	+\lambda\sum_{x=1}^V n_x(n_x-1).
\end{equation}
Introduce a gauge breaking source $\nu \in \CC$ to both Hamiltonians (\ref{I-R2}) and (\ref{Happrox}), by setting
$H_\sV(\nu) \vcentcolon= H_\sV - \sum_{x=1}^V (\bar{\nu} a\astr_x + \nu a^\ast_x)$ and 
$H^{\textrm{APP}}_\sV(r,\nu) \vcentcolon= H^{\text{APP}}_\sV(r) - \sum_{x=1}^V (\bar{\nu} a\astr_x + \nu a^\ast_x)$.
Then for all $\mu$ (with some constraints on $\nu$),
one finds that the pressures for these Hamiltonians are equivalent in the thermodynamic limit, i.e.
for large $V$ one obtains the estimate
\[
	0 \le p_\sV[H_\sV(\nu)] 
	- \sup_{r\in \CC} p_\sV[H^{\textrm{APP}}_\sV(r,\nu)] \le O(V^{-1/2})
\]
where $p_\sV[ \, \cdot \, ] \vcentcolon= p_\sV[ \, \cdot \, ](\beta, \mu)$ is the grand-canonical pressure. Henceforth 
the $\beta$ and $\mu$ dependencies are assumed unless explicitly given.

The exactness of this approximation is discussed in \cite{Ginibre:1968sf, LiebSeiringerC-nos}.
Recent applications of this approximation to a continuum Bose gas model appears 
in \cite{BruZagrebnov} and \cite{BruZagrebnov01}. This model in the presence of an additional random
potential is considered in \cite{DPZ}.
With this approximation technique, Bru and Dorlas managed to obtain the limiting pressure
and showed that in some regimes Bose-Einstein condensation occurs. 
They proved the following result:

\begin{theorem}							\label{thm_soln}
The pressure in the thermodynamic limit for the Infinite-Range-Hopping Bose-Hubbard Model,
$p(\beta, \mu) \vcentcolon= \thermlim p_\sV [H_\sV]$, is given by
\begin{equation}							\label{pressure}
	p(\beta, \mu) = \sup_{r \ge 0} \bigg\{  -r^2 + \frac{1}{\beta} \ln \trace_{\mathcal{F}_+(\CC)} 
	\e^{-\beta h(r) } \bigg\}
\end{equation}
where 
\[
	h(r) \vcentcolon= (1 - \mu)n + \lambda n(n-1) - r ( a + a^\ast)
\]
is a single site Hamiltonian with creation and annihilation operators $a^\ast$ and $a$,
and with number operator $n=a^\ast a$. Note that now the supremum is attained over the set of
non-negative real numbers.

The Euler-Lagrange equation for the variational principle is
\begin{equation}							\label{euler-lagrange}
    2r
=  \big\la a + a^\ast \big\ra_{h(r)}
=  \frac{\trace_{\mathcal{F}(\CC)} (a+a^\ast) \e^{-\beta h(r)}}
      {\trace_{\mathcal{F}(\CC)} \e^{-\beta h(r)}}.
\end{equation}
Moreover the density of the condensate is exactly given by
\[
	\rhocond \vcentcolon= \thermlim \frac{1}{V^2} \sum_{x,y=1}^V \big\la a^\ast_x a\astr_y \big\ra_{H_\sV} = r_\mu^2.
\]
where $r_\mu$ is the largest solution of $(\ref{euler-lagrange})$.
\end{theorem}

\begin{figure}[hbt]
  \begin{center}
    \begin{tabular}{cc}
      \resizebox{80mm}{!}{\includegraphics{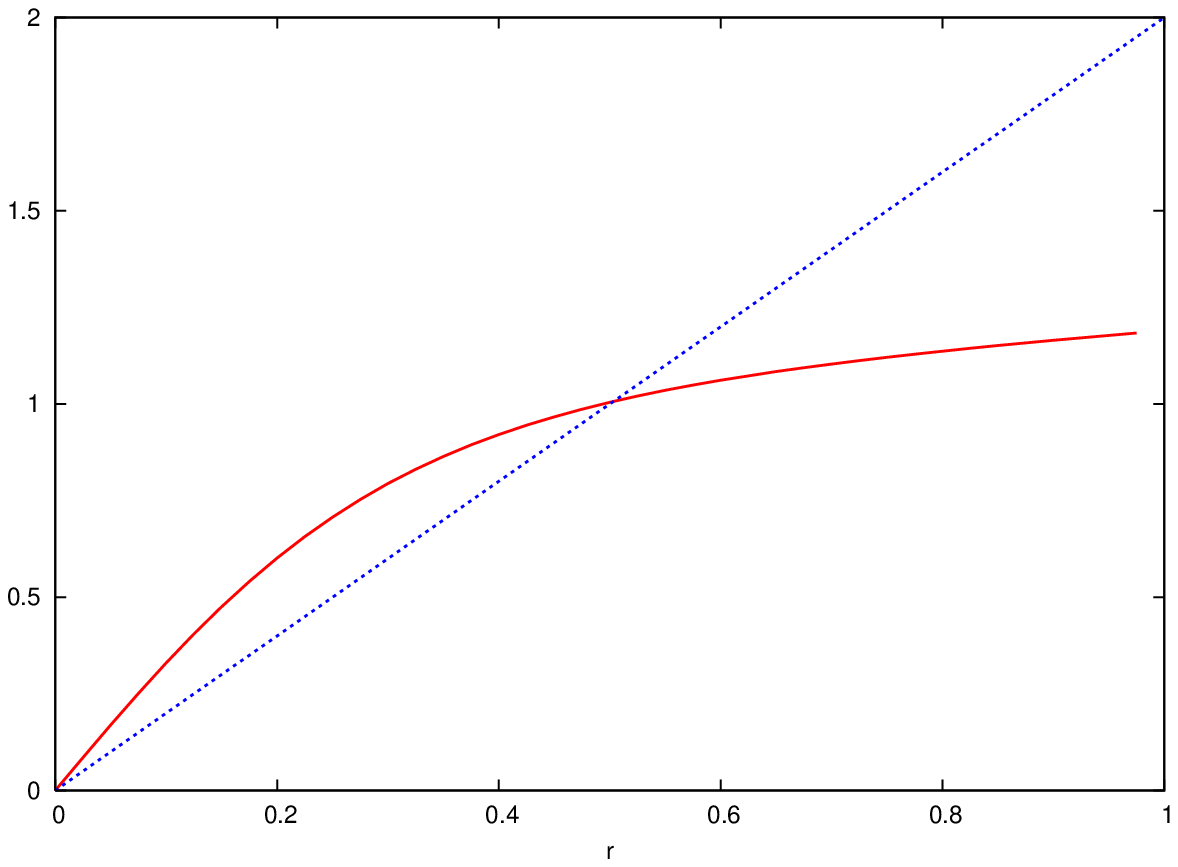}} &
      \resizebox{80mm}{!}{\includegraphics{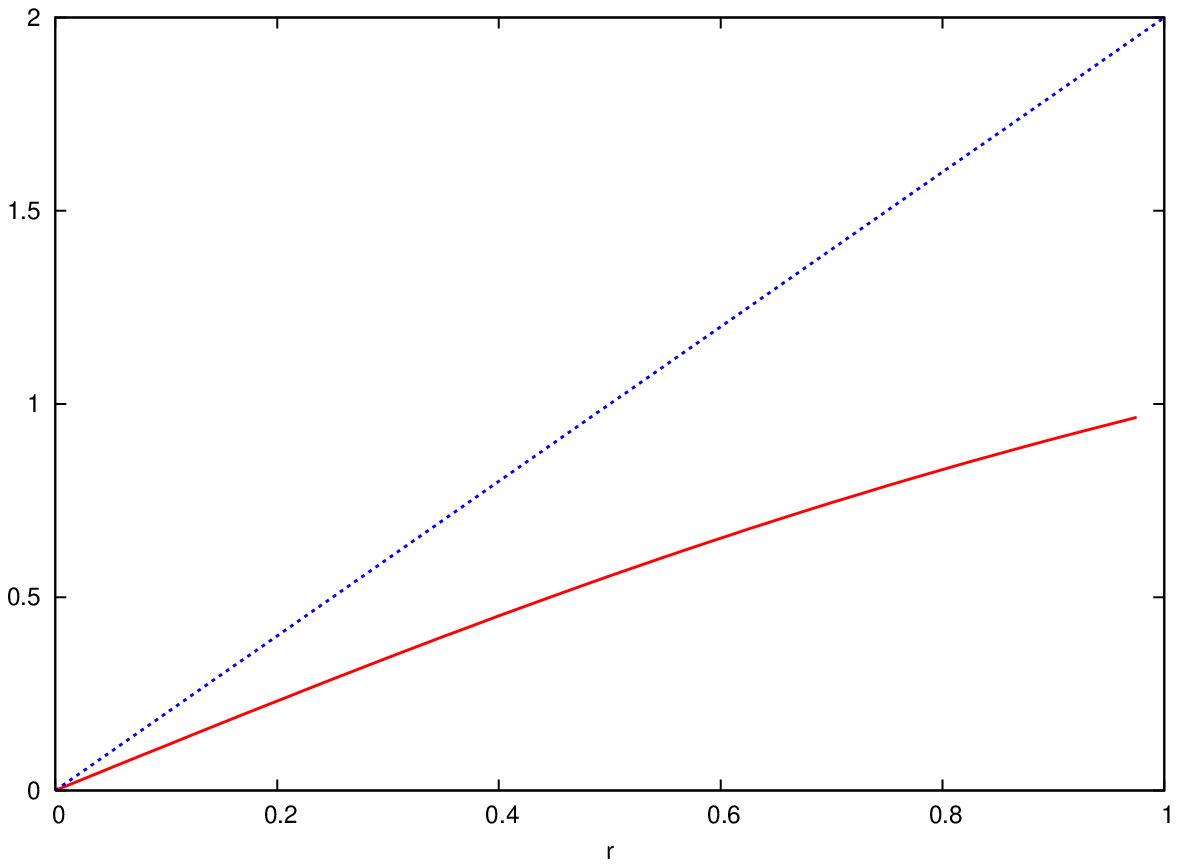}} \\
    \end{tabular}
    \caption[Comparison of $2r$ with $\la a+a^\ast\ra_{h(r)}$ with $\beta=4$, $\lambda=5$ 
    for the cases $\mu=1.5$ (condensation) and $\mu=5$ (no condensation)]{\it Comparison of 
    $2r$ with $\la a+a^\ast\ra_{h(r)}$ with $\beta=4$, $\lambda=5$ 
    for the cases $\mu=1.5$ (condensation) and $\mu=5$ (no condensation)}
    \label{figs2}
  \end{center}
\end{figure}

Equation (\ref{euler-lagrange}) can have at most two solutions. 
Clearly $r=0$ is always a solution. When $\beta$ is large enough, for certain values
of $\mu$ a second non-zero solution may appear (see Figure \ref{figs2}). So $r_\mu \vcentcolon= 0$
unless a second solution $r>0$ exists, in which case $r_\mu \vcentcolon= r$. 

Bru and Dorlas then obtained the properties of this model by using some numerical techniques
to find this maximal solution to
the Euler-Lagrange equation and use it to evaluate the pressure using (\ref{pressure}).

\begin{figure}[hbt]
\begin{center}
\includegraphics[width=16cm]{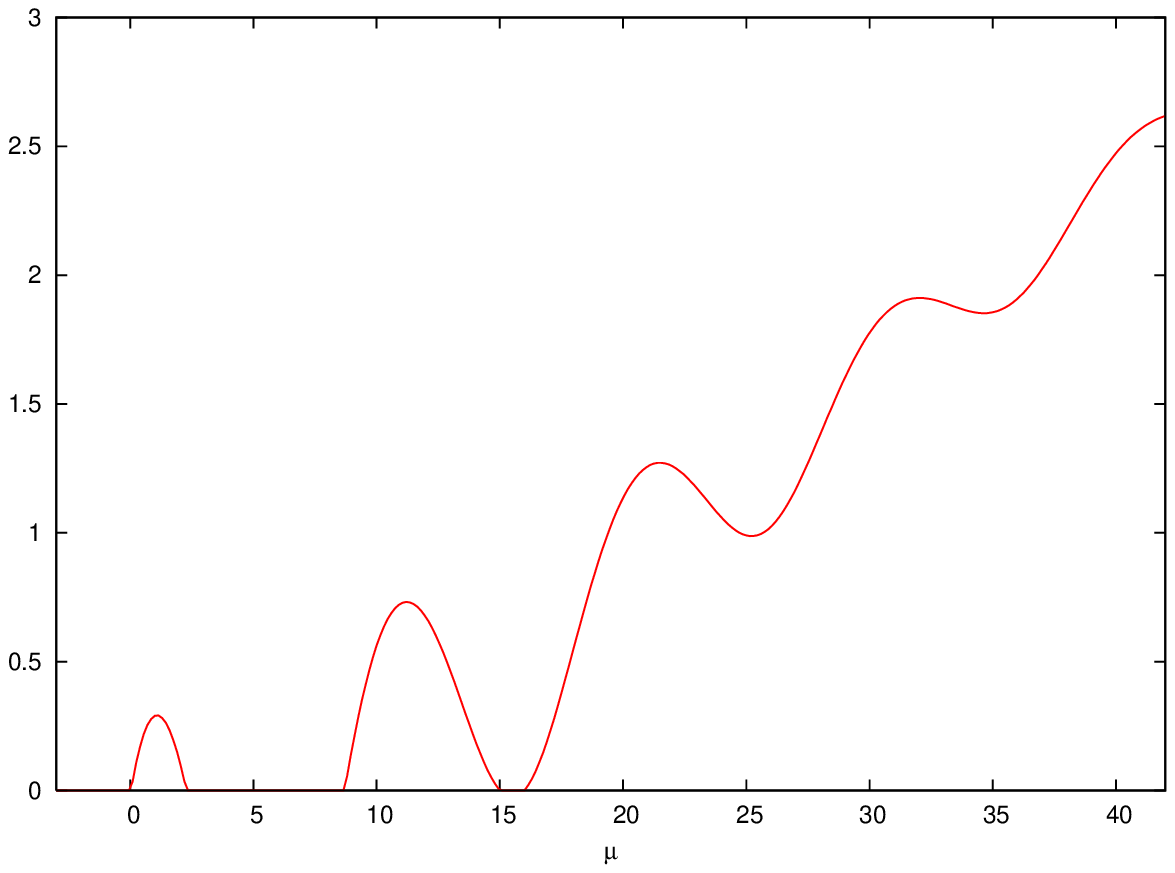}
\end{center}
\caption[Plot of $r_\mu^2 ( =\rhocond)$ versus $\mu$, for $\beta=4$, $\lambda=5$]{\it Plot of 
$r_\mu^2 ( =\rhocond)$ versus $\mu$, for $\beta=4$, $\lambda=5$}
\label{04fig1}
\end{figure}

As may be seen from Figure \ref{04fig1}, for sufficiently large $\beta$, there may exist several critical 
values of $\mu$ which correspond to intervals of $r_\mu = 0$ and $r_\mu > 0$. 

In addition Bru and Dorlas showed that Theorem \ref{thm_soln} holds in the presence of the gauge-symmetry 
breaking term. In that case, the corresponding Euler-Lagrange equation has a unique non-zero solution 
$r_\mu(\nu)$.


\section{Cycle Statistics and the \texorpdfstring{$q$}{q}-particle Hop}				\label{Section2}

From Chapter \ref{chapter2}, we have established, in the grand canonical ensemble,
that the density of particles on cycles of infinite length is
\[
	\rholong = \lim_{Q\to\infty} \sum_{q=Q+1}^\infty \thermlim \gC(q)
\]
where $\gC(q)$ is the density of particles on cycles of length $q$.
We intend to evaluate this expression for the Infinite-Range-Hopping Bose-Hubbard model
and compare it to the (already known) condensate density $\rhocond$.

In order to proceed, it is best to reaffirm the notation used in this chapter.
As before, the Hilbert space for a single particle
on a lattice of $V$ sites is $\Hone\vcentcolon=\CC^\sV$ and upon it we define the operator
\[
	h_\sV=I-P_\sV
\]
where $P_\sV$ is the orthogonal projection onto the ground state $\mathbf{g}_\sV$.
We can define the free Hamiltonian acting on the unsymmetrised Fock space $\FF$
by the second quantization of the operator $h_\sV$:
\[
	\gHam = d\Gamma(h_\sV).
\]
For bosons we consider the symmetric subspace of $\FF$, denoted $\FFsym$.

The operator which counts the number of particles at site labelled $x$, denoted $n_x$, is defined by
$n_x = d\Gamma(P_x) \vcentcolon= d\Gamma( |\mathbf{e}_x \ra\la \mathbf{e}_x |)$. Then
the total number operator is $N_\sV \vcentcolon= \sum_{x=1}^V n_x \equiv d\Gamma(I)$.

Then the Bose-Hubbard Hamiltonian acting upon $\FFsym$ may be expressed as
\[
	\gHam = d\Gamma(h_\sV) + \lambda \sum_{x=1}^V n_x (n_x - 1).
\]
Note $\gHam$ is defined for all particles (not just bosons), i.e. $\gHam$ may also
act upon $\FF$.


Recall we fixed $I^{(q)}$ as the identity operator on $\Hcan{q}$, and $\mathbb{I}$ as that for $\FFsym$.
By Theorem \ref{thm1-cycledens}, we have the following expression for the $q$-cycle density:
\begin{equation}							\label{05qcycle}
	\gC(q) 
=	\frac{1}{\gPart V} \; \trace_{\qFFsym}
	\left[ ( U_q \otimes \II ) \e^{-\beta(\gHam -\mu N_\sV)} \right]
\end{equation}
where $\qFFsym \vcentcolon= \Hcan{q} \otimes \FFsym$.

Note that partitioning the partition function into its cycle structure is only possible if
(\ref{Cycle_Cond1}) holds, i.e.
\[
	\trace_{\Hcan{n}} \left[ U_\pi \e^{-\beta \gHam} \right] \ge 0,
\]
however here by the random walk formulation one can see that the kernel of $\e^{-\beta \gHam}$
is positive and therefore the left-hand side of this expression is positive.

As is evident from equation (\ref{05qcycle}), the cycle statistics technique of Chapter \ref{chapter2}
results in splitting the symmetric Fock space $\FFsym$ into the tensor product of two spaces, an 
unsymmetrised $q$-particle space $\Hcan{q}$ and a symmetrised Fock space $\FFsym$. 
Similarly to the previous chapter, it is convenient to have a notation to 
split the action of an operator on $\FFsym$ to those upon $\Hcan{q}$ and $\FFsym$ individually.
Write
\begin{alignat*}{4}
	A^{(q)} &\vcentcolon= A^{(q)} \otimes \mathbb{I}
	\qquad&\text{and}&\qquad
	&d\Gamma'(A) &\vcentcolon= I^{(q)} \otimes d\Gamma(A)
\intertext{for any operator $A$ on $\Hone$. In this fashion, the number operators 
applied to $\mathcal{H}_{q, \sV}$ are defined as \vspace{-0.4cm}}
	N_x &\vcentcolon= P_x^{(q)} \otimes \mathbb{I}
	\qquad&\text{and}&
	&n_x &\vcentcolon= I^{(q)} \otimes d\Gamma(P_x).
\end{alignat*}
Then in this notation we may define $\gHam^{(q)}$ on $\mathcal{H}_{q, \sV}$ by
\[
	\gHam^{(q)} = h_\sV^{(q)} + d\Gamma'(h_\sV) + 
			\lambda \sum_{x=1}^V (n_x + N_x)(n_x + N_x - 1).
\]
As in the hard-core case, we intend to drop the contribution of the hopping of the $q$ 
unsymmetrised particles and see if this has any influence upon the cycle-density
expression in the thermodynamic limit. To do so define the modified Hamiltonian by the following:
\[
	\widetilde{H}_{\sV}^{(q)} \vcentcolon= I^{(q)} +
		d\Gamma'(h_\sV) + \lambda \sum_{x=1}^V (n_x + N_x)(n_x + N_x - 1)
\]
so that $\gHam^{(q)} = \widetilde{H}_{\sV}^{(q)} - P_\sV^{(q)}$, and then define the modified
cycle density, that which neglects the hopping of the $q$ distinguishable particles, by
\[
	\mgC(q) 
	= \frac{1}{\gPart} \frac{1}{V} \trace_{\qFFsym}
		\bigg[ U_q \e^{-\beta (\widetilde{H}_{\sV}^{(q)} - \mu N_\sV) } \bigg].
\]
We wish to prove for each $q$ that the cycle density $\gC(q)$ and the modified cycle density $\mgC(q)$
are equivalent in the thermodynamic limit, i.e.

\begin{theorem} 								\label{Prop2}
\[
	\thermlim \gC(q) = \thermlim \mgC(q)
\]
\end{theorem}
which implies that in the thermodynamic limit, we are able to disregard the hopping of the 
$q$-unsymmetrised particles in the cycle density.

\textbf{Proof:} 
The proof of this Theorem is structurally similar to that of Lemma \ref{c} from
the previous chapter when we considered the Hard-Core model. However 
there are sufficient differences between the two approaches to justify giving the complete proof.
We wish to obtain an upper bound for the following modulus:
\[
	\left |\trace_{\qFFsym}
	\Big[ U_q \e^{-\beta (\gHam - \mu N_\sV) } \Big]
	- \trace_{\qFFsym}
	\Big[ U_q \e^{-\beta (\widetilde{H}_\sV^{(q)} - \mu N_\sV) } \Big]\right |.
\]
Before proceeding, we first shall introduce a basis of the Hilbert space $\qFFsym$.
Let $\{ \phi_k \}_{k=0}^\infty$ be an orthonormal basis for $\FFsym$.

Let $\Lambda_{\sV}^{(q)}$ be the set of ordered $q$-tuples of (not necessarily distinct) indices 
of $\Lambda_\sV$ and for $\mathbf{i}\vcentcolon=(i_1, i_2, \dots , i_q) \in \Lambda_{\sV}^{(q)}$
let
\[
	|\ii\ra = | i_1, i_2, \dots , i_q \ra \vcentcolon= \mathbf{e}_{i_1} \otimes \mathbf{e}_{i_2}
	\otimes \dots \otimes \mathbf{e}_{i_q}.
\]
Then $\{|\ii \ra\,|\,\mathbf{i}\in \Lambda_{\sV}^{(q)}\}$ is an orthonormal basis for $\Hcan{q}$.

A basis for $\qFFsym$ may therefore be formed by taking the tensor product of the bases
of $\Hcan{q}$ and $\FFsym$, so the set
$\{|\mathbf{i} \ra \otimes \phi_k  \,|\, k=1,2,\dots ; \mathbf{i}\in \Lambda_{\sV}^{(q)} \}$
is an orthonormal basis for $\qFFsym$. For brevity we shall write
\begin{equation}									\label{basis-hvq}
	| \ii; k \ra \vcentcolon= |\mathbf{i} \ra \otimes \phi_k.
\end{equation}

For simplicity, denote $P \vcentcolon= P_\sV^{(q)}$ and $H \vcentcolon= \widetilde{H}_{\sV}^{(q)} - \mu N_\sV$.
We expand
\[
	\trace_{\qFFsym} \left[U_q \e^{-\beta (\gHam - \mu N_\sV )} \right]
=	\trace_{\qFFsym} \left[U_q \e^{-\beta (H - P) } \right]
\]
in a Dyson series in powers of $P$. If $m\ge1$, the $m^\text{th}$ term of this series is
\begin{multline}									\label{mth-term-dyson}
	X_m
\vcentcolon=	\beta^m \int_0^1\hskip -0.3cm ds_1 \int_0^{s_1} \hskip -0.4cm ds_2
	\dots  \int_0^{s_{m-1}}\hskip -0.8cm ds_m \;
	\trace_{\qFFsym} \bigg[ \e^{-\beta H (1-s_1)} P \e^{-\beta H (s_1-s_2)} P\cdots
\\
	\cdots P \e^{-\beta H (s_{m-1}-s_m)} P \e^{-\beta H s_m} U_q \bigg]. \qquad
\end{multline}
Let $P_r \vcentcolon= I \otimes \dots \otimes \underbrace{P_\sV}_{r^{\text{th}} \text{place}} \otimes \dots \otimes I$,
so that $P = \sum_{r=1}^q P_r$. Then  \vspace{-0.4cm}
\[
	X_m = \beta^m \sum_{r_1=1}^q \sum_{r_2=1}^q \dots \sum_{r_m=1}^q X_m( r_1, r_2, \dots , r_m )
\]
where 
\begin{multline}									\label{mth-term-dyson-r}
	X_m( r_1, r_2, \dots , r_m )
=	\int_0^1\hskip -0.3cm ds_1 \int_0^{s_1} \hskip -0.4cm ds_2
	\dots  \int_0^{s_{m-1}}\hskip -0.8cm ds_m \;
	\trace_{\qFFsym} \bigg[ \e^{-\beta H (1-s_1)} P_{r_1} \e^{-\beta H (s_1-s_2)} P_{r_2} \cdots
\\
	\cdots P_{r_{m-1}} \e^{-\beta H (s_{m-1}-s_m)} P_{r_m} \e^{-\beta H s_m} U_q \bigg]. \quad
\end{multline}
In terms of (\ref{basis-hvq}), the basis of $\qFFsym$, we may write
\begin{multline}									\label{04trace_term}
	X_m( r_1, r_2, \dots , r_m )
=	\int_0^1\hskip -0.3cm ds_1 \int_0^{s_1} \hskip -0.4cm ds_2
	\dots  \int_0^{s_{m-1}}\hskip -0.8cm ds_m \;\;
	\sum_{k^0, \dots\ ,k^m} \;\; \sum_{\ii^0} \cdots
	\sum_{\ii^m }
\\
	\qquad\qquad
	\la \ii^0; k^0 | \e^{-\beta H (1-s_1)} P_{r_1}
	| \ii^1; k^1 \ra
	\la \ii^1; k^1 | \e^{-\beta H (s_1-s_2)} P_{r_2}
	| \ii^2; k^2 \ra \cdots
\\
	\cdots
	\la \ii^{m-1}; k^{m-1} |\e^{-\beta H (s_{m-1} -s_m)} P_{r_m}
	| \ii^m; k^m \ra
	\la \ii^m; k^m |\e^{-\beta H s_m}  U_q  | \ii^0; k^0 \ra
\end{multline}
where it is understood that the $\ii$ summations are over $\Lambda_{\sV}^{(q)}$,
the set of ordered $q$-tuples (not necessarily distinct) of $\Lambda_\sV$,
and the $k$ summations are over the bases for $\FFsym$.

Notice that we may express
\[
	\e^{-\beta H s} | \ii; k \ra
= 	\e^{-\beta q (1-\mu) s} \; | \ii; \e^{-\beta H^\ii s } | k \ra
\]
where
\[
	H^\ii \vcentcolon= \gHam - \mu N_\sV + \lambda \sum_{x=1}^V ( N^\ii_x ( N^\ii_x - 1) + 2N^\ii_x n_x )
\]
and $N^\ii_x \vcentcolon = \sum_{j=1}^q \delta_{x, i_j}$ counts the number of particles at site $x$ which are in $q$-space.
Also, for any fixed $r$:
\[
	P_r | \ii; k \ra 
= 	\frac{1}{V} \sum_{j=1}^V | i_1,\dots,\widehat{i_r}, j,\dots,i_q ; k \ra
= 	\frac{1}{V} \sum_{j=1}^V | [r,j](\ii) ; k \ra
\]
using the same index replacement notation $[ \,\cdot\, , \,\cdot\, ]$ as in Chapter \ref{chapter4}.

Using these facts, a single inner product term of (\ref{04trace_term}) may be expressed as
\begin{align*}
	\la \ii ; k | \e^{-\beta H s} P_r | \jj; k' \ra
&=	\e^{-\beta q (1-\mu) s} \; \la k |
	\e^{-\beta H^\ii s } | k' \ra \; \la \ii | P_r| \jj \ra
\\
&=	\frac{\e^{-\beta q (1-\mu) s}}{V} \la k | \e^{-\beta H^\ii s } | k' \ra
	\sum_{m=1}^V \la \ii | j_1, \dots \widehat{j_r}, m, \dots, j_q \ra 
\\
&=	\frac{\e^{-\beta q (1-\mu) s}}{V} \la k | \e^{-\beta H^\ii s } | k' \ra
 	\sum_{m=1}^V \delta_{i_1, j_1} \dots \widehat{\delta_{i_r, j_r}} \delta_{i_r, m} \dots \delta_{i_q, j_q}
\\
&=	\frac{\e^{-\beta q (1-\mu) s}}{V} \la k | \e^{-\beta H^\ii s } | k' \ra \;
	\delta_{i_1, j_1} \dots \widehat{\delta_{i_r, j_r}} \dots \delta_{i_q, j_q}.
\end{align*}

Now if we sum over $\jj$
\begin{align*}
	\sum_{\jj} \la \ii ; k | \e^{-\beta H s} P_r | \jj; k' \ra \la \jj; k' |
&=	\frac{\e^{-\beta q (1-\mu) s}}{V} \la k | \e^{-\beta H^\ii s } | k' \ra 
	\sum_{j_r=1}^V \la i_1, \dots, \widehat{i_{r}}, j_r, \dots, i_q; k' |
\\
&=	\frac{\e^{-\beta q(1-\mu) s}}{V} \la k | \e^{-\beta H^\ii s } | k' \ra 
	\sum_{j_r=1}^V \la [r, j_r](\ii) ; k' |.
\end{align*}
Performing two summations for fixed $r_1$ and $r_2$ we get:
\begin{align*}
&	\sum_{\ii^1} \sum_{\ii^2} \la \ii^0 ; k^0 | \e^{-\beta H s} P_{r_1} | \ii^1; k^1 \ra \;
	\la \ii^1; k^1 | \e^{-\beta H t} P_{r_2} | \ii^2; k^2 \ra \; \la \ii^2; k^2 |
\\
&\quad= 	\frac{\e^{-\beta q (1-\mu) s}}{V} \sum_{i^1_{r_1}=1}^V \sum_{\ii^2} 
	\la k^0 | \e^{-\beta H^{\ii^0} s } | k^1 \ra \; \la [r_1, i^1_{r_1}](\ii^0) ; k^1 |
	 \e^{-\beta H t} P_{r_2} | \ii^2; k^2 \ra \; \la \ii^2; k^2 |
\\
&\quad=	\frac{\e^{-\beta q (1-\mu) (s+t)}}{V^2} \sum_{i^1_{r_1}=1}^V \sum_{i^2_{r_2}=1}^V 
	\la k^0 | \e^{-\beta H^{\ii^0} s } | k^1 \ra \; \la k^1 | \e^{-\beta H^{[r_1, i_{r_1}](\ii^0)} t } | k^2 \ra \;
	\la [r_2, i^2_{r_2} ; r_1, i^1_{r_1} ](\ii^0) ; k^2 |.
\end{align*}
Thus (\ref{mth-term-dyson-r}) looks like
{\allowdisplaybreaks
\begin{align*}
  	X_m&( r_1, r_2, \dots , r_m )
\\*
=& \;	\frac{ \e^{-\beta q(1-\mu)} }{V^m} 
	\int_0^1\hskip -0.3cm ds_1 \int_0^{s_1} \hskip -0.4cm ds_2
	\dots  \int_0^{s_{m-1}}\hskip -0.8cm ds_m \;\;
	\sum_{k^0 \dots k^m} \sum_{\ii^0}
	\sum_{i^1_{r_1} = 1}^V
	\sum_{i^2_{r_2} = 1}^V
	\cdots \sum_{i^m_{r_m} =1}^V
	\la k^0 |  \e^{-\beta(1-s_1) H^{\ii^0} }
	| k^1 \ra
\\*
&	\qquad \la k^1 |
	\e^{-\beta (s_1-s_2) H^{[r_1,i^1_{r_1}](\ii^0)}}
	| k^2 \ra
	\la k^2 | \e^{-\beta (s_2-s_3)
	H^{[r_2, i^2_{r_2} ; r_1, i^1_{r_1}](\ii^0)} }
	| k^3 \ra
	\cdots
\\*
&	\qquad \cdots
	\la k^m |
	\e^{-\beta s_m H^{ [r_m, i^m_{r_m}; \, \dots \, ; r_2, i^2_{r_2} ; r_1, i^1_{r_1}](\ii^0)} }
	| k^0 \ra
	\la [r_m, i^m_{r_m}; \, \dots \, ; r_2, i^2_{r_2} ; r_1, i^1_{r_1}](\ii^0) | U_q \ii^0 \ra
\\[0.4cm]
=&\;	\frac{\e^{-\beta q(1-\mu)} }{V^m}
	\int_0^1\hskip -0.3cm ds_1 \int_0^{s_1} \hskip -0.4cm ds_2
	\dots  \int_0^{s_{m-1}}\hskip -0.8cm ds_m
\\*
&	\qquad \sum_{\ii^0}
	\sum_{i^1_{r_1} = 1}^V
	\sum_{i^2_{r_2} = 1}^V
	\cdots \sum_{i^m_{r_m} = 1}^V
	\la [r_m, i^m_{r_m}; \, \dots \, ; r_2, i^2_{r_2} ; r_1, i^1_{r_1}](\ii^0) | U_q \ii^0 \ra
\\*
&	\qquad \trace_{\FFsym} \Bigg[ \e^{-\beta (1-s_1) H^{\ii^0} }
	\e^{-\beta (s_1-s_2) H^{[r_1, i^1_{r_1}](\ii^0)} }
	\cdots \cdots
	\e^{-\beta s_m H^{[r_m, i^m_{r_m}; \, \dots \, ; r_2, i^2_{r_2} ; r_1, i^1_{r_1}](\ii^0)} }
	\Bigg].
\end{align*}}
Using the H\"older inequality, for 
non-negative trace class operators $A_1, A_2, \dots , A_{m+1}$ we have that
\[
	\left| \trace \big( A_1 A_2 \dots A_{m+1} \big) \right|
\le	\trace \big| A_1 A_2 \dots A_{m+1} \big|
\le	\prod_{k=1}^{m+1} \big( \trace A_k^{p_k} \big)^{\tfrac{1}{p_k}}
\]
where $\sum_{k=1}^{m+1} \tfrac{1}{p_k} = 1$, $p_i > 0$.

Set $p_1 = \frac{1}{1-s_1},\ p_2 = \frac{1}{s_1 - s_2},\ \dots ,\ p_m = \frac{1}{s_{m-1}-s_m},
p_{m+1} = \frac{1}{s_m}$. Taking the modulus of the above trace
\begin{align*}
&	\Bigg| \trace_{\FFsym}  \Bigg[
	\e^{-\beta H^{\ii^0} (1-s_1)}
	\e^{-\beta H^{[r_1, i^1_{r_1}](\ii^0)} (s_1-s_2)}
	\; \cdots
	\cdots
	\e^{-\beta H^{[r_m, i^m_{r_m}; \, \dots \, ; r_2, i^2_{r_2} ; r_1, i^1_{r_1}](\ii^0)} (s_m)}
	\Bigg]  \Bigg|
\\
&\le	\quad \trace_{\FFsym} \bigg[
	\e^{-\beta H^{\ii^0}} \bigg]^{1-s_1}
	\trace_{\FFsym} \bigg[
	\e^{-\beta H^{[r_1, i^1_{r_1}](\ii^0)}} \bigg]^{s_1-s_2} \cdots
\\
&	\qquad\qquad\qquad \cdots \trace_{\FFsym}
	\bigg[
	\e^{-\beta H^{[r_m, i^m_{r_m}; \, \dots \, ; r_2, i^2_{r_2} ; r_1, i^1_{r_1}](\ii^0)}}
	\bigg]^{s_m} \!\!\!.
\end{align*}
But since the trace is independent of the $V-q$ sites 
$\{\ii^0, [r_1,i^1_{r_1}](\ii^0),[r_2, i^2_{r_2}; r_1,i^1_{r_1}](\ii^0) \,  \dots$ 
$\dots ,[r_m, i^m_{r_m}; \, \dots\ \, ; r_2, i^2_{r_2} ; r_1, i^1_{r_1}](\ii^0)\}$,
the product of all the trace terms above is equal to
\[
	\trace_{\FFsym} \left[ 
		\e^{-\beta H^{\boldl}} \right]
\]
with $\boldl = \{V-q+1, V-q+2, \dots ,V\}$.

This is independent of the $\ii^0$ and $i$ summations, so we need only consider
\[								
	\sum_{i^1_{r_1} = 1}^V
	\sum_{i^2_{r_2} = 1}^V
	\cdots \sum_{i^m_{r_m} = 1}^V
	\sum_{\ii^0}
	\la [r_m, i^m_{r_m}; \, \dots \, ; r_2, i^2_{r_2} ; r_1, i^1_{r_1}](\ii^0) | U_q \ii^0 \ra.
\]
Fix the values of $i^1_{r_1}, i^2_{r_2}, \dots , i^m_{r_m}$. We intend to show that
\[
	\sum_{\ii^0}
	\la [r_m, i^m_{r_m}; \, \dots \, ; r_2, i^2_{r_2} ; r_1, i^1_{r_1}](\ii^0) | U_q \ii^0 \ra = 1.
\]
If $\{r_1, r_2, \dots , r_m\} \ne \{1,2,\dots, q\} $, 
then $| [r_m, i^m_{r_m}; \, \dots \, ; r_2, i^2_{r_2} ; r_1, i^1_{r_1}](\ii^0) \ra$ is of the form\vspace{-0.2cm}
\[
	| j_1, j_2, \dots ,j_{n_1},i^0_{n_1+1}, \dots ,i^0_{n_2}, j_{n_2+1},
	\dots, j_{n_3}, i^0_{n_3+1}, \dots  ,
	i^0_{n_4}, j_{n_4+1}, \dots \dots\ra
\]
where $\{n_1, n_2, \dots \}$ is a non-empty ordered set of distinct integers between 0 and $q$.
This vector is clearly orthogonal to $U_q \ii^0$ except for the single choice of\vspace{-0.2cm}
\[
	\ii^0 = |  j_2, \dots , j_{n_1-1}, j_{n_1}, \dots , j_{n_1}, j_{n_2+1},
	\dots, j_{n_3-1}, j_{n_3}, \dots ,
	j_{n_3}, j_{n_4+1}, \dots \dots , j_1\ra.
\]
For the case $\{r_1, r_2, \dots , r_m\} = \{1,2,\dots, q\}$ 
notice that $| [r_m, i^m_{r_m}; \, \dots \, ; r_2, i^2_{r_2} ; r_1, i^1_{r_1}](\ii^0) \ra$ is independent of
$\ii^0$ so we may take it to be\vspace{-0.2cm}
\[
	|[r_m, i^m_{r_m}; \, \dots \, ; r_2, i^2_{r_2} ; r_1, i^1_{r_1}](\mathbf{s}^0) \ra\vspace{-0.2cm}
\]
where $\mathbf{s}^0 = (1,2,3,\dots,q)$. For each choice of $i^1_{r_1}, i^2_{r_2}, \dots , i^m_{r_m}$ 
there exists only one possible $\ii^0 \in \Lambda_\sV^{(q)}$ such that
\[
	\la [r_m, i^m_{r_m}; \, \dots \, ; r_2, i^2_{r_2} ; r_1, i^1_{r_1}](\mathbf{s}^0) | U_q \ii^0 \ra \ne 0.
\]
So we may conclude that
\begin{equation*}
	\sum_{\ii^0}
	\sum_{i^1_{r_1} = 1}^V
	\sum_{i^2_{r_2} = 1}^V
	\cdots \sum_{i^m_{r_m} = 1}^V
	\la [r_m, i^m_{r_m}; \, \dots \, ; r_2, i^2_{r_2} ; r_1, i^1_{r_1}](\ii^0) | U_q \ii^0 \ra
	= V^m
\end{equation*}
and by applying this, we see that the modulus of (\ref{mth-term-dyson-r}) may bounded above by
\begin{align*}
    |X_m( r_q, r_2, \dots, r_m)| & \le \trace_{\FFsym} \left[ 
		\e^{-\beta \Htilde^{\boldl}} \right]
		\e^{-\beta q(1-\mu)} \frac{1}{m! V^m} 
\\
&   \qquad \times
	\sum_{\ii^0}
	\sum_{i^1_{r_1} = 1}^V
	\sum_{i^2_{r_2} = 1}^V
	\cdots \sum_{i^m_{r_m} = 1}^V
	\la [r_m, i^m_{r_m}; \, \dots \, ; r_2, i^2_{r_2} ; r_1, i^1_{r_1}](\ii^0) | U_q \ii^0 \ra
\\[0.2cm]
&=	\trace_{\FFsym} \left[ 
		\e^{-\beta \Htilde^{\boldl}} \right]\e^{-\beta q(1-\mu)} \frac{1}{m!}.
\end{align*}
which is independent of $r_1, r_2, \dots, r_m$.
Hence the modulus of (\ref{mth-term-dyson}), the $m^\text{th}$ term of the Dyson series, may be bounded above by
\begin{align*}
	| X_m | 
\enspace\le\enspace
	\beta^m \sum_{r_1=1}^q \cdots \sum_{r_m=1}^q |X_m( r_q, r_2, \dots, r_m)|
\enspace\le\enspace
	\trace_{\FFsym} \left[ 
		\e^{-\beta \Htilde^{\boldl}} \right]\e^{-\beta q(1-\mu)} \frac{q^m \beta^m}{m!}.
\end{align*}
Noting that the zeroth term of the Dyson series is
\[
	X_0 = \trace_{\mathcal{H}_{q,V}} \left[U_q \e^{-\beta H} \right] =
	\trace_{\mathcal{H}_{q,V}} \left[U_q \e^{-\beta ( \widetilde{H}_\sV^{(q)} - \mu N_\sV ) } \right],
\]
we may re-sum the series to obtain
\begin{align*}
&\Bigg| \trace_{\mathcal{H}_{q,\sV}}
            \left[U_q \e^{-\beta (H_{\sV} - \mu N_\sV) } \right]
        - \trace_{\mathcal{H}_{q,\sV}}
            \left[U_q \e^{-\beta ( \widetilde{H}_\sV^{(q)} - \mu N_\sV ) } \right]  \Bigg| 
\\
& \quad \le \trace_{\FFsym} \left[
	\e^{-\beta \Htilde^{\boldl}}  \right]\e^{-\beta q(1-\mu)} 
	\sum_{m=1}^\infty \frac{q^m \beta^m}{m!}
\\[0.2cm]
& \quad = \trace_{\FFsym} \left[
	\e^{-\beta \Htilde^{\boldl}}  \right]
	\e^{\beta q\mu} ( 1 - \e^{-\beta q}).
\end{align*}
Thus
\begin{align*}
   	\left|\gC(q)-\mgC(q)\right |
 &= 	\frac{1}{V} \left| \frac{
	\trace_{\mathcal{H}_{q,\sV}}
		\big[U_q \e^{-\beta ( H_{\sV} - \mu N_\sV) } \big]
	- \trace_{\mathcal{H}_{q,\sV}}
		\big[U_q \e^{-\beta ( \widetilde{H}_\sV^{(q)} - \mu N_\sV ) } \big]}{\gPart} \right|
\\[0.2cm]
  &\le 	\frac{\e^{\beta q\mu} ( 1 - \e^{-\beta q})}{V} 
  	\frac{\trace_{\FFsym} \big[
		\e^{-\beta \Htilde^{\boldl}}  \big]}{\gPart}.
\end{align*}
Finally considering the ratio of the traces, since $H_{\sV} - \mu N_\sV - \Htilde^{\boldl} = \lambda 
\sum_{x=1}^\sV ( N^\boldl_x ( N^\boldl_x - 1) + 2N^\boldl_x n_x ) \ge 0$,
the second fraction is not greater than 1, implying\vspace{0.2cm}
\[
	\left|\gC(q)-\mgC(q)\right |
\le	\frac{\e^{\beta q\mu} ( 1 - \e^{-\beta q})}{V}\vspace{0.2cm}
\]
which goes to zero in the limit $V \to \infty$, as desired.\hfill $\square$
\pagebreak
\begin{corollary}							\label{thm-short-cycles}
The density of cycles of finite length in the IRH Bose-Hubbard model may be expressed as
\[
	\rhoshort
=	\sum_{q=1}^\infty \e^{-\beta(q-\mu)q} \thermlim 
	\frac{ \trace_{\FFsym} \exp\big\{ -\beta ( 2\lambda q n_1 + H_\sV - \mu N_\sV ) \big\} }
	{ \trace_{\FFsym} \exp\big\{ -\beta ( H_\sV - \mu N_\sV ) \big\} }.
\]
where $n_1$ is an operator which counts the number of bosons on the site labelled $1$.
\end{corollary}

\begin{proof}
Note that we may easily re-express the modified cycle density expression:
{\allowdisplaybreaks
\begin{align*}
	\mgC(q)
=&	\frac{\e^{-\beta (1-\mu) q}}{\gPart V}
   	\sum_{i_1=1}^V \dots \sum_{i_q=1}^V \sum_{k=1}^\infty
	\la i_1, i_2, \dots, i_q; k |
\\*
&	\quad\exp\left\{ -\beta \left( d\Gamma'(h_\sV)
		+ \lambda \sum_{x=1}^V (n_x + N_x)(n_x + N_x -1) - \mu \sum_{x=1}^V n_x
	\right) \right\}
\\*
&	\quad U_q | i_1, i_2, \dots, i_q; k \ra
\\[0.3cm]
=&	\frac{\e^{-\beta (1-\mu) q}}{\gPart V}
   	\sum_{i_1=1}^V \dots \sum_{i_q=1}^V \sum_{k=1}^\infty
	\la i_1, i_2, \dots, i_q |  U_q | i_1, i_2, \dots, i_q \ra
\\*
&	\quad \la k | \exp\left\{ -\beta \left( d\Gamma(h_\sV)
		+ \lambda \sum_{x=1}^V (n_x + \sum_{j=1}^q \delta_{i_j,x})
		(n_x + \sum_{j=1}^q \delta_{i_j,x} -1) - \mu \sum_{x=1}^V n_x
	\right) \right\} | k \ra
\intertext{and as $\la i_1, i_2, \dots, i_q | i_2, i_3, \dots, i_q, i_1 \ra \ne 0$ if and only if
$i_1 = i_2 = \dots = i_q \vcentcolon= i$ then}
=&	\frac{\e^{-\beta (1-\mu) q}}{\gPart V}
   	\sum_{i=1}^V \sum_{k=1}^\infty
	\la k |
\\*
&	\quad	\exp\left\{ -\beta \left( d\Gamma(h_\sV)
		+ \lambda \sum_{x=1}^V (n_x + q \delta_{i x})
		(n_x + q \delta_{i x}  -1) - \mu \sum_{x=1}^V n_x \right) \right\}
	|k \ra
\\[0.3cm]
=&	\frac{\e^{-\beta (1-\mu) q}}{\gPart V}
	\sum_{i=1}^V \trace_{\FFsym}
\\*
&	\quad \exp\Bigg\{ -\beta \Bigg( d\Gamma(h_\sV)
		+ \lambda (n_i + q)(n_i + q  -1)
		+ \lambda \sum_{\stackrel{x=1}{x \ne i}}^V n_x ( n_x - 1)
		- \mu \sum_{x=1}^V n_x \Bigg) \Bigg\}
\\[0.3cm]
=&	\frac{\e^{-\beta (1-\mu) q}}{\gPart}
	\trace_{\FFsym} \Bigg[
	\exp\Bigg\{ -\beta \Bigg( d\Gamma(h_\sV)
		+ \lambda (n_1 + q)(n_1 + q - 1)
\\*
&	\hspace{7cm}
		+ \lambda \sum_{x=2}^V n_x ( n_x - 1)
		 - \mu \sum_{x=1}^V n_x \Bigg) \Bigg\}
	\Bigg]
\end{align*}}
since the trace is independent of the basis chosen. Hence we obtain
\[
	\mgC(q) 
= 	\e^{-\beta (q-\mu) q}
	\frac{ \trace_{\FFsym} \exp\big\{-\beta (2\lambda q n_1 + H_\sV -\mu N_\sV) \big\} }
	{ \trace_{\FFsym} \exp\big\{-\beta (H_\sV -\mu N_\sV) \big\} }.
\]
where $n_1$ is an operator which counts the number of bosons on the site labelled 1 of the lattice.
The result follows by taking the thermodynamic limit and summing over all $q$.
\end{proof}

\section{The Approximating Hamiltonian Method}

In Corollary \ref{thm-short-cycles} we derive the following expression for the density of cycles of
finite length in the IRH Bose-Hubbard model:
\[
	\rhoshort
=	\sum_{q=1}^\infty \e^{-\beta(q-\mu)q} \thermlim 
	\frac{ \trace_{\FFsym} \exp\big\{ -\beta ( 2\lambda q n_1 + H_\sV - \mu N_\sV ) \big\} }
	{ \trace_{\FFsym} \exp\big\{ -\beta ( H_\sV - \mu N_\sV ) \big\} }.
\]
where $n_1$ is an operator which counts the number of bosons on the site labelled $1$.
Unfortunately we are unable to treat this expression analytically. Instead we apply
the Approximating Hamiltonian substitution to this expression and conjecture
that the approaches are equivalent in the limit.

If one were to substitute $H_\sV$ for the approximating Hamiltonian $H^{\text{APP}}_\sV(r_\mu)$ 
(where again $r_\mu$ is the maximal solution to the Euler-Lagrange equation (\ref{euler-lagrange})\!) 
into the right hand side of this expression, one would obtain:
\begin{equation}							\label{conj-short-cycles-approxed} 
	\rhoshort
=	\sum_{q=1}^\infty 
	\frac{  \trace_{\mathcal{F}_{+}(\CC)} \e^{ -\beta h_q(r_\mu) }}
	{ \trace_{\mathcal{F}_{+}(\CC)} \e^{ -\beta h_0(r_\mu)}}
\end{equation}
where 
\[
	h_q(r) \vcentcolon= (1-\mu)(n + q) + \lambda (n + q)(n + q - 1) - r (a + a^\ast)
\]
is another single-site Hamiltonian (note that $h_0(r) = h(r)$).
We conjecture that (\ref{conj-short-cycles-approxed}) gives the
correct expression for $\rhoshort$.

Moreover the fact that a state corresponding to $H_\sV$ in the thermodynamic limit may be shown 
to be a convex combination of one-site product states of the form
\[
	\omega(A) = \frac{\trace_{\mathcal{F}_{+}(\CC)} \e^{-\beta h_q(r_\mu) } A}
	{\trace_{\mathcal{F}_{+}(\CC)} \e^{-\beta h_0(r_\mu)}}
\]
supports this conjecture. We shall prove this conjecture for those values of $\mu$ such that $r_\mu=0$, 
but unfortunately are unable to do so when $r_\mu>0$. However we can prove 
an slightly weaker result with the addition of a gauge-breaking term.

Let $\gC(q,\nu)$ be the density of particles on cycles of length $q$ for 
the gauge-symmetry broken Hamiltonian $H_\sV(\nu)$ and set $c^\mu(q) \vcentcolon= \thermlim \gC(q)$.

\begin{theorem}								\label{prop3-equiv}
For $\mu$ such that $r_\mu=0$, we have 
\[
	c^\mu(q)
= 	\frac{  \trace_{\mathcal{F}_{+}(\CC)} \e^{ -\beta h_q(0) }}
	{ \trace_{\mathcal{F}_{+}(\CC)} \e^{ -\beta h_0(0) }}.
\]

More generally for any $\mu \in \RR$, for a fixed $\nu>0$ there exists a sequence $\nu_\sV \to \nu$ as $V \to \infty$,
independent of $q$ such that
\[
	\thermlim \gC(q,\nu_\sV)
=	\frac{ \trace_{\mathcal{F}_{+}(\CC)} \e^{ -\beta [ h_q(r_\mu(\nu)) - \nu (a + a^\ast) ] }}
	 { \trace_{\mathcal{F}_{+}(\CC)} \e^{ -\beta [ h_0(r_\mu(\nu)) - \nu (a + a^\ast) ] }}.
\]
\end{theorem}

\textbf{Proof:} 
For convenience, denote $\Hv{q} \vcentcolon= 2\lambda q n_1 + (q-\mu)q + H_\sV -\mu N_\sV$. 
Due to Theorem \ref{Prop2} we need to consider
\[
	c^\mu(q)
=	\thermlim \frac{\trace_{\FFsym} \big[\e^{- \beta \Hv{q}}\big]}
	{\trace_{\FFsym} \big[\e^{-\beta \Hv{0}}\big]}.
\]
Recall that
\begin{equation*}
	H_\sV
=	\frac{1}{2V} \sum_{x,y=1}^V (a^\ast_x-a^\ast_y)(a\astr_x-a\astr_y)
	+ \lambda \sum_{x=1}^V n_x (n_x-1).
\end{equation*}

Motivated by the occurrence of the site-specific operator $n_1$ in the numerator of the expression
of $\gC(q)$, we expand the expression for $H_\sV - \mu N_\sV$ to isolate the operators which apply to the site
labelled 1:
\begin{equation*}
	H_\sV - \mu N_\sV
	= (1-\mu) n_1 + \lambda n_1 (n_1 - 1) + \frac{n_1}{V}
		- \frac{a\astr_1}{V} \sum_{x\ne 1} a_x^\ast - \frac{a_1^\ast}{V} \sum_{x\ne 1} a\astr_x
		+ \widetilde{H}_\sV
\end{equation*}
where 
\[
	\widetilde{H}_\sV
\vcentcolon=	(1-\mu)\sum_{x\ne1} n\astr_x - \frac{1}{V} \sum_{x,y\ne1}  a^\ast_x a\astr_y 
	+ \lambda \sum_{x\ne1} n_x ( n_x - 1)
\]
is an Infinite-Range-Hopping Bose-Hubbard Hamiltonian for $\Lambda_\sV\setminus\{1\}$.
By denoting
\[
	h_{q,\sV} \vcentcolon= (1 - \mu)(n_1 + q) + \lambda (n_1 + q) (n_1 + q - 1) +\tfrac{n_1}{V} 
\]
we may then write
\begin{equation*}
	\Hv{q}
	= h_{q,\sV} - \frac{a\astr_1}{V} \sum_{x\ne 1} a_x^\ast - \frac{a_1^\ast}{V} \sum_{x\ne 1} a\astr_x
		+ \widetilde{H}_\sV.
\end{equation*}

Note that $h_{q,\sV} \to h_q$ on $\mathcal{F}_{+}(\CC)$ as $V \to \infty$.

We intend to completely segregate the Hamiltonian $\Hv{q}$ into two individual parts, one which operates solely upon the 
site labelled 1, and the other which applies only to the remaining $V-1$ sites. What prevents us from doing this 
immediately is of course the  ``cross-term''
\[
	\frac{a\astr_1}{V} \sum_{x\ne 1} a_x^\ast + \frac{a_1^\ast}{V} \sum_{x\ne 1} a\astr_x.
\]
Motivated by the Approximating Hamiltonian technique, we shall substitute this term with
\[
	a\astr_1 \bar{R} + a_1^\ast R
\]
for a certain $c_0$-number $R$. Without loss of generality, we may take $R$ to be a 
non-negative real number. Fixing
\[
	h_{q,\sV}(R) \vcentcolon= h_{q,\sV} - R ( a\astr_1 + a_1^\ast )
\]
then the resulting newly approximated Hamiltonian may be expressed as
\[
	\Happ{q}(R) = h_{q,\sV}(R) + \widetilde{H}_\sV.
\]
In the arguments that follow, we shall either take $R$ to equal $r_\mu$
in the variational principle, or a variable depending on $V$ which tends to $r_\mu$ in the limit.

\subsection{Case 1: \texorpdfstring{ Values of $\mu$ such that $r_\mu=0$ --}{} the absence of condensation}

First we shall state and prove the following:
\begin{proposition}								\label{PropB1}
For all $\mu\in\RR$ such that $r_\mu=0$,
\[
	\thermlim \frac{\trace_{\FFsym} \exp\big\{-\beta \Happ{q}(0) \big\}}
	{\trace_{\FFsym} \exp\big\{-\beta \Hv{q} \big\}} = 1.
\]
\end{proposition}

\begin{proof}
Using the Bogoliubov inequality:
\begin{equation} 								\label{bogoliubov}
	\la A-B \ra_B \le \ln \trace e^A - \ln \trace e^B \le \la A-B \ra_A
\end{equation}
for any $R$ we obtain
\begin{multline}								\label{sandwich}
	\beta \left\la a\astr_1 \left( \frac{\sum_{x\ne 1} a_x^\ast }{V} - R \right) 
	\right\ra_{\Happ{q}(R)} +\mathrm{h.c.}
\\	
\le	\ln \trace \exp\big\{-\beta \Hv{q} \big\} - \ln \trace \exp\big\{-\beta \Happ{q}(R) \big\}
\\	
\le	\beta \left\la a\astr_1 \left( \frac{\sum_{x\ne 1} a_x^\ast }{V} - R \right)
	\right\ra_{\Hv{q}} + \mathrm{h.c}.
\end{multline}
Taking the left-hand side, since $\Happ{q}(R)$ is a sum of two Hamiltonians which act upon different Hilbert spaces,
traces and therefore expectations may be easily de-coupled, so one may see that
\[
	\left\la a\astr_1 \left( \frac{\sum_{x\ne 1} a_x^\ast }{V}\right) \right\ra_{\Happ{q}(R)}
= 	\la a\astr_1 \ra_{h_{q,\sV}(R)}
	\left\la \frac{\sum_{x\ne 1} a_x^\ast}{V} \right\ra_{\widetilde{H}_\sV} = 0,
\]
which is zero since $\widetilde{H}_\sV$ is a gauge-invariant Hamiltonian:
$\left\la \sum_{x\ne 1} a_x^\ast \right\ra_{\widetilde{H}_\sV}=0$.

On the right-hand side, note that $\la a_1\ra_{\Hv{q}}=0$, also due to gauge invariance. Therefore we may simplify 
(\ref{sandwich}) to obtain
\begin{multline}							\label{sandwich1}
	-\beta R \big\la a\astr_1 + a^\ast_1 \big\ra_{h_{q,\sV}(R)}
\\
\le	\ln \trace \exp\big\{-\beta \Hv{q} \big\} - \ln \trace \exp\big\{-\beta \Happ{q}(R) \big\}
\\
\le	\beta \left\la a\astr_1 \left( \frac{\sum_{x\ne 1} a_x^\ast }{V}\right) \right\ra_{\Hv{q}} + \mathrm{h.c.}
\end{multline}

For this case we shall take $R=r_\mu = 0$. We therefore obtain
\begin{equation}\label{ineq}
0
\le	\ln \trace \exp\big\{-\beta \Hv{q} \big\} - \ln \trace \exp\big\{-\beta \Happ{q}(0) \big\}
\le	\beta \left\la a\astr_1 \left( \frac{\sum_{x\ne 1} a_x^\ast }{V}\right) \right\ra_{\Hv{q}} + \mathrm{h.c.}
\end{equation}
Now by the Schwarz inequality
\begin{align*}
	\left| \left\la a\astr_1 \left( \frac{\sum_{x\ne 1} a_x^\ast }{V} \right) \right\ra_{\Hv{q}} \right|
&=	\left| \left\la a\astr_1 \left( \frac{\sum_{x} a_x^\ast }{V}\right) - \frac{a\astr_1 a_1^\ast}{V} \right\ra_{\Hv{q}} \right|
\\
&\le	\la  n_1 \ra_{\Hv{q}}^{\tfrac{1}{2}} \left\la \frac{\sum a_x^\ast \sum a\astr_x}{V^2}\right\ra^{\tfrac{1}{2}}_{\Hv{q}}
	+ \frac{\la n_1 \ra_{\Hv{q}}+1}{V}.
\end{align*}

To consider this let $H^s_{q,\sV} \vcentcolon= \Hv{q} + s\frac{\sum a_x^\ast \sum a\astr_x}{V}$
and $\widehat{H}^s_{q,\sV} \vcentcolon= H^s_{q,\sV} - 2 \lambda q n_1$. Using the Bogoliubov inequality
(\ref{bogoliubov}), with $A = -\beta \widehat{H}^s_{q,\sV}$ and $B = -\beta H^s_{q,\sV}$, so that 
$A-B = 2 \beta \lambda q n_1$, one has
\begin{equation}										\label{upperbound}
	2 \beta \lambda q \la n_1 \ra_{H^s_{q,\sV}} 
\le	\ln \trace \e^{-\beta \widehat{H}^s_{q,\sV}} - \ln \trace \e^{-\beta H^s_{q,\sV}}
\le 	2 \beta \lambda q \la n_1 \ra_{\widehat{H}^s_{q,\sV}} 
=	2 \beta \lambda q \frac{ \la N \ra_{\widehat{H}^s_{q,\sV}} }{V}.
\end{equation}
The last equality may be seen due to the fact that the system $\widehat{H}^s_{q,\sV}$ is invariant under
permutation of the sites of the lattice. This identity implies the following:
\begin{enumerate}
\renewcommand{\labelenumi}{(\roman{enumi})}
\item With $s=0$ we get $\la n_1 \ra_{\Hv{q}} \le \la N \ra_{H_\sV} /V$, thus $\la n_1 \ra_{\Hv{q}}$ is bounded and 
	in the limit $\frac{\la n_1 \ra_{\Hv{q}}}{V} \to 0$.
\item \vspace{-0.3cm}\[
		0 \le \frac{1}{V} \ln \trace \e^{-\beta \widehat{H}^s_{q,\sV}} - \frac{1}{V} \ln \trace \e^{-\beta H^s_{q,\sV}}
		\le \frac{2 \beta \lambda q}{V} \frac{ \la N \ra_{\widehat{H}^s_{q,\sV}} }{V}
	\] indicating that in the limit, the pressures are the same for the two Hamiltonians. By using Griffith's Lemma 
	we may see that the condensate densities (the derivatives with respect to $s$ at zero) are both equal to zero 
	(since we are considering the case $r_\mu=0$ here).
	That is
	\[
		\thermlim \left\la \frac{\sum a_x^\ast \sum a\astr_x}{V^2}\right\ra_{\Hv{q}}
		= \thermlim \left\la \frac{\sum a_x^\ast \sum a\astr_x}{V^2}\right\ra_{H_\sV}
		\vcentcolon= 0.
	\]
\end{enumerate}
Using these facts, one may see that the right-hand side of (\ref{ineq}) goes to zero in the limit
and we can conclude that
\[
	\thermlim \frac{\trace_{\FFsym} \exp\big\{- \beta \Hv{q}\big\}}
		{\trace_{\FFsym} \exp\big\{-\beta \Happ{q}(0)\big\}}=1.\vspace{-0.1cm}
\]
\end{proof}\vspace{-0.2cm}
Using Proposition \ref{PropB1}, one immediately obtains the desired result:
\begin{align*}
	c^\mu(q)
&=	\thermlim c^\mu_\sV(q) = \thermlim \frac{\trace_{\FFsym} 
	\exp\big\{- \beta \Hv{q}\big\}}{\trace_{\FFsym} \exp\big\{-\beta \Hv{0}\big\}}
\\[0.1cm]
&=	\thermlim \frac{\trace_{\FFsym} 
	\exp\big\{- \beta \Happ{q}(0)\big\}}{\trace_{\FFsym} \exp\big\{-\beta \Happ{0}(0)\big\}}
\\[0.1cm]
& =	\thermlim \frac{\trace_{\mathcal{F}_+(\mathcal{H}_{\sV-1})} 
	\exp\big\{- \beta \widetilde{H}_\sV  \big\}
	}
	{\trace_{\mathcal{F}_+(\mathcal{H}_{\sV-1})} 
	\exp\big\{- \beta \widetilde{H}_\sV \big\}
	}
	\frac{\trace_{\mathcal{F}(\CC)} \exp\big\{- \beta h_{q,\sV} \big\}}
	{\trace_{\mathcal{F}(\CC)} \exp\big\{-\beta h_{0,\sV} \big\}} 
\\[0.1cm]
& = 	\frac{\trace_{\mathcal{F}(\CC)} 
	\exp\big\{- \beta h_{q}(0) \big\}}
	{\trace_{\mathcal{F}(\CC)} \exp\big\{-\beta h_{0}(0) \big\}}
\end{align*} 
in the limit since $h_{q,\sV} = h_{q,\sV}(0) \to h_q(0)$ on $\mathcal{F}(\CC)$.

\subsection{Case 2: \texorpdfstring{For any $\mu$}{absence or presence of condensation}}
Consider the case when $r_\mu > 0$ If we insert $R = r_\mu$ in the constraint inequality (\ref{sandwich1})
then its left-hand term is strictly negative, but its right-hand term is strictly positive, rendering the 
previous argument useless here.

We therefore introduce a gauge-breaking term $\bar{\nu} \sum_x a\astr_x+ \nu \sum_x a^\ast_x$ into
the Hamiltonians $\Hv{q}$ and $\Happ{q}(r_\mu)$. Without loss of generality we may assume $\nu$ to 
be real and positive, so denote
\[
	\Hv{q}(\nu) \vcentcolon= \Hv{q} - \nu \sum_{x=1}^V ( a\astr_x + a_x^\ast)
\]
and its corresponding approximation as
\[
	\Happ{q}(R,\nu) \vcentcolon= \Happ{q}(R) - \nu \sum_{x=1}^V ( a\astr_x + a_x^\ast).
\]
Again we wish to separate this Hamiltonian into parts, one acting upon the site labelled 1, the
other on the remaining $V-1$ sites. If we define:
\begin{align*}
	h_{q,\sV}(r,\nu) 
&\vcentcolon=	h_{q,\sV}(r) - \nu (a_1 + a^\ast_1)
\\	
&=	(1 - \mu)(n_1 + q) + \lambda (n_1 + q) (n_1 + q - 1) +\tfrac{n_1}{V} - (r+\nu)(a_1 + a^\ast_1).
\end{align*}
and
\[
	\widetilde{H}_\sV(\nu) \vcentcolon= \widetilde{H}_\sV - \nu \sum_{x\ne1} (a\astr_x + a^\ast_x)
\]
then we may write $\Happ{q}(R,\nu) = h_{q,\sV}(R,\nu) + \widetilde{H}_\sV(\nu)$. 
Denote $\thermlim h_{q,\sV}(r,\nu) = h_{q}(r,\nu)$ on $\mathcal{F}_{+}(\CC)$, i.e. obtain a 
``gauge-symmetry broken'' single site Hamiltonian
\[
	h_{q}(r,\nu) \vcentcolon= (1 - \mu)(n + q) + \lambda (n + q) (n + q - 1) - (r+\nu)(a + a^\ast).
\]
We shall first prove the following proposition:

\begin{proposition}									\label{PropB2}
For each $\nu>0$, there exists a sequence $\{ \nu_\sV \in \RR: \nu_\sV \in [\nu, \nu+1/\sqrt{V}]\}$
independent of $q$ such that
\begin{equation*}
	\thermlim \frac{\trace_{\FFsym} \exp\left\{-\beta \Hv{q}(\nu_\sV)\right\}}
	{\trace_{\FFsym} \exp\left\{-\beta \Happ{q}(r_\mu(\nu_\sV), \nu_\sV)\right\}} = 1
\end{equation*}
where $r_\mu(\nu)>0$ is the non-zero solution of
\begin{equation}									\label{r-nu-value}
	2 r
=	\la a + a^\ast \ra_{h_q(r,\nu) }.
\end{equation}
Note that $\lim_{\nu\to0} r_\mu(\nu) = r_\mu$, the maximal solution to $(\ref{euler-lagrange})$,
i.e. the positive square root of the condensate density. 
\end{proposition}

\begin{proof}
There is no immediate correlation between the chosen $R$ and $r_\mu$ as yet. For each $\nu>0$
take a sequence $\nu_\sV$ which tends to $\nu$ as $V \to \infty$.
Then using the Bogoliubov inequality again, we obtain
\begin{multline}					\label{gauge_invar_bogoulibov}
	\beta \left\la a\astr_1 \left( \frac{\sum_{x\ne 1} a_x^\ast }{V}
	- R\right) \right\ra_{\Happ{q}(R,\,\nu_\sV)} + \mathrm{h.c.}
\\
	\le
	\ln \trace \exp\left\{-\beta \Hv{q}(\nu_\sV) \right\} - \ln \trace \exp\left\{-\beta \Happ{q}(R,\nu_\sV)\right\}
\\
	\le
	\beta \left\la a\astr_1 \left( \frac{\sum_{x\ne 1} a_x^\ast }{V} 
	- R\right) \right\ra_{\Hv{q}(\nu_\sV)} + \mathrm{h.c}.
\end{multline}

As above, the left-hand side may be reduced to
\[
	\left\la a\astr_1 \left( \frac{\sum_{x\ne 1} a_x^\ast }{V}-R \right) \right\ra_{\Happ{q}(R,\,\nu_\sV)}
   = 	\la a\astr_1 \ra_{h_{q,\sV}(R,\,\nu_\sV)}
	\left\la \frac{\sum_{x\ne 1} a_x^\ast}{V} - R
	\right\ra_{\widetilde{H}_\sV(\nu_\sV)}.
\]
If we replace $R$ with the term
\begin{equation*}
	r^{-}_{\mu,\sV}(\nu_\sV) = \left\la \frac{\sum_{x\ne 1} a_x}{V} \right\ra_{\widetilde{H}_\sV(\nu_\sV)},
\end{equation*}
then the left-most side of (\ref{gauge_invar_bogoulibov}) is identically zero and we get that
\[
	0
\le	\ln \trace \exp\left\{-\beta \Hv{q}(\nu_\sV)\right\} - 
	\ln \trace \exp\left\{-\beta \Happ{q}(r^{-}_{\mu,\sV}(\nu_\sV),\nu_\sV)\right\}.
\]
Hence
\begin{equation}								\label{liminf-ineq}
	\liminf_{V\to\infty} \frac{\trace \exp\left\{-\beta \Hv{q}(\nu_\sV)\right\}}
	{\trace \exp\left\{-\beta \Happ{q}(r^{-}_{\mu,_\sV}(\nu_\sV), \nu\sV)\right\}} \ge 1.
\end{equation}

Now considering the right-hand side of (\ref{gauge_invar_bogoulibov}) for any $\mu$. 
Using the Schwarz inequality as before:
\begin{align}
&	\left| \left\la a\astr_1 \left( \frac{\sum_{x\ne 1} a_x^\ast }{V}
	- R \right) \right\ra_{\Hv{q}(\nu)} \right|
=	\left| \left\la a\astr_1 \left( \frac{\sum_{x} a_x^\ast }{V} - R \right)
	- \frac{a\astr_1 a_1^\ast}{V} \right\ra_{\Hv{q}(\nu)} \right|		\notag
\\[0.15cm]
&\qquad\le	\la n_1 \ra_{\Hv{q}(\nu)}^{\tfrac{1}{2}}
	\left\la \left( \frac{\sum_{x} a_x^\ast }{V} - R \right)\left( \frac{\sum_{x} a\astr_x }{V} - R \right)  
	\right\ra^{\tfrac{1}{2}}_{\Hv{q}(\nu)} + \frac{\la n_1 \ra_{\Hv{q}(\nu)}+1}{V}	\notag
\\[0.3cm]
&\qquad=	\left( \la n_1 \ra_{\Hv{q}(\nu)} \frac{1}{V} 
	\big\la \delta_0^\ast \delta\astr_0 \big\ra_{\Hv{q}(\nu)} \right)^{1/2} 
	+ \frac{\la n_1 \ra_{\Hv{q}(\nu)}+1}{V}					\label{ineq2}
\end{align}
where we have taken
\[	
	\delta\astr_0 \vcentcolon= \frac{1}{\sqrt{V}} \left( \sum_{x=1}^V a\astr_x - V R \right).
\]

Again in order to consider this, insert
$\Hv{q}(\nu)$ and $\widehat{H}_{q,\sV}(\nu) \vcentcolon= \Hv{q}(\nu) - 2 \lambda q n_1$ 
into the Bogoliubov inequality, to obtain
\begin{equation}									\label{bog-5}
	2 \beta \lambda q \la n_1 \ra_{\Hv{q}(\nu)} 
\le	\ln \trace \e^{-\beta \widehat{H}_{q,\sV}(\nu)} - \ln \trace \e^{-\beta \Hv{q}(\nu)}
\le	2 \beta \lambda q \la n_1 \ra_{\widehat{H}_{q,\sV}(\nu)} 
=	2 \beta \lambda q \frac{ \la N \ra_{\widehat{H}_{q,\sV}(\nu)} }{V}
\end{equation}
which implies the following facts:
\begin{enumerate}
\renewcommand{\labelenumi}{(\roman{enumi})}
\item $\la n_1 \ra_{\Hv{q}(\nu)} \le \la N \ra_{\widehat{H}_{q,\sV}(\nu)}/V$, and 
	hence $\frac{\la n_1 \ra_{\Hv{q}(\nu)}}{V} \to 0$ as $V\to\infty$.
\item Since $\widehat{H}_{q,\sV}(\nu) =  (q-\mu)q + H_\sV(\nu) - \mu N_\sV$,
	\[
		0 \le -\frac{ \beta(q-\mu)q }{V} + \frac{1}{V} \ln \trace \e^{-\beta ( H_\sV(\nu) - \mu N_\sV )} 
		- \frac{1}{V} \ln \trace \e^{-\beta \Hv{q}(\nu)}
		\le \frac{2 \beta \lambda q}{V} \frac{ \la N \ra_{H_\sV(\nu)} }{V}
	\] with which one may show that in the limit, the pressures are the same for $H_\sV(\nu)$ 
	and $\Hv{q}(\nu)$:
	\begin{equation}							\label{Pressures2}
		\thermlim p_\sV[ H_\sV(\nu) ] = \thermlim p_\sV[ \Hv{q}(\nu) ].
	\end{equation}
\end{enumerate}

Using fact (i) from above, we find that the only term we need yet be concerned with on the right hand side of 
(\ref{gauge_invar_bogoulibov}) is the first term of (\ref{ineq2}), whose behaviour in the thermodynamic 
limit is still unknown:
\[
	\frac{1}{V} \big\la \delta_0^\ast \delta\astr_0 \big\ra_{\Hv{q}(\nu)}.
\]

To deal with this we shall take $R$ to be the following:
\begin{equation*}
	r^{+}_{\mu,\sV}(\nu) = \frac{1}{V} \left\la \sum_{x=1}^V a_x \right\ra_{\Hv{q}(\nu)}
\end{equation*}
so that one has
\begin{equation}								\label{delta0}
	\delta_0 = \frac{1}{\sqrt{V}} \left( \sum_{x=1}^V a_x - \bigg\la \sum_{x=1}^V a_x \bigg\ra_{\Hv{q}(\nu)} \right).
\end{equation}

Now we shall state and use some lemmas, which are proved later: 

\begin{lemma} 								\label{lemma1}
For fixed $\nu>0$, a positive integer $q$, and $\delta_0$ is defined as $(\ref{delta0})$, then there exists a sequence 
$\{ \nu_\sV \in \RR : \nu_\sV \in [\nu, \nu+1/\sqrt{V}] \}$ independent of $q$
which tends to $\nu$ as $V \to \infty$, such that for large $V$ we have the approximation
\begin{equation*}
	\left\la \delta_0^\ast \delta\astr_0 \right\ra_{\Hv{q}(\nu_\sV)}
\le	\frac{\e^{q} \sqrt{M}}{\beta} \left(\sqrt{V} + \frac{1}{2\nu}\right) + 
	u\frac{q}{V} + w
\end{equation*}
for some constants $u$, $w>0$ and $M$, independent of $\nu$ and $q$.
\end{lemma}

Using this lemma, for large $V$ and fixed $q$, we obtain the following estimate
\begin{equation*}
	\ln \trace \exp\left\{-\beta \Hv{q}(\nu_\sV)\right\} - 
	\ln \trace \exp\left\{-\beta \Happ{q}(r^{+}_{\mu,\sV}(\nu_\sV),\nu_\sV)\right\}
	\le \frac{\text{const}}{V^{1/4}}
\end{equation*}
where $\nu_\sV \to \nu$ as $V \to \infty$, implying that 
\begin{equation}							\label{limsup-ineq}
	\limsup_{V\to\infty} \frac{\trace \exp\left\{-\beta \Hv{q}(\nu_\sV)\right\}}
	{\trace \exp\left\{-\beta \Happ{q}(r^{+}_{\mu,\sV}(\nu_\sV),\nu_\sV)\right\}} \le 1.
\end{equation}

\begin{lemma}							\label{lemma3-r_vals-equal}
For a fixed $\nu>0$ and for any sequence $\{ \nu_\sV \}$ which tends to $\nu$ as $V \to \infty$, then
\begin{equation*}
	\thermlim r^{-}_{\mu,\sV}( \nu_\sV) 
=	\thermlim r^{+}_{\mu,\sV}( \nu_\sV)
= 	r_\mu(\nu)
\end{equation*}
where $r_\mu(\nu)$ is the unique non-zero solution to the Euler-Lagrange equation $(\ref{r-nu-value})$.
\end{lemma}

For clarity, it is best to use the following short-hand to complete this argument:
\begin{align*}
	a_\sV &= \trace_{\mathcal{F}(\Hone)} \exp\big\{-\beta \Hv{q}(\nu_\sV) \big\}
\\
	b_\sV &= \trace_{\mathcal{F}(\Hone)} \exp\big\{-\beta \Happ{q}(r_{\mu}(\nu_\sV),\nu_\sV)\big\}
\\
	c_\sV &= \trace_{\mathcal{F}(\Hone)} \exp\big\{-\beta \Happ{q}(r^{-}_{\mu,\sV}(\nu_\sV),\nu_\sV)\big\}
\\
	d_\sV &= \trace_{\mathcal{F}(\Hone)} \exp\big\{-\beta \Happ{q}(r^{+}_{\mu,\sV}(\nu_\sV),\nu_\sV)\big\}.
\end{align*}
The penultimate step is to prove the following:
\begin{equation}									\label{limothers}
	\thermlim \frac{b_\sV}{c_\sV} = 1 \qquad \quad \text{and} \quad \qquad
	\thermlim \frac{b_\sV}{d_\sV} = 1.
\end{equation}
Considering the first, note that
\[
	\frac{b_\sV}{c_\sV} 
=	\frac{\trace_{\mathcal{F}(\CC)} \e^{-\beta h_q(r_{\mu}(\nu_\sV), \nu_\sV)}}
	{\trace_{\mathcal{F}(\CC)} \e^{-\beta h_q(r^{-}_{\mu,\sV}(\nu_\sV), \nu_\sV)}}.
\]
Once again using the Bogoliubov inequality (\ref{bogoliubov}) with 
$A=-\beta h_q(r^{-}_{\mu,\sV}(\nu_\sV), \nu_\sV)$ and 
$B=-\beta h_q(r_{\mu}(\nu_\sV), \nu_\sV)$,
then $A-B = \beta (r_\mu(\nu_\sV) - r^{-}_{\mu, \sV}(\nu_\sV) )( a + a^\ast )$ and we obtain
\begin{multline*}
	\beta (r_\mu(\nu_\sV) - r^{-}_{\mu, \sV}(\nu_\sV) ) 
	\left\la a + a^\ast \right\ra_{h_q(r_{\mu}(\nu_\sV), \nu_\sV)}
\\
\le	\ln b_\sV  - \ln c_\sV
\\
\le 	\beta (r_\mu(\nu_\sV) - r^{-}_{\mu, \sV}(\nu_\sV) ) 
	\left\la a + a^\ast \right\ra_{h_q(r^{-}_{\mu, \sV}(\nu_\sV), \nu_\sV)}
\end{multline*}
and as $V \to \infty$, both the left and right hand sides go to zero, implying the result. 
A similar procedure may be used to show the second.

To complete this proof, we wish to derive
\[
	\thermlim \frac{a_\sV}{b_\sV} = 1.
\]
with the use of the following information derived above (see (\ref{liminf-ineq}), (\ref{limsup-ineq})
and (\ref{limothers}))
\begin{align*}
	\liminf_{V \to \infty} \frac{a_\sV}{c_\sV} &\ge 1 &
	\limsup_{V \to \infty} \frac{a_\sV}{d_\sV} &\le 1
\\
	\thermlim \frac{b_\sV}{c_\sV} &= 1 &
	\thermlim \frac{b_\sV}{d_\sV} &= 1.
\end{align*}
Taking the supremum limit, one may see that:
\[
 	\thermlim \frac{a_\sV}{b_\sV}
\le	\limsup_{V \to \infty} \frac{a_\sV}{b_\sV}
=	\limsup_{V \to \infty} \frac{\frac{a_\sV}{d_\sV}}{\frac{b_\sV}{d_\sV}}
\le 	 \frac{ \limsup_{V \to \infty} \frac{a_\sV}{d_\sV}}{ \thermlim \frac{b_\sV}{d_\sV}}
\le 	1.
\]
The infimum limit follows similarly, proving the proposition.
\end{proof}

With the assistance of Proposition \ref{PropB2} we have our result:
\begin{align*}
	c^\mu(q, \nu) 
&=	\thermlim c^\mu_\sV(q, \nu_\sV)
=	\thermlim \frac{\trace_{\FFsym} \exp\big\{-\beta \Hv{q}(\nu_\sV)\big\}
	}{
	\trace_{\FFsym} \exp\big\{-\beta \Hv{0}(\nu_\sV)\big\}
	}
\\[0.1cm]
&=	\thermlim \frac{\trace_{\FFsym} \exp\big\{-\beta \Happ{q}(r_{\mu,\sV}(\nu_\sV),\nu_\sV)\big\}
	}{
	\trace_{\FFsym} \exp\big\{-\beta \Happ{0}(r_{\mu,\sV}(\nu_\sV),\nu_\sV)\big\}
	}
\\[0.1cm]
&=	\thermlim \frac{\trace_{\mathcal{F}_{+}(\mathcal{H}_{\sV-1})}
		\exp\big\{-\beta \widetilde{H}_\sV(\nu_\sV) \big\}
	}{
		\trace_{\mathcal{F}_{+}(\mathcal{H}_{\sV-1})} 
			\exp\big\{-\beta \widetilde{H}_\sV(\nu_\sV)\big\}
	}
	\frac{\trace_{\mathcal{F}(\CC)}
		\exp\big\{-\beta  h_{q,\sV}(r_{\mu,\sV}(\nu_\sV), \nu_\sV) \big\}
	}{
		\trace_{\mathcal{F}(\CC)} 
		\exp\big\{-\beta  h_{0,\sV}(r_{\mu,\sV}(\nu_\sV), \nu_\sV)\big\}
	}
\\[0.1cm]
&=	\frac{\trace_{\mathcal{F}(\CC)}
		\exp\big\{-\beta [h_q(r_\mu(\nu)) - \nu(a + a^\ast)]\big\}
	}{
		\trace_{\mathcal{F}(\CC)}
		\exp\big\{-\beta [h_0(r_\mu(\nu)) - \nu(a + a^\ast)]\big\}
	}.
\end{align*}

\pagebreak
\subsection{Proofs of Lemmas}
\begin{proof}
\textbf{of Lemma \ref{lemma1}}\\
Considering the term $\left\la \delta_0^\ast \delta\astr_0 \right\ra_{\Hv{q}(\nu)}$ in a similar fashion to
Appendix 1 of Dorlas and Bru \cite{BruDorlas}, note that since $[ \delta\astr_0, \delta_0^\ast ]=1$ we may write
\[
	\left\la \delta_0^\ast \delta\astr_0 \right\ra_{\Hv{q}(\nu)} 
	= \frac{1}{2} \left\la \{ \delta_0^\ast, \delta\astr_0 \} \right\ra_{\Hv{q}(\nu)} - 1
\]
where $\{ X,Y \} \vcentcolon= XY+YX$. Using the spectral decomposition of the Hamiltonian $\Hv{q}(\nu)$
\vspace{-0.2cm}:
\[
	\Hv{q}(\nu) \psi_n = E_n \psi_n \vspace{-0.2cm}
\]
and denoting $A_{mn} = (\psi_m, \delta\astr_0, \psi_n)$, one may get
\begin{align}
	\left\la \{\delta_0^\ast, \delta\astr_0 \} \right\ra_{\Hv{q}(\nu)} 
&=	\e^{-\beta V p[ \Hv{q}(\nu) ]} \sum_{m,n} |A_{mn}|^2 
	\left( \e^{-\beta E_n} - \e^{-\beta E_m} \right) 						\notag
\\
&\le	2 \e^{-\beta V p[ \Hv{q}(\nu) ]} \sum_{m,n} |A_{mn}|^2 
	\frac{ \left( \e^{-\beta E_n} - \e^{-\beta E_m} \right) }{ \beta (E_m - E_n ) } 		\notag
\\
& \qquad + \frac{1}{6} \e^{-\beta V p[ \Hv{q}(\nu) ]} \sum_{m,n} |A_{mn}|^2
	\left( \e^{-\beta E_n} + \e^{-\beta E_m} \right) \beta (E_m - E_n ) 			\notag
\\
&\le	2 (\delta\astr_0, \delta\astr_0)_{\Hv{q}(\nu)} 
	+ \frac{\beta}{6} \la [\delta_0^\ast , [ \Hv{q}(\nu) , \delta\astr_0]] \ra_{\Hv{q}(\nu)}. \label{decomp}
\end{align}
where the Duhamel inner product $( \, \cdot \, , \, \cdot \, )$ is defined as follows:
\begin{equation}											\label{duhamel}
	( A , B )_H = \frac{1}{\beta Z} \int_0^\beta \trace 
		\left[ A^\ast \e^{-(\beta - s) H} B \e^{-s H}\right] ds
\end{equation}
and $Z = \trace \e^{-\beta H}$.

Denote $c_0 = \tfrac{1}{\sqrt{V}} \sum a_x$. Consider the second term of (\ref{decomp})
while using the following identities (for some $x=1\dots V$)
\begin{align*}
	[ c_0^\ast, [n_x^2, c\astr_0]] = \frac{2}{V} (2 n_x + 1), \quad
	[ c_0^\ast, [n_x, c\astr_0]] &= \frac{1}{V}, \quad
	[ c_0^\ast, [ c_0^\ast  c\astr_0,  c\astr_0]] = 1, 
\\
	[ c_0^\ast, [ c\astr_0 , c\astr_0]] = [ c_0^\ast , [ c^\ast_0 , c\astr_0]] &= 0
\end{align*}
one may evaluate 
{\allowdisplaybreaks
\begin{align*}
	[\delta_0^\ast & , [ \Hv{q}(\nu) , \delta\astr_0]] = [ c_0^\ast , [ \Hv{q}(\nu) , c\astr_0]]
\\
&=	[ c_0^\ast , [ 2\lambda q n_1 + (q-\mu)q - c_0^\ast c\astr_0 + (1-\lambda - \mu) n_x 
	+ \lambda n_x^2 +\nu \sqrt{V} c^\ast_0 + \bar{\nu} \sqrt{V} c\astr_0, c\astr_0]]
\\
&=	2 \lambda q [ c_0^\ast , [ n_1 , c\astr_0]] - [ c_0^\ast , [ c_0^\ast c\astr_0 , c\astr_0]] 
	+ (1-\lambda - \mu) \sum_{x=1}^V [ c_0^\ast , [ n_x , c\astr_0]] 
	+ \lambda \sum_{x=1}^V [ c_0^\ast , [ n^2_x , c\astr_0]]
\\
&	\qquad + \nu \sqrt{V} [c_0^\ast , [ c^\ast_0 , c\astr_0]] 
	+ \bar{\nu} \sqrt{V} [c_0^\ast , [ c\astr_0 , c\astr_0]]
\\
&=	-\mu + \lambda + 2 \lambda \frac{2 N_\sV + q}{V}.
\end{align*}}

\textbf{Some useful inequalities and bounds}\\
Note the following operator inequalities:
\[
	\frac{1}{2V}\sum_{x,y} (a_x^\ast - a_y^\ast)(a_x - a_y) \ge 0
\quad\Rightarrow\quad
	\sum_x n_x - \frac{1}{V} \sum_{x,y} a_x^\ast a_y \ge 0
\quad\Rightarrow\quad
	c_0^\ast c_0 \le N_\sV
\]
\[
	\sum_{x} \left( n_x - \frac{N_\sV}{V} \right)^2 \ge 0
	\qquad \Rightarrow \qquad \sum_x n_x^2 \ge \frac{N_\sV^2}{V}
\]
\[
	\nu \sqrt{V} (c\astr_0 + c^\ast_0 ) \le \nu^2 c^\ast_0 c\astr_0 + V \le \nu^2 N_\sV + V.
\]
From these it is clear that the Hamiltonian with sources, $\Hv{q}(\nu)$, is superstable for fixed 
$q>0$ and $\lambda>0$, i.e.
\[
	\Hv{q}(\nu)
\ge	\lambda \frac{N_\sV^2}{V} - (\lambda + \mu + \nu^2) N_\sV - V.
\]
By the Bogoliubov inequality, since $H_\sV(\beta,\mu-2q\lambda,\nu) - \Hv{q}(\beta,\mu,\nu) - q(q - \mu)
= 2 q\lambda \sum_{i=2}^V n_i $, one has that:
\begin{equation}									\label{no-one-bound}
	\bigg\la \sum_{i=2}^V n_i \bigg\ra_{\Hv{q}(\beta,\mu,\nu)}
\le	\bigg\la \sum_{i=2}^V n_i \bigg\ra_{H_\sV(\beta,\mu-2q\lambda,\nu)}.
\end{equation}
We can find an upper bound for the expectation of the number operator with respect to $\Hv{q}(\nu)$
which is independent of $q$.
\[	
	\big\la N_\sV \big\ra_{\Hv{q}(\nu)} 
= 	\big\la n_1 \big\ra_{\Hv{q}(\nu)} + \bigg\la \sum_{i=2}^V n_i \bigg\ra_{\Hv{q}(\nu)}
\le 	\bigg\la \frac{N_\sV}{V} \bigg\ra_{H_\sV(\nu)}
	+ \bigg\la \sum_{i=2}^V n_i \bigg\ra_{H_\sV(\beta, \mu-2q\lambda, \nu)}
\]
using (\ref{bog-5}) and (\ref{no-one-bound}). Now
\[
	\bigg\la \sum_{i=2}^V n_i \bigg\ra_{H_\sV(\beta, \mu-2q\lambda, \nu)}
\le	\big\la N_\sV \big\ra_{H_\sV(\beta, \mu-2q\lambda, \nu)}
\le	\big\la N_\sV \big\ra_{H_\sV(\beta, \mu, \nu)}
\]
since $\big\la N_\sV \big\ra_{H_\sV(\beta, \mu, \nu)}$ is monotonically increasing in $\mu$.
Hence we have a bound independent of $q$:
\begin{equation}									\label{bound-indepq}
	\big\la N_\sV \big\ra_{\Hv{q}(\nu)} 
\le 	\bigg\la \frac{N_\sV}{V} \bigg\ra_{H_\sV(\nu)} 
	+ \big\la N_\sV \big\ra_{H_\sV(\beta, \mu, \nu)}
\le	2\big\la N_\sV \big\ra_{H_\sV(\nu)}
\end{equation}

The pressure $p_\sV[H_\sV(\nu)]$ is convex in $\mu \in \RR$ and $\nu \in \RR$. 
If one fixes the real numbers $\mu_0$ and $\nu_0 \ge 0$, then by superstability
there exists a uniform bound $M\vcentcolon= M(\beta,\mu) \ge 0$ such that (using $\la a\astr_x \ra_{\Hv{q}(\nu)} 
= \la a^\ast_x \ra_{\Hv{q}(\nu)}$ and (\ref{bound-indepq})):
\begin{equation}									\label{-404bound-2}
	\frac{1}{V} \big\la c\astr_0 \big\ra_{\Hv{q}(\nu)}^2 
\le	\frac{1}{V} \big\la c_0^\ast c_0 \big\ra_{\Hv{q}(\nu)}
\le	\bigg\la \frac{N_\sV}{V} \bigg\ra_{\Hv{q}(\nu)} 
\le	\bigg\la \frac{N_\sV}{V} \bigg\ra_{H_\sV(\nu)} 
\le 	M
\end{equation}
(c.f. (\ref{upperbound})\!) with $\mu \le \mu_0$ and $\nu < \nu_0$.

\textbf{Resuming the proof}\\
Then we have
\begin{align*}
	\left\la \delta_0^\ast \delta\astr_0 \right\ra_{\Hv{q}(\nu)}
&=	 (\delta_0, \delta_0 )_{\Hv{q}(\nu)}
	+ \frac{\beta}{3} \left( -\mu + \lambda + 2 \lambda \frac{2 \la N_\sV \ra_{\Hv{q}(\nu)} + q}{V} \right) - 1
\\
&\le	(\delta_0, \delta_0 )_{\Hv{q}(\nu)}
	+ \frac{\beta}{3} \left( -\mu + \lambda + 2 \lambda \left( 2M + \frac{q}{V} \right) \right).
\end{align*}
It remains to consider the inner product. For convenience denote $f_{q,\sV}(\nu) \vcentcolon= (\delta_0, \delta_0 )_{\Hv{q}(\nu)}$. 
If $\nu \in \CC$, then $\Hv{q}(\nu) = \Hv{q} - \sqrt{V} ( \bar{\nu} c\astr_0 + \nu c^\ast_0 )$ and so we 
would have the following:
\[
	f_{q,\sV}(\nu) = \frac{\partial}{\partial \nu} \frac{\partial}{\partial \bar{\nu}} p_\sV[\Hv{q}(\nu)].
\]
However as $\nu \in \RR$, we may use polar co-ordinates to write the following:
\begin{equation}									\label{04second-deriv-polar}
	f_{q,\sV}(\nu)
	= \frac{1}{4\beta\nu} \frac{\partial}{\partial \nu} 
	\left( \nu \frac{\partial}{\partial \nu} p_\sV[\Hv{q}(\nu)] \right).
\end{equation}
We first wish to show that $f_{q,\sV}(\nu)$ is bounded by a constant independent of $q$.
For fixed $V$ and $\nu$, recall we have
\begin{align*}
	4\beta f_{q,\sV}(\nu) 
&=	\frac{1}{\nu} \frac{\partial}{\partial \nu} p_\sV[H_{q,\sV}(\nu)]
	+  \frac{\partial^2}{\partial \nu^2} p_\sV[H_{q,\sV}(\nu)]
\\
&\le 	\frac{2 \sqrt{M}}{\nu} 
	+  \frac{\partial^2}{\partial \nu^2} p_\sV[H_{q,\sV}(\nu)]
\end{align*}
from (\ref{firstderiv-bound}). So it remains to consider the latter term:
\begin{align*}
	\beta^{-1} \frac{\partial^2}{\partial \nu^2} p_\sV[H_{q,\sV}(\nu)]
&=	\left( (c\astr_0 + c^\ast_0 )^\ast , c\astr_0 + c^\ast_0 \right)_{H_{q,\sV}(\nu)}
	- \big\la c\astr_0 + c^\ast_0 \big\ra_{H_{q,\sV}(\nu)}^2
\\
&\le	\left( c\astr_0 + c^\ast_0 , c\astr_0 + c^\ast_0 \right)_{H_{q,\sV}(\nu)}
\intertext{and using the fact that $(A,A) \le \tfrac{1}{2}\la A^\ast A + A A^\ast \ra$, we may obtain}
&\le	\big\la (c\astr_0 + c^\ast_0 )( c\astr_0 + c^\ast_0 ) \big\ra_{H_{q,\sV}(\nu)}
\\
&=	\big\la c^\ast_0 c\astr_0 +  c\astr_0 c^\ast_0  \big\ra_{H_{q,\sV}(\nu)}
	+ 2 \big|\big\la c\astr_0 c\astr_0 \big\ra_{H_{q,\sV}(\nu)} \big|
\\
&\le	2\big\la N_\sV \big\ra_{H_{q,\sV}(\nu)} + V 
	+ \sqrt{\big\la N_\sV \big\ra_{H_{q,\sV}(\nu)}^2 + V \big\la N_\sV \big\ra_{H_{q,\sV}(\nu)}}
\\
&\le	V (2M+1 + \sqrt{M(M+1)} )
\end{align*}
from (\ref{bound-indepq}). To summarize:
\begin{equation}									\label{f-bound-bad}
	f_{q,\sV}(\nu) \le \frac{\sqrt{M}}{2\beta\nu} + \frac{V}{4\beta} (4M+1 ).
\end{equation}
Note however that this procedure does not yield a satisfactory bound for our purposes, since we desire
a bound of order $V^\alpha$, where $\alpha<1$. Taking a different approach, fix $q$,
consider (\ref{04second-deriv-polar}), multiply both sides by $\nu$ and integrate:
\[
	\int_\nu^{\nu+\epsilon} \nu' f_{q,\sV}(\nu') d\nu' = \frac{1}{4\beta}  
	\left( \nu' \frac{\partial}{\partial \nu'} p_\sV[\Hv{q}(\nu')] \right) \bigg|_\nu^{\nu+\epsilon} 
\]
for $[\nu,\nu+\epsilon] \subset [0, \nu_0]$. 

By (\ref{-404bound-2}), we have for all $q \in \NN$, $\mu \in \RR$ and $\nu < \nu_0$ that
\begin{equation}								\label{firstderiv-bound}
	\frac{\partial}{\partial \nu} p_\sV[\Hv{q}(\nu)] 
= 	\frac{1}{\sqrt{V}}  \big\la c\astr_0 + c^\ast_0 \big\ra_{\Hv{q}(\nu)}
=	\frac{2}{\sqrt{V}} \left| \big\la c\astr_0 \big\ra_{\Hv{q}(\nu)} \right|
\le 	2 \sqrt{M}
\end{equation}
and so
\[
	\nu' \frac{\partial}{\partial \nu'} p_\sV[\Hv{q}(\nu')]  \bigg|_\nu^{\nu+\epsilon}
\le	\left| 2 (\nu+\epsilon)\sqrt{M}\right| + \left| 2\nu\sqrt{M}\right| 
=	2 (2\nu+\epsilon) \sqrt{M}
\]
with which we obtain
\begin{equation}								\label{int-f-bound}
	\int_\nu^{\nu+\epsilon} \nu' f_{q,\sV}(\nu') d\nu' 
\le 	\frac{\sqrt{M}}{\beta} \left(\nu+\frac{\epsilon}{2}\right) .
\end{equation}

From (\ref{f-bound-bad}), $| f_{q,\sV} (\nu) | < CV$ where $C$ is a constant independent of $q$
and $\nu' \in [\nu, \nu+\epsilon]$. Now let $\epsilon = V^{-1/2}$. 
Denote
\[
	F_\sV(\nu) \vcentcolon= \sum_{q=1}^\infty \e^{-q}  f_{q,\sV} (\nu).
\]
Since $f_{q,\sV}(\nu)$ is uniformly convergent in $\nu$ (see (\ref{f-bound-bad})\!) and each term 
is continuous, then $F_q(\nu)$ is continuous. From (\ref{int-f-bound}) may obtain:
\begin{equation}								\label{04bound-3}
	\int_\nu^{\nu+V^{-1/2}} \nu' F_\sV(\nu') d\nu' 
\le 	\frac{\sqrt{M}}{\beta} \left(\nu+\frac{1}{2\sqrt{V}}\right) \sum_{q=1}^{\infty} \e^{-q}.
\end{equation}
By the Mean-Value theorem, there exists some $\nu_\sV \in [\nu, \nu+V^{-1/2}]$ such that
\begin{equation}								\label{mvt}
	\int_\nu^{\nu+V^{-1/2}} \nu' F_\sV(\nu') d\nu' 
=	\frac{\nu_\sV}{\sqrt{V}} F_\sV(\nu_\sV)
\end{equation}
which in combination with (\ref{04bound-3}) yields
\[
	\frac{\nu_\sV}{\sqrt{V}} F_\sV(\nu_\sV) \,\,
\le 	\,\, \frac{\sqrt{M}}{\beta} \left(\nu+\frac{1}{2\sqrt{V}}\right)
	\frac{1}{\e-1} \,\,
\le 	\,\, \frac{\sqrt{M}}{\beta} \left(\nu+\frac{1}{2\sqrt{V}}\right).
\]
For any positive integer $q$, since $\e^{-q} f_{q,\sV}(\nu) \le F_\sV(\nu)$, then
\[
	\frac{\nu_\sV}{\sqrt{V}} \e^{-q} f_{q,\sV}(\nu_\sV)
\le 	\frac{\sqrt{M}}{\beta} \left(\nu+\frac{1}{2\sqrt{V}}\right)
\]
so that
\begin{align*}
	f_{q,\sV}(\nu_\sV)
&\le 	\frac{\e^{q} \sqrt{M}}{\beta} \left(\sqrt{V} \frac{\nu}{\nu_\sV} +\frac{1}{2\nu_\sV}\right).
\end{align*}
Then as $V$ increases, we have a sequence 
$\{ \nu_\sV  \in \RR: \nu_\sV \in [\nu, \nu+1/\sqrt{V}] \}$ satisfying (\ref{mvt})
which tends to $\nu>0$ as $V \to \infty$, so that for large $V$ we have the estimate:
\[
	f_{q,\sV}(\nu) \le \frac{\e^q \sqrt{M}}{\beta} \left(\sqrt{V}  + \frac{1}{2\nu}\right).
\]
\end{proof}

\begin{proof}
\textbf{of Lemma \ref{lemma3-r_vals-equal}}\\
Before proceeding, we need to show the following: for fixed $\nu >0$
\[
	\thermlim p_\sV[ \widetilde{H}_\sV(\nu) ] = \thermlim p_\sV[ H_\sV(\nu) ],
\]
i.e. a single site's contribution is irrelevant in the thermodynamic limit.
Recall that
\begin{align*}
	\widetilde{H}_\sV
&=	\frac{1}{2V} \sum_{x,y\ne1} ( a^\ast_x - a^\ast_y )( a\astr_x - a\astr_y ) 
	+ \lambda \sum_{x\ne1} n_x ( n_x - 1)
	+ \left(\frac{2}{V(V-1)} - \mu \right) \sum_{x\ne1} n_x.
\end{align*}
The corresponding pressure may be expressed as
\begin{align*}
	p_\sV[ \widetilde{H}_\sV ]
&=	\frac{1}{\beta V} \ln \trace_{\FFsym} \exp\left\{
	-\beta \widetilde{H}_\sV
	\right\}
\\
&=	\frac{1}{\beta V} \ln \trace_{\mathcal{F}_{+}(\mathcal{H}_{V-1})} \exp\bigg\{
	-\beta \bigg( \frac{1}{2V} \sum_{x,y=1}^{V-1} ( a^\ast_x - a^\ast_y )( a\astr_x - a\astr_y ) 
	+ \lambda \sum_{x=1}^{V-1} n_x ( n_x - 1)
\\
&	\qquad \qquad  + \left(\frac{2}{V(V-1)} - \mu \right) \sum_{x=1}^{V-1} n_x 
	\bigg)
	\bigg\}
\\
&=	\left( \frac{V-1}{V} \right)^2
	p_{\sV-1}\big[ H_{\sV-1}( \beta\left( \tfrac{V-1}{V} \right) , \lambda\left( \tfrac{V}{V-1} \right),
	\mu - \tfrac{2}{V(V-1)} )\big]
\end{align*}
where abusing notation temporarily we have explicitly included the parameters of the 
IRH Bose-Hubbard Hamiltonian,
i.e. we write the pressure of (\ref{I-R2}) as $p_\sV[ H_\sV ] \equiv p_\sV[ H_\sV(\beta,\lambda,\mu)]$.
Then in the limit, with the use of the Bogoliubov inequality, one may verify that
\begin{align*}
	\thermlim p_\sV[\widetilde{H}_\sV]
&=	\thermlim \left( \frac{V-1}{V} \right)^2
	p_{\sV-1}\big[ H_{\sV-1}( \beta\left( \tfrac{V-1}{V} \right) , \lambda\left( \tfrac{V}{V-1} \right),
	\mu - \tfrac{2}{V(V-1)} )\big]
\\[0.2cm]
&=	\thermlim p_\sV[H_\sV ] \equiv p(\beta,\mu).
\end{align*}

Now proceeding to prove this lemma, recall that we chose
\[
	r^{-}_{\mu,\sV}(\nu) = \left\la \frac{\sum_{x\ne 1} a_x}{V} \right\ra_{\widetilde{H}_\sV(\nu)}
\]
where $\widetilde{H}_\sV(\nu) = \widetilde{H}_\sV - \nu \sum_{x\ne1} ( a\astr_x + a_x^\ast )$
is the gauge-broken IRH Bose-Hubbard Hamiltonian on
all sites of the lattice barring the site $x=1$. Fixing a value of $\nu>0$,
there exists a unique $r_\mu(\nu)>0$ as the solution to the Euler-Lagrange equation (\ref{r-nu-value}), i.e.
\[
	2r_\mu(\nu) = \la a + a^\ast \ra_{h_q(r_\mu(\nu),\nu)}.
\]
The pressure $p_\sV[ \widetilde{H}_\sV(\nu) ]$ is convex in $\nu$ and its thermodynamic limit 
is differentiable for all $\nu > 0$. By Griffith's Lemma, we have
\begin{equation}								\label{griff1}
	\thermlim \frac{d}{d\nu} p_\sV[ \widetilde{H}_\sV( \nu_\sV) ]
=	\frac{d}{d\nu} \thermlim p_\sV[ \widetilde{H}_\sV(\nu) ].
\end{equation}
The left hand side of this evaluates to
\[
	\thermlim \frac{d}{d\nu_\sV} p_\sV[ \widetilde{H}_\sV( \nu_\sV) ]
=	\thermlim \frac{1}{V} \sum_{x\ne1} \la a_x + a^\ast_x \ra_{\widetilde{H}_\sV(\nu_\sV)}
=	\thermlim \frac{2}{V} \sum_{x\ne1} \la a_x \ra_{\widetilde{H}_\sV(\nu_\sV)}
=	2 \thermlim r^{-}_{\mu,\sV}(\nu_\sV).
\]
As shown above, we have that
\begin{multline}								\label{pressures2}
	\thermlim p_\sV[ \widetilde{H}_\sV(\nu) ]
=	\thermlim p_\sV [ H_\sV(\nu) ]
\\[0.2cm]
\vcentcolon=	-r_\mu(\nu)^2 + \frac{1}{\beta} \ln \trace \exp\big\{ \beta 
	\big[ (\mu-1) n -\lambda n(n-1) + (r_\mu(\nu)+\nu)(a+a^\ast)\big]\big\}.
\end{multline}
so the right-hand side of (\ref{griff1}) will become (also using (\ref{pressures2})\!)
{\allowdisplaybreaks
\begin{align*}
	\frac{d}{d\nu} & \thermlim p_\sV [ H_\sV(\nu) ]
\\[0.1cm]
&=	\frac{d}{d\nu} \left[-r_\mu(\nu)^2 + \frac{1}{\beta} \ln \trace \exp\big( \beta 
	\big[ (\mu-1) n -\lambda n(n-1) + (r_\mu(\nu)+\nu)(a+a^\ast)\big]\big) \right]
\\[0.1cm]
&=	-2r_\mu(\nu) \frac{dr_\mu(\nu)}{d\nu} + \left( \frac{dr_\mu(\nu)}{d\nu} + 1 \right) \la a 
	+ a^\ast \ra_{h_q(r_\mu(\nu),\nu)}
\\[0.1cm]
&=	\frac{dr_\mu(\nu)}{d\nu} \left( \la a + a^\ast \ra_{H(\nu)} - 2r_\mu(\nu) \right) + \la a 
	+ a^\ast \ra_{h_q(r_\mu(\nu),\nu)}
\\[0.1cm]
&=	\la a + a^\ast \ra_{h_q(r_\mu(\nu),\nu)} = 2r_\mu(\nu).
\end{align*}}
as desired.

Similarly, taking
\[
	r_{\mu,\sV}^{+}(\nu) = \frac{1}{V} \bigg\la \sum_{x=1}^V a_x \bigg\ra_{\Hv{q}(\nu)}
\]
where $\Hv{q}(\nu) = \Hv{q} - \nu \sum_{x=1}^V ( a\astr_x + a^\ast_x )$.
Label the corresponding pressure for this Hamiltonian as $p_\sV[ \Hv{q}(\nu)]$. Recall the expression 
(\ref{Pressures2}) that we previously derived: 
\[
	\thermlim p_\sV[ \Hv{q}(\nu)] = \thermlim p_\sV[ H_\sV(\nu)].
\]
As above by Griffith's Lemma, we have
\[
	\thermlim r^{+}_\sV(\nu_\sV) 
= 	\frac{1}{2} \thermlim \frac{d}{d\nu_\sV} p_\sV[ \Hv{q}( \nu_\sV) ]
=	\frac{1}{2} \frac{d}{d\nu} \thermlim p_\sV[ H_\sV(\nu) ]
=	r_\mu(\nu)
\]
as above.
\end{proof}

\section{The Absence of Condensation}
Using the first statement in Theorem \ref{prop3-equiv}
we can rigorously show in the absence of condensation that all particles are on short cycles.
\begin{lemma}								\label{no-long}
In the absence of condensation, i.e. for those $\mu$ such that $r_\mu=0$, the density of particles 
on short cycles equals the total density of the system, that is:
\[
	\rhoshort = \rhodens \quad \Rightarrow \quad \rholong = 0.
\]
\end{lemma}

\begin{proof}
For those values of $\mu$ such that $r_\mu = 0$, i.e. in the absence of condensation,
first note that the density (from \ref{pressure}) may be expressed as:
\begin{multline*}
	\rhodens
\vcentcolon= 	\frac{\partial}{\partial \mu} p(\mu)
=	\frac{\partial}{\partial \mu} 
	\frac{1}{\beta}\trace_{\mathcal{F}(\CC)} \exp \big\{ -\beta [
		 (1 - \mu)n +\lambda n(n - 1)
	] \big\}
\\
=	\frac{\trace_{\mathcal{F}(\CC)} \big[ n \exp \big\{ -\beta [
		(1 - \mu)n +\lambda n(n - 1)
	] \big\} \big]}
	{\trace_{\mathcal{F}(\CC)} \exp \big\{ -\beta [
		(1 - \mu)n +\lambda n(n - 1)
	] \big\}}.
\end{multline*}

Similarly, from Theorem \ref{prop3-equiv} one immediately obtains:
\[
	c^\mu(q,0)
   =	\frac{ \trace_{\mathcal{F}(\CC)} \exp \big\{ -\beta [
   		(1 - \mu)(n + q) + \lambda (n + q)(n + q - 1)
	] \big\} }
	{ \trace_{\mathcal{F}(\CC)} \exp \big\{ -\beta [
		(1 - \mu)n +\lambda n(n - 1)
	] \big\} }.
\]

Label the denominator $\Xi \vcentcolon= \trace_{\mathcal{F}(\CC)} \exp\{-\beta ( (1 - \mu)n + \lambda n(n - 1) )\}$.
The operator $n$ in this context counts the number of particles on the site, so in terms of a basis of 
occupation numbers, it has eigenvalues $k=0,1,2,\dots$. Summing over this basis
{\allowdisplaybreaks
\begin{align*}
	\sum_{q=1}^\infty c^\mu(q,0)
   =&	\frac{1}{\Xi} \sum_{q=1}^\infty \sum_{k=0}^\infty \exp \big\{ -\beta [
		\lambda (1 - \mu)(k + q) + \lambda (k + q)(k + q - 1)
	] \big\}
\intertext{and shifting the sum}
   =&	\frac{1}{\Xi} \sum_{q=1}^\infty \sum_{k=q}^\infty \exp \big\{ -\beta [
		(1 - \mu)k + \lambda k (k - 1)
	] \big\}
\\
   =&	\frac{1}{\Xi} \sum_{k=1}^\infty \sum_{q=1}^k \exp \big\{ -\beta [
		(1 - \mu)k + \lambda k (k - 1)
	] \big\}
\\
   =&	\frac{1}{\Xi} \sum_{k=1}^\infty k \exp \big\{ -\beta [
		(1 - \mu)k + \lambda k (k - 1)
	] \big\}
\\
   =&	\frac{1}{\Xi} \trace_{\mathcal{F}(\CC)} \big[ n \exp \big\{ -\beta [
		(1 - \mu)n +\lambda n(n - 1)
	] \big\} \big]
\\
   =&	\rhodens.
\end{align*}}
Therefore the absence of condensation implies that the sum of all finitely long cycle densities
equals the system density.
\end{proof}

\section{Numerical Analysis}
Using the second statement of Theorem \ref{prop3-equiv}, in the presence of condensation
we may obtain an expression for the short cycle density for a gauge-broken IRH Bose-Hubbard 
Model. 

\begin{figure}[hbt]
\begin{center}
\includegraphics[width=16cm]{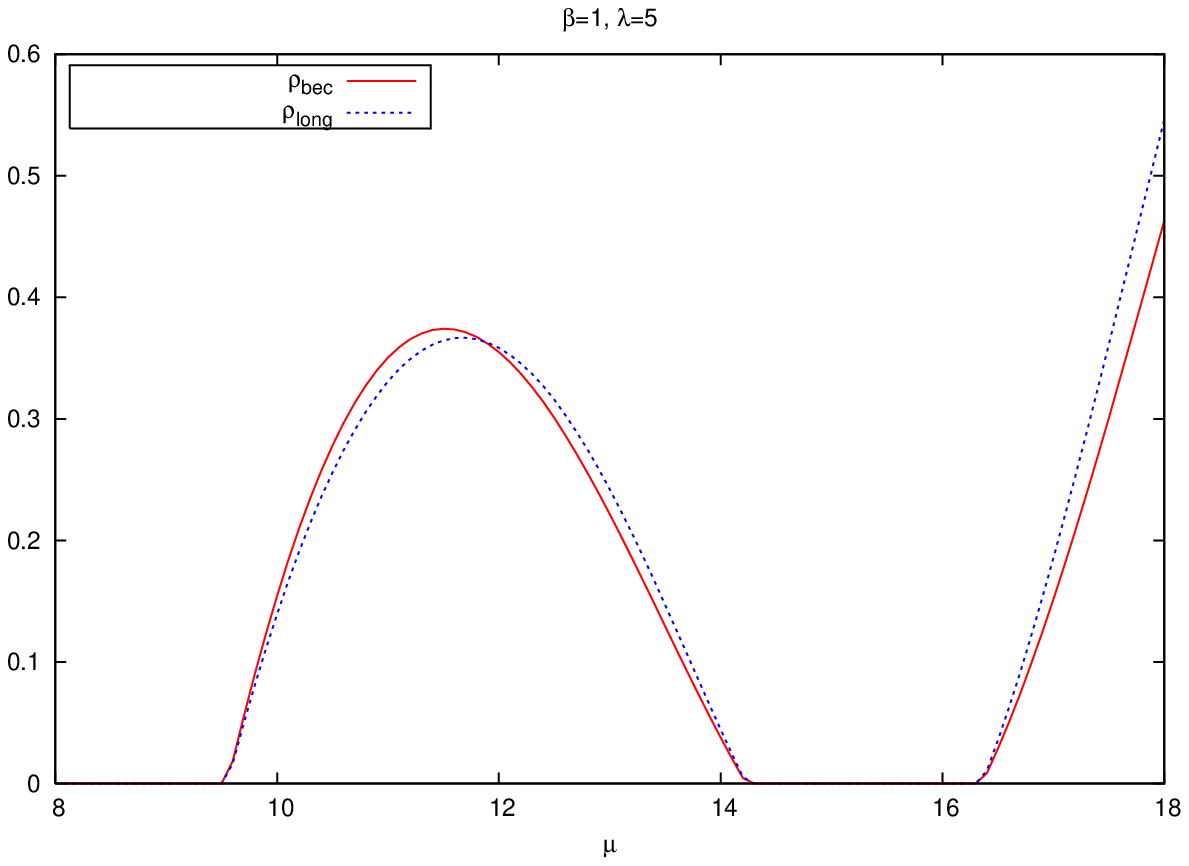}
\end{center}
\caption[Comparison of $\rhocond$ with $\rholong$]{\it Comparison of $\rhocond$ with $\rholong$}
\label{04fig2}
\end{figure}

\begin{figure}
  \begin{center}
    \begin{tabular}{cc}
      \resizebox{80mm}{!}{\includegraphics{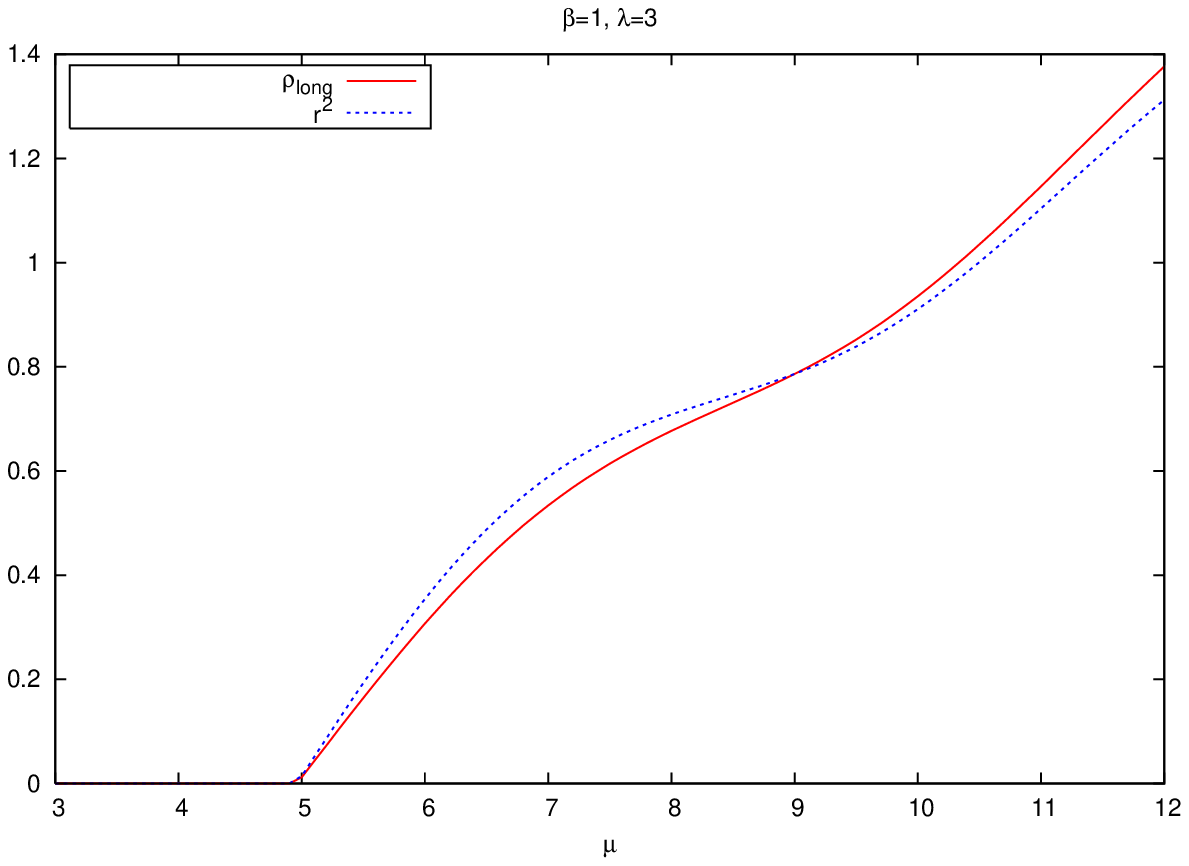}} &
      \resizebox{80mm}{!}{\includegraphics{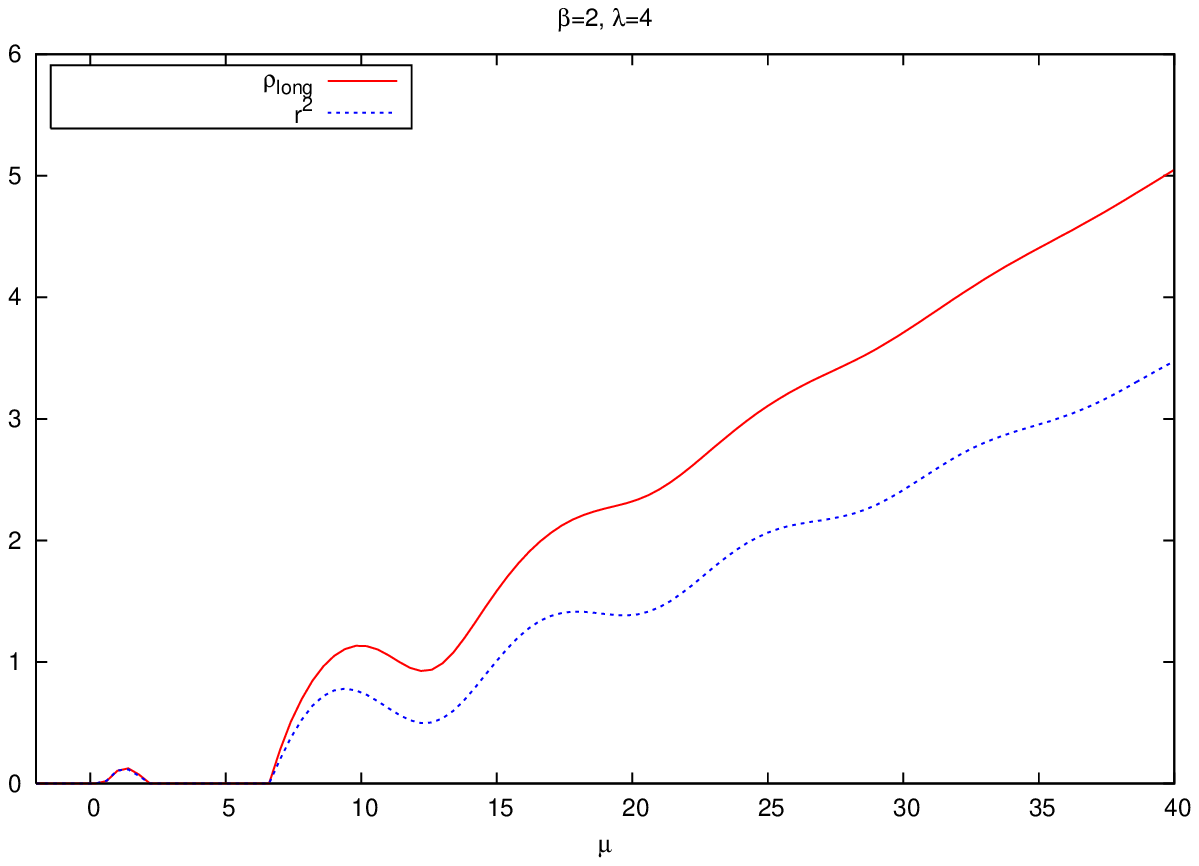}} \\
      \resizebox{80mm}{!}{\includegraphics{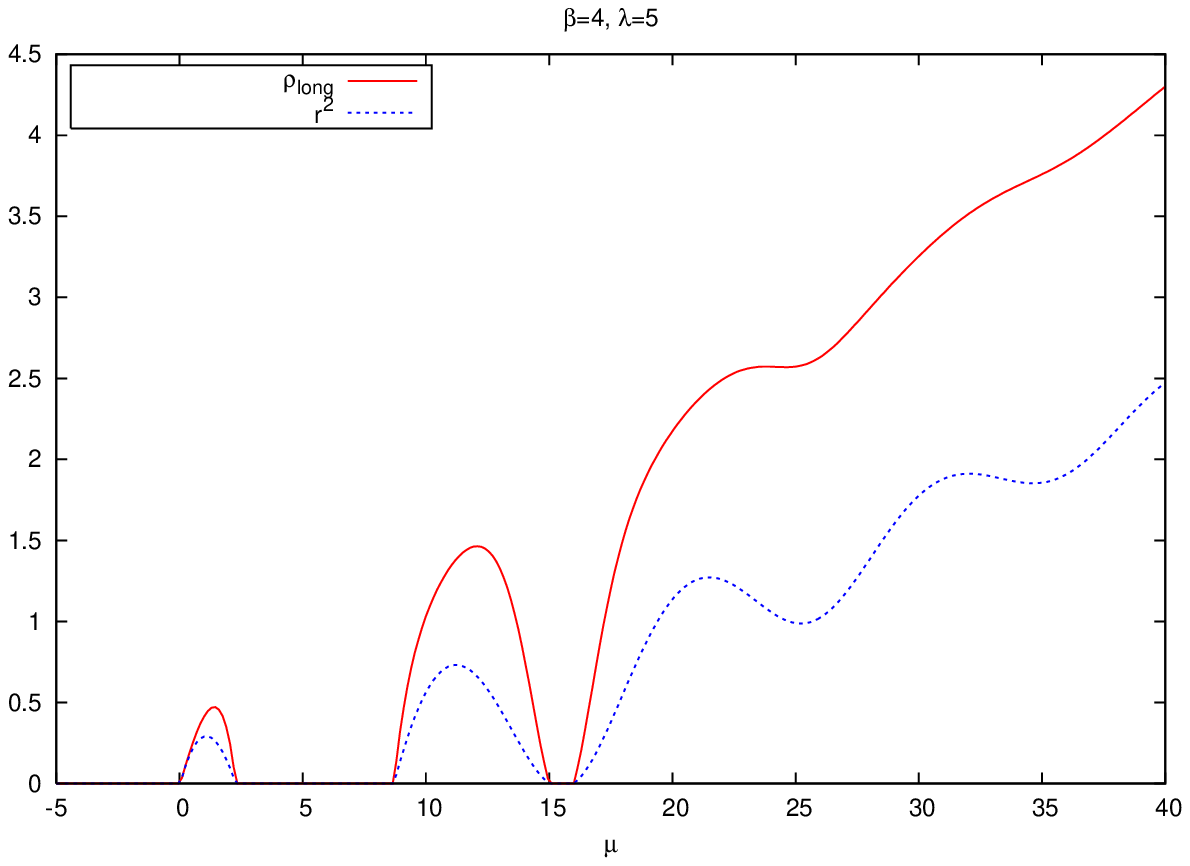}} &
      \resizebox{80mm}{!}{\includegraphics{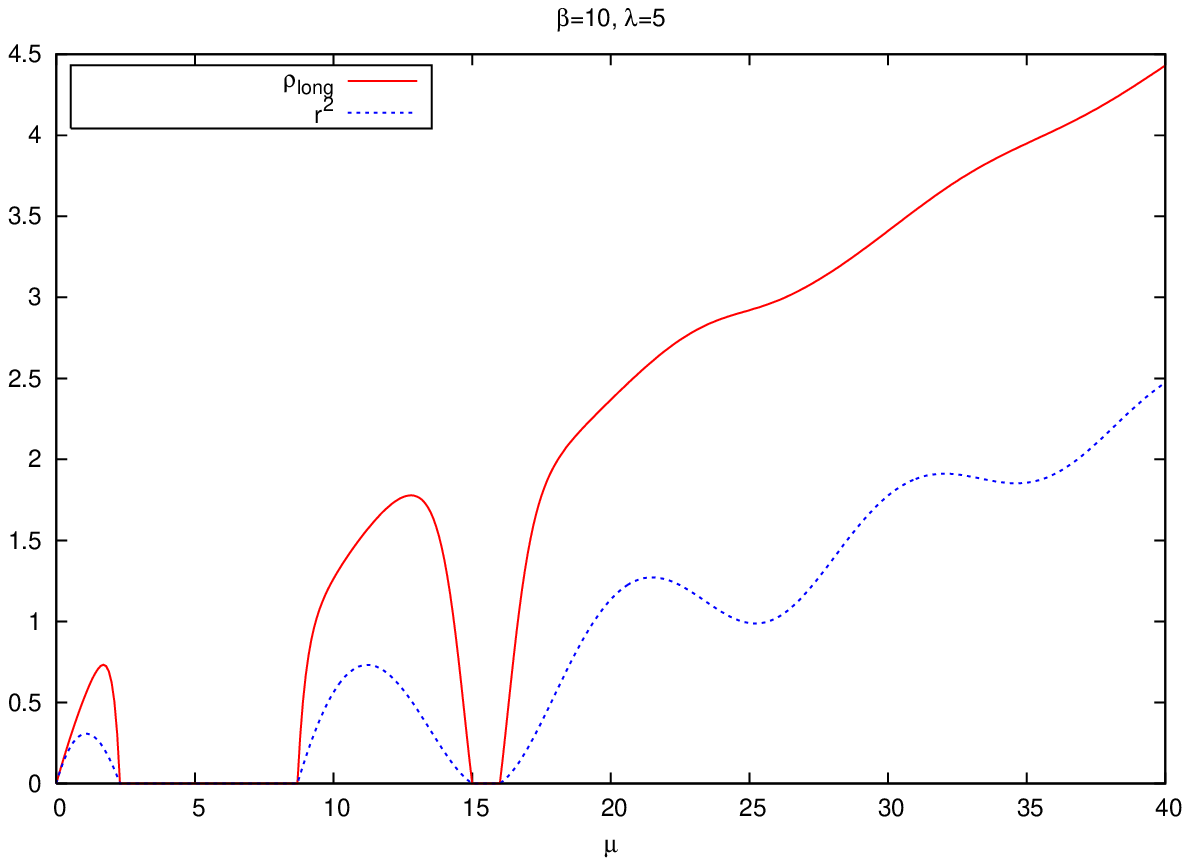}} \\
    \end{tabular}
    \caption[Comparison of long cycles with $\rhocond$ for various values of $\beta$ and $\lambda$]{\it Comparison 
    of long cycles with $\rhocond$ for various values of $\beta$ and $\lambda$}
    \label{test4}
  \end{center}
\end{figure}

Assuming our conjecture is correct, we apply some simple numerical techniques to this expression in order 
to compare long cycles with the Bose-Einstein condensate (see Appendix \ref{appendixC}
for a brief explanation of the algorithms used and the numerical source code).
As may be seen from Figures \ref{04fig2} and \ref{test4}
the calculations certainly agree with Lemma (\ref{no-long}), i.e. that the absence of condensation implies 
the lack of long cycles and visa versa. However more importantly they also indicate that while the 
presence of condensation coincides with the existence of long cycles, their respective densities are 
not necessarily equal. In fact, it is indicated that the long cycle density may be greater than 
or less than the condensate density for differing parameters.

\begin{appendix}

%
%
%

\setcounter{equation}{0}

\makeatletter
\renewcommand{\section}{\@startsection%
{section}%
{1}%
{0mm}%
{-\baselineskip}%
{0.5\baselineskip}%
{\normalfont\large\bfseries}%
} \makeatother

\renewcommand{\theequation}{\appendixletter .\arabic{equation}}

\renewcommand{\baselinestretch}{1}
\makeatletter
\renewcommand{\subsection}{\@startsection%
{subsection}%
{2}%
{0mm}%
{-\baselineskip}%
{0.5\baselineskip}%
{\normalfont\normalsize\bfseries}}%

\def\appendixletter{A}

\chapter[Penrose's Treatment of the Hard-Core Model]{Penrose's Treatment of the Infinite-Range-Hopping
Bose-Hubbard Model with Hard-Cores}
												\label{appendixA}
Here we shall briefly summarise the procedure in Penrose's work \cite{Penrose} in deriving an
expression for the condensate density for the Infinite-Range-Hopping
Bose-Hubbard model. We consider a collection of $n$ bosons
on a lattice of $V$ sites with the hard-core condition, i.e. no more than one particle per site.
As we are dealing with bosons, it is not necessary to specify which particle is where, only which sites are 
occupied and which are not. The allowed states of the system lie in a $n$-particle subspace $\mathcal{H}_n$ of
$\Hcan{n}$. Let $X$ be the set of lattice points which are occupied.
Then denote the corresponding quantum state $|X\ra \in \mathcal{H}_n$.
For a given $n$, there are $\binom{V}{n}$ such states, hence $\dim \mathcal{H}_n = \binom{V}{n}$.

Define creation and annihilation operators $A^\ast$ and $A$ by the following:
\begin{align*}
	A^\ast | X \ra &= \frac{1}{\sqrt{V}} \sum_{y \in X^c} | X \cup \{y\} \ra
\shortintertext{and}
	A | X \ra &=	\frac{1}{\sqrt{V}} \sum_{x \in X} | X \backslash \{x\} \ra.
\end{align*}
These operators satisfy the commutation relation $[A, A^\ast] = 1 - \tfrac{2}{V} N$, where
$N$ is the number operator defined by
\[
	N | X \ra = \#(X) | X \ra.
\]
Penrose finds that the Hamiltonian for this model may be expressed in terms of these operators
\[
	H = N - A^\ast A.
\]
where the $A^\ast A$ encompass the kinetic term of the Hamiltonian. 

From the commutation relation $[N,A^\ast] = A^\ast$ one may interpret the operator
$A^\ast$ as one which increases the number of particles in the system by one. However defining
the following operator
\begin{equation}									\label{apA-L}
	L = N - A^\ast A - V^{-1}\big( N^2 - N \big)
	= H - V^{-1}\big( N^2 - N \big)
\end{equation}
one may calculate that $[L, A^\ast] = 0$. Note also that $[H, L] = [H,N] = 0$. 
Hence we have two simultaneously
diagonalisable quantities with which we may be able to parameterise the 
eigenstates of $H$. By (\ref{apA-L}), we immediately have
\[
	H = L + V^{-1}\big( N^2 - N \big).
\]
Let $r$ be any integer satisfying $0 \le r \le V/2$ and construct the following subspace of $\mathcal{H}$:
\[
	\mathcal{H}_r = \big\{ | \phi \ra : N|\phi \ra = r | \phi \ra \big\},
\]
then by construction every state $|\phi\ra \in \mathcal{H}_r$ satisfies
\[
	N|\phi \ra = r | \phi \ra,
\qquad
	H|\phi \ra = r | \phi \ra,
\qquad
	L|\phi \ra = [r - V^{-1}(r^2 - r)] | \phi \ra.
\]
Then consider the further sequence of subspaces
\[
	\mathcal{H}_{r,m} = (A^\ast )^{m-r} \mathcal{H}_r
\]
for $m=r, r+1, \dots, V-r$ (with the consequence $\mathcal{H}_{r,r} = \mathcal{H}_r$).
One may then observe that for every state $|\phi\ra \in \mathcal{H}_{r,m}$ safisfies
\begin{equation}									\label{Penrose_N}
	N | \phi \ra = m | \phi \ra,
\qquad
	L | \phi \ra = [r - V^{-1}(r^2 - r)] | \phi \ra
\end{equation}
and consequently
\begin{equation}									\label{Penrose_Eigenvalues}
	H | \phi \ra = [r - V^{-1}(r^2 - r) + V^{-1}(m^2 - m)] | \phi \ra.
\end{equation}
The subspaces $\mathcal{H}_{r,m}$ for all $0 \le r \le V/2$ and $m=r, r+1, \dots V-r$
are all orthogonal, as they correspond to unique pairs of eigenvalues for the operators $N$ and $L$.
And thus they are eigenspaces of $H$.

Penrose then determines the dimensions of these subspaces to be
\[
	\dim \mathcal{H}_{r,m} = \dim \mathcal{H}_r = \binom{V}{r}\left( \frac{V-2r+1}{V-r+1} \right)
\]
with which we can write down an expression for the partition function of an $n$
particle system as a scaled sum of contributions from each subspace $\mathcal{H}_{r,n}$:
\[
	Z_\sbeta(n,V)=\hskip -0.3cm\sum_{r=0}^{\min(n, V-n)} z(r,n,V,\beta),
\]
where
\[
	z(r,n,V,\beta)
=	\left( \frac{V-2r+1}{V-r+1} \right) \binom{V}{r}
	\exp\left\{ -\frac{\beta}{V} \left[ Vr - r^2 + r + n^2 - n \right] \right\}.
\]


\section{Condensate Density}
To find the condensate density, we need to evaluate
\begin{equation}									\label{Penrose_dens}
	\rhobcond = \thermlim \frac{\la N_0 \ra}{V}
\end{equation}
where
\[
	\la N_0 \ra = \frac{\trace\big[ A^\ast A \e^{-\beta H}\big]}{\trace[ \e^{-\beta H} ]}
\]
Again we use the decomposition of the Hilbert space $\mathcal{H}_n$ into the
subspaces $\mathcal{H}_{r,m}$ as performed above. By (\ref{apA-L}) we have 
that $A^\ast A = N - H$, and using (\ref{Penrose_N}) and (\ref{Penrose_Eigenvalues})
we see that $\mathcal{H}_{r,m}$ is an eigenspace of $A^\ast A$ with eigenvalue
\[
	m - [r - V^{-1}(r^2 - r) + V^{-1}(m^2 - m)] = \frac{1}{V} \left[ (m-r)(V-m-r+1) \right].
\]
This gives us sufficient information to fully diagonalise the traces in (\ref{Penrose_dens}) to 
obtain
\begin{align*}
	\rhobcond 
&=	\thermlim \frac{\la N_0 \ra}{V}
\\
&=	\frac{1}{Z_\sbeta(n,V)} \sum_{r=0}^{\min(n, V-n)} \frac{ (N-r)(V-N-r+1)}{V^2} z(r,n,V,\beta).
\end{align*}
To evaluate this expression, we compare with Penrose's Large Deviation-like variational
formula from Section \ref{sub Z}. Set $x = r/V$ and $\rho = N/V$ and obtain
\[
	h_\sV( x ) = ( \rho - x)(1-\rho-x+1/V).
\]
This $h_\sV$ converges uniformly in $[0,\min(\rho,1-\rho)]$
as $V\to\infty$ to a continuous function $h:[0,\min(\rho,1-\rho)]\to \RR$ of the form
\[
	h(x) = ( \rho - x)(1-\rho-x).
\]
Inserting this into (\ref{LD}), we obtain the density of the condensate for this model:
\[
	\rhobcond =
	\begin{cases}
		    0		 & \mathrm{if}\ \  \rho\in[0,\rho_\beta] \cup [1-\rho_\beta,1],
    	\\
    		( \rho - \rho_\beta )(1-\rho-\rho_\beta) & \mathrm{if}\ \  \rho\in[\rho_\beta, 1-\rho_\beta]
	\end{cases}
\]
where $\rho_\beta$ is the critical temperature as stated in Chapter \ref{chapter4} (see 
equation (\ref{gbeta}) and the following paragraph for a definition).

\def\appendixletter{B}

\chapter{Counter-example to the Positivity of \texorpdfstring{$c(q,A)$}{the cycle expectation}}	\label{appendixE}

In Chapter \ref{chapter3} a series of three conditions was developed which when satisfied
by a particular model implies that the density of the Bose-Einstein condensate
equals the long cycle density in that model, i.e. $\rholong = \rhocond$. 
Here we wish to give an example of a model which breaks Condition \Rmnum{1}, which demanded 
that for any single-particle operator $A\ge0$, the respective
cycle expectation is non-negative, i.e. $c^n_\sV(q,A)\ge0$.
In this appendix we shall analyse more deeply
the Infinite-Range-Hopping Bose-Hubbard Model with Hard-Cores of Chapter \ref{chapter4} 
in order to evaluate the cycle statistics of this model for the particular case of three particles hopping 
on a lattice of $V$ sites. After 
explicitly finding the eigenvalues and eigenvectors for the three particle
Hamiltonian $H^{\rm hc}_{3,V}$ (using the notation of Chapter \ref{chapter4}), 
we derive expressions for projections upon this eigen-basis and
rewrite the partition function with respect to this diagonalisation.
Then we can obtain an analytical expression for the cycle density $c^3_\sV(q)$
for any $q$. We find that the addition of a 
the hard-core interaction actually causes a damping effect on the cycle density
of the particles. This damping can be exploited by choosing a particular
operator which projects onto a vector perpendicular to the ground state,
whose cycle expectation can be negative for certain temperatures.

\section{Diagonalising the 3-particle Hard-Core Hamiltonian}		\label{apE-sec1}
To begin, recall that the one particle Hamiltonian $h_\sV$ has eigenvalues 0 and 1. The eigenvector corresponding
to 0 is $\g=V^{-1/2}(1,1,1,\ldots 1)$ so then we may write $h_\sV = 1 - |\g\ra\la\g|$.
Let $\ee_i=(0,\ldots, 0,1,0,\ldots, 0)$ with 1 in the $i$-th place,
which represents a particle on the site labelled $i$ on the lattice. One may then evaluate that
\[
	h_\sV |\ee_i \ra = \ee_i - \tfrac{1}{V} \sum_j \ee_j.
\]
Note also that for any normalised $\cc \perp \g$ with components $c_i$, then
\begin{equation}									\label{apE-orthog}
	\sum_{\cc\perp \g} c_i \bar{c}_j=\delta_{ij}-1/V.
\end{equation}

$H^{\rm hc}_{3,V}$ is the  Hamiltonian for three particles with Hilbert space $\Hhccan{3}$.
Any state $\phi \in \Hhccan{3}$ may be expressed in the form
\begin{equation}									\label{apE-basicstate}
	\phi = \sum_{i,j,k=1}^V \phi_{ijk} \, \ee_i \otimes \ee_j \otimes \ee_k
\end{equation}
where $\phi_{ijk} \in \CC$ with the restriction that
$\phi_{ijk}=0$ when $i,j,k$ are not distinct, i.e. $\phi_{iij} = \phi_{iji} = \phi_{ijj} = 0$,
to satisfy the hard-core condition. 
Denote $\Phi = \sum_{i,j,k} \phi_{ijk}$.

We want to find such states (i.e. values of $\phi_{ijk}$) that solve the eigen-equation $H^{\rm hc}_{3,V}\phi = E\phi$,
where $E$ are the energy eigenvalues of the system. Inserting (\ref{apE-basicstate}) into
the eigen-problem, one obtains
\begin{align*}
	H^{\rm hc}_{3,V} \phi
&=	3 \phi - \frac{1}{V} \sum_{i,j,k=1}^V \phi_{ijk} \left( \sum_{i'\ne j,k} \ee_{i'} \right) \otimes \ee_j \otimes \ee_k
	- \frac{1}{V} \sum_{i,j,k=1}^V \phi_{ijk} \, \ee_i \otimes \left( \sum_{j'\ne i,k} \ee_{j'} \right) \otimes \ee_k
\\*
&\qquad\qquad\qquad
	- \frac{1}{V} \sum_{i,j,k=1}^V \phi_{ijk} \, \ee_i \otimes \ee_j \otimes \left( \sum_{k'\ne i,j} \ee_{k'} \right)
\\[0.2cm]
&=	3 \phi - \frac{1}{V} \sum_{i,j,k=1}^V \left[ 
	\sum_{i'\ne j,k} \phi_{i'jk} + \sum_{j'\ne i,k} \phi_{ij'k} + \sum_{k'\ne i,j} \phi_{ijk'}
	\right] \ee_i \otimes \ee_j \otimes \ee_k
\vcentcolon=	E \phi	
\end{align*}
which implies that
\begin{equation}									\label{c_ijk}
	\phi_{ijk} 
=	\begin{cases}
		\frac{1}{(3-E)V} \left[ 
		\sum_{i'\ne j,k} \phi_{i'jk} + \sum_{j'\ne i,k} \phi_{ij'k} + \sum_{k'\ne i,j} \phi_{ijk'}
		\right]
		&	\text{$i,j,k$ distinct},
	\\
		0	& \text{otherwise}.
	\end{cases}
\end{equation}
Suppose for the moment that $E \ne 3$, then for distinct indices $i,j,k$ denote
\[
	a_{jk} = \sum_{i'} \phi_{i'jk} \qquad b_{ki} = \sum_{j'} \phi_{ij'k} \qquad c_{ij} = \sum_{k'} \phi_{ijk'}
\]
(Note that we can drop the restrictions in the sums here since $\phi_{ijk}=0$ when $i,j,k$ are not distinct).
Then by definition $a_{ii} = b_{ii} = c_{ii} = 0$, $\sum_{jk} a_{jk} = \sum_{ki} b_{ki} = \sum_{ij} c_{ij} = \Phi$
and
\[
	\sum_j a_{jk} = \sum_i b_{ki} \qquad \sum_k b_{ki} = \sum_j c_{ij}
	\qquad \sum_k a_{jk} = \sum_i c_{ij}.
\]
Using this notation, (\ref{c_ijk}) becomes \vspace{-0.2cm}
\begin{equation}									\label{c_ijk_cleaner}
	\phi_{ijk} = \lambda^{-1} \big( a_{jk} + b_{ki} + c_{ij} \big)\vspace{-0.2cm}
\end{equation}
where $\lambda = (3-E) V$.

Sum (\ref{c_ijk_cleaner}) over $i (\ne j,k)$:
\begin{align*}
	\sum_{i (\ne j,k)} \phi_{ijk}
&=	\lambda^{-1} \sum_{i (\ne j,k)} \big( a_{jk} + b_{ki} + c_{ij} \big)
\\
\Rightarrow\quad a_{jk}
&=	\lambda^{-1} \left[ (V-2) a_{jk} + \sum_{i (\ne j,k)} b_{ki} + \sum_{i (\ne j,k)} c_{ij} \right]
\\
\Rightarrow\quad (\lambda - V + 2 ) a_{jk}
&=	\sum_i b_{ki} + \sum_i c_{ij} - b_{kj} - c_{kj}.
\end{align*}
By symmetry, summing  (\ref{c_ijk_cleaner}) over $j (\ne i,k)$ and over $k (\ne i,j)$, one can
obtain the following system of equations
\begin{align}
	(\lambda - V + 2 ) a_{jk} + b_{kj} + c_{kj}				\label{simul1}
&=	\sum_i b_{ki} + \sum_i c_{ij}
\\
	a_{ik} + (\lambda - V + 2 ) b_{ki} + c_{ik}				\label{simul2}
&=	\sum_j a_{jk} + \sum_j c_{ij}
\\
	a_{ji} + b_{ji} + (\lambda - V + 2 ) c_{ij}				\label{simul3}
&=	\sum_k a_{jk} + \sum_k b_{ki}
\end{align}
We want to find all possible solutions to this set of equations, so that to obtain values for $E$ and 
corresponding information on the coefficients $\phi_{ijk}$ (i.e. the $a_{jk},b_{ki},c_{ij}$).
Let 
\begin{align*}
	A_k = \sum_j a_{jk} \qquad B_k = \sum_j b_{jk} \qquad C_k = \sum_j c_{jk}
\\
	\tilde{A}_j = \sum_k a_{jk} \qquad \tilde{B}_j = \sum_k b_{jk} \qquad \tilde{C}_j = \sum_k c_{jk}
\end{align*}
where again $\sum_k A_k = \sum_k B_k = \sum_k C_k = \sum_j \tilde{A}_j = \sum_j \tilde{B}_j 
= \sum_j \tilde{C}_j = \Phi$. Also we have that
\[
	A_k = \tilde{B}_k, \qquad B_k = \tilde{C}_k, \qquad C_k = \tilde{A}_k.
\]
For brevity also denote $-\sigma = \lambda - V + 2$, then using these additional notations
the simultaneous equations (\ref{simul1})-(\ref{simul3}) may be written as
\begin{align}
	-\sigma a_{jk} + b_{kj} + c_{kj}				\label{simulB1}
&=	A_k + C_j
\\
	a_{ik} -\sigma b_{ki} + c_{ik}				\label{simulB2}
&=	B_k + A_j
\\
	a_{ji} + b_{ji} -\sigma c_{ij}				\label{simulB3}
&=	C_k + B_j
\end{align}
which we may rewrite in matrix form as
\begin{equation}							\label{simul_matrix}
	\begin{pmatrix}
		-\sigma & 0 & 0 & 0 & 1 & 1 \\
		0 & -\sigma & 0 & 1 & 0 & 1 \\
		0 & 0 & -\sigma & 1 & 1 & 0 \\
		0 & 1 & 1 & -\sigma & 0 & 0 \\
		1 & 0 & 1 & 0 & -\sigma & 0 \\
		1 & 1 & 0 & 0 & 0 & -\sigma
	\end{pmatrix}
	\begin{pmatrix}
		a_{jk} \\ b_{jk} \\ c_{jk} \\ a_{kj} \\ b_{kj} \\ c_{kj}
	\end{pmatrix}
	=
	\begin{pmatrix}
		A_k + C_j \\ B_k + A_j \\ C_k + B_j \\ A_j + C_k \\ B_j + A_k \\ C_j + B_k
	\end{pmatrix}.
\end{equation}

\textbf{Sum each row of (\ref{simul_matrix}) over both indices}\\
Begin by summing the first row of (\ref{simul_matrix}) (which is equation (\ref{simulB1})\!) over both $j$ 
and $k$ ($j \ne k$), then
{\allowdisplaybreaks
\begin{align*}
	-\sigma \sum_{\la j,k\ra} a_{jk} + \sum_{\la j,k\ra} b_{kj} + \sum_{\la j,k\ra} c_{kj}
&=	(V-1)\sum_{i,k} b_{ki} + (V-1) \sum_{i,j} c_{ij}
\\
\Rightarrow \qquad\qquad
	( -\sigma + 1 + 1 ) \Phi 
&=	2 (V-1) \Phi
\\
\Rightarrow\qquad\,\,
	( -\sigma - 2(V -2) ) \Phi 
&=	0
\end{align*}}
which implies either
\[
	\sigma = 2(2-V) \qquad \text{or} \qquad \Phi = 0.
\]
This result is independent of the row chosen, due to the symmetry of the system.

\textbf{Case I: $\sigma = 2(2-V) \Rightarrow E=\frac{6}{V}$}\\
In this case we obtain the following for (\ref{simul_matrix})
\[
	\begin{pmatrix}
		2(V-2) & 0 & 0 & 0 & 1 & 1 \\
		0 & 2(V-2) & 0 & 1 & 0 & 1 \\
		0 & 0 & 2(V-2) & 1 & 1 & 0 \\
		0 & 1 & 1 & 2(V-2) & 0 & 0 \\
		1 & 0 & 1 & 0 & 2(V-2) & 0 \\
		1 & 1 & 0 & 0 & 0 & 2(V-2)
	\end{pmatrix}
	\begin{pmatrix}
		a_{jk} \\ b_{jk} \\ c_{jk} \\ a_{kj} \\ b_{kj} \\ c_{kj}
	\end{pmatrix}
	=
	\begin{pmatrix}
		A_k + C_j \\ B_k + A_j \\ C_k + B_j \\ A_j + C_k \\ B_j + A_k \\ C_j + B_k
	\end{pmatrix}
\]
which has solution
\[
	a_{jk} = b_{jk} = c_{jk} = a_{kj} = b_{kj} = c_{kj} = 1
\]
since then $A_i = B_i = C_i = V-1$. Therefore using (\ref{c_ijk_cleaner}), the corresponding 
normalised eigenstate for the energy eigenvalue $6/V$ is
\[
	\mathbf{X}_3
=	\frac{1}{\sqrt{V(V-1)(V-2)}} \sum_{\la i,j,k \ra} \ee_i \otimes \ee_j \otimes \ee_k.
\]
This represents three bosons all in their lowest energy state.

\textbf{Case II: $\Phi=0$ where $A_i,B_i,C_i$ are not all zero}\\
We shall sum each of the three simultaneous equations (\ref{simulB1})--(\ref{simulB3}) over $k (\ne j)$.
Taking the first equation we get
\begin{equation*}
	-\sigma \tilde{A}_j + B_j + C_j = \underbrace{\sum_k \tilde{B}_k}_{=\Phi=0} - \tilde{B}_j + (V-1)C_j
\quad\Rightarrow\quad
	A_j + B_j + \tilde{\sigma}  C_j = 0
\end{equation*}
where $\tilde{\sigma} = -\sigma - V + 2 = V(1-E)+4$. Performing the same to the other two equations (\ref{simulB2})
and (\ref{simulB3}) one may form the system
\begin{equation}									\label{case2_matrix}
	\begin{pmatrix}
		\tilde{\sigma} & 1 & 1 \\
		1 & \tilde{\sigma} & 1 \\
		1 & 1 & \tilde{\sigma}
	\end{pmatrix}
	\begin{pmatrix} A_j \\ B_j \\ C_j \end{pmatrix}
=	0.
\end{equation}
Since we are assuming $A_j, B_j, C_j \ne0$ then this system of equations is solvable if and only
if the determinant of the matrix is zero, and therefore $\tilde{\sigma}$ must satisfy:\vspace{-0.2cm}
\begin{align*}
	(\tilde{\sigma} - 1)^2 (\tilde{\sigma} +& 2) = 0
\\
\Rightarrow \qquad E = 1 + 3/V \qquad &\text{or} \qquad E = 1+6/V.
\vspace{-0.2cm}\end{align*}
Each of these eigenvalues will be considered separately.

\textbf{Case II(a): $\Phi=0$ where $A_i,B_i,C_i$ are not all zero -- taking $E=1+3/V$}\\
In this case, $\tilde{\sigma} = 1$, and hence (\ref{case2_matrix}) becomes
\[
	\begin{pmatrix}
		1 & 1 & 1 \\
		1 & 1 & 1 \\
		1 & 1 & 1
	\end{pmatrix}
	\begin{pmatrix} A_j \\ B_j \\ C_j \end{pmatrix}
=	0
	\qquad \implies \qquad A_j + B_j + C_j = 0.
\]
The matrix has rank 1 and thus has a two dimensional solution space, which implies that
this eigenvalue is doubly-degenerate. Consider the following two orthogonal solutions\vspace{-0.1cm}
\[
	A_j = -B_j \text{ with } C_j =0, \qquad\text{ and }\qquad A_j = B_j = \tfrac{1}{2} C_j \vspace{-0.1cm}
\]
and take them one at a time:\newline
$\bullet\; A_j = -B_j$ with $C_j =0$\\
For this possibility, $\tilde{\sigma} = 1 \Rightarrow \sigma = 1-V$ and so (\ref{simul_matrix}) will become
\[
	\begin{pmatrix}
		V-1 & 0 & 0 & 0 & 1 & 1 \\
		0 & V-1 & 0 & 1 & 0 & 1 \\
		0 & 0 & V-1 & 1 & 1 & 0 \\
		0 & 1 & 1 & V-1 & 0 & 0 \\
		1 & 0 & 1 & 0 & V-1 & 0 \\
		1 & 1 & 0 & 0 & 0 & V-1
	\end{pmatrix}
	\begin{pmatrix}
		a_{jk} \\ b_{jk} \\ c_{jk} \\ a_{kj} \\ b_{kj} \\ c_{kj}
	\end{pmatrix}
	=
	\begin{pmatrix}
		A_k \\ A_j - A_k \\ -A_j \\ A_j \\ A_k - A_j \\ -A_k
	\end{pmatrix}
\]
which has solution
\[
	\frac{1}{V}
	\begin{pmatrix}
		\frac{A_j + (V-1)A_k}{V-2} \\
		A_j - A_k \\
		- \frac{A_k + (V-1)A_j}{V-2} \\
		\frac{A_k + (V-1)A_j}{V-2} \\
		- (A_j - A_k) \\
		- \frac{A_j + (V-1)A_k}{V-2} \\
	\end{pmatrix}
	\qquad \implies
	\begin{minipage}{6cm}
		\begin{align*}
			a_{jk} &= - c_{kj} \\
			c_{jk} &= - a_{kj} \\
			b_{jk} &= - b_{kj} \\
			\text{and } a_{jk} &+ b_{jk} + c_{jk} = 0
		\end{align*}
	\end{minipage}
\]
Therefore using (\ref{c_ijk_cleaner}), the corresponding eigenstate for the energy eigenvalue $1+3/V$
may be written as
\[
	\mathbf{X}_7 \propto
	\sum_{\la i,j,k\ra} \left[ \frac{A_j + (V-1)A_k}{V-2} + A_k - A_i - \frac{A_j + (V-1)A_i}{V-2}
		 \right]\,\ee_i \otimes \ee_j \otimes \ee_k
\]
which when simplified and normalised (i.e. taking $\sum_i |A_i|^2 = 1$ with $\sum_i A_i = 0$) becomes
\[
	\mathbf{X}_7
=	\frac{1}{\sqrt{2V(V-2)}} \sum_{\la i,j,k \ra} ( A_k - A_i )\,\ee_i \otimes \ee_j \otimes \ee_k.
\]

$\bullet\; A_j = B_j = \tfrac{1}{2} C_j$\\
By following the same procedure as above, one may find a second eigenvector of the 
degenerate eigenvalue $1+3/V$ of the form
\[
	\mathbf{X}_8
=	\frac{1}{\sqrt{6V(V-2)}} \sum_{\la i,j,k \ra} ( A_i + A_k - 2A_j )\,\ee_i \otimes \ee_j \otimes \ee_k
\]
which is orthogonal to $\mathbf{X}_7$.

\textbf{Case II(b): $\Phi=0$ where $A_i,B_i,C_i$ are not all zero -- taking $E=1+6/V$}\\
In this case, $\tilde{\sigma} = -2$, and hence (\ref{case2_matrix}) is
\[
	\begin{pmatrix}
		-2 & 1 & 1 \\
		1 & -2 & 1 \\
		1 & 1 & -2
	\end{pmatrix}
	\begin{pmatrix} A_j \\ B_j \\ C_j \end{pmatrix}
=	0.
\]
The matrix has rank 2 and thus has a single dimensional solution space defined by:
\[
	A_j = B_j = C_j.
\]
As $\tilde{\sigma} = -2 \Rightarrow \sigma = 4-V$ and so (\ref{simul_matrix}) will become
\[
	\begin{pmatrix}
		V-4 & 0 & 0 & 0 & 1 & 1 \\
		0 & V-4 & 0 & 1 & 0 & 1 \\
		0 & 0 & V-4 & 1 & 1 & 0 \\
		0 & 1 & 1 & V-4 & 0 & 0 \\
		1 & 0 & 1 & 0 & V-4 & 0 \\
		1 & 1 & 0 & 0 & 0 & V-4
	\end{pmatrix}
	\begin{pmatrix}
		a_{jk} \\ b_{jk} \\ c_{jk} \\ a_{kj} \\ b_{kj} \\ c_{kj}
	\end{pmatrix}
	=
	(A_j + A_k)
	\begin{pmatrix}
		1 \\ 1 \\ 1 \\ 1 \\ 1 \\ 1
	\end{pmatrix}
\]
which has solution
\[
	a_{jk} = b_{jk} = c_{jk} = a_{kj} = b_{kj} = c_{kj} = \frac{A_j + A_k}{V}.
\]
Hence we can construct the respective eigenvector of $1+6/V$:
\[
	\mathbf{X}_4=\frac{1}{\sqrt{3(V-2)(V-3)}} \sum_{\la i,j,k \ra} ( A_i + A_j + A_k )
					\,\ee_i \otimes \ee_j \otimes \ee_k.
\]
This is a bosonic state of two particles in their ground states with one excited particle.

\textbf{Case III: $\Phi=0$ where $A_i = B_i = C_i =0$}\\
In this case, (\ref{simul_matrix}) reduces to the following eigensystem
\[
	\begin{pmatrix}
		0\, & 0 & 0 & 0 & 1 & 1 \\
		0 & 0\, & 0 & 1 & 0 & 1 \\
		0 & 0 & 0\, & 1 & 1 & 0 \\
		0 & 1 & 1 & 0\, & 0 & 0 \\
		1 & 0 & 1 & 0 & 0\, & 0 \\
		1 & 1 & 0 & 0 & 0 & 0\,
	\end{pmatrix}
	\begin{pmatrix}
		a_{jk} \\ b_{jk} \\ c_{jk} \\ a_{kj} \\ b_{kj} \\ c_{kj}
	\end{pmatrix}
=
	\sigma \,\II_6\,
	\begin{pmatrix}
		a_{jk} \\ b_{jk} \\ c_{jk} \\ a_{kj} \\ b_{kj} \\ c_{kj}
	\end{pmatrix}
\]
where $\II_6$ is the $6 \times 6$ identity matrix.

The above matrix has the following eigenvalues and eigenvectors:
\begin{align*}
	\sigma&=2	&	&(1,1,1,1,1,1)^T &
\\
	\sigma&=-2	&	&(-1,-1,-1,1,1,1)^T &
\\
	\sigma&=1	&	&(1, -1, 0, -1, 1, 0)^T \text{ and } (1,1,-2,-1,-1,2)^T &
\\
	\sigma&=-1	&	&(-1, 1, 0, -1, 1, 0)^T \text{ and } (1,1,-2,1,1,-2)^T. &
\end{align*}
Note that the eigenvectors for the degenerate eigenvalues were chosen to be orthogonal.
Considering each eigen-solution in turn:

\textbf{Case III(a): $\sigma = 2 \Rightarrow E = 2 + 4/V$}\\*
Here $(a_{jk}, b_{jk}, c_{jk}, a_{kj}, b_{kj}, c_{kj})^T = (1,1,1,1,1,1)^T$ so we have
that
\[
	a_{jk} = b_{jk} = c_{jk} = a_{kj} = b_{kj} = c_{kj}
\]
and hence inserting these components into (\ref{c_ijk_cleaner}) we can construct a normalised state
\begin{equation*}	
	\mathbf{X}_5 = \frac{1}{\sqrt{3(V-4)}} \sum_{i,j,k} \big( a_{jk} + a_{ki} + a_{ij} \big)
				\,\ee_i \otimes \ee_j \otimes \ee_k
\end{equation*}
where $a_{ij} = a_{ji}$ and $A_j=0$ implies that $\sum_i a_{ij} = 0$ for all $j=1\dots V$, 
$\sum_{ij} | a_{ij} |^2=1$. This is a bosonic state of two excited particles and one 
in its ground state.

\textbf{Case III(b): $\sigma = -2 \Rightarrow E = 2$}\\*
Here by $(-1,-1,-1,1,1,1)^T$ one deduces the relations $a_{jk} = -a_{kj} = b_{jk} = -b_{kj} = c_{jk} = -c_{kj}$
with which the normalised state can be written as
\begin{equation}									\label{apE-A1}
	\mathbf{X}_1 = \frac{1}{\sqrt{3V}} \sum_{i,j,k} \big( a_{jk} + a_{ki} + a_{ij} \big)
				\,\ee_i \otimes \ee_j \otimes \ee_k
\end{equation}
where now $a_{ij} = -a_{ji}$ with $\sum_i a_{ij} = 0$ for all $j=1\dots V$, $\sum_{ij} | a_{ij} |^2=1$. This 
may be rewritten in a more intuitive form by setting $a_{ij} = c_i c'_j - c_j c'_i$ to obtain
\[
	\mathbf{X}_1=\frac{1}{\sqrt{6}} \big( \g \otimes \cc_1\otimes \cc_2 - \g \otimes \cc_2 \otimes \cc_1
	+ \cc_2 \otimes \g \otimes \cc_1- \cc_1\otimes \g \otimes \cc_2 
	+ \cc_1\otimes \cc_2 \otimes \g - \cc_2 \otimes \cc_1\otimes \g \big)
\]
where $\cc_1 \perp \cc_2 \perp \g$. This represents a fermionic state with a single particle in the ground
state.

\textbf{Case III(c): $\sigma = 1 \Rightarrow E = 2+3/V$}\\*
This eigenvalue is doubly degenerate, so we shall consider each eigenvector separately:\\
$\bullet$ By $(1,-1,0,-1,1,0)^T$, have that $a_{jk} = -a_{kj} = -b_{jk} = b_{kj}$ and $c_{jk} = c_{kj} = 0$
and hence
\[
	\mathbf{X}_{11}
=	\frac{1}{\sqrt{2(V-3)}} \sum_{\la i,j,k \ra} ( a_{jk} - a_{ki} )\,\ee_i \otimes \ee_j \otimes \ee_k
\]
with $a_{ij} = -a_{ji}$ where the  $a_{ij}$'s are scaled as above. 

$\bullet$ By $(1,1,-2,-1,-1,2)^T$, $a_{jk} = -a_{kj} = -b_{jk} = b_{kj} = \tfrac{1}{2} c_{jk} = -\tfrac{1}{2} c_{kj}$
and hence
\[
	\mathbf{X}_{12}
=	\frac{1}{\sqrt{6(V-3)}} \sum_{\la i,j,k \ra} ( a_{jk} + a_{ki} - 2a_{ij} )\,\ee_i \otimes \ee_j \otimes \ee_k.
\]

\textbf{Case III(d): $\sigma = -1 \Rightarrow E = 2+1/V$}\\*
$\bullet$ By $(-1,1,0,-1,1,0)^T$, $a_{jk} = a_{kj} = -b_{jk} = -b_{kj}$ and $c_{jk} = c_{kj} = 0$
and thus
\[
	\mathbf{X}_9
=	\frac{1}{\sqrt{2(V-1)}} \sum_{\la i,j,k \ra} ( a_{jk} - a_{ki} )\,\ee_i \otimes \ee_j \otimes \ee_k
\]
with $a_{ij} = a_{ji}$ and again the $a_{ij}$'s scaled as above. 

$\bullet$ By $(1,1,-2,1,1,-2)^T$, $a_{jk} = a_{kj} = b_{jk} = b_{kj} = -\tfrac{1}{2} c_{jk} = -\tfrac{1}{2} c_{kj}$
to form
\[
	\mathbf{X}_{10}
=	\frac{1}{\sqrt{6(V-1)}} \sum_{\la i,j,k \ra} ( a_{jk} + a_{ki} - 2a_{ij} )\,\ee_i \otimes \ee_j \otimes \ee_k.
\]
\textbf{Case IV: $E=3$}\\
Here we cannot apply the above method as the right-hand-side of (\ref{c_ijk_cleaner})
is not defined. However we can propose a fermionic state with all excited particles of the form:
\[
	\mathbf{X}_2=\frac{1}{\sqrt{6}} \sum_{\pi \in S_3} \mathrm{sgn}(\pi) 
		\cc_{\pi(1)} \otimes \cc_{\pi(2)} \otimes \cc_{\pi(3)}
\]
where $\cc_1 \perp \cc_2 \perp \cc_3 \perp \g$ and $\|\cc_i\|=1$. This can be quickly verified
by recalling that $h_\sV = 1 - |\g \ra\la \g |$ and $H^{\rm hc}_{3,V} = \Phc{3} h_\sV^{(3)} \Phc{3}$. 
Similarly for bosons:
\[
	\mathbf{X}_6
	\propto \sum_{\la i_1, i_2, i_3 \ra} \phi_{i_1 i_2 i_3} \sum_{\pi \in S_3} \ee_{i_{\pi(1)}} \otimes \ee_{i_{\pi(2)}} 
	\otimes \ee_{i_{\pi(3)}}
\]
where $\phi_{iii}=\phi_{ijj} = \phi_{jij} = \phi_{jji} = 0$ and $\sum_i \phi_{ijk} = \sum_j \phi_{ijk} = \sum_k \phi_{ijk} = 0$.

\subsection*{Summary of Eigenvalues and Eigenstates of $H^{\rm hc}_{3,V}$}
We split the Hilbert space $\Hhccan{3}$ into a symmetric subspace, an anti-symmetric subspace, and a 
subspace which is neither symmetric nor anti-symmetric.
On the odd (fermionic) space the three-particle Hamiltonian $H^{\rm hc}_{3,V}$ has eigenvalues 
2 and 3 with eigenvectors:
\newline
$\bullet$ eigenvalue 2:
\[
	\mathbf{X}_1
=	\frac{1}{\sqrt{3V}} \sum_{\la i,j,k \ra} ( a_{jk} + a_{ki} + a_{ij} )\,\ee_i \otimes \ee_j \otimes \ee_k
\]
where $a_{ij} = -a_{ji}$ with $a_{ii} = 0$, $\sum_{i,j} |a_{ij}|^2 = 1$ and $\sum_i a_{ij} = 0$ for all $j$.
$\bullet$ eigenvalue 3:
\[
	\mathbf{X}_2=\frac{1}{\sqrt{6}} \sum_{\pi \in S_3} \mathrm{sgn}(\pi) 
		\cc_{\pi(1)} \otimes \cc_{\pi(2)} \otimes \cc_{\pi(3)}
\]
where $\cc_1 \perp \cc_2 \perp \cc_3 \perp \g$ and $\|\cc_i\|=1$.

The eigenvectors of $H^{\rm hc}_{3,V}$ in the even (bosonic) space are of the form:
\newline
$\bullet$ eigenvalue $6/V$:
\[
	\mathbf{X}_3 =\frac{1}{\sqrt{V(V-1)(V-2)}} \sum_{\la i,j,k \ra}\,\ee_i \otimes \ee_j \otimes \ee_k.
\]
$\bullet$ eigenvalue $1+6/V$:
\[
	\mathbf{X}_4=\frac{1}{\sqrt{3(V-2)(V-3)}} \sum_{\la i,j,k \ra} 
		( A_i + A_j + A_k )\,\ee_i \otimes \ee_j \otimes \ee_k
\]
where $\sum_i A_i = 0$ and $\sum_i | A_i|^2 = 1$.

$\bullet$ eigenvalue $2+4/V$:
\[
	\mathbf{X}_5
	=\frac{1}{\sqrt{3(V-4)}} \sum_{\la i,j,k \ra} ( a_{jk} + a_{ki} + a_{ij} )\,\ee_i \otimes \ee_j \otimes \ee_k
\]
where $a_{ij} = a_{ji}$ with $a_{ii} = 0$, $\sum_{i,j} |a_{ij}|^2 = 1$ and $\sum_i a_{ij} = 0$ for all $j$.

$\bullet$ eigenvalue $3$:
\[
	\mathbf{X}_6
	\propto \sum_{\la i_1, i_2, i_3 \ra} c_{i_1 i_2 i_3} \sum_{\pi \in S_3} \ee_{i_{\pi(1)}} \otimes \ee_{i_{\pi(2)}} 
	\otimes \ee_{i_{\pi(3)}}
\]
where $c_{iii}=c_{ijj} = c_{jij} = c_{jji} = 0$ and $\sum_i c_{ijk} = \sum_j c_{ijk} = \sum_k c_{ijk} = 0$.

The following eigenvalues of $H^{\rm hc}_{3,V}$ are degenerate with eigenvectors that lie outside both 
the even and odd eigenspaces:
\newline
$\bullet$ eigenvalue $1+3/V$:
\begin{align*}
	\mathbf{X}_7
&=	\frac{1}{\sqrt{2V(V-2)}} \sum_{\la i,j,k \ra} ( A_k - A_i )\,\ee_i \otimes \ee_j \otimes \ee_k
\\
	\mathbf{X}_8
&=	\frac{1}{\sqrt{6V(V-2)}} \sum_{\la i,j,k \ra} ( A_i + A_k - 2A_j )\,\ee_i \otimes \ee_j \otimes \ee_k
\end{align*}
where $\sum_i A_i = 0$ and $\sum_i | A_i|^2 = 1$.

$\bullet$ eigenvalue $2+1/V$:
\begin{align*}
	\mathbf{X}_9
&=	\frac{1}{\sqrt{2(V-1)}} \sum_{\la i,j,k \ra} ( a_{jk} - a_{ki} )\,\ee_i \otimes \ee_j \otimes \ee_k
\\
	\mathbf{X}_{10}
&=	\frac{1}{\sqrt{6(V-1)}} \sum_{\la i,j,k \ra} ( a_{jk} + a_{ki} - 2a_{ij} )\,\ee_i \otimes \ee_j \otimes \ee_k
\end{align*}
where  $a_{ij} = a_{ji}$ with $a_{ii} = 0$, $\sum_{i,j} |a_{ij}|^2 = 1$ and $\sum_i a_{ij} = 0$ for all $j$.

$\bullet$ eigenvalue $2+3/V$:
\begin{align*}
	\mathbf{X}_{11}
&=	\frac{1}{\sqrt{2(V-3)}} \sum_{\la i,j,k \ra} ( a_{jk} - a_{ki} )\,\ee_i \otimes \ee_j \otimes \ee_k
\\
	\mathbf{X}_{12}
&=	\frac{1}{\sqrt{6(V-3)}} \sum_{\la i,j,k \ra} ( a_{jk} + a_{ik} - 2a_{ij} )\,\ee_i \otimes \ee_j \otimes \ee_k
\end{align*}
where  $a_{ij} = -a_{ji}$ with $a_{ii} = 0$,  $\sum_{i,j} |a_{ij}|^2 = 1$ and $\sum_i a_{ij} = 0$ for all $j$.

Note that these eigenvalues can also be acquired from equation (\ref{Penrose_Eigenvalues}) by Penrose 
(see Appendix \ref{appendixA}):
\[
	r - V^{-1}(r^2 - r) + V^{-1} ( m^2 - m)
\]
where here $m=3$ and $r = 0, 1, 2, 3$.

We shall next derive explicit expressions for projections upon these
eigenvectors. However in order to do this, one must consider the two particle case for some
expressions which will be required for the three particle calculation.


\section{The 2-particle Hard-Core Hamiltonian}
By performing the technique of the previous section for a system of two hard-core particles on a lattice of $V$ sites, 
one can diagonalise the Hamiltonian $H^{\rm hc}_{2,V}$ as follows:

In the odd space the two-particle Hamiltonian has eigenvalues 1 and 2 with eigenvectors:\vspace{0.3cm}
\\
$\bullet$ eigenvalue 1:
\[
	\mathbf{Y}_1
=	\frac{1}{\sqrt{2V}} \sum_{i,j=1}^V (c_i - c_j)\ \ee_i \otimes \ee_j
=	\frac{1}{\sqrt{2}}(\g\otimes \cc-\cc\otimes \g)
\]
where $\cc \perp \g$ and $\|\cc\|=1$.

$\bullet$ eigenvalue 2:
\[
	\mathbf{Y}_2
=	\frac{1}{2} \sum_{i,j=1}^V a_{ij} \ \ee_i \otimes \ee_j
=	\frac{1}{\sqrt{2}}(\cc\otimes \cc'-\cc'\otimes \cc)
\]
where  $a_{ij} = -a_{ji}$ with $a_{ii} = 0$,  $\sum_{i,j} |a_{ij}|^2 = 1$ and $\sum_i a_{ij} = 0$ for all $j$.
The second expression is found by taking $a_{ij} = 2^{-1/2}(c_i c'_j - c'_i c_j)$, in which case 
$\cc,\cc' \perp \g$ and $\cc \perp \cc'$ with unit norm $\|\cc\|=\|\cc'\|=1$.

The eigenvectors in the even space are of the form: \vspace{0.3cm}
\\
$\bullet$ eigenvalue $2/V$:
\[
	\mathbf{Y}_3
=	\frac{1}{\sqrt{V(V-1)}} \sum_{i,j=1}^V (1-\delta_{ij})\ \ee_i \otimes \ee_j
=	\frac{1}{\sqrt{V(V-1)}} \left(V\,\g \otimes \g-\sum_{i=1}^V \ee_i\otimes \ee_i \right).
\]
$\bullet$ eigenvalue $1+2/V$:
\begin{align*}
	\mathbf{Y}_4
&=	\frac{1}{\sqrt{2(V-2)}} \sum_{i,j=1}^V (c_i + c_j)(1-\delta_{ij})\,\ee_i \otimes \ee_j
\\
&=	\frac{1}{\sqrt{2(V-2)}} \left(\sqrt{V}( \g\otimes \cc+\cc\otimes \g )
	-2\sum_{i=1}^V c_i\ \ee_i\otimes \ee_i \right) \vspace{0.2cm}
\end{align*}
where
$\cc \perp \g$, $\|\cc\|=1$. Note that $\mathbf{X}_4(\cc)\perp \mathbf{X}_4(\cc')$ if
$\cc\perp \cc'$.

$\bullet$ eigenvalue 2:
\[
	\mathbf{Y}_5
=	\frac{1}{\sqrt{2}}\sum_{i,j=1}^V a_{ij} (\ee_i\otimes \ee_j+\ee_j\otimes \ee_i)
\]
where $a_{ij}=a_{ji}$ with $a_{ii}=0$, $\sum_{i}^V a_{ij}=0\,\forall\,j$ and $\sum_{i,j=1}^V |a_{ij}|^2=1$.

Denote $P^{[2]}_i$ as the projection onto $\mathbf{Y}_i$. We may evaluate these five projections as
follows:
\begin{align*}
	P^{[2]}_1(i_1,i_2;j_1,j_2)
&=	\sum_{\cc : \sum c_i = 0} Y_1(i_1, i_2) \overline{Y_1(j_1, j_2)}
=	\frac{1}{2V} \sum_{\cc \perp \g} \left( c_{i_1} \bar{c_{j_1}} - c_{i_1} \bar{c}_{j_2} 
	- c_{i_2} \bar{c}_{j_1} + c_{i_2} \bar{c}_{j_2} \right)
\\
&=	\frac{1}{2V} (\delta_{i_1j_1}-\delta_{i_1j_2}-\delta_{i_2j_1}+\delta_{i_2j_2} )
\end{align*}
using (\ref{apE-orthog}).
\begin{align}
	P^{[2]}_2(i_1,i_2;j_1,j_2)
&=	\sum_{a} a_{i_1 i_2} \bar{a}_{j_1 j_2}
=	\frac{1}{4} \mathop{\sum_{\cc, \cc'}}_{\cc \perp \cc' \perp \g} (c_{i_1} c'_{i_2} - c'_{i_1} c_{i_2} )
	(\bar{c}_{j_1} \bar{c}'_{j_2} - \bar{c}'_{j_1} \bar{c}_{j_2} )				\notag
\\
&=	\frac{1}{4} \big[ 
	D(i_1, i_2, j_1, j_2) + D(i_2, i_1, j_2, j_1) - D(i_1, i_2, j_2, j_1) - D(i_2, i_1, j_1, j_2)
	\big]												\label{ap-aij}
\end{align}
where
\begin{align*}
	D(i_1, i_2, j_1, j_2) 
&= 	\sum_{\cc: \,\cc \perp \g} c_{i_1} \bar{c}_{j_1}
	\sum_{\cc': \,\cc' \perp \cc,\g} c'_{i_2} \bar{c}'_{j_2}
= 	\sum_{\cc: \,\cc \perp \g} c_{i_1} \bar{c}_{j_1}\left( 
	\sum_{\cc': \,\cc' \perp \g} c'_{i_2} \bar{c}'_{j_2} - c_{i_2} \bar{c}_{j_2} \right)
\\
&= 	\left(\delta_{i_1 j_1} - \frac{1}{V} \right) \left(\delta_{i_2 j_2} - \frac{1}{V} \right)
	- \sum_{\cc: \,\cc \perp \g} c_{i_1} \bar{c}_{j_1} c_{i_2} \bar{c}_{j_2}.
\end{align*}
When we reinsert this into (\ref{ap-aij}), the latter terms of the above expression will all cancel
leaving 
\begin{align}
	P^{[2]}_2(i_1,i_2;j_1,j_2) 							\label{P2}
&= 	\frac{1}{2}(\delta_{i_1 j_1} \delta_{i_2 j_2} - \delta_{i_1 j_2} \delta_{i_2 j_1})
	- \frac{1}{2V}(\delta_{i_1 j_1} - \delta_{i_1 j_2} - \delta_{i_2 j_1} + \delta_{i_2 j_2}).
\shortintertext{Similarly we have}
	P^{[2]}_3(i_1,i_2;j_1,j_2)
&=	\frac{1}{V(V-1)}(1-\delta_{i_1i_2})(1-\delta_{j_1j_2})			\notag
\\
	P^{[2]}_4(i_1,i_2;j_1,j_2)
&=	\frac{1}{2(V-2)}(\delta_{i_1j_1}+\delta_{i_1j_2}+\delta_{i_2j_1}
	+\delta_{i_2j_2}-4/V) (1-\delta_{i_1i_2})(1-\delta_{j_1j_2}).	\notag
\end{align}
We cannot derive this projection from the eigenvector $\mathbf{Y}_5$ itself due to lack of information,
but since $P^{[2]}_5=I^{[2]}-P^{[2]}_1-P^{[2]}_2-P^{[2]}_3-P^{[2]}_4$ and the matrix for $I^{[2]}$ has kernel
$\delta_{i_1j_1}\delta_{i_2j_2}(1-\delta_{i_1i_2})$, we can deduce 
\begin{multline}									\label{P5}
	P^{[2]}_5(i_1,i_2;j_1,j_2)
=	\Bigg\{ \frac{1}{2} (\delta_{i_1j_1}\delta_{i_2j_2} +\delta_{i_1j_2}\delta_{i_2j_1})
	-\frac{1}{2(V-2)}(\delta_{i_1j_1}+\delta_{i_1j_2}+\delta_{i_2j_1} +\delta_{i_2j_2})
\\ 
	+ \frac{1}{(V-1)(V-2)}\Bigg \}  (1-\delta_{i_1i_2})(1-\delta_{j_1j_2}).
\end{multline}
Consequently the partition function for the two particle hard-core Hamiltonian may be written as
\begin{align*}
	\e^{-\beta H^{\rm hc}_{2,V}}
&= 	\e^{-\beta} P^{[2]}_1+\e^{-2\beta} P^{[2]}_2+\e^{-2\beta/V} P^{[2]}_3+\e^{-\beta(1+2/V)} P^{[2]}_4+
	\e^{-2\beta} P^{[2]}_5
\\
&=	\e^{-2\beta} I^{[2]} + (\e^{-\beta} - \e^{-2\beta}) P^{[2]}_1 + (\e^{-2\beta/V} - \e^{-2\beta}) P^{[2]}_3 
	+ ( \e^{-\beta(1+2/V)}  - \e^{-2\beta}) P^{[2]}_4.
\end{align*}
With this one can evaluate, for instance, the 2-cycle density
\[
	c^2_\sV(q\!=\!2) = \sum_{i,j=1}^V c_2(i,j)
\quad\text{ where }\quad
	c_2(i,j)
=	\sum_{k=1}^V \e^{-\beta H^{\rm hc}_{2,V}}(i,k;k,j)
=	 f_\sbeta(V)+g_\sbeta(V)\delta_{ij}.
\]
Observe that the term $g_\sbeta(V) \delta_{ij}$ is a consequence of the hard-core interaction, so deleting it 
returns the cycle density of the two free particles. It can be shown that for two particles the functions 
$f_\sbeta$ and $g_\sbeta$ are both non-negative for all possible cycle lengths (i.e. one and two). 
However in the three particle case, the Hilbert space can be divided into subspaces which are not only even and 
odd, but also those which are neither even nor odd. These additional spaces
can result in negative contributions to the cycle density expression, as we shall shortly see.

\section{Cycle Density Expression for the 3-particle Hard-Core Model}
Returning to the three particle case, let $P_i$ denote the projection onto the 
eigenvector $\mathbf{X}_i$. It is useful to set the operator 
\[
	\Delta^{\pm}(i, k; j, l) = \delta_{i j} +\delta_{k l} \pm ( \delta_{i l} + \delta_{j k} )
\]
and also to abuse notation slightly by defining 
\[
	\delta_{ijk} = \begin{cases} 1 & \text{ if } i \ne j \ne k \\ 0 & \text{ otherwise.}\end{cases}
\]

Taking these projection in order of complexity. First one may easily see that
\[
	P_3(i_1,i_2,i_3;j_1,j_2, j_3)
=	\frac{1}{V(V-1)(V-2)} ( 1 - \delta_{i_1 i_2 i_3} )( 1 - \delta_{j_1 j_2 j_3} )
\]
where we introduce $1-\delta_{i_1 i_2 i_3}$ terms to enforce the summation's conditions
(and hence the hard-core condition) that $i_1\ne i_2\ne i_3$.

Secondly, the eigenvectors $\mathbf{X}_4$, $\mathbf{X}_7$ and $\mathbf{X}_8$ 
all contain terms of the form $A_i$ where $\sum_i A_i = 0$ and $\sum_i |A_i|^2 = 1$.
Then choosing $\mathbf{X}_4$ we have
\[
	P_4(i_1,i_2,i_3;j_1,j_2, j_3) 
= 	\sum_{\mathbf{A} :\sum_i A_i = 0} (A_{i_1} + A_{i_2} + A_{i_3} )(\bar{A}_{j_1} + \bar{A}_{j_2} + \bar{A}_{j_3} ).
\]
But since $\sum_i A_i = 0$ one can say that the vector $\mathbf{A} \perp \g$ (as we do in the 
two-particle case) and hence
\[
	\sum_{\mathbf{A} \perp \g} A_i \bar{A}_j = \delta_{ij} - \frac{1}{V}.
\]
With this we may calculate the following (with $P_{l,m} \vcentcolon= P_l + P_m$):
\begin{align*}
	P_4(i_1,i_2,i_3;j_1,j_2, j_3)
&=	\frac{1}{3(V-2)(V-3)} \bigg[ \delta_{i_1 j_1} + \delta_{i_1 j_2} + \dots + \delta_{i_3 j_3} - \frac{9}{V} \bigg]
\\*
&\qquad\qquad
	\times ( 1 - \delta_{i_1 i_2 i_3} )( 1 - \delta_{j_1 j_2 j_3} )
\\[0.2cm]
	P_{7,8}(i_1,i_2,i_3;j_1,j_2, j_3)
&=	\frac{1}{3V(V-2)}
	\bigg[ \Delta^{-}(i_1 ,i_2 ; j_1, j_2) + \Delta^{-}(i_2 ,i_3 ; j_2, j_3) + \Delta^{-}(i_3 ,i_1 ; j_3, j_1) \bigg]
\\*
&\qquad\qquad
	\times ( 1 - \delta_{i_1 i_2 i_3} )( 1 - \delta_{j_1 j_2 j_3} ).
\end{align*}

Thirdly consider the eigenvectors $\mathbf{X}_1$, $\mathbf{X}_{11}$ and $\mathbf{X}_{12}$. 
These all have the property that $a_{ij} = -a_{ji}$.
From the expression (\ref{apE-A1}) for the eigenvector $\mathbf{X}_1$
we may construct the corresponding projection $P_1$:
\[
	P_1(i_1,i_2,i_3;j_1,j_2, j_3) 
=	\frac{1}{3V} \sum_a ( a_{i_2 i_3} + a_{i_3 i_1} + a_{i_1 i_2} )
	( \bar{a}_{j_2 j_3} + \bar{a}_{j_3 j_1} + \bar{a}_{j_1 j_2} ).
\]
In two particle calculation above for odd $a_{ij}$ we derived a projection (\ref{P2}) of the form:
\[
	P^{[2]}_2(i_1,i_2;j_1,j_2)
=	\sum_a a_{i_1 i_2} \bar{a}_{j_1 j_2}
= 	\frac{1}{2}(\delta_{i_1 j_1} \delta_{i_2 j_2} - \delta_{i_1 j_2} \delta_{i_2 j_1})
	- \frac{1}{2V}(\delta_{i_1 j_1} - \delta_{i_1 j_2} - \delta_{i_2 j_1} + \delta_{i_2 j_2})
\]
which we can use here to obtain
\begin{multline*}
	P_1(i_1,i_2,i_3;j_1,j_2, j_3) 
=	\frac{1}{3V} \Big[ P^{[2]}_2(i_2,i_3;j_2,j_3) + P^{[2]}_2(i_2,i_3;j_3,j_1) + P^{[2]}_2(i_2,i_3;j_1,j_2) 
\\
\hspace{3cm} +  P^{[2]}_2(i_3,i_1;j_2,j_3) + P^{[2]}_2(i_3,i_1;j_3,j_1) + P^{[2]}_2(i_3,i_1;j_1,j_2)
\\
	+  P^{[2]}_2(i_1,i_2;j_2,j_3) + P^{[2]}_2(i_1,i_2;j_3,j_1) + P^{[2]}_2(i_1,i_2;j_1,j_2) \Big].
\end{multline*}
This can be simplified (all the $\tfrac{1}{2V}$ terms cancel) to obtain
\begin{multline*}
	P_1(i_1,i_2,i_3;j_1,j_2, j_3)
=	\frac{1}{6V} \bigg\{
	\delta_{i_1 j_1} \Delta^{-}(i_2, i_3; j_2, j_3) + \delta_{i_1 j_2} \Delta^{-}(i_3, i_2; j_1, j_3)
	+ \delta_{i_1 j_3} \Delta^{-}(i_3, i_2; j_2, j_1) 
\\
	+ \delta_{i_2 j_1} (\delta_{i_3 j_2} - \delta_{ i_3 j_3} )
	+ \delta_{i_2 j_2} ( \delta_{i_3 j_3} - \delta_{i_3 j_1}) + \delta_{i_2 j_3} ( \delta_{i_3 j_1} - \delta_{i_3 j_2} )
	\bigg\}.
\end{multline*}
In a similar fashion, one can find expressions for the projections $P_{11}(i_1,i_2,i_3;j_1,j_2, j_3)$ and 
$P_{12}(i_1,i_2,i_3;j_1,j_2, j_3)$, which we shall combine to obtain
\begin{align*}
	P_{11,12}(i_1,i_2,i_3;j_1,j_2, j_3)
&=	\frac{1}{6(V-3)} \bigg\{ 
	\delta_{i_1j_1} \Delta^{+}(i_2, i_3; j_2, j_3) + \delta_{i_2 j_2} \Delta^{+}(i_1, i_3; j_1, j_3)
\\*
	+ \delta_{i_3j_3} \Delta^{+}(i_1, & i_2; j_1, j_2)
	+ \delta_{i_3j_2} (-2\delta_{i_2j_3} - \delta_{i_1j_3} - \delta_{i_2j_1})
	+ \delta_{i_3j_1} (-2\delta_{i_1j_3} - \delta_{i_1j_2} - \delta_{i_2j_3} )
\\*
	- 2\delta_{i_2j_1} \delta_{i_1j_2} -& \delta_{i_2j_3} \delta_{i_1j_2}
	- \delta_{i_2j_1} \delta_{i_1j_3}
	- \frac{3}{V}  \big[ \Delta^{-}(i_1 ,i_2 ; j_1, j_2) + \Delta^{-}(i_2 ,i_3 ; j_2, j_3)
\\*&\qquad\qquad
	+ \Delta^{-}(i_3 ,i_1 ; j_3, j_1) \big] \bigg\} ( 1 - \delta_{i_1 i_2 i_3} )( 1 - \delta_{j_1 j_2 j_3} ).
\end{align*}

Finally the remaining vectors we shall consider, $\mathbf{X}_5$, $\mathbf{X}_{9}$ and $\mathbf{X}_{10}$,
all have that $a_{ij} = a_{ji}$. As above, these three-particle projections can be
expressed in terms of a sum of a particular two-particle projection with different arguments.
In this case the projection is $P^{[5]}_2(i_1,i_2;j_1,j_2)$ for which we have an explicit expression (\ref{P5}).
With this one can evaluate
{\allowdisplaybreaks
\begin{align*}
	P_5(i_1,i_2,i_3;j_1,j_2, j_3)
&=	\frac{1}{3(V-4)} \bigg\{ \frac{9}{(V-1)(V-2)}
	 - \frac{2}{V-2}\big( \delta_{i_1j_1} + \delta_{i_1 j_2} + \dots + \delta_{i_3 j_3}\big) 
\\*
	+ \frac{1}{2} \bigg[ \delta_{i_1j_1} \Delta^{+}( i_2,i_3;&j_2,j_3)  + \delta_{i_1j_2} \Delta^{+}(i_2,i_3;j_1,j_3)
	+ \delta_{i_1j_3} \Delta^{+}(i_2,i_3;j_1,j_2) 
\\*
	+ \delta_{i_2j_1} ( \delta_{i_3j_2} + \delta_{i_3j_3})&
	+ \delta_{i_2j_2} (\delta_{i_3j_3} + \delta_{i_3j_1}) 
	+ \delta_{i_2j_3} (\delta_{i_3j_1} + \delta_{i_3j_2})  \bigg]\bigg\}
	( 1 - \delta_{i_1 i_2 i_3} )( 1 - \delta_{j_1 j_2 j_3} )
\\[0.3cm]
	P_{9,10}(i_1,i_2,i_3;j_1,j_2, j_3)
&=	\frac{1}{6(V-1)} \bigg\{ 
	\delta_{i_1j_1} \Delta^{-}(i_2, i_3; j_2, j_3) + \delta_{i_2 j_2} \Delta^{-}(i_1, i_3; j_1, j_3)
\\*
	+ \delta_{i_3j_3} \Delta^{-}(i_1, & i_2; j_1, j_2)
	+ \delta_{i_3j_2} (2\delta_{i_2j_3} - \delta_{i_1j_3} - \delta_{i_2j_1})
	+ \delta_{i_3j_1} (2\delta_{i_1j_3} - \delta_{i_1j_2} - \delta_{i_2j_3} )
\\*
	+ 2\delta_{i_2j_1} \delta_{i_1j_2} -& \delta_{i_2j_3} \delta_{i_1j_2}
	- \delta_{i_2j_1} \delta_{i_1j_3}
	- \frac{1}{V-2} \big[ \Delta^{-}(i_1 ,i_2 ; j_1, j_2) + \Delta^{-}(i_2 ,i_3 ; j_2, j_3)
\\*&\qquad\qquad
	+ \Delta^{-}(i_3 ,i_1 ; j_3, j_1) \big] \bigg\} ( 1 - \delta_{i_1 i_2 i_3} )( 1 - \delta_{j_1 j_2 j_3} ).
\end{align*}}

What remains is 
$P_{2,6}$, the projection onto the space spanned by $\mathbf{X}_2$ and $\mathbf{X}_6$.
This can be written as $P_{2,6}=I-P_1-P_3-P_4-P_5-P_{7,8} - P_{9,10} - P_{11,12}$ where we note that
$I$ can be written as $\delta_{i_1j_1}\delta_{i_2j_2}\delta_{i_3j_3} (1-\delta_{i_1i_2i_3} )$.

\begin{figure}[hbt]
\begin{center}
\includegraphics[width=12cm]{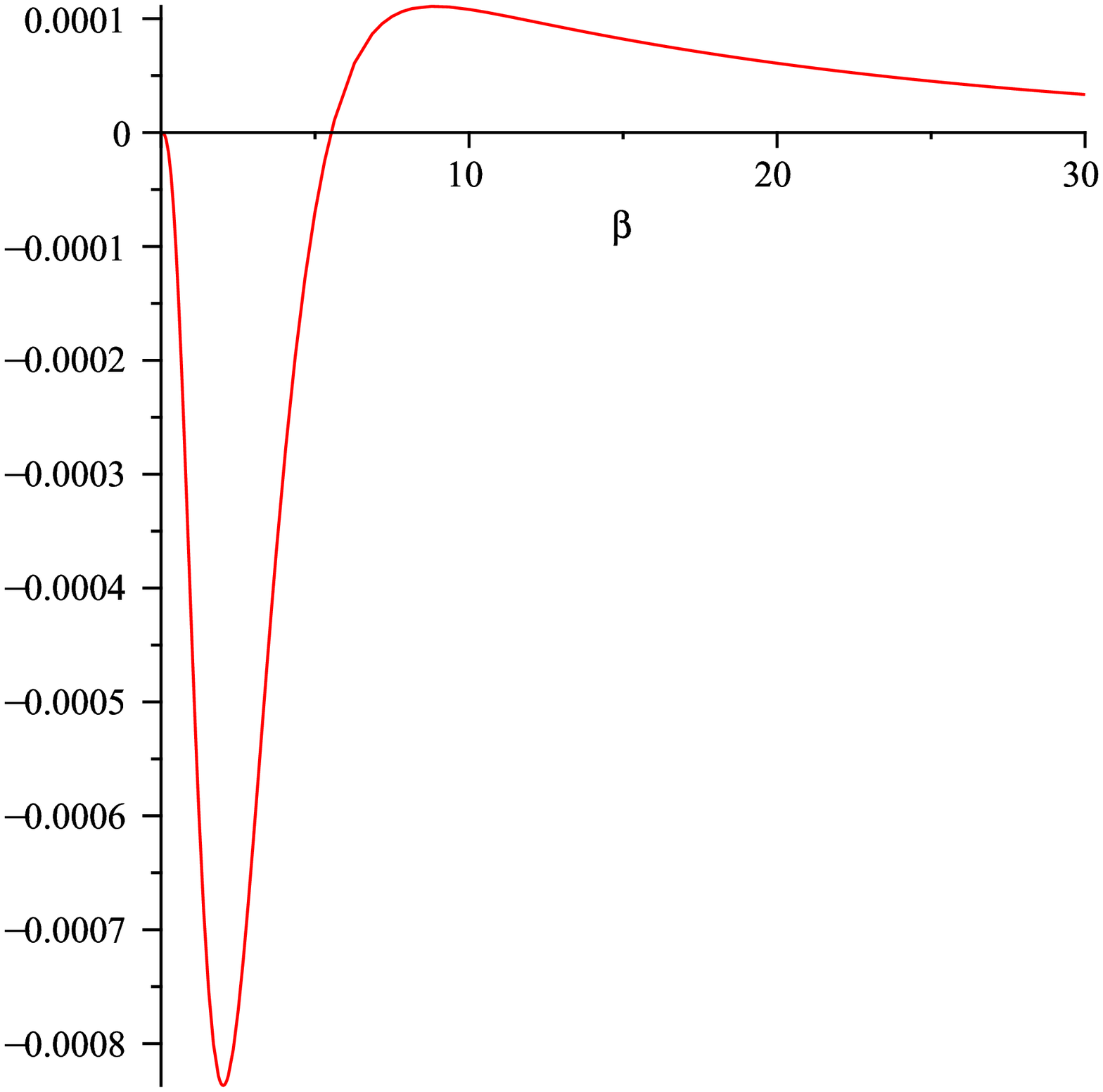}
\end{center}
\caption[Plot of the function $g_\sbeta(V)$ versus $\beta$ for $V=100$]{\it Plot of the function 
$g_\sbeta(V)$ versus $\beta$ for $V=100$}
\label{fig5}
\end{figure}

The Hamiltonian $H^{\rm hc}_{3,V}$ is diagonal with respect to these vectors $\mathbf{X}_i$
and hence it can be written in terms of these induced projections:
\begin{align*}
	\e^{-\beta H^{\textsf hc}_{3,V}}
&= 	\e^{-2\beta} P_1 + \e^{-3\beta} P_{2,6} + \e^{-6\beta/V} P_3 + \e^{-\beta(1+6/V)} P_4 
	+ \e^{-2\beta} P_5 + \e^{-\beta(1+3/V)} P_{7,8}
\\&\qquad
	+ \e^{-\beta(2+1/V)} P_{9,10} + \e^{-\beta(2+3/V)} P_{11,12}
\\[0.2cm]
&= 	\e^{-3\beta} I + (\e^{-2\beta} - \e^{-3\beta}) P_1 + (\e^{-6\beta/V} - \e^{-3\beta}) P_3
	+ (\e^{-\beta(1+6/V)} - \e^{-3\beta}) P_4
\\&\qquad
	+ (\e^{-2\beta} - \e^{-3\beta}) P_5 + (\e^{-\beta(1+3/V)} - \e^{-3\beta}) P_{7,8} 
	+ (\e^{-\beta(2+1/V)} - \e^{-3\beta}) P_{9,10} 
\\&\qquad
	+ (\e^{-\beta(2+3/V)} - \e^{-3\beta}) P_{11,12}.
\end{align*}
Thus we can find an explicit expression for the kernel of the operator $\e^{-\beta H^{\rm hc}_{3,V}}$.
Let us consider the case of a cycle of length three, i.e. 
\[
	\sum_{\la k,l \ra} \e^{-\beta H^{\rm hc}_{3,V}}(i,k,l; k,l,j).
\]
Then one can evaluate the three-cycled projections to write
{\allowdisplaybreaks
\begin{align*}
	\sum_{\la k,l \ra} I(i,k,l; k,l,j) &= 0
\\[0.1cm]
	\sum_{\la k,l \ra} P_1(i,k,l; k,l,j) &= \frac{(V-2)(V-3)}{6V} + \delta_{ij} \frac{V-2}{3}
\\[0.1cm]
	\sum_{\la k,l \ra} P_3(i,k,l; k,l,j) &= \frac{V-3}{V(V-1)} + \delta_{ij} \frac{2}{V(V-1)}
\\[0.1cm]
	\sum_{\la k,l \ra} P_4(i,k,l; k,l,j) &= \frac{2V-9}{3V} + \delta_{ij} \frac{V+6}{3V}
\\[0.1cm]
	\sum_{\la k,l \ra} P_5(i,k,l; k,l,j) &= \frac{(V-3)(V-7)}{6(V-1)} + \delta_{ij} \frac{(V-3)(V+2)}{3(V-1)}
\\[0.1cm]
	\sum_{\la k,l \ra} P_{7,8}(i,k,l; k,l,j) &= -\frac{2(V-3)}{3V} - \delta_{ij} \frac{V+3}{3V}
\\[0.1cm]
	\sum_{\la k,l \ra} P_{9,10}(i,k,l; k,l,j) &= -\frac{(V-3)(V-4)}{6(V-1)} - \delta_{ij} \frac{(2V+1)(V-3)}{6(V-1)}
\\[0.1cm]
	\sum_{\la k,l \ra} P_{11,12}(i,k,l; k,l,j) &= -\frac{(V-2)(V-6)}{6V} - \delta_{ij} \frac{(2V+3)(V-2)}{6V}
\end{align*}}
which returns the following expression
\[
	\sum_{\la k,l \ra} \e^{-\beta H^{\rm hc}_{3,V}}(i,k,l; k,l,j) = f_\sbeta(V) + \delta_{ij} g_\sbeta(V)
\]
where $f_\sbeta(V)$ is a non-negative expression of the form
\begin{multline*}
	f_\sbeta(V) = (\e^{-2\beta} - \e^{-3\beta}) \frac{(V-2)(V-3)}{6V}
	+ (\e^{-6\beta/V} - \e^{-3\beta}) \frac{V-3}{V(V-1)}
	+ (\e^{-\beta(1+6/V)} - \e^{-3\beta}) \frac{2V-9}{3V}
\\[0.1cm]
	+ (\e^{-2\beta} - \e^{-3\beta}) \frac{(V-3)(V-7)}{6(V-1)}
	- (\e^{-\beta(1+3/V)} - \e^{-3\beta}) \frac{2(V-3)}{3V}
\\[0.1cm]
	- (\e^{-\beta(2+1/V)} - \e^{-3\beta}) \frac{(V-3)(V-4)}{6(V-1)}
	- (\e^{-\beta(2+3/V)} - \e^{-3\beta}) \frac{(V-2)(V-6)}{6V}
\end{multline*}
and $g_\sbeta(V)$ is a not-always-positive function as may be seen from Figure \ref{fig5},
\begin{multline*}
	g_\sbeta(V) = (\e^{-2\beta} - \e^{-3\beta}) \frac{V-2}{3}
	+ (\e^{-6\beta/V} - \e^{-3\beta}) \frac{2}{V(V-1)}
	+ (\e^{-\beta(1+6/V)} - \e^{-3\beta}) \frac{V+6}{3V}
\\[0.1cm]
	+ (\e^{-2\beta} - \e^{-3\beta}) \frac{(V-3)(V+2)}{3(V-1)}
	- (\e^{-\beta(1+3/V)} - \e^{-3\beta}) \frac{V+3}{3V}
\\[0.1cm]
	- (\e^{-\beta(2+1/V)} - \e^{-3\beta}) \frac{(2V+1)(V-3)}{6(V-1)}
	- (\e^{-\beta(2+3/V)} - \e^{-3\beta}) \frac{(2V+3)(V-2)}{6V}.
\end{multline*}
	
Recall from equation (\ref{cycle-expt-can}) that $c^n_\sV(q,A)$ is an expression for the expectation of a single
particle operator $A\ge0$ in terms of cycle statistics. In the hard-core model, the expansion
for three particles on a lattice of $V$ sites can be written as (see (\ref{hc_cycle_expectation})\!)
\[
	c^3_\sV(q,A)
= 	\frac{1}{ Z_\sbeta(3,V) V}\trace_{\Phc{3} (\HHcan{q} \otimes \HHcansym{3-q})} \bigg[
	\left(  A \otimes I^{(q-1)} \right) U_q \e^{-\beta H^{\textsf hc}_{3,V} } \bigg]
\]
for $q=1,2,3$. Taking the case $q=3$, in terms of the kernel of $H^{\textsf hc}_{3,V}$, we may expand this to
obtain
{\allowdisplaybreaks
\begin{align*}
	c^3_\sV(3,A)
&=	\frac{1}{ Z_\sbeta(3,V) V} \trace_{\Phc{3} \HHcan{3}} \bigg[
	\left( A \otimes I \otimes I \right) U_3 \e^{-\beta H^{\textsf hc}_{3,V} } \bigg]
\\[0.2cm]
&=	\frac{1}{ Z_\sbeta(3,V) V}
	\sum_{\la j,k,l\ra} \bigg[ \left( A \otimes I \otimes I \right) 
	U_3 \e^{-\beta H^{\rm hc}_{3,V}} \bigg](j,k,l; j,k,l)
\\
&=	\frac{1}{ Z_\sbeta(3,V) V}
	\sum_{\la j,k,l\ra} \sum_{m=1}^V A(j,m) \e^{-\beta H^{\rm hc}_{3,V}}(m,k,l; k,l,j)
\\
&=	\frac{1}{ Z_\sbeta(3,V) V}
	\sum_{ j,m=1}^V A(j,m) \left( f_\sbeta(V) + \delta_{mj} g_\sbeta(V) \right).
\end{align*}}
Let $A$ be a projection onto states orthogonal to the ground state $\g$, for 
example let $\cc = 2^{-1/2}(1, -1, 0, \dots, 0)$ and fix $A = | \cc \ra\la \cc |$. 
It has kernel $A(j,m) = 2^{-1}(\delta_{1j} - \delta_{2j})(\delta_{1m} - \delta_{2m})$.
Then
\begin{equation*}
	c^3_\sV(3,| \cc \ra\la \cc |)
=	\frac{1}{2 Z_\sbeta(3,V) V}
	\sum_{ j,m=1}^V (\delta_{1j} - \delta_{2j})(\delta_{1m} - \delta_{2m})
	\left( f_\sbeta(V) + \delta_{mj} g_\sbeta(V) \right)
=	\frac{g_\sbeta(V)}{ Z_\sbeta(3,V) V}
\end{equation*}
which is not necessarily a non-negative number (see Figure \ref{fig5}).

\def\appendixletter{C}

\chapter{Numerical Source Code}							\label{appendixC}
Contained within this appendix is the C++ source code to generate Figures \ref{fig1}, \ref{fig2} and \ref{test4} 
of Chapter \ref{chapter5} on the Infinite-Range-Hopping Bose-Hubbard Model, which compare the density of the Bose 
condensate with the long cycle density. The algorithm represents the Approximated Hamiltonian,
equation (\ref{Happrox}), as a finite dimensional symmetric tridiagonal matrix with respect to the occupation number 
basis, and diagonalises the matrix to find its eigenvalues and eigenvectors.

The matrix is diagonal except in the case of $r>0$, i.e. when BEC occurs. So for each $\mu$, one needs to calculate
the solution of $p'(r) = 2r$, done using a simple binary search-like convergence algorithm (see lines 143-210).
This value of $r$ is then used to evaluate the pressure, density and the cycle-densities. 

This code uses the open source \texttt{GMP} \cite{GMP} and \texttt{CLN} \cite{CLN} libraries to boost the accuracy of the 
calculations beyond that of standard double precision arithmetic. This was done for two reasons,
first so that the algorithm was capable of summing the exponents of the eigenvalues without overflowing,
and secondly to zoom in on the critical values of $\mu$ to check for the discontinuity in $r$.
The matrix diagonalisation algorithm is a modified \texttt{tqli} method from Numerical Recipes in 
C \cite{NumericalRecipes}, enhanced for higher precision.
The code automatically generates a plotdata file and a command file and calls \texttt{gnuplot} \cite{GNUPLOT} to
render two plots automatically, one comparing short cycles with the total density, the other long cycles and BEC.
Both these plots are saved in two appropriately named EPS files.

\definecolor{lightgrey}{gray}{0.97}
\definecolor{darkgrey}{gray}{0.35}
\begin{spacing}{1}
\lstset{
language=c++,
basicstyle=\scriptsize\ttfamily,
linewidth=\textwidth,
tabsize=4,
upquote=true,	
breaklines=true,
backgroundcolor=\color{lightgrey},
numbers=left,
numberstyle=\tiny,
stepnumber=5, 
rulecolor=\color{darkgrey},
frame=tb,
numbersep=10pt,
commentstyle=\textit, 
stringstyle=\upshape,
showstringspaces=false}
\begin{lstlisting}
/*
 *  soft-core-cycles.cpp
 *  
 * To test hypothesis that \sum_{q=1}^\infty c_V^\mu(q) = \rho(\beta, \mu) - r_0^2
 * And so r_0^2 is the density of BEC in the model
 */

#include <iostream>
#include <fstream>
#include <sstream>
#include <string>
#include <exception>
#include <iomanip>
#include <vector>
#include <cstdlib>
#include <cmath>
#include <cln/cln.h>

using namespace std;
using namespace cln;

/* To enable Debugging output, activate the line below */
//#define DEBUG

/* Common functions */
cl_F f(int, double);
cl_F f_approx(int, int, double);
cl_F partition_of_approx(double, double);
cl_F cycle_density_numerator(int, double, double);
double density(double, double);
void deriv_of_pressure(double, double, cl_F*, cl_F*);
void tqli(vector<cl_F>&, vector<cl_F>&, int, vector<cl_F>&, bool);
cl_F pythag( const cl_F, const cl_F);

/* Advanced variables for precision purposes */
const int decimal_places = 40;			//How many decimal places the arithmetic uses
										//(as algorithm is arbitrary precision)
const cl_F TINY_T = "1.0e-40_30";		//What constitutes negligibly small
										//Check "deciman_places" before setting
const double TINY = 1.0e-15;			//set how accurate r value should be, e-12 is 
										//good enough for 7-8decimal places
const int datapoints = 100;				//Number of datapoints to take between mu_min 
										//and mu_max

/* Initial defaults, usually overridden via command line */
double beta = 100;
double mu_min = 0;
double mu_max = 10;
double lambda = 3;

/* No need to change anything from here on */
float_format_t precision = float_format(decimal_places);

const cl_F zero = cl_float(0.0, precision);

/* Handy macros */
#define SIGN(a,b) ((b)<zero ? -abs(a) : abs(a))
#ifdef DEBUG
#define verbose if (0) ; else cerr
#else
#define verbose if (1) ; else cerr
#endif

//The matrix will be a symmetric tridiagonal. This is most efficiently stored in two arrays,  
//one with the diagonal entries, one with the off-diagonal entries.
vector <cl_F> diag;				//tqli will place eigenvalues in this later
vector <cl_F> offdiag;
vector <cl_F> eigenvectors;		//to store generated eigenvalues of matrix
int number_of_occupations = -1;


int main (int argc, char * const argv[])
{	
	cl_F cycles=zero, expectation=zero, partition=zero;
	double r_found=0.0, diff=0.0, density_found=0.0, short_cycles=0.0;
	char barspin[4] = {'\\', '|', '/', '-'};
	int iteration=0, bariterator=0;
	double mu_iterator=0.1;
	ofstream plotdata, gnuplot_commands;
	
	//Check command line variables before starting
	if( argc != 6 ){
		cout << endl << "Usage: " << argv[0] 
		<< " [number of occupations] [beta] [lambda] [mu_min] [mu_max]" << endl;
		return 1;
	}
	
	number_of_occupations = atoi(argv[1]);
	if( number_of_occupations < 0 ){
		cout << endl << "Usage: " << argv[0] 
		<< " [number of occupations] [beta] [lambda] [mu_min] [mu_max]" << endl 
		<< "where you must supply a positive integer for occupations" << endl;
		return 1;
	}
	
	beta = atof(argv[2]);
	if( beta < 0){
		cout << endl << "Usage: " << argv[0] 
		<< " [number of occupations] [beta] [lambda] [mu_min] [mu_max]" << endl 
		<< "where you must supply a positive integer for beta" << endl;
		return 1;
	}
	
	lambda = atof(argv[3]);
	if( lambda < 0){
		cout << endl << "Usage: " << argv[0] 
		<< " [number of occupations] [beta] [lambda] [mu_min] [mu_max]" << endl 
		<< "where you must supply a positive number for lambda" << endl;
		return 1;
	}
	
	mu_min = atof(argv[4]);
	mu_max = atof(argv[5]);
	if( mu_min > mu_max){
		cout << endl << "Usage: " << argv[0] 
		<< " [number of occupations] [beta] [lambda] [mu_min] [mu_max]" << endl 
		<< "where mu_min <= mu_max!" << endl;
		return 1;
	}
	
	//Construct filename for plotdata
	stringstream s;
	s << "plotdata-beta" << beta << "-lambda" << lambda << "-mu" 
		<< mu_min << '~' << mu_max << "-n" << number_of_occupations;
	string filename = s.str();

	
	//Open "plotdata" file and prepare it
	plotdata.open(filename.c_str(), ios::out);

	plotdata << "# Dataset for beta=" << beta << ", lambda=" << lambda 
				<< " with " << number_of_occupations << " occupations" << endl 
				<< "# Data format: mu   density   r   shortcycles" << endl;
	
	//Initialize the matrix
	diag.resize(number_of_occupations);
	offdiag.resize(number_of_occupations);
	eigenvectors.resize(number_of_occupations*number_of_occupations);
	
	//Set mu_iterator so that we evaluate correct number of datapoints
	mu_iterator = (mu_max-mu_min)/datapoints;
	
	for(double mu=mu_min; mu <= mu_max; mu+=mu_iterator){
		bariterator=0;
		cl_F partition = zero;
		cl_F expectation = zero;
#ifndef DEBUG
		//Progress meter
		iteration = floor(((mu-mu_min)*100)/(mu_max-mu_min));
		cout << '\r' << iteration << "% complete";
		cout.flush();
#endif	
		//Check to see if p'(r) will ever intersect 2*r. Do this by evaluating both at TINY 
		//and if p'(TINY) > 2*TINY => r>0 is a solution. Otherwise no. This is due to
		//concavity of p'(r) as postulated (not proven, but looks likely).
		
		r_found = 0.0;   //set to lowest guaranteed solution
		
		//Find expectation of (a+a*) i.e. p'(r)
		deriv_of_pressure(0.00001, mu, &partition, &expectation);	
		if( expectation > 0.00002 ){
			//otherwise skip mu loop to next iteration, since r=0 is only possibility
			
			//Now know r>0. Need to iterate over r to locate it.
			double r_increment = 0.2;	//Set initial r_increment for r loop
			double r_initial = 0.002;	//Set initial point of interval to seek for r in
			double r_final = 20;		//Set final point of interval to seek r in
			
			// For this mu, find the max value of r s.t. 2*r = p'(r)
			for(double r_val=r_initial; r_val < r_final; r_val+=r_increment){
#ifndef DEBUG
				cout << '\r' << iteration << "% complete " << barspin[(bariterator++)%4]; 
				cout.flush();
#endif
				//Find expectation of (a+a*) i.e. p'(r)
				deriv_of_pressure(r_val, mu, &partition, &expectation);
				
				diff = double_approx(expectation) - 2*r_val; //difference
				
				if(fabs(diff) <= TINY ){  //if expectation==2*r, set r_found and break loop
					r_found = r_val;
					verbose << '!';
					verbose.flush();
					break;
				}
				else{
					//Has p'(r) crossed 2*r yet?
					if( diff > 0 ){  //No, so leave all else as is
						if(r_increment < 1.0e-15){ //Unfortunate possibility
							r_found = r_val;
							break; //Almost gone beyond precision
						}
						verbose << '+'; 
						verbose.flush();
					}
					else{ //Yes we've intersected the line 2*r, somewhere in interval 
						  //(r_val-r_interval, r_val+r_interval)
						r_final = r_val+r_increment;
						r_val -= r_increment;	//need to double back, iterate & converge 
												//to value of r at intersection
						r_initial = r_val;
						r_increment /= 2; //use binary search as most reliable
						verbose << "->" << mu << ", diff: " << diff << ", r_val: " << r_val 
							<< ", r_increment: " << r_increment << endl;
						verbose.flush();
					}
				}
				
			}//r loop
		}//end if(i.e. no intersection)
		verbose << "Found for mu=" << mu << "\t r=" << r_found << endl;
		
		//Have found value for supremum of r (r_found), evaluate the pressure at r
		cycles = zero;
		try{
			//Eval sum of cycles - short cycle sum
			partition = partition_of_approx(mu, r_found);

			for(int q=1; q<=number_of_occupations; q++){
				verbose << q << '\t' << cycle_density_numerator(q, mu, r_found) 
					/ partition << endl;
				cycles = cycles + cycle_density_numerator(q, mu, r_found); 
			}
			cycles = cycles / partition; // cycles/partition = cycle_density

			//print values to file
			density_found = density(r_found, mu);
			short_cycles = double_approx( cycles );
			plotdata << mu << '\t' << density_found << '\t' << r_found << '\t' << short_cycles << endl;
		} catch( int a ){ //Divide by zero error, report, skip this iteration & continue
			verbose << "Divide by zero in TQLI, skipping evaluation for mu=" << mu << endl;
			continue;
		}
		
	}//end mu loop
	
	cout << '\r' << "100% complete" << endl; //just tidy the output
	plotdata.close();
	
	//Generate a plot_command file for gnuplot so graphs will be generated on the data found
	s.str("");
	s << "plot_commands-beta" << beta << "-lambda" << lambda << "-mu" 
		<< mu_min << '~' << mu_max << "-n" << number_of_occupations;
	string plot_cmd = s.str();
	gnuplot_commands.open(plot_cmd.c_str(), ios::out);
	gnuplot_commands << "set term postscript eps enhanced colour" << endl;
	
	s.str("");
	s << "density-r^2_vs_short_cycles-beta" << beta << "-lambda" << lambda << "-mu" 
		<< mu_min << '~' << mu_max << "-n" << number_of_occupations;
	
	//Print plot_command file
	gnuplot_commands << "set out \"" << s.str() << ".eps\"\n"
		"set key spacing 1.3\n"
		"set key box\n"
		"set style line 5 lt 1 lw 2\n"
		"set style line 6 lt 3 lw 2\n"
		"set title \"{/Symbol b}=" << beta << ", {/Symbol l}=" << lambda
		<< " with " << number_of_occupations << " occupations\"\n"
		"set xrange [" << mu_min << ":" << mu_max << "]\n"
		"set xlabel \"{/Symbol m}\"\n"
		"set key right bottom\n"
		"plot \"" << filename << "\" using 1:($2-($3*$3)) title '{/Symbol r}-r^2' "
		"with lines linestyle 5, \\\n\"" << filename << "\" using 1:4 title "
		"'{/Symbol r}_{short}' with lines linestyle 6" << endl;

	s.str("");	//reset s for a new filename
	s << "r^2_vs_long_cycles-beta" << beta << "-lambda" << lambda << "-mu" 
		<< mu_min << '~' << mu_max << "-n" << number_of_occupations;
	
	gnuplot_commands << "set out \"" << s.str() << ".eps\"\n"
		"set key left top\n"
		"set title \"{/Symbol b}=" << beta << ", {/Symbol l}=" << lambda
		<< " with " << number_of_occupations << " occupations\"\n"
		"plot \"" << filename << "\" using 1:($2-$4) title '{/Symbol r}_{long}' "
		"with lines linestyle 5, \\\n\"" << filename << "\" using 1:($3*$3) title "
		"'r^2' with lines linestyle 6;" << endl;
	
	gnuplot_commands.close();
	plot_cmd = "gnuplot " + plot_cmd;
	system(plot_cmd.c_str());		//run gnuplot and point it to the plot_command file.
	return 0; 
}



/*  
 * Evaluates d/dr p(r) 
 */
void deriv_of_pressure(double r, double mu, cl_F *partition, cl_F *expectation)
{	
	cl_F temp = zero;
	*partition = zero;
	
	for(int n=0; n<number_of_occupations; n++){
		diag[n] = f(n, mu);
		if(n!=0)  //note: offdiag[0] not used
			offdiag[n] = cl_float(r*beta,precision)*sqrt(cl_float(n,precision));
		//Create identity matrix here, needed for tqli to work correctly
		for(int m=0; m<number_of_occupations; m++){
			eigenvectors[m+n*number_of_occupations] 
				= (m==n) ? cl_float(1.0,precision) : zero;
		}
	}
	
	/* Execute the "TriDiagonal QL Implicit" Algorithm - from Numerical Receipes in C */
	tqli(diag, offdiag, number_of_occupations, eigenvectors, true);
	
	//Evaluate the partition function
	for(int n=0; n<number_of_occupations; n++){
		*partition = *partition + exp(cl_float(diag[n], precision));
	}
	
	*expectation = zero;
	
	//Evaluate the numerator, Tr[ (a+a*)exp{f(n)+r*beta*(a+a*)} ]
	for(int n=0; n<number_of_occupations; n++){
		for(int k=0; k<number_of_occupations-1; k++){ //less one
			temp = temp + cl_float(2*sqrt(k+1), precision)
				*cl_float(eigenvectors[k+n*number_of_occupations], precision)
				*cl_float(eigenvectors[k+1+n*number_of_occupations], precision);
		}
		*expectation = *expectation + temp*exp(cl_float(diag[n], precision));
		temp = zero;
	}
	*expectation = *expectation / *partition;
}



/*
 * Find value of partition function for the approximating hamiltonian - hence r dependence.
 *   -> could also be known as "cycle_density_denominator" to match below function
 *
 * Evals: Tr[ -\beta( \lambda n(n-1) + (1-\mu)n + r(a+a*)]
 */
cl_F partition_of_approx(double mu, double r)
{	
	cl_F partition = zero;
	
	//Construct the matrix - first with q=0
	for(int n=0; n<number_of_occupations; n++){
		diag[n] = f_approx(n, 0, mu);
		if(n!=0)  //offdiag[0] not used
			offdiag[n] = cl_float(r*beta,precision)*sqrt(cl_float(n,precision));
	}
	
	/* Execute the "TriDiagonal QL Implicit" Algorithm - from Numerical Receipes in C */
	tqli(diag, offdiag, number_of_occupations, eigenvectors, false);
	
	//Sum the exponentiated eigenvalues
	for(int n=0; n<number_of_occupations; n++){
		partition = partition + exp(cl_float(diag[n],precision));
	}
	return partition;
}	



/*
 * Evaluate the density of cycles of length q in the approximating hamiltonian case
 *
 * Evals: Tr[ -\beta( \lambda (n+q)(n+q-1) + (1-\mu)(n+q) + r(a+a*)]
 */
cl_F cycle_density_numerator(int q, double mu, double r)
{	
	cl_F numerator = zero;
	
	//Construct the matrix - with q non-zero!!!
	for(int n=0; n<number_of_occupations; n++){
		diag[n] = f_approx(n, q, mu);
		if(n!=0)  //offdiag[0] not used
			offdiag[n] = cl_float(r*beta,precision)*sqrt(cl_float(n,precision));
	}
	
	/* Execute the "TriDiagonal QL Implicit" Algorithm - from Numerical Receipes in C */
	tqli(diag, offdiag, number_of_occupations, eigenvectors, false);
	
	//Sum the exponentiated eigenvalues
	for(int n=0; n<number_of_occupations; n++){
		numerator = numerator + exp(cl_float(diag[n],precision));
	}
	return numerator;
}



/*
 * Diagonal values of B-H hamiltonian (off-diagonal values generated separately)
 */
cl_F f(int n, double mu)
{
	return cl_float(beta,precision)*( cl_float((mu + lambda - 1)*n, precision) 
									  - cl_float(lambda*n*n,precision) );
}



/*
 * Diagonal values of Approximator of the B-H hamiltonian - with cycles
 */
cl_F f_approx(int n, int q, double mu)
{
	return cl_float(-beta,precision)*( cl_float(lambda*(n+q)*(n+q-1), precision) 
									   + cl_float((1-mu)*(n+q),precision) );
}
/* 
 * Density evaluator
 */
double density(double r, double mu)
{	
	cl_F e_vec, temp;
	cl_F numerator = zero;
	cl_F denominator = zero;
	
	for(int n=0; n<number_of_occupations; n++){
		diag[n] = f(n, mu);
		if(n!=0)  //offdiag[0] not used
			offdiag[n] = cl_float(r*beta,precision)*sqrt(cl_float(n,precision));
		//Create identiy matrix
		for(int m=0; m<number_of_occupations; m++){
			eigenvectors[m+n*number_of_occupations] 
				= (m==n) ? cl_float(1.0,precision) : zero;
		}	
	}
	
	/* Execute the "TriDiagonal QL Implicit" Algorithm - from Numerical Receipes in C */
	//Only want eigenvalues, so set last argument in tqli as true
	tqli(diag, offdiag, number_of_occupations, eigenvectors, true);
	
	for(int k=0; k<number_of_occupations; k++){
		e_vec = exp(cl_float(diag[k],precision));
		denominator = denominator + cl_float(e_vec,precision);
		temp = zero;
		for(int n=1; n<number_of_occupations; n++){ //n==0 always zero so skip
			temp = temp + n*cl_float(eigenvectors[n+k*number_of_occupations],precision)
							*cl_float(eigenvectors[n+k*number_of_occupations],precision);
		}
		numerator = numerator + cl_float(e_vec*temp,precision);
	}
	
	return double_approx(numerator/denominator);
}



/* Based on the 'tqli' algorithm from Numerical Recipes in C
* Finds the eigenvalues and eigenvectors of a symmetric tridiagonal matrix
* --enhanced for arbitrary precision arithmetic - reliable up to 10 decimal places it appears--
*/
void tqli(vector<cl_F> &d, vector<cl_F> &e, int n, vector<cl_F> &z, bool eval_eigenvecs)
{
	int m,l,iter,i,k;
	cl_F s=zero,r=zero,p=zero,g=zero,f=zero,dd=zero,c=zero,b=zero;
	
	for(i=1;i<n;i++) e[i-1]=e[i];	//shift off-diagonal entries by one
	e[n-1]=zero;
	
	for(l=0;l<n;l++){	//main loop
		iter=0;
		do{
			for(m=l;m<n-1;m++){
				dd=abs(d[m])+abs(d[m+1]);
				if (abs(e[m]) < TINY*dd) break; //this offdiagonal negligible, go onto next
			}
			if(m != l){
				if (iter++ == 30){	//This algorithm quickly converges, but if not
					verbose << "Too many iterations in TQLI" << endl;
					throw 1;
				}
				if( e[l] == zero ) throw 1; //Divide-by-zero catch
				g=(d[l+1]-d[l])/(2.0*e[l]);
				r=pythag(g,1.0);  //pythag
				g=d[m]-d[l]+e[l]/(g+SIGN(r,g));
				s=c=cl_float(1,precision);
				p=zero;
				for(i=m-1;i>=l;i--){
					f=s*e[i];
					b=c*e[i];
					e[i+1]=(r=pythag(f,g));
					if(r == zero){
						d[i+1] = d[i+1]-p;
						e[m]=zero;
						break;
					}
					s=cl_float(f/r,precision);  //helps catch underflows
					c=cl_float(g/r,precision);
					g=d[i+1]-p;
					r=(d[i]-g)*s+2.0*c*b;
					d[i+1]=g+(p=s*r);
					g=c*r-b;
					// Next loop can be omitted if eigenvectors not wanted
					if(eval_eigenvecs){
						for(k=0;k<n;k++){
							f=z[k+(i+1)*n];
							z[k+(i+1)*n]=s*z[k+i*n]+c*f;
							z[k+i*n]=c*z[k+i*n]-s*f;
						}
					}
				}
				if(r == zero && i >=l) continue;
				d[l]=d[l]-p;
				e[l]=g;
				e[m]=zero;
			}
		}while(m != l);
	}
}



/*
 *	pythag - finds sqrt(a*a + b*b) in best way possible to avoid overflows
 *	Needed by tqli()
 */
cl_F pythag( const cl_F a, const cl_F b)
{
	cl_F absa=abs(a), absb=abs(b);
	return( (absa > absb) ? absa*sqrt(1.0+(absb*absb)/(absa*absa)) : 
			(absb==zero ? zero : absb*sqrt(1.0+(absa*absa)/(absb*absb))));
}
\end{lstlisting}
\end{spacing}

\end{appendix}

\phantomsection
\listoffigures
\addcontentsline{toc}{chapter}{List of Figures}

\pagebreak
\phantomsection

\end{spacing}
\bibliography{main-small-combined}
\addcontentsline{toc}{chapter}{Bibliography}

   \end{document}